\documentclass[JHEP,11pt]{article}
\usepackage{amsmath,amssymb,amsthm,amsfonts}
\usepackage{array}
\usepackage{booktabs}
\usepackage{cancel}
\usepackage{caption,subcaption}
\usepackage{enumitem}
\usepackage{fancybox}
\usepackage{float}
\usepackage[T1]{fontenc}
\usepackage[utf8]{inputenc}
\usepackage{graphicx}
\usepackage{jheppub}\usepackage{cleveref}
\usepackage{longtable}
\usepackage{makecell}
\usepackage{mathrsfs}
\usepackage{mdframed}
\usepackage{multirow}
\usepackage{placeins}
\usepackage{tabu}
\usepackage{tikz,tikz-3dplot}
\usepackage[compat=1.0.0]{tikz-feynman}
\usepackage{slashed}
\usepackage{soul}
\usepackage{subcaption}
\usepackage[usestackEOL]{stackengine}
\usepackage{verbatim}
\usepackage{wrapfig}
\usepackage[dvipsnames]{xcolor}
\usetikzlibrary{calc,arrows.meta,shapes.geometric}

\def\x{\boldsymbol{x}}
\def\s{\boldsymbol{s}}
\def\r{\boldsymbol{r}}

\newcommand{\leftrarrows}{\mathrel{\raise.75ex\hbox{\oalign{%
  $\scriptstyle\leftarrow$\cr
  \vrule width0pt height.5ex$\hfil\scriptstyle\relbar$\cr}}}}
\newcommand{\lrightarrows}{\mathrel{\raise.75ex\hbox{\oalign{%
  $\scriptstyle\relbar$\hfil\cr
  $\scriptstyle\vrule width0pt height.5ex\smash\rightarrow$\cr}}}}
\newcommand{\Rrelbar}{\mathrel{\raise.75ex\hbox{\oalign{%
  $\scriptstyle\relbar$\cr
  \vrule width0pt height.5ex$\scriptstyle\relbar$}}}}

\newcommand\greencheckmark[1][]{%
  \tikz[scale=0.4,#1]{\fill(0,.35) -- (.25,0) -- (1,.7) -- (.25,.15) -- cycle;}%
}
\newcommand\crossmark[1][]{%
  \tikz[scale=0.4,#1]{
    \fill(0,0)--(0.1,0) .. controls (0.5,0.4) .. (1,0.7)--(0.9,0.7) ..  controls (0.5,0.5) ..(0,0.1) --cycle;
    \fill(1,0.1)--(0.9,0.1) .. controls (0.5,0.3) .. (0,0.7)--(0.1,0.7) .. controls (0.5,0.4) ..(1,0.2) --cycle;
  }%
}

\makeatletter
\def\leftrightarrowsfill@{\arrowfill@\leftrarrows\Rrelbar\lrightarrows}
\newcommand{\xleftrightarrows}[2][]{\ext@arrow 3399\leftrightarrowsfill@{#1}{#2}}
\makeatother

\newtheorem{theorem}{Theorem}[section]
\newtheorem{corollary}{Corollary}[theorem]
\newtheorem{lemma}[theorem]{Lemma}

\theoremstyle{definition}

\theoremstyle{remark}

\title{\boldmath All-order prescription for facet regions in massless wide-angle scattering}

\author[]{Yao Ma}

\affiliation[]{Institute for Theoretical Physics, ETH Zürich,\\ 8093 Zürich, Switzerland}

\emailAdd{yaomay@phys.ethz.ch}

\abstract{
We take a step toward answering a long-standing question in the asymptotic expansion of Feynman integrals: how to systematically determine and interpret the regions in the Expansion-by-Regions technique for multiscale processes? Focusing on generic massless wide-angle scattering, we provide an all-order momentum-space prescription for facet regions, which generally dominate---and in most cases exhaust---the contributions in a given asymptotic expansion. This extends the Euclidean-space picture, where regions correspond to specific subgraphs, to the complexities of Minkowski space.

Our results are derived from a novel analytical approach combining graph theory and convex geometry, and they yield several key byproducts. Theoretically, we uncover for the first time the algebraic structure underlying momentum modes (collinear, soft, and their hierarchies): they form a non-distributive lattice in general. Phenomenologically, we show that ``cascading modes'' are always absent in our scope. Algorithmically, we implement this prescription in ``\texttt{Facet Region Interpreter}'', a computer program to systematically derive regions in momentum space, by which we validate our prescription through 18,000+ examples.
}

\begin{document}
\maketitle
\flushbottom

\section{Introduction}
\label{section-introduction}

Evaluating multiscale Feynman integrals remains one of the central challenges in precision calculations for high-energy physics, particularly at particle colliders. These integrals often arise in processes involving widely separated energy scales, where exact analytical evaluation is often intractable, necessitating approximate methods, such as asymptotic expansions, which capture the dominant contributions across different kinematic regimes.

Among various techniques for deriving asymptotic expansions for Feynman integrals, the \emph{Expansion by Regions} (EbR, also referred to as the ``Method of Regions'')~\cite{BnkSmn97} has become a standard tool for obtaining such expansions in a wide variety of kinematic limits. The basic idea of EbR is to decompose the integration domain into distinct ``regions'', each characterized by a specific homogeneous scaling of the integration variables (momenta or parameters). In each region, the integrand is expanded accordingly in powers of the small expansion parameters, and the resulting contributions are integrated over the full space after an appropriate analytic continuation. Although a rigorous mathematical proof of this method is still lacking, its reliability is well established by numerous examples. A crucial step in applying EbR is the identification of all relevant regions contributing to the asymptotic expansion.

For asymptotic expansions in Euclidean space, such as large-mass or large-momentum expansions, the structure of these regions is well understood from a graph-theoretical perspective. Specifically, each relevant region corresponds to a particular assignment of large loop momenta within an ``asymptotically irreducible subgraph''~\cite{Pivovarov:1986xz,Smn90,Smn94,Smirnov:2002pj}. Moreover, a mathematical proof has been established that summing the expanded contributions over all such subgraphs exactly reproduces the original Feynman integral. In this sense, EbR is equivalent to ``expansion by subgraphs''.

In Minkowski space, however, the richer infrared structure renders the identification of regions considerably more nontrivial. For a long time, this task was performed on a case-by-case basis, often guided by heuristic arguments derived from explicit calculations and physical intuition. Only in recent years has a more systematic geometric approach emerged, which associates a multiscale $N$-propagator Feynman graph $\mathcal{G}$ with an $(N+1)$-dimensional polytope $\Delta(\mathcal{G})$ in Lee-Pomeransky parameter space and identifies regions with certain codimension-1 faces---called ``lower facets''---of this polytope~\cite{PakSmn11,JtzSmnSmn12,AnthnrySkrRmn19,SmnvSmnSmv19}. The entries of the vector normal to each lower facet encode the scaling exponents of the corresponding region. Building on this, several computer programs, such as Asy2~\cite{JtzSmnSmn12}, ASPIRE~\cite{AnthnrySkrRmn19}, and pySecDec~\cite{HrchJnsSlk22}, have been developed to automate region identification.

Nevertheless, our understanding of region structures remains incomplete. A key limitation is that certain contributing regions can lie in the interior (rather than on the boundary) of the polytope and are therefore invisible to the geometric method above. We refer to these as \emph{hidden regions}~\cite{GrdHzgJnsMa24,Ma25,Becher:2025igg}, in contrast to the \emph{facet regions} that reside on the boundaries. Identifying hidden regions is technically demanding and can involve sophisticated Landau analysis. Although refs.~\cite{JtzSmnSmn12,AnthnrySkrRmn19} showed that some hidden regions can be unveiled through suitable variable changes, this technique relies on the linear cancellation structure in the $\mathcal{F}$ polynomial, and is not generally applicable. For instance, in massless two-to-two scattering, hidden regions first appear at three loops and involve quadratic cancellations in $\mathcal{F}$, requiring more intricate treatments~\cite{GrdHzgJnsMa24}. A recent proposal for a general and systematic approach to hidden regions can be found in ref.~\cite{Gardi:2026mcx}.

For massless wide-angle scattering processes, hidden regions are generally rare~\cite{GrdHzgJnsMa24}. In most cases, facet regions alone exhaust the contributions in the expansion, and the geometric approach to regions is fully reliable. Even so, a complete understanding of facet regions has been lacking. First, the geometric approach yields regions in terms of scalings of the Lee-Pomeransky parameters, leaving the physical interpretation in momentum space obscure, which often requires nontrivial translation. Second, as the polytope dimension increases with the number of propagators, existing computational tools provide regions only to a finite order, preventing direct access to a formal, all-order prescription of the full region structure.

\begin{figure}[t]
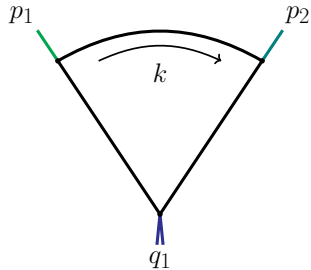

\centering
\include{figs/form_factor_expansion}
\vspace{-3em}\caption{The one-loop form factor with all the edges massless. Even for expanding this relatively simple graph near the same infrared limit ($p_1^2,p_2^2\ll 1$ and $q_1^2\sim 1$), the region structure depends strongly on the precise form of the expansion.}
\label{figure-form_factor_expansions}
\end{figure}

The second limitation identified above is particularly subtle, because the region structure depends sensitively on both the external kinematics and the graph topology. For example, consider the massless form factor in figure~\ref{figure-form_factor_expansions} in the following infrared limit: the external momentum $q_1$ is off shell ($q_1^2\sim 1$), while $p_1$ and $p_2$ are on shell ($p_1^2,p_2^2\ll 1$), pointing along distinct lightcones. Near this kinematic point, different asymptotic expansions yield distinct region structures, as we list below.\footnote{In this paper, the virtuality of the soft mode is $\mathcal{O}(\lambda^2)$ while that of the collinear is $\mathcal{O}(\lambda)$, consistent with the terminology in ref.~\cite{Stm95book,ClsSprStm04,Cls11book,BchBrgFrl15book}. In some other literature, this soft mode is referred to as the ``infrared mode''~\cite{Kcmsk89} or the ``ultrasoft mode''~\cite{Smn02book,SmnRkmt99,Jtz11,Smn99,KuhnPeninSmn00}.

To present the regions, here we have chosen the lightcone coordinate $k^\mu = (k^+, k^-, \boldsymbol{k}_\perp)$ and the center-of-mass frame, such that $p_1$ and $p_2$ are close to the ``$+$'' and ``$-$'' directions respectively. Later in section~\ref{section-kinematic_setup} where generic wide-angle kinematics is considered, we will use multiple lightcone coordinates for the external momenta (see eq.~\eqref{eq:virtuality_expansion_p_scaling} therein).}
\begin{itemize}
    \item $p_1^2/q_1^2\sim p_2^2/q_1^2\sim \lambda$. There are four regions (where $Q\sim \mathcal{O}(1)$ is the hard scale):
    \begin{flalign*}
    &\qquad\textup{hard region }(H):\ \ k^\mu\sim Q(1,1,1);&\\
    &\qquad\textup{collinear-1 region }(C_1):\ k^\mu\sim Q(1,\lambda,\sqrt\lambda);&\\
    &\qquad\textup{collinear-2 region }(C_2):\ k^\mu\sim Q(\lambda,1,\sqrt\lambda);&\\
    &\qquad\textup{soft region }(S):\ k^\mu\sim Q(\lambda,\lambda,\lambda).&
    \end{flalign*}    
    \item $p_1^2/q_1^2\sim \lambda$ and $p_2^2/q_1^2\sim \lambda^2$. There are four regions:
    \begin{flalign*}
    &\qquad\textup{hard region }(H):\ \ k^\mu\sim Q(1,1,1);&\\
    &\qquad\textup{collinear-1 region }(C_1):\ k^\mu\sim Q(1,\lambda,\sqrt\lambda);&\\
    &\qquad(\textup{collinear-2)}^2\textup{ region }(C_2^2):\ k^\mu\sim Q(\lambda^2,1,\lambda);&\\
    &\qquad\textup{soft}\cdot\textup{collinear-2 region }(SC_2):\ k^\mu\sim Q(\lambda^2,\lambda,\lambda^{3/2}).&
    \end{flalign*}    
    \item $p_1^2/q_1^2\sim \lambda$ and $p_2^2/q_1^2=0$. There are two regions:
    \begin{flalign*}
    &\qquad\textup{hard region }(H):\ \ k^\mu\sim Q(1,1,1);&\\
    &\qquad\textup{collinear-1 region }(C_1):\ k^\mu\sim Q(1,\lambda,\sqrt\lambda).&
    \end{flalign*}
\end{itemize}
These examples illustrate that even for this relatively simple graph, the region structure depends highly on the external kinematics: changing the hierarchical relations among the virtualities can alter the scalings of individual regions, introduce hierarchical modes (e.g., $C_2^2$ or $SC_2$), or cause certain regions to disappear. In particular, we note that in the third kinematics above ($p_1^2/q_1^2\sim \lambda$ and $p_2^2/q_1^2=0$), the $C_2$ and $S$ regions are both absent, because their corresponding expanded integrals are \emph{scaleless} and thus vanishing.

\begin{figure}[t]
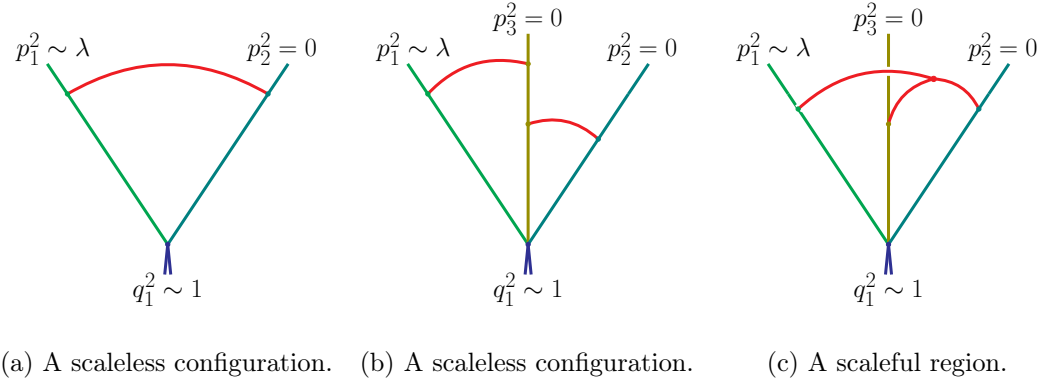

\centering
\begin{subfigure}[b]{0.3\textwidth}
\centering
\include{figs/form_factor_like_nonregion_example1}
\vspace{-2em}
\caption{A scaleless configuration.}
\label{form_factor_like_nonregion_example1}
\end{subfigure}
\begin{subfigure}[b]{0.3\textwidth}
\centering
\include{figs/form_factor_like_nonregion_example2}
\vspace{-2em}
\caption{A scaleless configuration.}
\label{form_factor_like_nonregion_example2}
\end{subfigure}
\begin{subfigure}[b]{0.3\textwidth}
\centering
\include{figs/form_factor_like_region_example}
\vspace{-2em}
\caption{A scaleful region.}
\label{form_factor_like_region_example}
\end{subfigure}
\caption{Three examples where the loop momenta are all soft. The expanded integrals corresponding to (a) and (b) are scaleless, and should not be taken into account in EbR. In contrast, (c) leads to a scaleful integral and should be considered as one region.}
\label{figure-form_factor_like_examples}
\end{figure}

Even for fixed external kinematics, whether a given loop-momentum configuration contributes as a region depends crucially on the graph topology. As we have just explained, the configuration in figure~\ref{form_factor_like_nonregion_example1} is scaleless, where the {\color{Red}\bf red} curve represents the soft propagator while the {\color{Green}\bf green} and {\color{teal}\bf teal} lines represent the collinear-1 and collinear-2 propagators, respectively. Similarly, the configuration in figure~\ref{form_factor_like_nonregion_example2}, which can be seen as a two-loop generalization, is also scaleless. Note that in these examples, we have set $q_1^2\sim 1$, $p_1^2\sim \lambda$, and $p_2,p_3$ strictly on shell ($p_2^2=p_3^2=0$). However, under the same kinematics, if we change the topology of the two-loop graph into figure~\ref{form_factor_like_region_example}, the obtained configuration is scaleful, and must therefore be included as a region.

The primary goal of this paper is to resolve these subtleties by extending the expansion-by-subgraphs paradigm in Euclidean space to generic massless wide‑angle scattering with non-exceptional external momenta in Minkowski space. To this end, we unify the various asymptotic expansions illustrated above (and others arising in massless wide‑angle scattering) under a single framework which we call the \emph{virtuality expansion} (defined in section~\ref{section-kinematic_setup}). Within this framework, we systematically analyze the region structure for arbitrary loop orders, numbers of external legs, and orders in the expansion parameter. Specifically, we address two central questions: (1) \emph{Given a virtuality expansion, what are the allowed momentum modes (collinear, soft, and their hierarchies)?} (2) \emph{What requirements must subgraphs associated with these modes satisfy, in order to yield scaleful contributions?} Answering these questions enables us to derive the complete set of facet regions directly from a Feynman graph and its external kinematics, providing an all-order prescription in momentum space while circumventing the limitation of computer codes.

Our approach combines convex geometry with graph theory. On the one hand, for each region, i.e., for a given scaling of the Lee-Pomeransky parameters $x_1,\dots,x_N$, the leading terms in the expansion all correspond to points on a ``lower facet'' of the polytope $\Delta(\mathcal{G})$. On the other hand, these leading terms can be characterized by the minimum spanning trees and 2-trees of $\mathcal{G}$. Consequently, the problem of identifying regions can be reformulated as follows:
\\
\makebox[\textwidth]{%
  \fbox{%
    \begin{minipage}{\dimexpr\linewidth-2\fboxsep-2\fboxrule\relax}
      \emph{we seek all scalings of the Lee-Pomeransky parameters $x_1,\dots,x_N$, such that the parameter-space points, which correspond to the minimum spanning trees and 2-trees of $\mathcal{G}$, span a lower facet of $\Delta(\mathcal{G})$.}
    \end{minipage}%
  }%
}

This approach was introduced in the author's previous work~\cite{Ma23}, where it was applied to two special cases of the virtuality expansion: the ``on-shell expansion'' and the ``soft expansion''. There, all-order prescriptions for the region structure were derived, with the on-shell case rigorously proven and the soft case resting on a conjecture about the allowed momentum modes. The general prescription presented here will unify and extend those results to arbitrary virtuality hierarchies, from which the earlier conjecture for the soft expansion can be automatically validated as a special case.

As a key byproduct of this analysis, we uncover, for the first time, the algebraic structure underlying the infinitely many momentum modes that arise in generic virtuality expansions. These modes admit a natural partial order by their softness; defining join and meet operations on this partially ordered set (poset) endows it with the structure of a lattice. We explore further algebraic properties of this lattice, which are essential for deriving region structures.

With that said, however, we emphasize the \emph{limitations of this work} to avoid potential misunderstandings of its scope.
\begin{enumerate}
    \item This paper only considers massless Feynman integrals in wide-angle kinematics. Massive propagators or collinear external momenta would require future extensions.
    \item Hidden regions are not taken into account. Although they are very few and mostly absent~\cite{GrdHzgJnsMa24}, it is imperative to first check their existence in advance. Notably, a recent work of the ``Hidden Region Finder'' (HRF)~\cite{Gardi:2026mcx} can be used as a first check: for a given Feynman integral, if HRF indicates that no hidden regions exist, all regions are described by our prescription; otherwise, HRF may output the hidden regions, and our prescription describes the remaining regions.
    \item The derivation of the all-order prescription starts from a \emph{fundamental pattern}, which rests on two assumptions: (1) infrared regions originate from solutions of the Landau equations, and (2) the hard scattering of each region occurs at a single position. We will explain why these assumptions are natural in section~\ref{section-fundamental_facet_region_structure}, and see how they are supported by the validation in section~\ref{section-validation_through_codes}. Nevertheless, we emphasize that a proof of the fundamental pattern is still lacking.
\end{enumerate}

This paper is organized as follows. In section~\ref{section-basic_concepts_notations}, we introduce the basic concepts and notation that form the foundation of our analysis. In section~\ref{section-algebraic_graphic_mode_structure}, we study the structures of momentum modes from both algebraic and graphic perspectives. These structures are not merely mathematical curiosities but also provide key relations used in the later proofs. We then derive the general prescription for facet regions: section~\ref{section-fundamental_facet_region_structure} establishes a fundamental pattern under two natural assumptions, while section~\ref{section-further_requirements_mode_subgraphs} presents the subgraph requirements, which act as a \emph{necessary and sufficient} condition for the fundamental pattern to yield scaleful contributions, i.e., regions. These constitute the main results of this paper. Notably, these requirements constrain not only the connections among subgraphs, but also the mode structure: given an expansion, one can determine which modes can appear in arbitrary graphs; in particular, the phenomenon of ``cascading modes'' (the number of momentum modes grows unbounded as the loop order increases) is always absent in the virtuality expansion. The proofs of these requirements, which form the most technical part of the paper, are provided in section~\ref{section-proof_subgraph_requirements}. In section~\ref{section-multiloop_examples}, we check our prescription through two multiloop graphs, show its results completely match the output of \texttt{pySecDec}, and point out the potential of this work to extend to massive Feynman integrals by connecting it with a recent study on quark mass corrections to $gg\to HH$~\cite{Jaskiewicz:2024xkd}. Furthermore, in section~\ref{section-validation_through_codes} we validate our prescription by implementing it in a computer program called ``\texttt{Facet Region Interpreter}'', and check its correctness against 18,000+ multiloop examples. We conclude in section~\ref{section-conclusion_outlook} with a summary and an outlook for future research. Additional technical details are collected in appendix~\ref{appendix-details_proof}.

\section{Basic concepts and notation}
\label{section-basic_concepts_notations}

In this section, we lay the foundation for our region analysis. In section~\ref{section-kinematic_setup}, we define the virtuality expansion, which encompasses a wide range of expansions for massless wide‑angle kinematics. In section~\ref{section-momentum_modes}, we formulate the momentum modes that typically appear in the virtuality expansion, define their virtuality degrees, and introduce a partial order on them. In section~\ref{section-parametric_representations_spanning_trees}, we demonstrate, within the Lee–Pomeransky representation of Feynman integrals, how the leading terms of the expansion, certain boundaries of a constructed polytope, and the minimum spanning (2‑)trees of $\mathcal{G}$ are interrelated.

\subsection{Kinematic setup}
\label{section-kinematic_setup}

For the massless wide-angle kinematics considered in this paper, the external momenta can be classified into three sets. The first set consists of momenta $p_1,\dots,p_K$, each $p_i$ collinear to a specific lightlike vector $\beta_i^\mu$, with the component $p_i\cdot\overline{\beta}_i$\footnote{Let $\beta_i^\mu\!\equiv\!\frac{1}{\sqrt{2}}(1,\widehat{\boldsymbol{v}}_i)$, where $\widehat{\boldsymbol{v}}_i$ is a unit three‑vector. Then $\overline{\beta}_i^\mu\! \equiv\! \frac{1}{\sqrt{2}} (1, -\widehat{\boldsymbol{v}}_i)$ is the null vector opposite to~$\beta_i^\mu$, satisfying $\beta_i \cdot \overline{\beta}_i =1$.} at the hard scale $Q\sim \mathcal{O}(1)$. The second set consists of hard external momenta $q_1,\dots,q_L$ whose components are all $\mathcal{O}(1)$. The third set consists of generalized soft momenta $l_k^\mu$, whose components are all approaching zero (possibly at different rates). For simplicity, we require these momenta to be non-exceptional: any nonempty proper subset of them does not sum to zero.

More precisely, let $\lambda$ denote a small scaling parameter. By omitting the hard scale $Q$ from now on, the external momentum scalings are then:
\begin{subequations}
\label{eq:virtuality_expansion}
\begin{align}
    & p_i^\mu=(p_i\cdot\overline{\beta}_i,\ p_i\cdot \beta_i,\ \boldsymbol{p}_{i\perp}^{})\sim (1,\lambda^{a_i},\lambda^{a_i/2})&& \text{for }i=1,\dots,K; \label{eq:virtuality_expansion_p_scaling}\\
    & q_j^\mu\sim 1\ \ \textup{(all components)}&& \text{for }j=1,\dots,L; \label{eq:virtuality_expansion_q_scaling}\\
    & l_k^\mu=(l_k\cdot\overline{\beta}'_k,\ l_k\cdot \beta'_k,\ \boldsymbol{l}_{k\perp}^{})\sim \lambda^{b_k}(1,\lambda^{c_k},\lambda^{c_k/2})&& \text{for }k=1,\dots,M; \label{eq:virtuality_expansion_l_scaling}\\
    & p_{i_1}\cdot p_{i_2}\sim 1, \quad l_{k_1}\cdot l_{k_2}\sim \lambda^{b_{k_1}+b_{k_2}} ,\quad p_{i_1}\cdot l_{k_2}\sim \lambda^{b_{k_2}} &&\text{for }i_1\neq i_2,\ k_1\neq k_2. \label{eq:virtuality_expansion_wide_angle}
\end{align}
\end{subequations}
Note that each mode is written in the lightcone frame adapted to its own jet direction, which is different from the notation in section~\ref{section-introduction} where a unique lightcone coordinate is used. The wide‑angle nature is encoded in eq.~(\ref{eq:virtuality_expansion_wide_angle}), indicating that the lightlike vectors $\beta_1,\dots,\beta_K,\beta'_1,\dots,\beta'_M$ are all along distinct directions. The exponents $a_i,b_k,c_k$ are taken from the following sets:
\begin{subequations}
\label{eq:virtuality_expansion_exponents_range}
    \begin{align}
        & a_i\in \mathbb{N}_+^\infty\equiv \mathbb{N}_+\cup \{+\infty\}=\{1,2,\dots,+\infty\};\\
        & b_k\in \mathbb{N}_+=\{1,2,\dots\};\\
        & c_k\in \mathbb{N}^\infty\equiv \mathbb{N}\cup \{+\infty\}=\{0,1,\dots,+\infty\}.
    \end{align}
\end{subequations}
Note that the value $+\infty$ implies that $p_i^\mu$ and $l_k^\mu$ can be exactly lightlike, i.e., $p_i^2=0$ and $l_k^2=0$ for some $i,k$.

\begin{figure}[t]
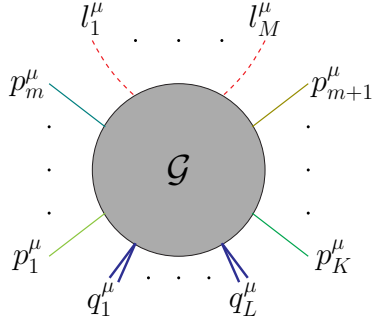

\centering
\include{figs/problem_to_study}
\vspace{-2em}\caption{General wide-angle scattering graph $\mathcal{G}$, with external momenta~$\{p_i^\mu\}_{i=1,\dots,K}$, $\{q_j^\mu\}_{j=1,\dots,L}$, and $\{l_k^\mu\}_{k=1,\dots,M}$ subject to eq.~(\ref{eq:virtuality_expansion}).}
\label{figure-problems_to_study}
\end{figure}

We define the \emph{virtuality expansion} as the expansion in the parameter $\lambda$. All the examples discussed in section~\ref{section-introduction} are special cases of the virtuality expansion. The general configuration is illustrated in figure~\ref{figure-problems_to_study}.

\subsection{Momentum modes}
\label{section-momentum_modes}

We now formalize the momentum modes that are relevant in the virtuality expansion.
\bigbreak
\noindent\underline{\emph{Notation of Modes}}.

Unless otherwise specified, we use the roman letter $X$ to denote a general momentum mode. The full set of modes in the virtuality expansion follows certain patterns. To describe them, we begin with three basic types:
\begin{subequations}
\label{eq:three_basic_modes}
    \begin{align}
        & \textup{Hard mode (}H\textup{):}&& k_H^\mu\sim 1\ \ \textup{(all its components)};\\
        & \textup{Collinear-}i\textup{ mode (}C_i\textup{):}&&k_{C_i}^\mu=(k_{C_i}\cdot\overline{\beta}_i,\ k_{C_i}\cdot \beta_i,\ \boldsymbol{k}_{C_i\perp}^{})\sim (1,\lambda,\lambda^{1/2});\\
        & \textup{Soft mode (}S\textup{):}&&k_{S}^\mu\sim \lambda\ \ \textup{(all its components)}.
    \end{align}
\end{subequations}
As we will propose in section~\ref{section-fundamental_facet_region_structure} and justify in section~\ref{section-implication_infrared_compatibility_requirement}, all momentum modes considered in this paper can be denoted by $S^m C_i^n$, where $(m,n)\in \mathbb{N}^2$. The mode $S^m C_i^n$ can be regarded as the Hadamard product (componentwise product) of $m$ soft modes and $n$ collinear-$i$ modes.\footnote{The special case $(m,n)=(0,0)$ corresponds to the hard mode $H$.} A momentum $k_X^\mu$ in the mode $X=S^m C_i^n$ scales as
\begin{eqnarray}
    k_X^\mu = (k_X\cdot \overline{\beta}_i,\, k_X\cdot \beta_i,\, \boldsymbol{k}_{\perp}) \sim (\lambda^m,\lambda^{m+n},\lambda^{m+n/2}).
\end{eqnarray}

For a given momentum $k$, we use $\mathscr{X}(k)$ to denote its corresponding momentum mode. For a given edge $e\in \mathcal{G}$, we use $\mathscr{X}(e)$ to denote the mode of its line momentum. Later in section~\ref{section-mode_subgraphs}, we will use the same notation to denote the modes of certain subgraphs of $\mathcal{G}$.

Under this construction, the external momenta in (\ref{eq:virtuality_expansion}) belong to the following modes:
\begin{align}
\begin{split}
    &\mathscr{X}(p_i):\ C_i,\ C_i^2,\ C_i^3,\ \dots,\ C_i^\infty;\\
    &\mathscr{X}(q_j):\ H;\\
    &\mathscr{X}(l_k):\ S,\ SC_k,\ SC_k^2,\ S^2,\ \dots,\ SC_k^\infty,\ \dots.
\end{split}
\end{align}
Note that the mode $C_i^\infty$ means that $p_i^\mu$ is precisely parallel to $\beta_i^\mu$. In this way, all modes from the external kinematics can be expressed in the form $S^mC_i^n$, with $(m,n)\in (\mathbb{N}^\infty)^2$.

\bigbreak
\noindent\underline{\emph{Virtuality Degree of a Mode}}.

For a given mode $X$, we denote its \emph{virtuality degree} by $\mathscr{V}(X)$, defined as the power of $\lambda$ in $k_X^2$, where $k_X$ is an $X$-mode momentum:
\begin{eqnarray}
    \mathscr{V}(X)=a,\quad\text{if }\ k_X^2\sim \lambda^a.
\end{eqnarray}
It follows that $\mathscr{V}(S^m C_i^n) = 2m+n$.

Note that if the line momentum of an edge $e$, $k_e^\mu$, is in the $X$ mode, then its corresponding Lee-Pomeransky parameter $x_e$ (see section~\ref{section-parametric_representations_spanning_trees}) scales as:
\begin{eqnarray}
    x_e\sim \lambda^{-\mathscr{V}(X)},
\end{eqnarray}
which is due to $x_e\sim (k_e^2)^{-1}$, a fact we will emphasize again in section~\ref{section-parametric_representations_spanning_trees}.

\bigbreak
\noindent\underline{\emph{Harder and Softer Modes}}.

We define a partial order of the modes as follows. Given two distinct modes $X_1$ and $X_2$, we say $X_1$ is \emph{harder than} $X_2$, or $X_2$ is \emph{softer than} $X_1$, if for some vectors $v$ we have
\begin{align}
    |k_{X_1}\cdot v| \gg |k_{X_2}\cdot v|,
\end{align}
meanwhile for \emph{all} vectors $w$, we have
\begin{align}
    k_{X_1}\cdot w \gtrsim k_{X_2}\cdot w,
\end{align}
where ``$\gtrsim$'' means either ``$\gg$'' or ``$\sim$''.
In momentum space, this implies that the range of $k_{X_1}^\mu$ is a proper superset of the range of $k_{X_2}^\mu$. If $X_1$ is neither harder nor softer than $X_2$, we call $X_1$ and $X_2$ \emph{overlapping}.

For example, $C_i$ is harder than $S$, which is clear from their scalings:
\begin{align}
\label{eq:partial_order_example_soft_collinear}
    k_{C_i}^\mu\sim (1,\lambda,\lambda^{1/2}),\quad k_S^\mu \sim (\lambda,\lambda,\lambda).
\end{align}
More generally, if two distinct modes $X_1 = S^{m_1} C_i^{n_1}$ and $X_2 = S^{m_2} C_i^{n_2}$ with the same $i$, then $X_1$ is harder than $X_2$ if and only if $m_1 \leqslant m_2$ and $m_1 + n_1 \leqslant m_2 + n_2$. If instead $X_1 = S^{m_1} C_i^{n_1}$ and $X_2 = S^{m_2} C_j^{n_2}$ with $i \neq j$, then $X_1$ is harder than $X_2$ if and only if $m_1 + n_1 \leqslant m_2$.

A crucial property is that if $X_1$ is harder than $X_2$, then $\mathscr{V}(X_1) < \mathscr{V}(X_2)$. Note that the converse does not hold: for example, $\mathscr{V}(S) = 2 < \mathscr{V}(C_i^3) = 3$, but $S$ is not harder than $C_i^3$.

\subsection{Parametric representation, polytope, and spanning (2-)trees}
\label{section-parametric_representations_spanning_trees}

We consider a massless Feynman integral dimensionally regularized in $D=4-2\epsilon$ dimensions:
\begin{eqnarray}
\label{eq:Feynman_integarl_dim_reg}
    \mathcal{I}(\mathcal{G})\equiv \mathcal{C}\cdot \int \prod_{i=1}^L \frac{d^Dk_i}{i\pi^{D/2}} \prod_{e\in \mathcal{G}} \frac{1}{D_e^{\nu_e}},
\end{eqnarray}
where $\mathcal{C}$ is an overall factor, $L$ is the number of independent loop momenta, and $e$ represents the edges of the graph $\mathcal{G}$. As all the propagators are massless here, we have $D_e = l_e^2+i0$ with $l_e^\mu$ the line momentum of $e$, associated with power $\nu_e$. Below, we discuss the Lee-Pomeransky representation of $\mathcal{I}$, which allows us to relate the leading terms in the expansion to
\begin{enumerate}
    \item [(1)] points on the ``lower facets'' of a constructed polytope $\Delta(\mathcal{G})$,
    \item [(2)] the minimum spanning (2-)trees of $\mathcal{G}$.
\end{enumerate}
These relations are the foundation of our later analyses.

\bigbreak
\noindent\underline{\emph{Lee-Pomeransky Representation}}.

For convenience, we use $\s$ to denote the set of scalar products formed among the external momenta of $\mathcal{G}$, so $\mathcal{I}$ is a function of $\s$. The Lee-Pomeransky representation of $\mathcal{I}(\s)$ reads~\cite{LeePmrsk13}
\begin{equation}
\mathcal{I}(\s) = \mathcal{C}'\cdot \int_0^\infty \left( \prod_{e\in \mathcal{G}} \frac{dx_e}{x_e}\right)\cdot \left(\prod_{e\in \mathcal{G}} x_e^{\nu_e} \right) \cdot \Big( \mathcal{P}(\x;\s) \Big)^{-D/2},
\label{lee_pomeransky_definition}
\end{equation}
where the overall factor $\mathcal{C}'\equiv \frac{\Gamma(D/2)}{\Gamma((L+1)D/2-\sum_{e\in \mathcal{G}}\nu_e) \prod_{e\in \mathcal{G}}\Gamma(\nu_e)}\!\cdot\!\mathcal{C}$. In the integrand above, we have used $\x$ to denote the set of Lee-Pomeransky parameters $x_1,\dots,x_N$, with $N$ being the number of edges of $\mathcal{G}$. $\mathcal{P}(\x;\s)$ is called the Lee-Pomeransky polynomial, defined as
\begin{eqnarray}
\label{lee_pomeransky_integrand_definition}
    \mathcal{P}(\x;\s)\equiv \mathcal{U}(\x)+\mathcal{F}(\x;\s),
\end{eqnarray}
where $\mathcal{U}(\x)$ and $\mathcal{F}(\x;\s)$ are the first and second Symanzik polynomials, respectively:
\begin{eqnarray}
\label{eq:UFterm_massless_general_expression}
    \mathcal{U}(\x)=\sum_{T^1}^{}\prod_{e\notin T^1}^{}x_e,\qquad \mathcal{F}(\x;\s) = \sum_{T^2}^{} (-Q_{T^2}^{2}) \prod_{e\notin T^2}^{}x_e.
\end{eqnarray}
The notations $T^1$ and $T^2$ denote a \emph{spanning tree} and a \emph{spanning 2-tree} of the graph $\mathcal{G}$, respectively. The momentum $Q_{T^2}^{\mu}$ is the total momentum flowing between the components of $T^2$, and $Q_{T^2}^{2}$ is referred to as the \emph{kinematic factor} of $T^2$.

In contrast to the momentum representation, where each region is characterized by the scaling of loop momenta, in the Lee–Pomeransky representation each region $R$ is characterized by the scaling of the parameters:\footnote{More precisely, (\ref{eq:facet_region_general_parameter_scaling}) describes the scaling behavior of facet regions. For hidden regions, one generally has $x_e=\mathcal{O}(\lambda^{v_e})$ with additional constraints on $\x$~\cite{GrdHzgJnsMa24,Ma25}, which we do not investigate in this work.}
\begin{align}
\label{eq:facet_region_general_parameter_scaling}
    x_e\sim \lambda^{v_e}\quad \forall e\in\mathcal{G}.
\end{align}
Let us remark that a key relation between the Lee–Pomeransky parameter $x_e$ and the line momentum $k_e$ (for massless $e$) is~\cite{GrdHzgJnsMaSchlk22}
\begin{align}
\label{eq:region_scaling_relation_momentum_LPparameter}
    x_e\sim (k_e^2)^{-1}\sim \lambda^{-\mathscr{V}(X)},
\end{align}
where $X$ is the momentum mode of $k_e$ and $\mathscr{V}(X)$ is its virtuality degree defined in section~\ref{section-momentum_modes}. This relation enables a translation between the momentum and parametric descriptions of a given region. If the scalings of $k_e^\mu$ are known, the scaling of $x_e$ follows immediately from the above. Conversely, given the scalings of $x_e$, the scaling of $k_e^2$ is obtained directly; the scalings of the components of $k_e^\mu$ can then be deduced from momentum conservation, Landau equations, etc., a process that is sometimes nontrivial~\cite{Ma23} but always feasible.

We now return to the Lee-Pomeransky-parameter description of a region, (\ref{eq:facet_region_general_parameter_scaling}). Each term of $\mathcal{P}(\x;\s)$, denoted by $s\cdot x_1^{r_1}x_2^{r_2}\cdots x_N^{r_N}$ with $s$ being the kinematic coefficient, scales as
\begin{align}
\label{eq:LP_term_scaling}
    s\cdot x_1^{r_1}x_2^{r_2}\cdots x_N^{r_N}\sim \lambda^{v_1r_1+\dots+v_Nr_N+r_{N+1}},\quad\textup{if }s\sim \lambda^{r_{N+1}}.
\end{align}
The leading terms in the expansion then correspond to those with the minimum value of $v_1r_1+\dots+v_Nr_N+r_{N+1}$. Below we will construct a polytope $\Delta$ corresponding to the Feynman integral and see that the leading terms are located on the ``lower facets'' of $\Delta(\mathcal{G})$.

\bigbreak
\noindent\underline{\emph{Polytope and Lower Facets}}.\footnote{This part provides a brief sketch of the Newton polytope approach. One may refer to ref.~\cite{HrchJnsSlk22} for pedagogical examples and more details.}

Let us map each individual term $s\cdot x_1^{r_1}x_2^{r_2}\cdots x_N^{r_N}$ to a specific point $\r$ in the $(N+1)$-dimensional space as follows:
\begin{eqnarray}
\label{eq:map_LP_term_point}
    s\cdot x_1^{r_1}x_2^{r_2}\cdots x_N^{r_N} \rightarrow \r\equiv (r_1,r_2,\dots,r_N,r_{N+1})\quad\textup{if }s\sim \lambda^{r_{N+1}}.
\end{eqnarray}
Next, we construct a polytope $\Delta(\mathcal{G})$, defined as the \emph{convex hull} of all these points obtained from above.

A crucial observation is the correspondence between the leading terms in a region $R$ and a certain \emph{lower facet} of $\Delta(\mathcal{G})$, denoted by $f_R$. By ``facet'', we refer to those $N$-dimensional boundaries of $\Delta(\mathcal{G})$, and ``lower facets'' refer to those facets whose inward-pointing normal vectors have a positive $(N+1)$-th entry. To see why this correspondence holds, first note that the set of points corresponding to the leading terms, denoted by $\{ \r_1,\r_2,\dots,\r_m \}$, must span an $N$-dimensional space; otherwise, the leading terms do not involve all $N$ parameters, and scaleless integrals would appear in the expansion. We denote this space by $f_R$. To see why $f_R$ must be a lower facet of $\Delta(\mathcal{G})$, observe from (\ref{eq:LP_term_scaling}) that $\r_1,\r_2,\dots,\r_m$ must all have the same minimum value of
$$\boldsymbol{v}\cdot\r_i^{} = v_1^{}r_{i,1}^{}+v_2^{}r_{i,2}^{}+\dots+v_N^{}r_{i,N}^{}+r_{i,N+1}^{},\quad i=1,\dots,m,$$
where $\boldsymbol{v}$ has been defined as the following $(N+1)$-dimensional vector\footnote{Strictly speaking, $\boldsymbol{v}$ is a covector, and in this sense we should have $v^1r_{i,1}+\dots+v^Nr_{i,N}+r_{i,N+1}$ above. Here we work with the Euclidean inner product and do not distinguish index positions.}:
\begin{align}
\label{eq:region_vector_def}
    \boldsymbol{v}\equiv (v_1,v_2,\dots,v_N;1).
\end{align}
On the one hand, for a convex polytope $\Delta(\mathcal{G})$, minimizing a linear functional $\boldsymbol{v}\cdot\r$ over $\r\in\Delta(\mathcal{G})$ always yields points on a supporting hyperplane that touches $\Delta(\mathcal{G})$ at its boundary. Hence $f_R$ forms a face of $\Delta(\mathcal{G})$, and since it spans $N$ dimensions, it is a facet, with its normal vector $\boldsymbol{v}$ pointing into the polytope. On the other hand, the $(N+1)$-th entry of $\boldsymbol{v}$ is positive (equal to $1$). Thus $f_R$ is a lower facet.

\bigbreak
\noindent\underline{\emph{Weighted Graph and Minimum Spanning (2-)trees}}.

The leading terms in the expansion can also be related to the minimum spanning (2-)trees of $\mathcal{G}$. Let us begin by defining a weighted graph in general. We associate each edge $e\in \mathcal{G}$ with a real number $w(e)$, which we refer to as the \emph{weight} of~$e$. We then define the weight of a spanning tree $T^1$ or a spanning 2‑tree $T^2$ as follows:
\begin{itemize}
    \item $w(T^1)$ is the sum over $w(e)$ where $e$ is \emph{not} in $T^1$;
    \item $w(T^2)$ is the sum over $w(e)$ where $e$ is \emph{not} in $T^2$, plus a \emph{kinematic contribution} $b$ if the kinematic factor $Q_{T^2}^{2}$ is $\mathcal{O}(\lambda^b)$.
\end{itemize}
Note that we exclude those spanning 2‑trees with $Q_{T^2}^{2}=0$, as they do not appear in the $\mathcal{F}$ polynomial. Summarizing the above, for each term of $\mathcal{P}(\x;\s)$ the weight of its corresponding spanning (2‑)tree $T$ is
\begin{align}
\label{eq:definition_spanning_1and2_trees_weights}
    w(T)\equiv \sum_{e\notin T} w(e) + b\quad \text{ if }\ \  s\sim \lambda^b.
\end{align}
Among all such $T$, we select those with the \emph{smallest} weight and call them the \emph{minimum spanning (2-)trees} of $\mathcal{G}$.

Recall that in the massless case, each power $r_e$ in a term $s\!\cdot\! x_1^{r_1}x_2^{r_2}\cdots x_N^{r_N}$ of $\mathcal{P}(\x;\s)$ is either $0$ or $1$. Setting $w(e)=v_e$ in eq.~(\ref{eq:definition_spanning_1and2_trees_weights}), we see that the minimum spanning (2‑)trees of $\mathcal{G}$ all minimize $\boldsymbol{v}\cdot\r$, the exact quantity that characterizes leading terms as discussed earlier. In this sense, every leading $\mathcal{U}$ term corresponds to a minimum spanning tree of $\mathcal{G}$, and every leading $\mathcal{F}$ term corresponds to a minimum spanning 2‑tree of $\mathcal{G}$.

To clarify the notation above: for each edge $e\in \mathcal{G}$, we have let $\mathscr{V}(\mathscr{X}(e))$ denote the virtuality degree of its line‑momentum mode, $v_e$ the corresponding component of the region vector or the power of $\lambda$ in the scaling of $x_e$, and $w(e)$ the weight assigned to it. These quantities are simply related by:
\begin{align}
    \boxed{\ v_e = w(e) =-\mathscr{V}(\mathscr{X}(e)).\ }
\end{align}

With this construction, the identification of regions can then be formulated as follows. \emph{We aim to find all possible scalings of the Lee–Pomeransky parameters $x_1,\dots,x_N$ such that the points associated with the minimum spanning (2‑)trees of $\mathcal{G}$ are sufficient to span a lower facet of the polytope $\Delta(\mathcal{G})$.} This is the core methodology of the region analysis in this paper.

\section{Algebraic and graphical structures of modes}
\label{section-algebraic_graphic_mode_structure}

This section provides a more comprehensive examination of the structures of momentum modes from two perspectives. In section~\ref{section-wedge_vee_operations}, we begin by introducing two fundamental operations, $\wedge$ and $\vee$, defined on any two given modes, and explore several of their algebraic properties. Then, in section~\ref{section-mode_subgraphs}, we assign to each edge and vertex in a given region a specific mode and develop the concepts of mode subgraphs and mode components, which can be seen as fundamental building blocks of each region. These notions will play essential roles in the technical proofs in the later sections.

\subsection{\texorpdfstring{The $\wedge$ and $\vee$ operations}{The wedge and vee operations}}
\label{section-wedge_vee_operations}

In section~\ref{section-momentum_modes}, we defined a partially ordered set (poset) of momentum modes ordered by the ``softer than'' (or ``harder than'') relation. We will now define two operations on these modes, which correspond to the mathematical concepts of the \emph{join} and \emph{meet} within this poset.

For any two given modes $X_1$ and $X_2$, we define
\begin{itemize}
    \item $X_1\wedge X_2$: the hardest mode that is softer than or equal to both $X_1$ and $X_2$;
    \item $X_1\vee X_2$: the softest mode that is harder than or equal to both $X_1$ and $X_2$.
\end{itemize}
These operations are well-defined: among all modes simultaneously softer than or equal to both $X_1$ and $X_2$, there exists a unique hardest one; similarly, among all modes simultaneously harder than or equal to both, there exists a unique softest one. Explicit expressions confirming this are derived below.

Two properties follow immediately from the definition. First, both the operations $\wedge$ and $\vee$ satisfy \emph{commutativity}:
\begin{eqnarray}
\label{eq:operation_commutativity}
    X_1\wedge X_2 = X_2\wedge X_1,\qquad X_1\vee X_2 = X_2\vee X_1.
\end{eqnarray}
Second, if $X_1$ is softer (harder) than $X_2$, we have $X_1\wedge X_2 = X_1$ ($X_2$), and if $X_1$ and $X_2$ are overlapping, then $X_1\wedge X_2$ is softer than both $X_1$ and $X_2$, while $X_1\vee X_2$ is harder than both $X_1$ and $X_2$.

We now derive explicit expressions for overlapping $X_1$ and $X_2$.

We start from the case where the modes $X_1$ and $X_2$ are possibly (soft-)collinear to the same direction $\beta_i$, i.e., $X_1=S^{m_1}C_i^{n_1}$ and $X_2=S^{m_2}C_i^{n_2}$. In this case, one can assume $m_1>m_2$ and $m_1+n_1<m_2+n_2$ without loss of generality. Let us further denote $X_1\wedge X_2 = S^{m_3}C_i^{n_3}$. From the text below~(\ref{eq:partial_order_example_soft_collinear}), we must have $m_3 \geqslant m_1$ and $m_3 + n_3 \geqslant m_2 + n_2$, yielding the following candidates:
\begin{align}
    &(m_3,n_3)=(m_1,m_2+n_2-m_1), &&\rightarrow &&S^{m_1}C_i^{m_2+n_2-m_1};\nonumber\\
    &(m_3,n_3)=(m_1+1,m_2+n_2-m_1-1), &&\rightarrow &&S^{m_1+1}C_i^{m_2+n_2-m_1-1};\nonumber\\
    &\quad\dots&&&&\quad\dots.\nonumber
\end{align}
The hardest mode from above is $S^{m_1}C_i^{m_2+n_2-m_1}$, so we have $X_1\wedge X_2 = S^{m_1}C_i^{m_2+n_2-m_1}$. Similarly, $X_1\vee X_2 = S^{m_2}C_i^{m_1+n_1-m_2}$. Note that these results hold when $m_1=m_2$ and/or $m_1+n_1=m_2+n_2$ as well.

We then consider the case where $X_1$ and $X_2$ are (soft-)collinear to distinct directions, i.e., $X_1=S^{m_1}C_i^{n_1}$ and $X_2=S^{m_2}C_j^{n_2}$ with $i\neq j$. Here, without loss of generality, we suppose $m_1+n_1\leqslant m_2+n_2$. If we further have $m_1+n_1\leqslant m_2$, then $X_1$ is harder than $X_2$, and we automatically have $X_1\wedge X_2 = S^{m_2}C_j^{n_2}$ and $X_1\vee X_2 = S^{m_1}C_i^{n_1}$.

If $m_2\leqslant m_1 + n_1$ instead, let us denote $X_1\wedge X_2 = S^{m_3}C_k^{n_3}$, which must be simultaneously softer than $X_1$ and $X_2$. If $k=j$, then we need $m_1+n_1\leqslant m_3$ and $m_2+n_2\leqslant m_3+n_3$ (from the text below~(\ref{eq:partial_order_example_soft_collinear})), whose solutions and corresponding modes are:
\begin{align}
    &(m_3,n_3)=(m_1+n_1,m_2+n_2-m_1-n_1), &&\rightarrow &&S^{m_1+n_1}C_j^{m_2+n_2-m_1-n_1};\nonumber\\
    &(m_3,n_3)=(m_1+n_1+1,m_2+n_2-m_1-n_1-1), &&\rightarrow &&S^{m_1+n_1+1}C_j^{m_2+n_2-m_1-n_1-1};\nonumber\\
    &\quad\dots&&&&\quad\dots\nonumber
\end{align}
among which the hardest mode is $S^{m_1+n_1}C_j^{m_2+n_2-m_1-n_1}$. If $k\neq j$ otherwise, then the only requirement we need is $m_3\geqslant m_2+n_2$ ($\geqslant m_1+n_1$ automatically). The hardest mode satisfying this condition is $S^{m_2+n_2}$, which is softer than (or equal to, if $m_1+n_1=m_2+n_2$) the mode $S^{m_1+n_1}C_j^{m_2+n_2-m_1-n_1}$. Therefore, we can always write $X_1\wedge X_2 = S^{m_1+n_1}C_j^{m_2+n_2-m_1-n_1}$ in the case of $m_2\leqslant m_1+n_1$.

It remains to work out the mode $X_1\vee X_2$ under the condition $m_2\leqslant m_1+n_1$, which further depends on the values of $m_1$ and $m_2$. Let us consider the subcase $m_1\leqslant m_2$ first and denote $X_1\vee X_2 \equiv S^{m_4}C_\ell^{n_4}$. For $\ell=i$, requiring that $S^{m_4}C_\ell^{n_4}$ is harder than both $S^{m_1}C_i^{n_1}$ and $S^{m_2}C_j^{n_2}$ leads to $m_4+n_4\leqslant \min\{m_2,m_1+n_1\} = m_2$ and $m_4\leqslant m_1$ (from the text below~(\ref{eq:partial_order_example_soft_collinear})). The solutions and corresponding modes are then:
\begin{align}
    &(m_4,n_4)=(m_1,m_2-m_1), &&\rightarrow &&S^{m_1}C_i^{m_2-m_1};\nonumber\\
    &(m_4,n_4)=(m_1-1,m_2-m_1+1), &&\rightarrow &&S^{m_1-1}C_i^{m_2-m_1+1};\nonumber\\
    &\quad\dots&&&&\quad\dots\nonumber
\end{align}
among which the softest mode is $S^{m_1}C_i^{m_2-m_1}$. For $\ell\neq i$, requiring $S^{m_4}C_\ell^{n_4}$ to be harder than both $S^{m_1}C_i^{n_1}$ and $S^{m_2}C_j^{n_2}$ leads to only $m_4+n_4\leqslant m_1$. The softest mode allowed is then $S^{m_1}$, which is harder than (or equal to, if $m_1=m_2$) the mode $S^{m_1}C_j^{m_2-m_1}$. As a result, we have $X_1\vee X_2 = S^{m_1}C_i^{m_2-m_1}$ for $m_1\leqslant m_2$. Similarly, $X_1\vee X_2 = S^{m_2}C_j^{m_1-m_2}$ for $m_1\geqslant m_2$.

The results are summarized in the following theorem.
\begin{theorem}
\label{theorem-overlapping_modes_intersection_union_rules}
    Any two overlapping modes $X_1 = S^{m_1}C_i^{n_1}$ and $X_2 = S^{m_2}C_j^{n_2}$ fall into one of the following cases, with the corresponding expressions of $X_1\wedge X_2$ and $X_1\vee X_2$, up to swapping $1\leftrightarrow2$:
    \begin{enumerate}
        \item $i=j$, $m_2<m_1\leqslant m_1+n_1<m_2+n_2$:\\
        $X_1\wedge X_2 = S^{m_1}C_i^{m_2+n_2-m_1}$ and $X_1\vee X_2 = S^{m_2}C_i^{m_1+n_1-m_2}$.
        \item $i\neq j$, $m_1\leqslant m_2\leqslant m_1+n_1\leqslant m_2+n_2$:\\
        $X_1\wedge X_2 = S^{m_1+n_1}C_j^{m_2+n_2-m_1-n_1}$ and $X_1\vee X_2 = S^{m_1}C_i^{m_2-m_1}$.
        \item $i\neq j$, $m_2\leqslant m_1\leqslant m_1+n_1\leqslant m_2+n_2$:\\
        $X_1\wedge X_2 = S^{m_1+n_1}C_j^{m_2+n_2-m_1-n_1}$ and $X_1\vee X_2 = S^{m_2}C_j^{m_1-m_2}$.
    \end{enumerate}
\end{theorem}
Theorem~\ref{theorem-overlapping_modes_intersection_union_rules} can be regarded as an alternative definition of the $\wedge$ and $\vee$ operations for overlapping modes. Tables~\ref{table-mode_intersection_examples} and~\ref{table-mode_union_examples} list the results of these operations for several typical examples of $X_1$ and $X_2$. The tables are only half filled, as the remaining entries follow from the commutativity of the operations, eq.~(\ref{eq:operation_commutativity}).
\begin{table}[t]
\centering
\begin{tabular}{|c||*{10}{c|}}
\hline
\makecell{$\boldsymbol{\bigwedge}$} & $\boldsymbol{H}$ & $\boldsymbol{C_i}$ & $\boldsymbol{C_j}$ & $\boldsymbol{S}$ & $\boldsymbol{C_i^2}$ & $\boldsymbol{C_j^2}$ & $\boldsymbol{SC_i}$ & $\boldsymbol{SC_j}$ & $\boldsymbol{C_i^3}$ & $\boldsymbol{C_j^3}$ \\ \hline\hline
$\boldsymbol{H}$ & $H$ & $C_i$ & $C_j$ & $S$ & $C_i^2$ & $C_j^2$ & $SC_i$ & $SC_j$ & $C_i^3$ & $C_j^3$ \\ \hline
$\boldsymbol{C_i}$ &  & $C_i$ & $S$ & $S$ & $C_i^2$ & $SC_j$ & $SC_i$ & $SC_j$ & $C_i^3$ & $SC_j^2$ \\ \hline
$\boldsymbol{C_j}$ &  &  & $C_j$ & $S$ & $SC_i$ & $C_j^2$ & $SC_i$ & $SC_j$ & $SC_i^2$ & $C_j^3$ \\ \hline
$\boldsymbol{S}$ &  &  &  & $S$ & $SC_i$ & $SC_j$ & $SC_i$ & $SC_j$ & $SC_i^2$ & $SC_j^2$ \\ \hline
$\boldsymbol{C_i^2}$ &  &  &  &  & $C_i^2$ & $S^2$ & $SC_i$ & $S^2$ & $C_i^3$ & $S^2C_j$ \\ \hline
$\boldsymbol{C_j^2}$ &  &  &  &  &  & $C_j^2$ & $S^2$ & $SC_j$ & $S^2C_i$ & $C_j^3$ \\ \hline
$\boldsymbol{SC_i}$ &  &  &  &  &  &  & $SC_i$ & $S^2$ & $SC_i^2$ & $S^2C_j$ \\ \hline
$\boldsymbol{SC_j}$ &  &  &  &  &  &  &  & $SC_j$ & $S^2C_i$ & $SC_j^2$ \\ \hline
$\boldsymbol{C_i^3}$ &  &  &  &  &  &  &  &  & $C_i^3$ & $S^3$ \\ \hline
$\boldsymbol{C_j^3}$ &  &  &  &  &  &  &  &  &  &  $C_j^3$ \\ \hline
\end{tabular}
\caption{Examples of the $\wedge$ operation of momentum modes, where we assume $i\neq j$. Only the diagonal and upper-right portion of the table is shown; the remaining entries can be obtained using the commutativity of $\wedge$.}
\label{table-mode_intersection_examples}
\end{table}
\begin{table}[t]
\centering
\begin{tabular}{|c||*{10}{c|}}
\hline
\makecell{$\boldsymbol{\bigvee}$} & $\boldsymbol{H}$ & $\boldsymbol{C_i}$ & $\boldsymbol{C_j}$ & $\boldsymbol{S}$ & $\boldsymbol{C_i^2}$ & $\boldsymbol{C_j^2}$ & $\boldsymbol{SC_i}$ & $\boldsymbol{SC_j}$ & $\boldsymbol{C_i^3}$ & $\boldsymbol{C_j^3}$ \\ \hline\hline
$\boldsymbol{H}$ & $H$ & $H$ & $H$ & $H$ & $H$ & $H$ & $H$ & $H$ & $H$ & $H$ \\ \hline
$\boldsymbol{C_i}$ &  & $C_i$ & $H$ & $C_i$ & $C_i$ & $H$ & $C_i$ & $C_i$ & $C_i$ & $H$ \\ \hline
$\boldsymbol{C_j}$ &  &  & $C_j$ & $C_j$ & $H$ & $C_j$ & $C_j$ & $C_j$ & $H$ & $C_j$ \\ \hline
$\boldsymbol{S}$ &  &  &  & $S$ & $C_i$ & $C_j$ & $S$ & $S$ & $C_i$ & $C_j$ \\ \hline
$\boldsymbol{C_i^2}$ &  &  &  &  & $C_i^2$ & $H$ & $C_i^2$ & $C_i$ & $C_i^2$ & $H$ \\ \hline
$\boldsymbol{C_j^2}$ &  &  &  &  &  & $C_j^2$ & $C_j$ & $C_j^2$ & $H$ & $C_j^2$ \\ \hline
$\boldsymbol{SC_i}$ &  &  &  &  &  &  & $SC_i$ & $S$ & $C_i^2$ & $C_j$ \\ \hline
$\boldsymbol{SC_j}$ &  &  &  &  &  &  &  & $SC_j$ & $C_i$ & $C_j^2$ \\ \hline
$\boldsymbol{C_i^3}$ &  &  &  &  &  &  &  &  & $C_i^3$ & $H$ \\ \hline
$\boldsymbol{C_j^3}$ &  &  &  &  &  &  &  &  &  &  $C_j^3$ \\ \hline
\end{tabular}
\caption{Examples of the $\vee$ operation of momentum modes, where we assume $i\neq j$. Only the diagonal and upper-right portion of the table is shown; the remaining entries can be obtained using the commutativity of $\vee$.}
\label{table-mode_union_examples}
\end{table}

So far we have shown that $X_1\wedge X_2$ and $X_1\vee X_2$ are well and uniquely defined for any two modes $X_1$ and $X_2$. With this construction, the poset of modes is indeed a \emph{lattice}, a crucial mathematical concept in order theory~\cite{Bkhf40book,BrsSkpnv81book,DvyPrstl02book}.

One immediate corollary is the \emph{associativity} of $\wedge$ and $\vee$, respectively:
\begin{eqnarray}
    (X_1\wedge X_2)\wedge X_3 = X_1\wedge (X_2\wedge X_3),\qquad (X_1\vee X_2)\vee X_3 = X_1\vee (X_2\vee X_3).
\end{eqnarray}
To see this, let us denote $X_a = (X_1\wedge X_2)\wedge X_3$ and $X_b = X_1\wedge (X_2\wedge X_3)$. By definition, $X_a$ is simultaneously softer than or equal to $X_1$, $X_2$, and $X_3$, thus it is also softer than or equal to $X_2\wedge X_3$. Since $X_b$ is the hardest mode that is simultaneously softer than or equal to both $X_1$ and $X_2\wedge X_3$, we know that $X_a$ is softer than or equal to $X_b$. The same logic in reverse shows $X_b$ is softer than or equal to $X_a$. Thus $X_a = X_b$. The proof for $\vee$ is analogous.

Associativity implies that $X_1\wedge\dots \wedge X_n$ ($X_1\vee\dots \vee X_n$) is exactly the hardest (softest) mode that is softer (harder) than or equal to all of $X_1,\dots,X_n$ simultaneously.

One may wonder whether $\wedge$ and $\vee$ satisfy the distributive laws. In general, the answer is \emph{no}: taking $X_1=C_1$, $X_2=C_2$, and $X_3=C_3$, we have
\begin{align}
\begin{split}
    &(X_1\wedge X_2)\vee X_3 = C_3 \neq H = (X_1\vee X_3)\wedge (X_2\vee X_3),\\
    &(X_1\vee X_2)\wedge X_3 = C_3 \neq S= (X_1\wedge X_3)\vee  (X_2\wedge X_3).
\end{split}    
\end{align}
The lattice is then non-distributive (more precisely, it contains the ``diamond $M_3$'' as a sub-lattice). However, in the special cases involving only one or two lightlike directions ($\beta$ in eq.~\eqref{eq:virtuality_expansion}), the distributive law still holds. This implies an interesting physical corollary: non-distributivity is a genuine multi-jet ($\geqslant3$ directions) phenomenon, whereas the Sudakov-type (two directions) case is distributive.

The observation above shows one difference between ($\wedge$,$\vee$) and ($\cap$,$\cup$) from set theory. The following theorem, meanwhile, describes a more striking distinction.
\begin{theorem}
\label{theorem-mode_algebra_two_representatives}
    For any set of modes $\{X_1,\dots,X_n\}$ with $n\geqslant3$, there exist $a,b,c,d\in \{1,\dots,n\}$ such that
    \begin{eqnarray}
    X_a\wedge X_b = X_1\wedge\cdots\wedge X_n ,\qquad X_c\vee X_d = X_1\vee\cdots\vee X_n .
    \label{eq:intersection_union_reduction_property}
    \end{eqnarray}
\end{theorem}
In other words, to obtain $X_1\wedge\dots \wedge X_n$ ($X_1\vee\dots \vee X_n$) we can simply take two representatives of this set, $X_a$ and $X_b$ ($X_c$ and $X_d$), and $X_a\wedge X_b$ ($X_c\vee X_d$) is the same mode.

\begin{proof}
    To begin, note that it suffices to prove the theorem if we show the existence of $X_a$--$X_d$ for $n=3$, because the corresponding modes can be identified recursively for any larger $n$.
    
    We next focus on the first equation of (\ref{eq:intersection_union_reduction_property}) and consider the special case where $X_i=C_i^{n_i}$ for $i=1,2,3$.\footnote{Here we have taken all the collinear directions distinct, because if two modes are collinear to the same direction, then one of them is either softer or harder than the other. The mode $X_1\wedge X_2 \wedge X_3$ then automatically reduces to the $\wedge$ operation of two representatives that are collinear to distinct directions.} Among the three integers $\{n_1,n_2,n_3\}$, we take the largest two of them, $n_a$ and $n_b$ with $a,b\in\{1,2,3\}$. For example,
    \begin{itemize}
        \item $X_1=C_1$, $X_2=C_2^2$, $X_3=C_3^2 \qquad \Rightarrow \qquad (a,b) = (2,3)$\,;
        \item $X_1=C_1$, $X_2=C_2$, $X_3=C_3^2 \qquad \Rightarrow \qquad (a,b) = (1,3)\,\textup{ or }\,(2,3)$\,.
    \end{itemize}
    Without loss of generality, we can further assume $n_a\leqslant n_b$. It then follows that
    \begin{eqnarray}
        X_a \wedge X_b = S^{n_a} C_b^{n_b-n_a} = X_1\wedge X_2 \wedge X_3.
    \end{eqnarray}
    The second equality is because $S^{n_a} C_b^{n_b-n_a}$ is already softer than the third mode.
    
    For any generic mode $X = S^m C_i^n$, there exist two collinear modes $C_i^{m+n}$ and $C_j^m$ ($j\neq i$), such that $X= C_i^{m+n}\wedge C_j^m$. We can thus use this decomposition to rewrite $X_1\wedge X_2\wedge X_3$ as the $\wedge$ operation of 6 collinear modes, which equals the $\wedge$ of two of them (denoted as $C_a^{n_a}$ and $C_b^{n_b}$) from the analysis above. This mode $C_a^{n_a} \wedge C_b^{n_b}$, meanwhile, also equals the $\wedge$ of two original modes generating $C_a^{n_a}$ and $C_b^{n_b}$. We thus complete proving the first equation of (\ref{eq:intersection_union_reduction_property}).

    We finally show the existence of $c,d\in \{1,2,3\}$ such that $X_c\vee X_d = X_1\vee\dots \vee X_n$. We first assume $X_i = S^{m_i}C_i^{n_i}$ with distinct $i=1,2,3$. Similar as above, we take $c$ and $d$ such that $m_c$ and $m_d$ are the smallest among $\{m_1,m_2,m_3\}$, with $m_c\geqslant m_d$. This leads to
    \begin{eqnarray}
    \label{eq:theorem_mode_algebra_two_representatives_vee_distinct_directions}
        X_c\vee X_d = S^{m_d} C_d^{m_c-m_d},
    \end{eqnarray}
    which is harder than or equal to the third mode. So $X_c\vee X_d = X_1\vee X_2\vee X_3$, and the second equation of (\ref{eq:intersection_union_reduction_property}) is justified under this assumption.

    It remains to consider the possibility that some overlapping modes in $\{X_1,\dots,X_n\}$ are (soft-)collinear along the same direction, i.e., $X_1=S^{m_1}C_i^{n_1}$ and $X_2=S^{m_2}C_i^{n_2}$ with the relation $m_2<m_1\leqslant m_1+n_1<m_2+n_2$ (case \emph{1} of theorem~\ref{theorem-overlapping_modes_intersection_union_rules}). In this case, however, one can always modify $X_1$ into a direction-independent mode $S^{m_1+n_1}$, because
    \begin{align}
        S^{m_1+n_1}\vee S^{m_2}C_i^{n_2} = S^{m_1}C_i^{n_1}\vee S^{m_2}C_i^{n_2}\ (=S^{m_2}C_i^{m_1+n_1-m_2}),
    \end{align}
    as verified directly from the explicit expressions. The directions are now effectively distinct, so eq.~(\ref{eq:theorem_mode_algebra_two_representatives_vee_distinct_directions}) applies.
\end{proof}

We remark that the choice of $X_a$ and $X_b$ in eq.~(\ref{eq:intersection_union_reduction_property}) may not be unique. For example,
\begin{align}
    C_i\wedge C_j = C_1\wedge\cdots\wedge C_n \ (=S),\qquad C_i\vee C_j = C_1\vee\cdots\vee C_n \ (=H),
\end{align}
for any $i,j\in\{1,\dots,n\}$ with $i\neq j$.

The following theorem relates the virtuality degrees of modes under the $\wedge$ and $\vee$ operations, revealing a particular elegance of the mode algebra.
\begin{theorem}
\label{theorem-virtualities_relation}
    Given any two modes $X_1$ and $X_2$, their virtuality degrees can be related to those of $X_1\wedge X_2$ and $X_1\vee X_2$ simply as follows:
    \begin{eqnarray}
        \mathscr{V}(X_1)+\mathscr{V}(X_2) = \mathscr{V}(X_1\wedge X_2) + \mathscr{V}(X_1\vee X_2).
    \label{eq:virtualities_relation}
    \end{eqnarray}
\end{theorem}
\begin{proof}
    The equality reduces to a trivial relation $\mathscr{V}(X_1)+\mathscr{V}(X_2) = \mathscr{V}(X_1)+\mathscr{V}(X_2)$ if $X_1$ is softer/harder than $X_2$. For overlapping $X_1$ and $X_2$, eq.~(\ref{eq:virtualities_relation}) can be justified directly from theorem~\ref{theorem-overlapping_modes_intersection_union_rules}. For instance, in case \emph{1} there, we have $\mathscr{V}(X_1) = 2m_1+n_1$, $\mathscr{V}(X_2) = 2m_2+n_2$, $\mathscr{V}(X_1\wedge X_2) = m_1+m_2+n_2$, and $\mathscr{V}(X_1\vee X_2) = m_1+m_2+n_1$. Clearly eq.~(\ref{eq:virtualities_relation}) holds. The remaining two cases are verified similarly.
\end{proof}
A simple nontrivial example is $X_i=C_i$ for $i=1,2$. Here $X_1\wedge X_2 = S$ and $X_1\vee X_2 = H$, and eq.~(\ref{eq:virtualities_relation}) reduces to $1+1=2+0$.

The relation (\ref{eq:virtualities_relation}) can be generalized to three or more modes, as stated in the following corollary.
\begin{corollary}
\label{theorem-virtualities_relation_corollary1}
    Given any modes $X_1,\dots,X_n$ ($n\geqslant2$), we have
    \begin{eqnarray}
        \mathscr{V}(X_1)+\dots+\mathscr{V}(X_n) \leqslant \mathscr{V}(X_1\wedge X_2)+\dots+\mathscr{V}(X_{n-1}\wedge X_n) + \mathscr{V}(X_1\vee \dots\vee X_n).
    \label{ineq:virtualities_relation}
    \end{eqnarray}
\end{corollary}
\begin{proof}
    We show (\ref{ineq:virtualities_relation}) by induction. This inequality at the special case of $n=2$ reduces to eq.~(\ref{eq:virtualities_relation}). Suppose it holds at $n=k$, then for $n=k+1$, we have
    \begin{align}
        &\mathscr{V}(X_1)+\dots+\mathscr{V}(X_{k+1})\leqslant \mathscr{V}(X_1\!\wedge\! X_2)+\dots+\mathscr{V}(X_{k-1}\!\wedge\! X_k) + \mathscr{V}(X_1\!\vee \dots\vee\! X_k) + \mathscr{V}(X_{k+1})\nonumber\\
        &\ =\mathscr{V}(X_1\wedge X_2)+\dots+\mathscr{V}(X_{k-1}\wedge X_k)+\mathscr{V}\big((X_1\vee\dots\vee X_k)\wedge X_{k+1}\big)+\mathscr{V}(X_1\vee\dots\vee X_{k+1})\nonumber\\
        &\ \leqslant\mathscr{V}(X_1\wedge X_2)+\dots+\mathscr{V}(X_{k-1}\wedge X_k)+\mathscr{V}(X_k\wedge X_{k+1})+\mathscr{V}(X_1\vee\dots\vee X_{k+1}).
    \end{align}
    In the first line above, we have applied the induction hypothesis. In the second line, we have applied theorem~\ref{theorem-virtualities_relation} by regarding $X_1\vee\dots\vee X_k$ as a single mode. In the third line, we have used the defining property of $\wedge$ and $\vee$ operations: since the mode $X_1\vee\dots\vee X_k$ is harder than or equal to $X_k$, the mode $(X_1\vee\dots\vee X_k)\wedge X_{k+1}$ is then harder than or equal to $X_k\wedge X_{k+1}$, thus $\mathscr{V}\big((X_1\vee\dots\vee X_k)\wedge X_{k+1}\big)\leqslant \mathscr{V}(X_k\wedge X_{k+1})$.
\end{proof}
For general $n$, the mathematical induction above can be viewed as the following recursive procedure.
\begin{enumerate}
    \item [1,] Apply theorem~\ref{theorem-virtualities_relation} to two selected modes $X_1,X_2$ and obtain
    \begin{align}
    \label{eq:virtualities_ineq_relation_recursive_step1}
        \mathscr{V}(X_1)+\mathscr{V}(X_2) = \mathscr{V}(X_1\wedge X_2) + \mathscr{V}(X_1\vee X_2).
    \end{align}
    \item [2,] Apply theorem~\ref{theorem-virtualities_relation} to $X_1\vee X_2$ and $X_3$,
    $$\mathscr{V}(X_1\vee X_2)+\mathscr{V}(X_3) = \mathscr{V}((X_1\vee X_2)\wedge X_3) + \mathscr{V}(X_1\vee X_2\vee X_3).$$
    Since $\mathscr{V}((X_1\vee X_2)\wedge X_3)\leqslant \mathscr{V}(X_2\wedge X_3)$, we have, in combination with eq.~(\ref{eq:virtualities_ineq_relation_recursive_step1}),
    \begin{align}
    \label{eq:virtualities_ineq_relation_recursive_step2}
        \mathscr{V}(X_1)+\mathscr{V}(X_2)+\mathscr{V}(X_3) \leqslant \mathscr{V}(X_1\wedge X_2) + \mathscr{V}(X_2\wedge X_3) + \mathscr{V}(X_1\vee X_2\vee X_3).
    \end{align}
    \item [3,] Apply theorem~\ref{theorem-virtualities_relation} to $X_1\vee X_2\vee X_3$ and $X_4$, \dots.
\end{enumerate}
It follows from above that the inequality in (\ref{ineq:virtualities_relation}) comes purely from
\begin{align}
    \mathscr{V}((X_1\vee X_2)\wedge X_3)\leqslant \mathscr{V}(X_2\wedge X_3),\quad \mathscr{V}((X_1\vee X_2\vee X_3)\wedge X_4)\leqslant \mathscr{V}(X_3\wedge X_4),\quad\dots.\nonumber
\end{align}
Therefore, the equality in (\ref{ineq:virtualities_relation}) holds if and only if all the equalities above hold. In fact, this is equivalent to a more general condition, which we state as another corollary of theorem~\ref{theorem-virtualities_relation}.
\begin{corollary}
\label{theorem-virtualities_relation_corollary2}
    The equality in (\ref{ineq:virtualities_relation}) holds if and only if all the following conditions hold:
    \begin{align}
    \begin{split}
        &(X_{i_1}\vee X_{i_2})\wedge X_{i_3} = X_{*_3}\wedge X_{i_3},\\
        &(X_{i_1}\vee X_{i_2}\vee X_{i_3})\wedge X_{i_4} = X_{*_4}\wedge X_{i_4},\\
        &\qquad\dots\\
        &(X_{i_1}\vee \dots\vee X_{i_{n-1}})\wedge X_{i_n} = X_{*_n}\wedge X_{i_n},
    \end{split}
    \label{eq:virtuality_equation_condition}
    \end{align}
with $\{i_1,\dots,i_n\}$ being any permutation of $\{1,\dots,n\}$ such that for each $k\in \{1,\dots,n\}$, the set $\{i_1,\dots,i_k\}$ consists of $k$ consecutive integers. The indices $*_3,\dots,*_n$ are uniquely determined such that each $*_j$ is from $\{i_1,\dots,i_{j-1}\}$ and adjacent to $i_j$.
\end{corollary}
\begin{proof}
    Following the same procedure of (\ref{eq:virtualities_ineq_relation_recursive_step1}) and (\ref{eq:virtualities_ineq_relation_recursive_step2}), for any permutation $\{1,\dots,n\}\rightarrow \{i_1,\dots,i_n\}$ we have
    \begin{align}
        &\mathscr{V}(X_{i_1})+\mathscr{V}(X_{i_2}) = \mathscr{V}(X_{i_1}\wedge X_{i_2}) + \mathscr{V}(X_{i_1}\vee X_{i_2}),\nonumber\\
        &\mathscr{V}(X_{i_1})+\mathscr{V}(X_{i_2})+\mathscr{V}(X_{i_3}) \leqslant \mathscr{V}(X_{i_1}\wedge X_{i_2}) + \mathscr{V}(X_{*_3}\wedge X_{i_3}) + \mathscr{V}(X_{i_1}\vee X_{i_2}\vee X_{i_3}),\nonumber\\
        &\mathscr{V}(X_{i_1})+\dots+\mathscr{V}(X_{i_4}) \leqslant \mathscr{V}(X_{i_1}\!\wedge\! X_{i_2}) + \mathscr{V}(X_{*_3}\!\wedge\! X_{i_3}) + \mathscr{V}(X_{*_4}\!\wedge\! X_{i_4}) + \mathscr{V}(X_{i_1}\!\vee \dots\vee\! X_{i_4}),\nonumber\\
        &\dots,\nonumber
    \end{align}
    where $*_j$ can be any element of $\{i_1,\dots,i_{j-1}\}$ for $j=3,\dots,n$. In the end, we have
    \begin{align}
    \label{eq:proof_virtualities_relation_corollary2_step1}
        \sum_{j=1}^n \mathscr{V}(X_{i_j}) \leqslant \mathscr{V}(X_{i_1}\vee X_{i_2}) + \sum_{j=3}^n \mathscr{V}(X_{*_j}\wedge X_{i_j}) + \mathscr{V}(X_{i_1}\vee\dots\vee X_{i_n}).
    \end{align}
    Since we aim to obtain \eqref{ineq:virtualities_relation}, the following pairs $(i_1,i_2)$, $(*_3,i_3)$, \dots, $(*_n,i_n)$ must be $(1,2)$, \dots, $(n-1,n)$, each once. This follows precisely the index constraints in the statement ($\{i_1,i_2\}$ consists of two consecutive integers, and each $*_j\in \{i_1,\dots,i_{j-1}\}$ is adjacent to $i_j$).

    The equality in (\ref{eq:proof_virtualities_relation_corollary2_step1}) is achieved if and only if
    \begin{align}
    \label{eq:proof_virtualities_relation_corollary2_step2}
    \begin{split}
        &\mathscr{V}((X_{i_1}\vee X_{i_2})\wedge X_{i_3}) = \mathscr{V}(X_{*_3}\wedge X_{i_3}),\\
        &\mathscr{V}((X_{i_1}\vee X_{i_2}\vee X_{i_3})\wedge X_{i_4}) = \mathscr{V}(X_{*_4}\wedge X_{i_4}),\\
        &\qquad\dots\\
        &\mathscr{V}((X_{i_1}\vee \dots\vee X_{i_{n-1}})\wedge X_{i_n}) = \mathscr{V}(X_{*_n}\wedge X_{i_n}).
    \end{split}
    \end{align}
    These equations are equivalent to those in eq.~(\ref{eq:virtuality_equation_condition}), because if $X_a$ is harder than or equal to $X_b$, then $\mathscr{V}(X_a) = \mathscr{V}(X_b)$ immediately leads to $X_a=X_b$. Conversely, if any equality above becomes ``$<$'' instead, then the mode on its left-hand side would be harder than the mode on the right-hand side, and the strict inequality would hold in \eqref{ineq:virtualities_relation}.
\end{proof}

Later in section~\ref{section-leading_terms}, we will see the essential role theorem~\ref{theorem-virtualities_relation} and its corollaries play in characterizing general leading $\mathcal{F}$ terms in any given expansion, as a necessity in deriving the region structure.

\subsection{Mode subgraphs}
\label{section-mode_subgraphs}

We begin with the notion of graph adjacency. Two graphs $\Gamma_1$ and $\Gamma_2$ are called \emph{adjacent} if there exists a vertex $v\in \Gamma_1$ and an edge $e\in \Gamma_2$ (or vice versa) such that $v$ is an endpoint of $e$.

\begin{figure}[t]
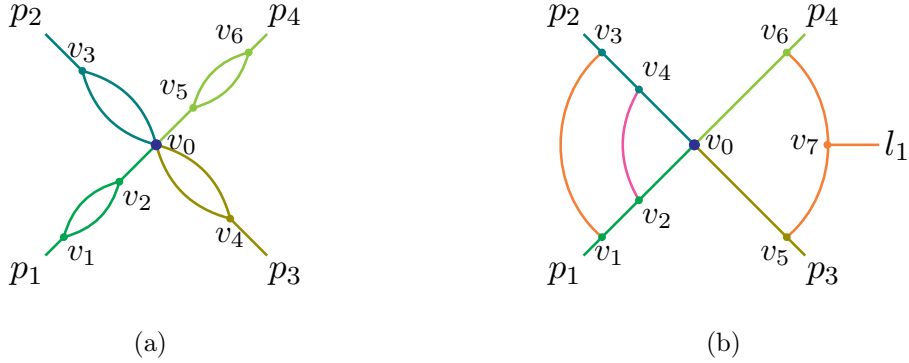

\centering
\begin{subfigure}[b]{0.4\textwidth}
\centering
\include{figs/mode_subgraphs_example1}
\vspace{-1em}
\caption{}
\label{mode_subgraphs_example1}
\end{subfigure}
\hspace{3em}
\begin{subfigure}[b]{0.4\textwidth}
\centering
\include{figs/mode_subgraphs_example2}
\vspace{-1em}
\caption{}
\label{mode_subgraphs_example2}
\end{subfigure}
\caption{Two examples illustrating some special cases of mode subgraphs, where the external momenta and internal vertices are explicitly specified. (a): the $H$ subgraph $\Gamma_H$ consists of the vertex $v_0$ only. (b): $\Gamma_H$ consists of the vertex $v_0$ only; additionally, the $S^2$ subgraph $\Gamma_{S^2}$ consists of two connected components.}
\label{figure-mode_subgraphs_examples}
\end{figure}

Mode subgraphs are defined as subgraphs formed by edges of a unique mode. For each given mode $X$, the corresponding \emph{$X$-mode subgraph}---or simply \emph{$X$ subgraph}, denoted $\Gamma_X$---is defined to consist of the following edges and vertices:
\begin{itemize}
    \item edges whose line momenta are in the $X$ mode;
    \item vertices whose adjacent line momenta $k_a^\mu$ satisfy $\bigvee_a \mathscr{X}(k_a) = X$.
\end{itemize}
Note that an $X$ subgraph may consist of a single vertex, and it may have several disconnected components. Two examples are shown in figure~\ref{figure-mode_subgraphs_examples}.
In figure~\ref{mode_subgraphs_example1} we require each edge to be collinear to a certain $p_i$, with $p_i^2\sim \lambda$ for $i=1,2,3,4$. In figure~\ref{mode_subgraphs_example2} we require the {\color{Rhodamine}\bf rhodamine} edges to be in the $S$ mode, the {\color{Orange}\bf orange} edges in the $S^2$ mode, and each solid edge collinear to a certain $p_i$, with $p_i^2=0$ for $i=1,2,3,4$ and $l_1^\mu\sim \lambda^2$. The mode subgraphs for these examples are listed below (where we use $e_{ij}$ to indicate the edge connecting $v_i$ and $v_j$):
\begin{eqnarray}
    \textup{Figure~\ref{mode_subgraphs_example1}: }&& \Gamma_H=(\{v_0\},\varnothing),\quad \Gamma_{C_1}=(\{v_1,v_2\}, \{e'_{12},e''_{12},e_{02}\}), \quad \dots.\nonumber\\
    \textup{Figure~\ref{mode_subgraphs_example2}: }&& \Gamma_H=(\{v_0\},\varnothing),\quad \Gamma_{C_1^2}=(\{v_1\}, \{e_{12}\}), \quad \Gamma_{C_1}=(\{v_2\}, \{e_{02}\}), \quad\dots,\nonumber\\
    &&\Gamma_S=(\varnothing,\{e_{24}\}),\quad \Gamma_{S^2}= (\varnothing,\{e_{13}\})\ \cup\ (\{v_7\}, \{e_{57},e_{67}\}).\nonumber
\end{eqnarray}
Note that the edges $e'_{12}$ and $e''_{12}$ denote the two distinct edges connecting $v_1$ and $v_2$ in figure~\ref{mode_subgraphs_example1}. In both examples, $\Gamma_H$ consists of a single vertex $v_0$. In figure~\ref{mode_subgraphs_example2}, $\Gamma_{S^2}$ has two connected components, which are aside ``$\cup$'' above.

For a given mode subgraph $\Gamma$, we denote its corresponding mode by $\mathscr{X}(\Gamma)$ or $\mathscr{X}(\Gamma')$ for any subgraph $\Gamma' \subset \Gamma$. By assigning edges and vertices into different mode subgraphs as described, we obtain the following two properties.
\begin{enumerate}
    \item $\Gamma_1\cap\Gamma_2 = \varnothing$ for any two mode subgraphs $\gamma_1$ and $\gamma_2$. In other words, the entire graph $\mathcal{G}$ is a \emph{disjoint union} of mode subgraphs.
    \item Two adjacent mode subgraphs cannot have overlapping modes.
\end{enumerate}
The first property follows by construction. To prove the second, assume the contrary: a vertex $v\in \Gamma_1$ is an endpoint of an edge $e\in \Gamma_2$, with $\mathscr{X}(\Gamma_1)$ overlapping with $\mathscr{X}(\Gamma_2)$. By definition, $\mathscr{X}(v) = \mathscr{X}(\Gamma_1)$. At the same time, $\mathscr{X}(v)$ must be equal to or harder than $\mathscr{X}(\Gamma_1)\vee \mathscr{X}(\Gamma_2)$, which is (strictly) harder than either $\mathscr{X}(\Gamma_1)$ or $\mathscr{X}(\Gamma_2)$ because these modes overlap. This yields a contradiction. Hence, adjacent mode subgraphs must have non‑overlapping modes.

From this, we can make another observation: for a mode subgraph $\Gamma_X$, any vertex $v\notin \Gamma_X$ that is adjacent to $\Gamma_X$ must correspond to a mode harder than $X$. This observation motivates the construction of \emph{contracted mode subgraphs}. A \emph{contracted $X$ subgraph}, denoted $\widetilde{\Gamma}_X$, is obtained from $\Gamma_X$ by introducing an auxiliary vertex and identifying it with all vertices of other mode subgraphs that are adjacent to $\Gamma_X$.

Let us illustrate this concept with a few examples. In figure~\ref{mode_subgraphs_example1}, the only vertex not belonging to $\Gamma_{C_1}$ but adjacent to it is the hard vertex $v_0$. To obtain $\widetilde{\Gamma}_{C_1}$, we simply replace $v_0$ by the auxiliary vertex of $\widetilde{\Gamma}_{C_1}$. So we have
$$\widetilde{\Gamma}_{C_1} = \Gamma_{C_1}\cup \{\textup{auxiliary vertex}\},$$
and the same conclusion for the $C_2,C_3,C_4$ subgraphs.

As a slightly more involved example, consider figure~\ref{mode_subgraphs_example2}. The vertices not belonging to $\Gamma_{S}$ but adjacent to it are $v_2$ and $v_4$. To construct $\widetilde{\Gamma}_{S}$, we identify these vertices with an auxiliary vertex. Similarly, $\widetilde{\Gamma}_{S^2}$ can be obtained. The graphs $\widetilde{\Gamma}_{S}$ and $\widetilde{\Gamma}_{S^2}$ are depicted below, with $v_*$ representing their auxiliary vertices.
\begin{equation}
\label{eq:contracted_mode_subgraph_illustration}
\widetilde{\Gamma}_{S}\ =\ 
\begin{tikzpicture}[baseline=(current bounding box.center), scale=0.45]
\node [draw, Rhodamine, circle, minimum size=3pt, fill=Rhodamine, inner sep=0pt, outer sep=0pt] () at (7,5) {};
\draw (7,5) .. controls (5,7) and (3,7) .. (3,5) [Rhodamine, very thick];
\draw (7,5) .. controls (5,3) and (3,3) .. (3,5) [Rhodamine, very thick];
\node at (7.7,5) {$v_*$};
\end{tikzpicture}\ ;
\qquad
\widetilde{\Gamma}_{S^2}\ =\ 
\begin{tikzpicture}[baseline=(current bounding box.center), scale=0.45]
\node [draw, Rhodamine, circle, minimum size=3pt, fill=Orange, inner sep=0pt, outer sep=0pt] () at (5,5) {};
\node [draw, Rhodamine, circle, minimum size=3pt, fill=Orange, inner sep=0pt, outer sep=0pt] () at (7,5) {};
\draw (5,5) .. controls (4,7) and (3,7) .. (3,5) [Orange, very thick];
\draw (5,5) .. controls (4,3) and (3,3) .. (3,5) [Orange, very thick];
\draw (5,5) .. controls (6,7) and (7,7) .. (7,5) [Orange, very thick];
\draw (5,5) .. controls (6,3) and (7,3) .. (7,5) [Orange, very thick];
\draw [very thick, color=Orange] (7,5) -- (8,5);
\node at (5.7,5) {$v_*$};
\node at (8.5,5) {$l_1$};
\end{tikzpicture}\ .
\end{equation}

For a given mode $X$, an \emph{$X$-mode component} (or for simplicity, \emph{$X$ component}) is a subgraph of $\Gamma_X$ that becomes a one‑vertex irreducible (1VI) component of $\widetilde{\Gamma}_X$. We use $\gamma_X$ to denote general mode components. For example, figure~\ref{mode_subgraphs_example2} contains one $S$-mode component and two $S^2$-mode components, as a direct conclusion from~(\ref{eq:contracted_mode_subgraph_illustration}).

Let us understand this key concept from the following example:
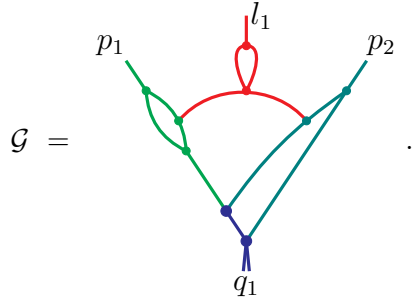
\begin{equation}
\mathcal{G}\ =\ 
\begin{tikzpicture}[baseline=(current bounding box.center), scale=0.4]
\draw [very thick, color=Green] (4.33,3) -- (3,5);
\draw [very thick, color=Green, bend left = 30] (3,5) to (1.67,7);
\draw [very thick, color=Green, bend right = 30] (3,5) to (1.67,7);
\draw [very thick, color=Green] (1.67,7) -- (1,8);
\draw [very thick, color=teal] (5,2) -- (8.33,7);
\draw [very thick, color=teal] (8.33,7) -- (9,8);
\draw [very thick, color=teal, bend left = 10] (4.33,3) to (8.33,7);
\draw [very thick, color=Blue] (5,2) -- (4.9,1);
\draw [very thick, color=Blue] (5,2) -- (5.1,1);
\draw [very thick, color=Blue] (5,2) -- (4.33,3);
\draw [very thick, color=Red, bend left = 50] (2.75,6) to (7,6);
\draw [very thick, color=Red, bend left = 60, in = 90, out = 30] (5,7) to (5,8.5);
\draw [very thick, color=Red, bend right = 60, in = -90, out = -30] (5,7) to (5,8.5);
\draw [very thick, color=Red] (5,8.5) to (5,9.5);
\node [draw, Blue, circle, minimum size=4pt, fill=Blue, inner sep=0pt, outer sep=0pt] () at (5,2) {};
\node [draw, Blue, circle, minimum size=4pt, fill=Blue, inner sep=0pt, outer sep=0pt] () at (4.33,3) {};
\node [draw, Green, circle, minimum size=3pt, fill=Green, inner sep=0pt, outer sep=0pt] () at (3,5) {};
\node [draw, Green, circle, minimum size=3pt, fill=Green, inner sep=0pt, outer sep=0pt] () at (1.67,7) {};
\node [draw, Green, circle, minimum size=3pt, fill=Green, inner sep=0pt, outer sep=0pt] () at (2.75,6) {};
\node [draw, teal, circle, minimum size=3pt, fill=teal, inner sep=0pt, outer sep=0pt] () at (8.33,7) {};
\node [draw, teal, circle, minimum size=3pt, fill=teal, inner sep=0pt, outer sep=0pt] () at (7,6) {};
\node [draw, Red, circle, minimum size=3pt, fill=Red, inner sep=0pt, outer sep=0pt] () at (5,7) {};
\node [draw, Red, circle, minimum size=3pt, fill=Red, inner sep=0pt, outer sep=0pt] () at (5,8.5) {};
\node at (0.5,8.5) {\large $p_1$};
\node at (9.5,8.5) {\large $p_2$};
\node at (5,0.5) {\large $q_1$};
\node at (5.5,9.5) {\large $l_1$};
\end{tikzpicture}.\nonumber
\end{equation}
In this graph, the mode subgraphs are marked in distinct colors. The $H$ subgraph is in {\color{Blue}\bf blue}, the $C_1$ subgraph is in {\color{Green}\bf green}, the $C_2$ subgraph is in {\color{teal}\bf teal}, and the $S$ subgraph is in {\color{Red}\bf red}. The mode components are as follows.
\begin{equation}
    H\textup{ component }(\gamma_H^{}):\quad\qquad\qquad\qquad\qquad
    \begin{tikzpicture}[baseline=(current bounding box.center), scale=0.5]
    \draw [very thick, color=Blue] (5,2) -- (4.9,1);
    \draw [very thick, color=Blue] (5,2) -- (5.1,1);
    \draw [very thick, color=Blue] (5,2) -- (4.33,3);
    \node [draw, Blue, circle, minimum size=4pt, fill=Blue, inner sep=0pt, outer sep=0pt] () at (5,2) {};
    \node [draw, Blue, circle, minimum size=4pt, fill=Blue, inner sep=0pt, outer sep=0pt] () at (4.33,3) {};
    \node at (5,0.5) {$q_1$};
    \end{tikzpicture}\quad.
\end{equation}
\begin{equation}
    C_1\textup{ component }(\gamma_{C_1}^{}):\qquad
    \begin{tikzpicture}[baseline=(current bounding box.center), scale=0.5]
    \draw [very thick, color=Green, bend left = 30] (3,5) to (1.67,7);
    \draw [very thick, color=Green, bend right = 30] (3,5) to (1.67,7);
    \draw [very thick, color=Green] (1.67,7) -- (1,8);
    \node [draw, Green, circle, minimum size=3pt, fill=Green, inner sep=0pt, outer sep=0pt] () at (3,5) {};
    \node [draw, Green, circle, minimum size=3pt, fill=Green, inner sep=0pt, outer sep=0pt] () at (1.67,7) {};
    \node [draw, Green, circle, minimum size=3pt, fill=Green, inner sep=0pt, outer sep=0pt] () at (2.75,6) {};
    \node at (0.5,8.5) {$p_1$};
    \end{tikzpicture},
    \qquad
    \begin{tikzpicture}[baseline=(current bounding box.center), scale=0.5]
    \node [draw, Green, circle, minimum size=3pt, fill=Green, inner sep=0pt, outer sep=0pt] () at (3,5) {};
    \draw [very thick, color=Green] (4.33,3) -- (3,5);
    \end{tikzpicture}\quad.
\end{equation}
\begin{equation}
    C_2\textup{ component }(\gamma_{C_2}^{}):\qquad\qquad\ \ 
    \begin{tikzpicture}[baseline=(current bounding box.center), scale=0.5]
    \draw [very thick, color=teal] (6.33,4) -- (8.33,7);
    \draw [very thick, color=teal] (8.33,7) -- (9,8);
    \draw [very thick, color=teal, bend left = 10] (5.5,4.5) to (8.33,7);
    \node [draw, teal, circle, minimum size=3pt, fill=teal, inner sep=0pt, outer sep=0pt] () at (8.33,7) {};
    \node [draw, teal, circle, minimum size=3pt, fill=teal, inner sep=0pt, outer sep=0pt] () at (7,6.1) {};
    \node at (9.5,8.5) {$p_2$};
\end{tikzpicture}\quad.
\end{equation}
\begin{equation}
    S\textup{ component }(\gamma_S^{}):\quad
    \begin{tikzpicture}[baseline=(current bounding box.center), scale=0.5]
    \draw [very thick, color=Red, bend left = 60, in = 90, out = 30] (5,7) to (5,8.5);
    \draw [very thick, color=Red, bend right = 60, in = -90, out = -30] (5,7) to (5,8.5);
    \draw [very thick, color=Red] (5,8.5) to (5,9.5);
    \node [draw, Red, circle, minimum size=3pt, fill=Red, inner sep=0pt, outer sep=0pt] () at (5,7) {};
    \node [draw, Red, circle, minimum size=3pt, fill=Red, inner sep=0pt, outer sep=0pt] () at (5,8.5) {};
    \node at (5.5,9.5) {$l_1$};
    \end{tikzpicture},
    \qquad
    \begin{tikzpicture}[baseline=(current bounding box.center), scale=0.5]
    \node [draw, Red, circle, minimum size=3pt, fill=Red, inner sep=0pt, outer sep=0pt] () at (5,6.9) {};
    \draw [very thick, color=Red, bend left = 50] (3,6) to (7,6);
    \end{tikzpicture}\quad.
\end{equation}
One can verify that each of the mode components above does correspond to a 1VI component of the contracted mode subgraph.

Note that each individual mode component must be connected (a property we will use in section~\ref{section-proof_subgraph_requirements}), because distinct connected components of $\Gamma_X$ do not lie in the same 1VI component of $\widetilde{\Gamma}_X$.

To end this section, we note a theorem that relates contracted mode subgraphs to minimum spanning trees of $\mathcal{G}$, as proved in ref.~\cite{Ma23}.
\begin{theorem}
\label{theorem-weight_hierarchical_partition_tree_structure}
Let $\mathcal{G}$ be a weighted graph such that for every $n\in \mathbb{N}$ the graph $\bigcup_{w(X)\geqslant -n} \Gamma_X$ (where $w(X)=-\mathscr{V}(X)$) is connected. Then, for any $T^1$ being a spanning tree of $\mathcal{G}$,
\begin{eqnarray}
\widetilde{\gamma}_{X}\cap T^1\text{ is a spanning tree of }\widetilde{\gamma}_{X}\ (\forall\text{ mode } X) \ \ \Leftrightarrow \ \ T^1\text{ is a minimum spanning tree of }\mathcal{G}. \nonumber
\end{eqnarray}
\end{theorem}
This theorem will be used as follows. In section~\ref{section-connectivity_requirement}, we will see that the condition of the theorem, $\bigcup_{w(X)\geqslant -n} \Gamma_X$ for any $n\in \mathbb{N}$, is indeed satisfied as a consequence of the ``First Connectivity Requirement''. Then, because every leading $\mathcal{U}$ term corresponds to a minimum spanning tree of $\mathcal{G}$, it follows immediately that $\widetilde{\gamma}_{X}\cap T^1$ is a spanning tree of $\widetilde{\gamma}_{X}$ for each $X$. This gives a graph‑theoretical perspective for characterizing leading $\mathcal{U}$ terms and paves the way for a similar characterization of leading $\mathcal{F}$ terms, which will be detailed in section~\ref{section-leading_terms}.

\section{The fundamental pattern}
\label{section-fundamental_facet_region_structure}

This section introduces the fundamental pattern that governs facet regions at all orders in the virtuality expansion of wide-angle kinematics, eq.~\eqref{eq:virtuality_expansion}. The pattern rests on two natural assumptions, which will be stated and explained.

\begin{figure}[t]
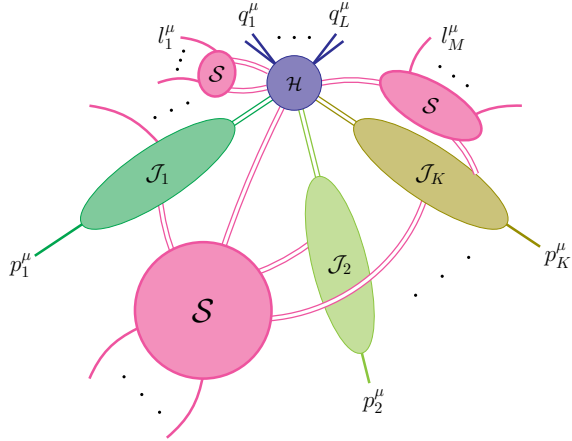

\centering
\include{figs/virtuality_expansion_fundamental_pattern}
\vspace{-3em}\caption{The fundamental pattern for facet regions in the virtuality expansion of massless wide-angle scattering. The graph comprises a connected hard subgraph $\mathcal{H}$, jet subgraphs $\mathcal{J}_1,\dots,\mathcal{J}_K$, and a soft subgraph $\mathcal{S}$. Off-shell external momenta $q_1,\dots,q_L$ attach exclusively to $\mathcal{H}$. Each on-shell external momentum $p_i$ attaches to its jet $\mathcal{J}_i$ (if nonempty) or directly to $\mathcal{H}$ (otherwise). Doubled lines represent any number of edges.}
\label{figure-virtuality_expansion_fundamental_pattern}
\end{figure}

In any such expansion, the relevant regions include a \emph{hard region} where all loop momenta are hard, and several \emph{infrared regions} where some loop momenta are collinear, soft, or in other infrared modes (as defined in section~\ref{section-momentum_modes}). All these regions conform to the fundamental pattern shown in figure~\ref{figure-virtuality_expansion_fundamental_pattern}.

In this pattern, the hard scattering of external momenta $\{p_i\}_{i=1}^K$ and $\{q_j\}_{j=1}^L$ occurs within a single connected hard subgraph $\mathcal{H}$. Collinear line momenta (those parallel to a given $p_i$) flow through the corresponding jet subgraph $\mathcal{J}_i$. In other words, large ($\sim \mathcal{O}(1)$) momentum components traverse $\mathcal{H}$ and the jets, while small momenta (vanishing in the infrared limit) are exchanged via the soft subgraph $\mathcal{S}$. The off-shell legs $\{q_j\}$ attach only to $\mathcal{H}$; on-shell legs $\{p_i\}$ attach to $\mathcal{J}_i$ (if $\mathcal{J}_i\neq \varnothing$) or $\mathcal{H}$ directly (otherwise); soft external momenta $\{l_k\}_{k=1}^M$ may attach to $\mathcal{H}$, the jets $\mathcal{J}\equiv \bigcup_{i=1}^K \mathcal{J}_i$, and $\mathcal{S}$.
We emphasize the notational distinction: calligraphic letters ($\mathcal{H}$, $\mathcal{J}_i$, $\mathcal{S}$) denote subgraphs, while roman letters ($H$, $C_i$, $S$, \dots) denote momentum modes.

At this stage, we provisionally allow hard loops within $  \mathcal{J} \cup \mathcal{S}  $, although they will later be excluded by the ``First Connectivity Theorem'' (theorem~\ref{theorem-mode_subgraphs_connectivity1}). The subgraphs $\mathcal{H}$, $\mathcal{J}$, and $\mathcal{S}$ are thus related to mode components as follows. $\mathcal{H}$ consists solely of $H$ components; each $\mathcal{J}_i$ is the union of $C_i^{n_i}$ components ($n_i\in\mathbb{N}_+$) and possibly some $H$ components (hard loops in $\mathcal{J}_i$); $S$ is the union of $S^mC_k^{n}$ components ($m\in \mathbb{N}_+$, $n\in \mathbb{N}$, $k$ unrestricted) and possibly $H$ components.

From the description above, the fundamental pattern relies on the following \emph{assumptions}.\footnote{Besides these assumptions, one may wonder why it is legitimate to consider only $m,n\in\mathbb{N}$ in $S^mC_i^n$; for example, why the ``semihard mode'' $(\lambda^{1/2},\lambda^{1/2},\lambda^{1/2})$ is not relevant. To see this, first note that one can redefine $\lambda$ to modify its power from a rational number to an integer; for example, the semihard mode can be included in the fundamental pattern with expansion parameter $\lambda'\equiv \lambda^{1/2}$. (As we will explain later in section~\ref{section-implication_infrared_compatibility_requirement}, those cases are actually ruled out due to the ``infrared-compatibility requirement''.) Therefore, one can safely take $m,n\in\mathbb{N}$ in the region structure subject to the external kinematics~(\ref{eq:virtuality_expansion}). Another natural question is why ultraviolet-mode momenta (whose components go to infinity as $\lambda\to 0$) cannot contribute. The reason is that an ultraviolet-mode line momentum $k_e^\mu$ would imply a positive weight $w_e$, which is forbidden for Feynman integrals with only propagators that are quadratic in momenta, as proved in lemma~2.3 of ref.~\cite{Ma23}. For Feynman integrals where momenta can be linear in some propagators (e.g., Wilson lines), ultraviolet modes can appear~\cite{Gardi:2025ule}, but that is beyond the scope of this paper.}
\begin{enumerate}
    \item Infrared regions all correspond to singularities of the Feynman integrand, i.e., solutions of the Landau equations, whose configurations can be described by figure~\ref{figure-virtuality_expansion_fundamental_pattern}.
    \item Configurations where the hard scattering occurs at multiple positions (Landshoff scattering) are not relevant to facet regions.
\end{enumerate}
These assumptions are empirically well supported by the EbR literature~\cite{Smn90,Smn94,Smirnov:2002pj} and proofs for some special cases~\cite{Ma23,GrdHzgJnsMa24}, though a general rigorous proof remains elusive. In the remainder of this section, we explain why both are natural from physical and mathematical perspectives.

\bigbreak
\noindent\emph{Infrared regions and Landau equations.}

Let us demonstrate why the configuration in figure~\ref{figure-virtuality_expansion_fundamental_pattern} is natural for infrared regions from the perspective of the Landau equations. If the Feynman integrand remains entirely regular as $\lambda\to 0$ in (\ref{eq:virtuality_expansion}), expanding in the hard region alone would suffice to reproduce the full asymptotic series. However, when the integrand possesses singularities that cannot be avoided by contour deformation, contributions from neighborhoods of those singularities become essential, meaning additional infrared regions must be included.

It is known that all such singularities are captured by the Landau equations~\cite{Lnd59,EdenLdshfOlvPkhn02book,Cls20}. The physical solutions of the Landau equations (those with $\alpha_e\in [0,1]$) admit a classical picture known as the \emph{Coleman--Norton interpretation}~\cite{ClmNtn65}. For wide‑angle scattering, this classical picture is precisely the structure shown in figure~\ref{figure-virtuality_expansion_fundamental_pattern}, provided the hard subgraph is connected---a point we will justify later. Therefore, figure~\ref{figure-virtuality_expansion_fundamental_pattern} is a natural pattern for infrared regions.

\begin{figure}[t]
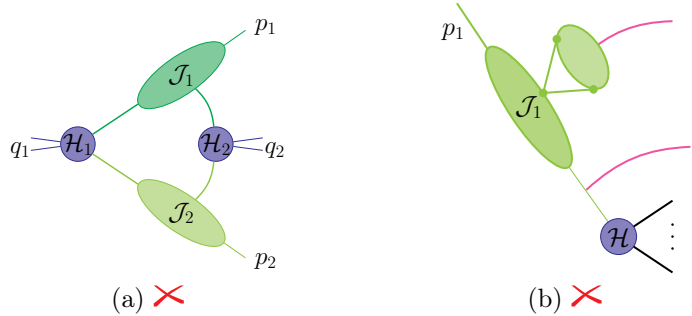

\centering
\begin{subfigure}[b]{0.25\textwidth}
\centering
\include{figs/configurations_violating_Landau_equations_I}
\vspace{-3em}\caption{$\crossmark[Red]$}
\label{configurations_violating_Landau_equations_I}
\end{subfigure}
\qquad\qquad
\begin{subfigure}[b]{0.25\textwidth}
\centering
\include{figs/configurations_violating_Landau_equations_II}
\vspace{-3em}\caption{$\crossmark[Red]$}
\label{configurations_violating_Landau_equations_II}
\end{subfigure}
\caption{Configurations that cannot be solutions of the Landau equations because they do not conform with the Coleman--Norton interpretation. (a) short-distance propagators (those within $\mathcal{H}_2$) connecting vertices from two jets after scattering at $\mathcal{H}_1$ in advance. (b) the existence of a ``vacuum bubble'' within the jet $\mathcal{J}_1$.}
\label{figure-configurations_violating_Landau_equations}
\end{figure}

Note that the Coleman--Norton interpretation, as a consequence of the Landau equations, already imposes constraints on the region structure. For example, once two jets depart from a hard subgraph, the vertices belonging to distinct jets are widely separated, so no short‑distance propagators can connect them. Consequently, the structure in figure~\ref{configurations_violating_Landau_equations_I} cannot correspond to a solution of the Landau equations and thus does not represent a region. As another example, configurations containing a ``vacuum bubble'' within a jet (possibly connected by soft edges) also violate the Coleman--Norton interpretation. Therefore, figure~\ref{configurations_violating_Landau_equations_II} cannot be part of any region. These pathological configurations have been implicitly excluded in the fundamental pattern.

Another implication from the Landau equations is the absence of the Glauber mode (i.e., $(\lambda^a,\lambda^b,\lambda^c)$ with $a+b>2c$) in $\mathcal{S}$. This is because in the wide‑angle kinematics, one can always deform the integration contour to avoid pinches in the Glauber scaling~\cite{ClsStm81}.

\bigbreak
\noindent\emph{Absence of Landshoff-scattering configurations.}

\begin{figure}[t]
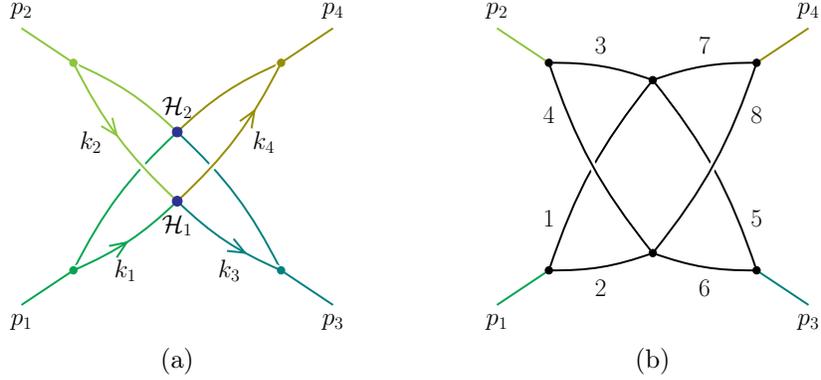

\centering
\begin{subfigure}[b]{0.3\textwidth}
\centering
\include{figs/2to2_Landshoff_scattering_configuration}
\vspace{-3em}\caption{}
\label{2to2_Landshoff_scattering_configuration}
\end{subfigure}
\qquad\qquad
\begin{subfigure}[b]{0.3\textwidth}
\centering
\include{figs/2to2_Landshoff_scattering_parameterization}
\vspace{-3em}\caption{}
\label{2to2_Landshoff_scattering_parameterization}
\end{subfigure}
\caption{A three-loop $2\to 2$ Landshoff scattering graph in the wide‑angle kinematics, where the two hard‑scattering components are $\mathcal{H}_1$ and $\mathcal{H}_2$. (a) the momentum configuration, with different types of green representing different jets. (b) the parameterization of the edges for (\ref{eq:additional_constraint_parameter_3loop_Landshoff}).}
\label{figure-2to2_Landshoff_scattering_example}
\end{figure}

The Landau equations do not force the hard subgraph $\mathcal{H}$ to be connected in general. Actually, hard scattering can occur at multiple disconnected locations, a pattern known as \emph{Landshoff scattering}~\cite{Ldsf74}. A three-loop example of two-to-two Landshoff scattering is shown in figure~\ref{2to2_Landshoff_scattering_configuration}. Here, the incoming momenta $p_1$ and $p_2$ split, scatter separately at the hard subgraphs $\mathcal{H}_1$ and $\mathcal{H}_2$, and then merge into the outgoing momenta $p_3$ and $p_4$. On top of this graph, one can also add soft momenta mediating between the $\mathcal{H}$ and $\mathcal{J}$ subgraphs.
At higher loops, patterns with three or more connected components of $\mathcal{H}$ can also arise.

We emphasize that Landshoff scattering \emph{does} correspond to singularities of the Feynman integrand and can yield valid regions in the asymptotic expansion. For example, figure~\ref{2to2_Landshoff_scattering_configuration} has been verified in ref.~\cite{GrdHzgJnsMa24} as a hidden regions in the ``on-shell expansion'' (see (\ref{eq:wideangle_onshell_kinematics}) for definition). Our statement here, by contrast, is that these configurations are \emph{never} facet regions.

As the first evidence, this has been proved to all orders for the on-shell expansion in ref.~\cite{Ma23}. In that work, no fundamental-pattern-like pictures are assumed. The hard subgraph $\mathcal{H}$, defined upon purely graph-theoretical concepts, is proved to be connected by showing that there exists a special vertex $v_0\in \mathcal{H}$, such that any other vertex of $\mathcal{H}$ can be reached from $v_0$ through a path in $\mathcal{H}$ (lemma 3.5 there).

For more general expansions, a rigorous proof is lacking, but there is an intuitive way to see this. When the hard scattering occurs at a single $\mathcal{H}$, the $\mathcal{O}(1)$ components of line momenta from distinct jets are independent of each other. In contrast, in the Landshoff-scattering pattern, those components are correlated, imposing extra constraints on the scaling of the Lee–Pomeransky parameters. As a result, the Lee-Pomeransky-parameter description of facet regions, eq.~(\ref{eq:facet_region_general_parameter_scaling}), is not sufficient here.

For example, in figure~\ref{2to2_Landshoff_scattering_configuration}, the $\mathcal{O}(1)$ components of all the $k_i$, namely $k_i\cdot\overline{\beta}_i$, must approach the same value in the infrared limit~\cite{BottsStm89}. Assuming $p_i^2\sim\lambda$, we have
\begin{align}
    \frac{k_i\cdot\overline{\beta}_i}{p_i\cdot\overline{\beta}_i}-\frac{k_j\cdot\overline{\beta}_j}{p_j\cdot\overline{\beta}_j}\ \sim\ \sqrt{\lambda}\ \ \textup{for any }i\neq j.
\end{align}
In the parametric representation, this additional constraint becomes~\cite{GrdHzgJnsMa24}
\begin{align}
\label{eq:additional_constraint_parameter_3loop_Landshoff}
    \frac{x_1}{x_2}\!-\!\frac{x_3}{x_4}\ \sim\ \frac{x_3}{x_4}\!-\!\frac{x_5}{x_6}\ \sim\ \frac{x_5}{x_6}\!-\!\frac{x_7}{x_8}\ \sim\ \sqrt{\lambda},
\end{align}
with the parameterization of edges shown in figure~\ref{2to2_Landshoff_scattering_parameterization}. Note that each $x_i$ is $\mathcal{O}(\lambda^{-1})$, so each individual ratio in (\ref{eq:additional_constraint_parameter_3loop_Landshoff}) is $\mathcal{O}(1)$, while their differences scale as $\sqrt{\lambda}$. This explains why (\ref{eq:facet_region_general_parameter_scaling}) alone is insufficient to characterize the region.

Based on the above, we propose that Landshoff scatterings cannot be characterized by the facet-region scaling behavior (\ref{eq:facet_region_general_parameter_scaling}) in general. Therefore, each facet region must feature a single connected hard subgraph, as in figure~\ref{figure-virtuality_expansion_fundamental_pattern}.

\section{Further requirements of the mode subgraphs}
\label{section-further_requirements_mode_subgraphs}

The fundamental pattern (figure~\ref{figure-virtuality_expansion_fundamental_pattern}) alone is far from sufficient to determine the complete list of regions. Typically, a given graph admits infinitely many configurations consistent with the fundamental pattern, because loops in $\mathcal{S}$ have modes of the form $S^mC_i^n$, and we can make $m$ arbitrarily large without violating momentum conservation. Even when all relevant modes in a graph are specified in advance, the number of configurations may still vastly exceed the number of actual regions, as many configurations are actually scaleless and therefore do not contribute to the asymptotic expansion. Hence, a crucial task is to derive a set of graph‑theoretic rules that impose further constraints on the mode subgraphs, thereby providing a necessary and sufficient condition for a configuration to be a valid region.

As we will see, these constraints fall into two categories: \emph{connectivity} and \emph{infrared compatibility}---concepts that will be explained in detail. In section~\ref{section-connectivity_requirement}, we present the connectivity requirements, which comprise the First and Second Connectivity Theorems. In section~\ref{section-infrared_compatibility_requirements_overall}, we describe the infrared‑compatibility requirement. These requirements impose strong constraints not only on the connection of mode subgraphs, but also on the momentum modes allowed for a given external kinematics, as we demonstrate in section~\ref{section-implication_infrared_compatibility_requirement}. In particular, we point out that the virtuality expansion \emph{never features ``cascading modes''}. Finally, in section~\ref{section-summary_matching_results_known_expansions} we summarize the subgraph requirements and illustrate how they match the known results for some special cases of virtuality expansion.

\subsection{The connectivity requirements}
\label{section-connectivity_requirement}

As the name suggests, the connectivity requirements govern how the mode subgraphs in the fundamental pattern (figure~\ref{figure-virtuality_expansion_fundamental_pattern}) are interconnected.

\begin{figure}[t]
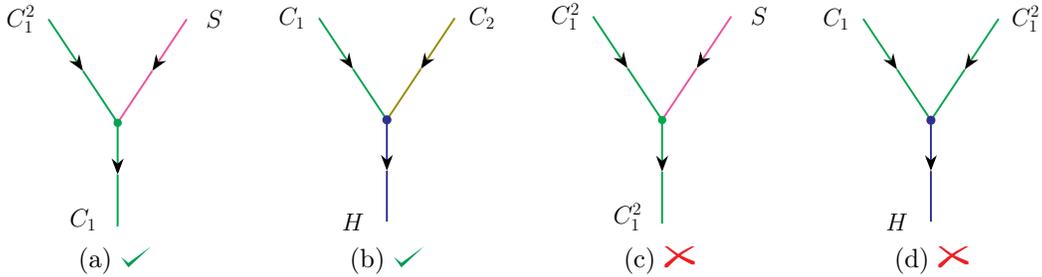

\centering
\begin{subfigure}[b]{0.2\textwidth}
\centering
\include{figs/momentum_conservation_allowed_config1}
\vspace{-3em}\caption{$\greencheckmark[ForestGreen]$}
\label{momentum_conservation_allowed_config1}
\end{subfigure}
\quad
\begin{subfigure}[b]{0.2\textwidth}
\centering
\include{figs/momentum_conservation_allowed_config2}
\vspace{-3em}\caption{$\greencheckmark[ForestGreen]$}
\label{momentum_conservation_allowed_config2}
\end{subfigure}
\quad
\begin{subfigure}[b]{0.2\textwidth}
\centering
\include{figs/momentum_conservation_violated_config1}
\vspace{-3em}\caption{$\crossmark[Red]$}
\label{momentum_conservation_violated_config1}
\end{subfigure}
\quad
\begin{subfigure}[b]{0.2\textwidth}
\centering
\include{figs/momentum_conservation_violated_config2}
\vspace{-3em}\caption{$\crossmark[Red]$}
\label{momentum_conservation_violated_config2}
\end{subfigure}
\caption{Some examples where the connection of mode subgraphs satisfies momentum conservation (in (a) and (b), marked with $\greencheckmark[ForestGreen]$) or violates momentum conservation (in (c) and (d), marked with $\crossmark[Red]$).}
\label{figure-momentum_conservation_configurations}
\end{figure}

To begin, we note that momentum conservation already imposes implicit constraints on the connectivity of mode subgraphs. At any vertex~$v$, we partition the incident line momenta into two nonempty sets $\{k_1,\dots,k_a\}$ (incoming) and $\{k_{a+1},\dots,k_b\}$ (outgoing). Their modes then satisfy
$$\mathscr{X}(k_1)\vee\dots\vee\mathscr{X}(k_a) = \mathscr{X}(k_{a+1})\vee\dots\vee\mathscr{X}(k_b).$$
Consequently, the configurations in figures~\ref{momentum_conservation_allowed_config1} and~\ref{momentum_conservation_allowed_config2} are allowed, whereas those in figures~\ref{momentum_conservation_violated_config1} and~\ref{momentum_conservation_violated_config2} are forbidden and thus cannot appear in any region. We have incorporated these necessary constraints implicitly into the fundamental pattern rather than stating them as an independent requirement.

We now investigate a more subtle connectivity property. We observe that the union of all $X$-mode subgraphs with $\mathscr{V}(X)$ below any specific value is connected. Likewise, the union of all $X$-mode subgraphs with $X$ harder than any specific mode is connected. As an explicit example, consider the following region of the graph $\mathcal{G}$:
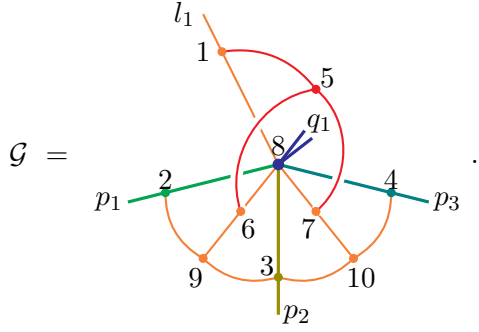
\begin{equation}
\mathcal{G}\ =\ 
\begin{tikzpicture}[baseline=(current bounding box.center), scale=0.5]
\draw [thick, color=Orange] (5,5) -- (3,9);
\draw [very thick, color=Green] (5,5) -- (1,4);
\draw [very thick, color=olive] (5,5) -- (5,1);
\draw [very thick, color=teal] (5,5) -- (9,4);
\draw [very thick, color=Blue] (5,5) -- (5.7,5.9);
\draw [very thick, color=Blue] (5,5) -- (5.9,5.7);
\path (2,4.25) edge [thick, Orange, bend right=30] (3,2.5) {};
\path (3,2.5) edge [thick, Orange, bend right=30] (5,2) {};
\path (5,2) edge [thick, Orange, bend right=30] (7,2.5) {};
\path (7,2.5) edge [thick, Orange, bend right=30] (8,4.25) {};
\path (3,2.5) edge [thick, Orange] (4,3.75) {};
\path (5,5) edge [thick, Orange] (4,3.75) {};
\path (7,2.5) edge [thick, Orange] (6,3.75) {};
\path (5,5) edge [thick, Orange] (6,3.75) {};

\draw (6,7) edge [thick, draw=white, double=white, double distance=3pt, bend right=50] (4,3.75) node [] {};\path (6,7) edge [thick, Red, bend right=50] (4,3.75) {};
\draw (6,7) edge [thick, draw=white, double=white, double distance=3pt, bend left=50] (6,3.75) node [] {};\path (6,7) edge [thick, Red, bend left=50] (6,3.75) {};
\path (6,7) edge [thick, Red, bend right=30] (3.5,8) {};

\node [draw, Blue, circle, minimum size=4pt, fill=Blue, inner sep=0pt, outer sep=0pt] () at (5,5) {};
\node [draw, Orange, circle, minimum size=3pt, fill=Orange, inner sep=0pt, outer sep=0pt] () at (3,2.5) {};
\node [draw, Orange, circle, minimum size=3pt, fill=Orange, inner sep=0pt, outer sep=0pt] () at (7,2.5) {};
\node [draw, Orange, circle, minimum size=3pt, fill=Orange, inner sep=0pt, outer sep=0pt] () at (4,3.75) {};
\node [draw, Orange, circle, minimum size=3pt, fill=Orange, inner sep=0pt, outer sep=0pt] () at (6,3.75) {};
\node [draw, Orange, circle, minimum size=3pt, fill=Orange, inner sep=0pt, outer sep=0pt] () at (3.5,8) {};
\node [draw, Green, circle, minimum size=3pt, fill=Green, inner sep=0pt, outer sep=0pt] () at (2,4.25) {};
\node [draw, olive, circle, minimum size=3pt, fill=olive, inner sep=0pt, outer sep=0pt] () at (5,2) {};
\node [draw, teal, circle, minimum size=3pt, fill=teal, inner sep=0pt, outer sep=0pt] () at (8,4.25) {};
\node [draw, Red, circle, minimum size=3pt, fill=Red, inner sep=0pt, outer sep=0pt] () at (6,7) {};

\node at (3,8) {$1$};
\node at (2,4.6) {$2$};
\node at (4.7,2.3) {$3$};
\node at (8,4.6) {$4$};
\node at (6.3,7.3) {$5$};
\node at (4.2,3.3) {$6$};
\node at (5.8,3.3) {$7$};
\node at (5,5.5) {$8$};
\node at (2.8,2) {$9$};
\node at (7.2,2) {$10$};
\node at (2.5,9) {$l_1$};
\node at (0.5,4) {$p_1$};
\node at (5.5,1) {$p_2$};
\node at (9.5,4) {$p_3$};
\node at (6.1,6.1) {$q_1$};
\end{tikzpicture}.\nonumber
\end{equation}
Here, the external kinematics consist of a $C_1^2$-mode momentum $p_1$, a $C_2$-mode momentum $p_2$, a $C_3^2$-mode momentum $p_3$, an off-shell (hard-mode) momentum $q_1$, and an $SC_4$-mode momentum (soft emission) $l_1$. The vertices are labeled $1$–$10$. The mode subgraphs are colored as follows: $H$ mode ({\color{Blue}\bf blue}), $C_1^2$ mode ({\color{Green}\bf green}), $C_2$ mode ({\color{olive}\bf olive}), $C_3^2$ mode ({\color{teal}\bf teal}), $SC_i$ modes with $i=1,3,4$ ({\color{Orange}\bf orange}), and $S^2$ mode ({\color{Red}\bf red}).
In other words,
\begin{eqnarray}
    && \Gamma_H=(\{v_8\},\varnothing),\quad \Gamma_{C_1^2}=(\{v_2\}, \{e_{28}\}),\quad \Gamma_{C_2}=(\{v_3\}, \{e_{38}\}),\quad \Gamma_{C_3^2}=(\{v_4\}, \{e_{48}\}), \nonumber\\
    && \Gamma_{SC_1}=(\{v_6,v_9\},\{e_{29},e_{69},e_{39},e_{68}\}),\quad \Gamma_{SC_3}=(\{v_7,v_{10}\},\{e_{4;10},e_{7;10},e_{3;10},e_{78}\}), \nonumber\\
    &&\Gamma_{SC_4}=(\{v_1\},\{e_{18}\}),\quad \Gamma_{S^2}=  (\{v_5\}, \{e_{51},e_{56},e_{57}\}).\nonumber
\end{eqnarray}
The notation $e_{ij}$ above denotes the edge connecting $v_i$ and $v_j$. Let us now consider certain types of subgraphs of $\mathcal{G}$. The first type is $\bigcup_{\mathscr{V}(X)\leqslant n} \Gamma_X$:
\begin{eqnarray}
&&\bigcup_{\mathscr{V}(X)\leqslant 0} \Gamma_X =
\begin{tikzpicture}[scale=0.5]
\draw [very thick, color=Blue] (5,5) -- (5.7,5.9);
\draw [very thick, color=Blue] (5,5) -- (5.9,5.7);
\node [draw, Blue, circle, minimum size=4pt, fill=Blue, inner sep=0pt, outer sep=0pt] () at (5,5) {};
\end{tikzpicture},
\qquad
\bigcup_{\mathscr{V}(X)\leqslant 1} \Gamma_X =
\begin{tikzpicture}[baseline=(current bounding box.center), yshift=-2pt, scale=0.5]
\draw [very thick, color=olive] (5,5) -- (5,1);
\draw [very thick, color=Blue] (5,5) -- (5.7,5.9);
\draw [very thick, color=Blue] (5,5) -- (5.9,5.7);
\node [draw, Blue, circle, minimum size=4pt, fill=Blue, inner sep=0pt, outer sep=0pt] () at (5,5) {};
\node [draw, olive, circle, minimum size=3pt, fill=olive, inner sep=0pt, outer sep=0pt] () at (5,2) {};
\end{tikzpicture},
\qquad
\bigcup_{\mathscr{V}(X)\leqslant 2} \Gamma_X =
\begin{tikzpicture}[baseline=(current bounding box.center), yshift=-2pt, scale=0.5]
\draw [very thick, color=Green] (5,5) -- (1,4);
\draw [very thick, color=olive] (5,5) -- (5,1);
\draw [very thick, color=teal] (5,5) -- (9,4);
\draw [very thick, color=Blue] (5,5) -- (5.7,5.9);
\draw [very thick, color=Blue] (5,5) -- (5.9,5.7);
\node [draw, Blue, circle, minimum size=4pt, fill=Blue, inner sep=0pt, outer sep=0pt] () at (5,5) {};
\node [draw, Green, circle, minimum size=3pt, fill=Green, inner sep=0pt, outer sep=0pt] () at (2,4.25) {};
\node [draw, olive, circle, minimum size=3pt, fill=olive, inner sep=0pt, outer sep=0pt] () at (5,2) {};
\node [draw, teal, circle, minimum size=3pt, fill=teal, inner sep=0pt, outer sep=0pt] () at (8,4.25) {};
\end{tikzpicture},\nonumber\\
&&\bigcup_{\mathscr{V}(X)\leqslant 3} \Gamma_X =
\begin{tikzpicture}[baseline=(current bounding box.center), scale=0.5]
\draw [thick, color=Orange] (5,5) -- (3,9);
\draw [very thick, color=Green] (5,5) -- (1,4);
\draw [very thick, color=olive] (5,5) -- (5,1);
\draw [very thick, color=teal] (5,5) -- (9,4);
\draw [very thick, color=Blue] (5,5) -- (5.7,5.9);
\draw [very thick, color=Blue] (5,5) -- (5.9,5.7);
\path (2,4.25) edge [thick, Orange, bend right=30] (3,2.5) {};
\path (3,2.5) edge [thick, Orange, bend right=30] (5,2) {};
\path (5,2) edge [thick, Orange, bend right=30] (7,2.5) {};
\path (7,2.5) edge [thick, Orange, bend right=30] (8,4.25) {};
\path (3,2.5) edge [thick, Orange] (4,3.75) {};
\path (5,5) edge [thick, Orange] (4,3.75) {};
\path (7,2.5) edge [thick, Orange] (6,3.75) {};
\path (5,5) edge [thick, Orange] (6,3.75) {};
\node [draw, Blue, circle, minimum size=4pt, fill=Blue, inner sep=0pt, outer sep=0pt] () at (5,5) {};
\node [draw, Orange, circle, minimum size=3pt, fill=Orange, inner sep=0pt, outer sep=0pt] () at (3,2.5) {};
\node [draw, Orange, circle, minimum size=3pt, fill=Orange, inner sep=0pt, outer sep=0pt] () at (7,2.5) {};
\node [draw, Orange, circle, minimum size=3pt, fill=Orange, inner sep=0pt, outer sep=0pt] () at (4,3.75) {};
\node [draw, Orange, circle, minimum size=3pt, fill=Orange, inner sep=0pt, outer sep=0pt] () at (6,3.75) {};
\node [draw, Orange, circle, minimum size=3pt, fill=Orange, inner sep=0pt, outer sep=0pt] () at (3.5,8) {};
\node [draw, Green, circle, minimum size=3pt, fill=Green, inner sep=0pt, outer sep=0pt] () at (2,4.25) {};
\node [draw, olive, circle, minimum size=3pt, fill=olive, inner sep=0pt, outer sep=0pt] () at (5,2) {};
\node [draw, teal, circle, minimum size=3pt, fill=teal, inner sep=0pt, outer sep=0pt] () at (8,4.25) {};
\end{tikzpicture},
\qquad
\bigcup_{\mathscr{V}(X)\leqslant 4} \Gamma_X = \mathcal{G}.
\label{eq:connected_mode_subgraphs_type1}
\end{eqnarray}
The second type is $\bigcup_{X'\textup{ not softer than }X} \Gamma_{X'}$ (note that $X$ is also included in this union):
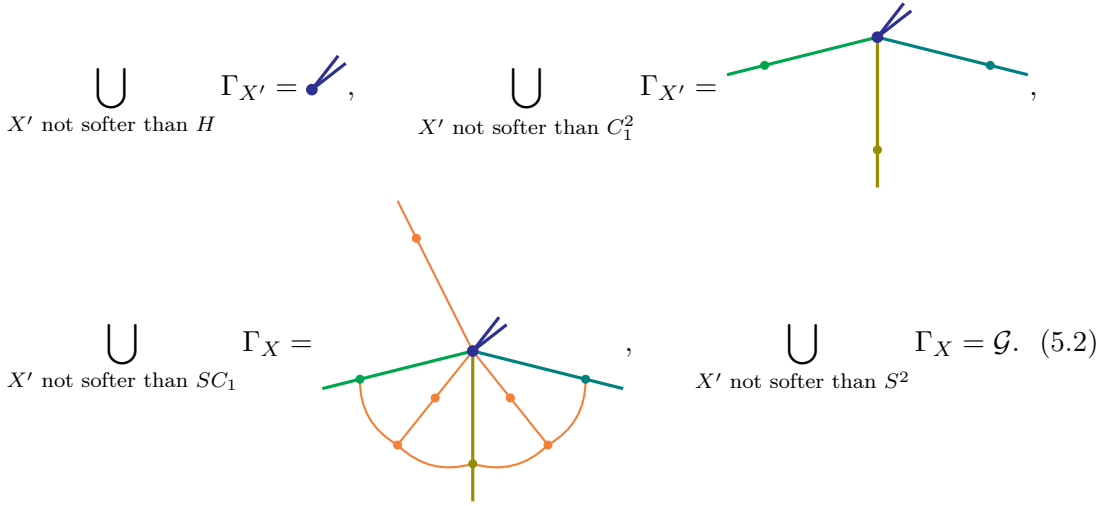
\begin{eqnarray}
&&\bigcup_{X'\textup{ not softer than }H} \Gamma_{X'} =
\begin{tikzpicture}[scale=0.5]
\draw [very thick, color=Blue] (5,5) -- (5.7,5.9);
\draw [very thick, color=Blue] (5,5) -- (5.9,5.7);
\node [draw, Blue, circle, minimum size=4pt, fill=Blue, inner sep=0pt, outer sep=0pt] () at (5,5) {};
\end{tikzpicture},
\qquad
\bigcup_{X'\textup{ not softer than }C_1^2} \Gamma_{X'} =
\begin{tikzpicture}[baseline=(current bounding box.center), yshift=-2pt, scale=0.5]
\draw [thick, color=Orange] (5,5) -- (3,9);
\draw [very thick, color=Green] (5,5) -- (1,4);
\draw [very thick, color=olive] (5,5) -- (5,1);
\draw [very thick, color=teal] (5,5) -- (9,4);
\draw [very thick, color=Blue] (5,5) -- (5.7,5.9);
\draw [very thick, color=Blue] (5,5) -- (5.9,5.7);
\path (5,2) edge [thick, Orange, bend right=30] (7,2.5) {};
\path (7,2.5) edge [thick, Orange, bend right=30] (8,4.25) {};
\path (7,2.5) edge [thick, Orange] (6,3.75) {};
\path (5,5) edge [thick, Orange] (6,3.75) {};
\node [draw, Blue, circle, minimum size=4pt, fill=Blue, inner sep=0pt, outer sep=0pt] () at (5,5) {};
\node [draw, olive, circle, minimum size=3pt, fill=olive, inner sep=0pt, outer sep=0pt] () at (5,2) {};
\node [draw, teal, circle, minimum size=3pt, fill=teal, inner sep=0pt, outer sep=0pt] () at (8,4.25) {};
\node [draw, Orange, circle, minimum size=3pt, fill=Orange, inner sep=0pt, outer sep=0pt] () at (7,2.5) {};
\node [draw, Orange, circle, minimum size=3pt, fill=Orange, inner sep=0pt, outer sep=0pt] () at (3.5,8) {};
\node [draw, Green, circle, minimum size=3pt, fill=Green, inner sep=0pt, outer sep=0pt] () at (2,4.25) {};
\node [draw, Orange, circle, minimum size=3pt, fill=Orange, inner sep=0pt, outer sep=0pt] () at (6,3.75) {};
\end{tikzpicture},\nonumber\\
&&\bigcup_{X'\textup{ not softer than }SC_1} \Gamma_X =
\begin{tikzpicture}[baseline=(current bounding box.center), scale=0.5]
\draw [thick, color=Orange] (5,5) -- (3,9);
\draw [very thick, color=Green] (5,5) -- (1,4);
\draw [very thick, color=olive] (5,5) -- (5,1);
\draw [very thick, color=teal] (5,5) -- (9,4);
\draw [very thick, color=Blue] (5,5) -- (5.7,5.9);
\draw [very thick, color=Blue] (5,5) -- (5.9,5.7);
\path (2,4.25) edge [thick, Orange, bend right=30] (3,2.5) {};
\path (3,2.5) edge [thick, Orange, bend right=30] (5,2) {};
\path (5,2) edge [thick, Orange, bend right=30] (7,2.5) {};
\path (7,2.5) edge [thick, Orange, bend right=30] (8,4.25) {};
\path (3,2.5) edge [thick, Orange] (4,3.75) {};
\path (5,5) edge [thick, Orange] (4,3.75) {};
\path (7,2.5) edge [thick, Orange] (6,3.75) {};
\path (5,5) edge [thick, Orange] (6,3.75) {};
\node [draw, Blue, circle, minimum size=4pt, fill=Blue, inner sep=0pt, outer sep=0pt] () at (5,5) {};
\node [draw, Orange, circle, minimum size=3pt, fill=Orange, inner sep=0pt, outer sep=0pt] () at (3,2.5) {};
\node [draw, Orange, circle, minimum size=3pt, fill=Orange, inner sep=0pt, outer sep=0pt] () at (7,2.5) {};
\node [draw, Orange, circle, minimum size=3pt, fill=Orange, inner sep=0pt, outer sep=0pt] () at (4,3.75) {};
\node [draw, Orange, circle, minimum size=3pt, fill=Orange, inner sep=0pt, outer sep=0pt] () at (6,3.75) {};
\node [draw, Orange, circle, minimum size=3pt, fill=Orange, inner sep=0pt, outer sep=0pt] () at (3.5,8) {};
\node [draw, Green, circle, minimum size=3pt, fill=Green, inner sep=0pt, outer sep=0pt] () at (2,4.25) {};
\node [draw, olive, circle, minimum size=3pt, fill=olive, inner sep=0pt, outer sep=0pt] () at (5,2) {};
\node [draw, teal, circle, minimum size=3pt, fill=teal, inner sep=0pt, outer sep=0pt] () at (8,4.25) {};
\end{tikzpicture},
\qquad
\bigcup_{X'\textup{ not softer than }S^2} \Gamma_X = \mathcal{G}.
\label{eq:connected_mode_subgraphs_type2}
\end{eqnarray}
The third type is $\bigcup_{X'\textup{ harder than }X} \Gamma_{X'}$. By considering all possible modes $X$, we obtain the same four graphs as above. It is straightforward to verify that all these graphs are connected. In fact, this connectedness is a general requirement for any region of any graph to yield a scaleful expanded integral. This property is formalized as the \emph{First Connectivity Theorem}.
\begin{theorem}[\textbf{First Connectivity Theorem}]
\label{theorem-mode_subgraphs_connectivity1}
    For any region, the following subgraphs of $\mathcal{G}$ must all be connected:
    \begin{enumerate}
        \item [1,] $\bigcup_{\mathscr{V}(X)\leqslant n} \Gamma_X$ for any $n\in \mathbb{N}$,
        \item [2,] $\bigcup_{X'\textup{ not softer than }X} \Gamma_{X'}$ for any mode $X$,
        \item [3,] $\bigcup_{X'\textup{ harder than }X} \Gamma_{X'}$ for any mode $X$.
    \end{enumerate}
\end{theorem}
The significance of this theorem is that once the connectedness of $\mathcal{H}$ is confirmed (as part of the assumptions of the fundamental region pattern in figure~\ref{figure-virtuality_expansion_fundamental_pattern}), all other graphs in statement \emph{1} with $n \in \mathbb{N}_+$ are automatically connected as well. With this said, configurations such as a hard loop inside a jet (figure~\ref{configurations_violating_first_connectivity_theorem1}), or a soft loop surrounded by softer ones (figure~\ref{configurations_violating_first_connectivity_theorem2}), etc., are all excluded from the regions by this theorem.
\begin{figure}[t]
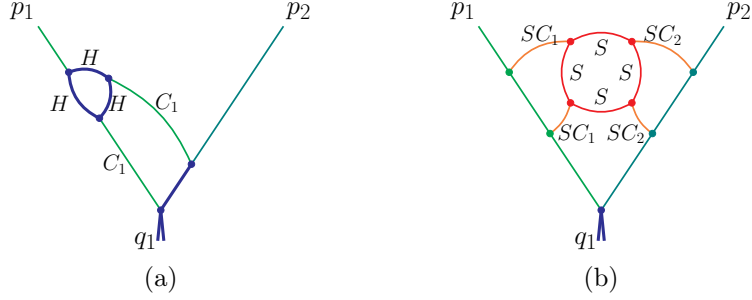

\centering
\begin{subfigure}[b]{0.27\textwidth}
\centering
\include{figs/configurations_violating_first_connectivity_theorem1}
\vspace{-3em}\caption{}
\label{configurations_violating_first_connectivity_theorem1}
\end{subfigure}
\qquad\qquad
\begin{subfigure}[b]{0.27\textwidth}
\centering
\include{figs/configurations_violating_first_connectivity_theorem2}
\vspace{-3em}\caption{}
\label{configurations_violating_first_connectivity_theorem2}
\end{subfigure}
\caption{Examples of configurations that are scaleless and thus not facet regions due to the First Connectivity Theorem. (a): a hard loop ({\color{Blue}\bf blue}) inside a jet. (b): an $S$-mode loop ({\color{Red}\bf red}) surrounded by softer edges.}
\label{figure-configurations_violating_first_connectivity_theorem}
\end{figure}

To prove theorem~\ref{theorem-mode_subgraphs_connectivity1}, a key observation is that the connectedness of the graphs in \emph{1} automatically implies that of the graphs in \emph{2} and \emph{3}. We formulate it as the following lemma.
\begin{lemma}
\label{lemma-first_connectivity_theorem_1to23}
    If all the graphs $\bigcup_{\mathscr{V}(X)\leqslant n} \Gamma_X$ are connected ($n=0,1,2,\dots$), then the following two types of graphs are also connected.
    \begin{itemize}
        \item $\bigcup_{X'\textup{ not softer than }X} \Gamma_{X'}$ for any mode $X$,
        \item $\bigcup_{X'\textup{ harder than }X} \Gamma_{X'}$ for any non-hard mode $X$.
    \end{itemize}
\end{lemma}
\begin{proof}
We prove the connectedness of both types by contradiction.

For the first type, suppose there is a mode $X_0$ such that $\mathcal{G}_0:={\bigcup}_{X'\textup{ not softer than }X_0} \Gamma_{X'}$ has multiple connected components, say, $\mathcal{G}_0^{(1)}$, $\mathcal{G}_0^{(2)}$, and possibly others. That is, the edges that are external to $\mathcal{G}_0^{(1)}$ and $\mathcal{G}_0^{(2)}$ are all softer than $X_0$. For $i=1,2$, let us denote the edges external to $\mathcal{G}_0^{(i)}$ by $e_1^{(i)},\dots,e_{n_i}^{(i)}$, and define
$$\mathscr{V}^{(i)}:= \textup{min}\{\mathscr{V}(e_1^{(i)}),\dots,\mathscr{V}(e_{n_i}^{(i)})\},\quad \mathscr{V}_0 = \textup{max}\{\mathscr{V}^{(1)},\mathscr{V}^{(2)}\}.$$
By construction, any edge external to $\mathcal{G}_0^{(i)}$ that is incident to a vertex of $\mathcal{G}_0^{(i)}$ has a mode softer than $X_0$, while that vertex's mode is not softer than $X_0$; hence the vertex's mode is harder than the edge's mode (recall that adjacent modes subgraphs cannot have overlapping modes, as established in section~\ref{section-mode_subgraphs}). From this, $\mathcal{G}_0^{(1)}$ and $\mathcal{G}_0^{(2)}$ both contain some vertices with virtuality degrees smaller than $\mathscr{V}_0$.

Meanwhile, from the definition of $\mathscr{V}_0$, either $\mathcal{G}_0^{(1)}$ or $\mathcal{G}_0^{(2)}$ has external edges whose virtuality degrees are all $\geqslant \mathscr{V}_0$. So the graph $\bigcup_{\mathscr{V}(X)\leqslant \mathscr{V}_0-1} \Gamma_X$ would be disconnected, contradicting the assumption. Therefore, all the first-type graphs in the lemma statement must be connected.

For the second type, we start by collecting all the $X$s leading to disconnected graphs $\bigcup_{X'\textup{ harder than }X} \Gamma_{X'}$, and focus on one with the smallest virtuality degree. Namely, the graph
$$\mathcal{G}_*:={\bigcup}_{X'\textup{ harder than }X_*} \Gamma_{X'}$$
consists of at least two connected components. One contains the hard subgraph $\mathcal{H}$. The other, denoted as $\gamma_*$, is necessarily a mode component (defined in section~\ref{section-mode_subgraphs}) with $\mathscr{X}(\gamma_*)$ harder than $X_*$.
Let us pick a vertex $v_* \in \gamma_*$ and denote $n_* = \mathscr{V}(\mathscr{X}(\gamma_*))$. Since the graph $\bigcup_{\mathscr{V}(X)\leqslant n_*} \Gamma_X$ is connected, there exists a path $P_*$ in this graph connecting $v_*$ to $\mathcal{H}$. Meanwhile, since the two subgraphs $\gamma_*$ and $\mathcal{H}$ are disconnected from each other, the path must at some point leave the component $\gamma_*$. The first edge on $P_*$ that lies outside $\gamma_*$ is then an edge $e_*$ not in $\mathcal{G}_*$ but adjacent to $\gamma_*$ (i.e., $e_*\in P_*\setminus \mathcal{G}_*$).

We now examine the mode of $e_*$:
\begin{itemize}
    \item because $e_* \in P_*$, we have $\mathscr{V}(\mathscr{X}(e_*)) \leqslant n_* = \mathscr{V}(\mathscr{X}(\gamma_*))$;
    \item because $e_*\notin \mathcal{G}_*$, the mode $\mathscr{X}(e_*)$ is not harder than $X_*$, thus not harder than $\mathscr{X}(\gamma_*)$.
\end{itemize}
These relations force $\mathscr{X}(e_*)$ and $\mathscr{X}(\gamma_*)$ to be overlapping with each other. However, this contradicts the previous conclusion that adjacent mode components cannot have overlapping modes. Therefore, all the second-type graphs in the lemma statement are connected.
\end{proof}

Given these arguments, proving theorem~\ref{theorem-mode_subgraphs_connectivity1} reduces to establishing the connectedness of the graphs in \emph{1}, which still remains technically involved. A detailed demonstration of this fact is provided in section~\ref{section-proof_connectivity_requirement_theorem}.

\begin{figure}[t]
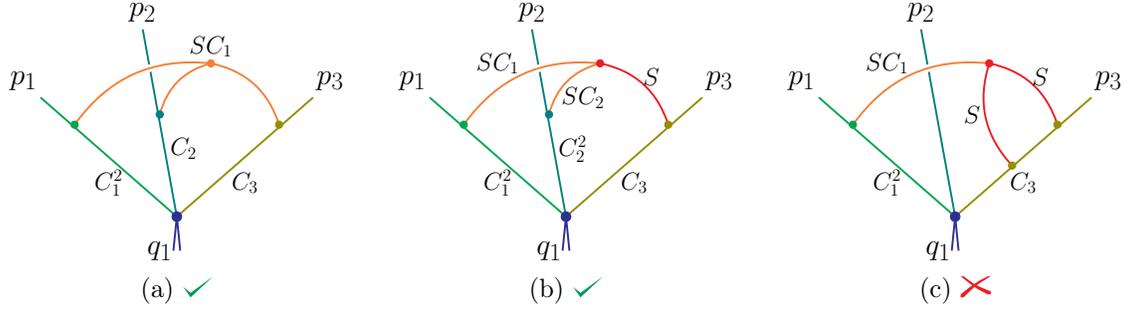

\centering
\begin{subfigure}[b]{0.3\textwidth}
\centering
\include{figs/mode_neighboring_allowed_config1}
\vspace{-3em}\caption{$\greencheckmark[ForestGreen]$}
\label{mode_neighboring_allowed_config1}
\end{subfigure}
\quad
\begin{subfigure}[b]{0.3\textwidth}
\centering
\include{figs/mode_neighboring_allowed_config2}
\vspace{-3em}\caption{$\greencheckmark[ForestGreen]$}
\label{mode_neighboring_allowed_config2}
\end{subfigure}
\quad
\begin{subfigure}[b]{0.3\textwidth}
\centering
\include{figs/mode_neighboring_violated_config1}
\vspace{-3em}\caption{$\crossmark[Red]$}
\label{mode_neighboring_violated_config1}
\end{subfigure}
\caption{Two regions (a) and (b) of the same graph, and a non-region configuration (c) of a similar graph. In all these configurations, the external momenta satisfy $p_1^2\sim p_2^2\sim \lambda^2$, $p_3^2\sim \lambda$, and $q_1^2\sim 1$.}
\label{figure-mode_neighboring_configurations}
\end{figure}

\bigbreak
We now consider another essential observation: for each mode subgraph $\gamma$, its mode $\mathscr{X}(\gamma)$ should be ``derivable'' from either the harder neighbors or softer neighbors (including external momenta) of $\gamma$. That is, we expect $\mathscr{X}(\gamma)$ to be equal to either of the following:
\begin{enumerate}
    \item [(1)] $\bigwedge_i\mathscr{X}(\gamma_i)$, where $\gamma_i$ are those adjacent to $\gamma$ with harder weights,
    \item [(2)] $\bigvee_i\mathscr{X}(k_i)$, where $k_i$ are those external momenta entering $\gamma$, or line momenta of softer-mode subgraphs adjacent to $\gamma$.
\end{enumerate}
To illustrate, consider the three examples in figure~\ref{figure-mode_neighboring_configurations}, where the mode of each edge is specified. Their external momenta all satisfy $p_1^2\sim p_2^2\sim \lambda^2$, $p_3^2\sim \lambda$, and $q_1^2\sim 1$.
All these configurations satisfy momentum conservation and theorem~\ref{theorem-mode_subgraphs_connectivity1}. One can further verify that figures~\ref{mode_neighboring_allowed_config1} and \ref{mode_neighboring_allowed_config2} also meet the requirement stated above. For example, the $SC_1$ subgraph in figure~\ref{mode_neighboring_allowed_config1} is adjacent to the harder‑mode subgraphs $\gamma_{C_1^2}^{}$, $\gamma_{C_2}^{}$, and $\gamma_{C_3}^{}$. By definition,
$$SC_1 = C_1^2\wedge C_2\wedge C_3.$$
The $S$ subgraph in figure~\ref{mode_neighboring_allowed_config2} is adjacent to softer-mode subgraphs $\gamma_{SC_1}^{}$ and $\gamma_{SC_2}^{}$, and by definition,
$$S = SC_1\vee SC_2.$$
In contrast, figure~\ref{mode_neighboring_violated_config1} contains a mode subgraph $S$ that satisfies neither condition (1) nor (2) above, so it does not correspond to a valid region.

We now make two remarks. First, by theorem~\ref{theorem-mode_algebra_two_representatives}, in either $\bigwedge_i\mathscr{X}(\gamma_i)$ or $\bigvee_i\mathscr{X}(k_i)$, one can always choose exactly two representatives from the index set. Second, the requirement that $\gamma$ is adjacent to $\gamma_1,\dots$, as described above, can actually be loosened to a generalized concept ``relevant'' (to be defined in section~\ref{section-formal_construction_statements}). We now formulate this relaxed connectivity requirement as the \emph{Second Connectivity Theorem}.
\begin{theorem}[\textbf{Second Connectivity Theorem}]
\label{theorem-mode_subgraphs_connectivity2}
    Each mode subgraph $\gamma$ must satisfy at least either of the following conditions:
    \begin{enumerate}
        \item [1,] $\gamma$ is \emph{relevant} to two harder-mode subgraphs $\gamma_1$ and $\gamma_2$, with $\mathscr{X}(\gamma) = \mathscr{X}(\gamma_1)\wedge \mathscr{X}(\gamma_2)$;
        \item [2,] two momenta $k_1$ and $k_2$, from the external kinematics and/or relevant softer-mode subgraphs, enter $\gamma$, with $\mathscr{X}(\gamma) = \mathscr{X}(k_1)\vee \mathscr{X}(k_2)$.
    \end{enumerate}
\end{theorem}
The proof of this theorem will be deferred to the end of section~\ref{section-leading_terms}.

\subsection{The infrared-compatibility requirement}
\label{section-infrared_compatibility_requirements_overall}

For any given region, its mode components must not only satisfy the connectivity requirements but also acquire their corresponding infrared scalings from the external kinematics, either directly or indirectly, otherwise the expanded integral would remain scaleless. This additional requirement, called the \emph{infrared-compatibility requirement}, exhibits a particularly interesting ``message-delivery pattern''. In the following, we first introduce the basic idea and then formulate the requirement precisely.

\subsubsection{Basic idea: delivery of infrared scalings}
\label{section-basic_idea_delivery_infrared_message}

Below we discuss possible mechanisms by which a mode component $\gamma$ acquires its infrared scaling. We shall collate the mechanisms with some typical examples, which motivates the message-delivery pattern.

\bigbreak\noindent\emph{Mechanism 1.}

The most straightforward way for a mode component $\gamma$ to obtain its infrared scaling is from the external momenta directly. This happens when the total external momentum entering $\gamma$ is precisely in the mode $\mathscr{X}(\gamma)$. In this case, one can construct a unitarity cut through $\gamma$, with the total incoming external momenta on one side. The momentum flowing across the cut is then of the mode $\mathscr{X}(\gamma)$.

\begin{figure}[t]
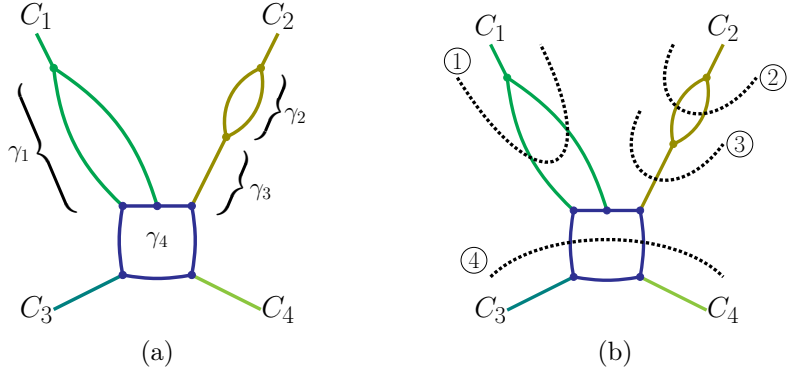

\centering
\begin{subfigure}[b]{0.27\textwidth}
\centering
\include{figs/infrared_scaling_delivery_external_components}
\vspace{-3em}\caption{}
\label{infrared_scaling_delivery_external_components}
\end{subfigure}
\qquad\qquad
\begin{subfigure}[b]{0.3\textwidth}
\centering
\include{figs/infrared_scaling_delivery_external_cuts}
\vspace{-3em}\caption{}
\label{infrared_scaling_delivery_external_cuts}
\end{subfigure}
\caption{An example showing the case where all the mode components acquire their infrared scalings from the external kinematics. (a): the mode components $\gamma_1$--$\gamma_4$. (b): the unitarity cuts across the mode components, each isolating some external momentum.}
\label{figure-infrared_scaling_delivery_external}
\end{figure}

An example is shown in figure~\ref{figure-infrared_scaling_delivery_external}, where the external momenta modes are $C_1$--$C_4$. The mode subgraph $\Gamma_{H}$ is colored {\color{Blue}\bf blue}, $\Gamma_{C_1}$ {\color{Green}\bf green}, and $\Gamma_{C_2}$ {\color{olive}\bf olive}.
There are four mode components labeled in figure~\ref{infrared_scaling_delivery_external_components}: one $C_1$ component ($\gamma_1$), two $C_2$ components ($\gamma_2$ and $\gamma_3$), and one $H$ component ($\gamma_4$). Each $\gamma_i$ obtains its momentum scaling directly from the external kinematics, with the corresponding unitarity cut indicated in figure~\ref{infrared_scaling_delivery_external_cuts}. One can verify that the external momentum flowing across each cut has the same mode as the corresponding component. In other words, $\gamma_1$--$\gamma_4$ acquire their infrared scalings from the external kinematics.

\bigbreak\noindent\emph{Mechanism 2.}

Suppose, for instance, that $\gamma_1$ and $\gamma_2$ have already acquired their infrared scalings from the external kinematics as above, i.e., external momenta entering $\gamma_1$ and $\gamma_2$ have the modes $\mathscr{X}(\gamma_1)$ and $\mathscr{X}(\gamma_2)$, respectively. Then any $\gamma$ that is adjacent to both $\gamma_1$ and $\gamma_2$ and satisfying $\mathscr{X}(\gamma) = \mathscr{X}(\gamma_1) \wedge \mathscr{X}(\gamma_2)$ would also acquire its infrared scaling. In other words, such a $\gamma$ acquires its scaling from two ``harder-mode neighbors'', rather than from any $\mathscr{X}(\gamma)$-mode external momentum.

The previously discussed figure~\ref{mode_neighboring_allowed_config1} is an example, where the $C_1^2$ and $C_3$ components have already obtained their scalings from the external kinematics (recall $\mathscr{X}(p_1)=C_1^2$ and $\mathscr{X}(p_3)=C_3$ there), and the $SC_1$ component (note that $SC_1 = C_1^2\wedge C_3$), which is adjacent to both, thereby obtains its infrared scaling as well.

\bigbreak\noindent\emph{Mechanism 3.}

In certain situations, $\gamma$ can acquire its infrared scaling from only a single (rather than multiple as in Mechanism 2) ``harder-mode neighbor''. The simplest example illustrating this is the region in figure~\ref{form_factor_like_region_example}, where $p_1^2\sim \lambda$ while $p_2^2 = p_3^2 = 0$. Consider $\gamma$ as the soft component exchanged among $\mathcal{J}_1$, $\mathcal{J}_2$, and $\mathcal{J}_3$. Only $\mathcal{J}_1$ acquires its infrared scaling directly from the external momentum $p_1$. The soft component $\gamma$, however, can still acquire its own scaling in this topology. Afterwards, it can in turn deliver scaling to the other jets ($\mathcal{J}_2$ and $\mathcal{J}_3$) adjacent to it, effectively acting as a ``messenger''. Thus, the initial scaling from $p_1$ is propagated through the graph via $\gamma$.

Let us further recall that figures~\ref{form_factor_like_nonregion_example1} and \ref{form_factor_like_nonregion_example2} both correspond to scaleless integrals. From this, one can tentatively conclude that soft components adjacent to three or more jets can act as messengers, while those adjacent to only two or fewer jets cannot. In fact, for configurations involving only modes $H$, $C$, and $S$, this has already been justified in section 4 of ref.~\cite{Ma23}.

\begin{figure}[t]
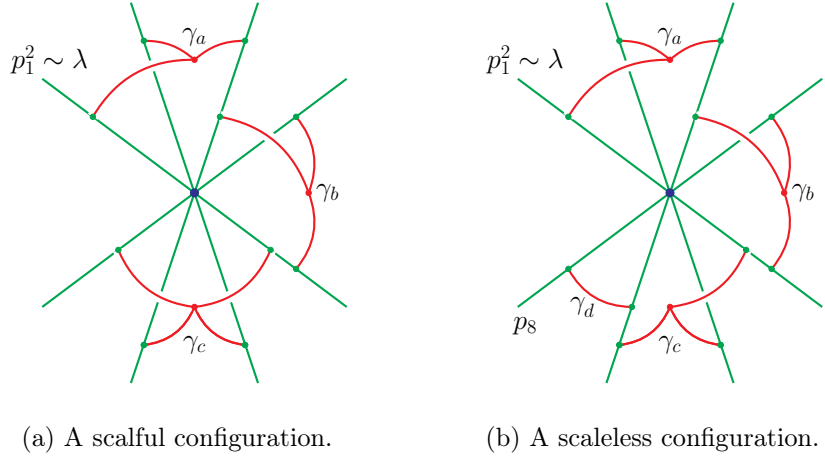

\centering
\begin{subfigure}[b]{0.3\textwidth}
\centering
\include{figs/infrared_compatibility_example1}
\vspace{-2em}
\caption{A scaleful configuration.}
\label{infrared_compatibility_example1}
\end{subfigure}
\qquad\qquad
\begin{subfigure}[b]{0.3\textwidth}
\centering
\include{figs/infrared_compatibility_example2}
\vspace{-2em}
\caption{A scaleless configuration.}
\label{infrared_compatibility_example2}
\end{subfigure}
\caption{Two examples demonstrating the ``message-delivery'' mechanism. In these $4 \to 4$ scattering configurations, the external momentum $p_1$ has a small virtuality ($p_1^2 \sim \lambda$) while all other $p_i$ are strictly on-shell ($p_i^2 = 0$ for $i\neq 1$). Green lines represent distinct collinear modes, and red curves represent the soft mode. The left example (a) corresponds to a scaleful expanded integral, whereas the right example (b) is scaleless. This difference arises because the ``infrared message'', i.e., the small virtuality $p_1^2$, can be delivered by the ``messengers'' to all other jets in (a), but not to $\mathcal{J}_8$ in (b).}
\label{figure-infrared_compatibility_examples}
\end{figure}

In general, the infrared scaling of a mode component $\gamma$ can also be acquired after several rounds of the delivery of infrared scaling. This is why we call it a ``message-delivery pattern''. Consider, for example, the two configurations in figure~\ref{figure-infrared_compatibility_examples}, where $p_1^2 \sim \lambda$ while $p_i^2 = 0$ for all $i \neq 1$. As one can verify, figure~\ref{infrared_compatibility_example1} yields a scaleful integral while figure~\ref{infrared_compatibility_example2} is scaleless. From the perspective of the mechanisms above, in figure~\ref{infrared_compatibility_example1} the infrared scaling from $p_1$ can be delivered to all mode components via messengers (which are the three soft components $\gamma_a$, $\gamma_b$, and $\gamma_c$, each adjacent to three or more jets). In contrast, in figure~\ref{infrared_compatibility_example2} one soft component $\gamma_d$ is adjacent to exactly two jets, thus cannot acquire their infrared scalings through the message-delivery pattern.

As we will see, the preceding examples illustrate three distinct mechanisms by which a mode component can acquire its infrared scaling: (1) from the external kinematics directly, (2) from two harder-mode neighbors directly, and (3) from one harder-mode neighbor via a messenger. In the following, we formalize them on a rigorous footing for all-order analysis.

\subsubsection{Formal constructions and statement}
\label{section-formal_construction_statements}

Motivated by the examples above, we now introduce a series of definitions to formulate the infrared-compatibility requirement.

\bigbreak\noindent\emph{Marginal Softness.}

We start with a key relation between momentum modes.
$X_1$ is called \emph{marginally softer than} $X_2$, if $X_1$ is softer than $X_2$ yet some components of the momentum $k_{X_1}$ are of the same scale as the corresponding components of $k_{X_2}$. Alternatively, this means the leading contribution of $(k_{X_1} + k_{X_2})^2$ retains a dependence on $k_{X_1}$.

For example, the mode $S$ is marginally softer than $C_1$, whereas $S^2$ is not, as an evident fact from the expansion of $(k_S+k_{C_1})^2$ and $(k_{S^2}+k_{C_1})^2$:
\begin{subequations}
    \begin{align}
        &(k_S+k_{C_1})^2= \underset{\mathcal{O}(\lambda)}{\underbrace{k_{C_1}^2}} + \underset{\mathcal{O}(\lambda)}{\underbrace{2k_{C_1}\cdot k_S}} + \underset{\mathcal{O}(\lambda^2)}{\underbrace{k_S^2}}\sim k_{C_1}^2 + 2k_{C_1}\cdot k_S,&\rightarrow\text{ dependent on }k_S;\\
        &(k_{S^2}+k_{C_1})^2= \underset{\mathcal{O}(\lambda)}{\underbrace{k_{C_1}^2}} + \underset{\mathcal{O}(\lambda^2)}{\underbrace{2k_{C_1}\cdot k_{S^2}}} + \underset{\mathcal{O}(\lambda^4)}{\underbrace{k_{S^2}^2}}\sim k_{C_1}^2,&\rightarrow\text{ independent of }k_{S^2}.
    \end{align}
\end{subequations}
A given mode can be marginally softer than several others. For example, $S^2$ is marginally softer than both $SC_i$ and $C_i^2$ for any $i$. More generally, the modes for which $S^m$ is marginally softer are precisely those of the form $S^{m-m'}C_i^{m'}$ with $m'\in\{1,\dots,m\}$.

\bigbreak\noindent\emph{Relevance of Mode Components.}

We now extend this idea from modes to mode components $\gamma_1$ and $\gamma_2$. We say $\gamma_1$ \emph{is relevant to} $\gamma_2$ if the following conditions hold:
\begin{enumerate}
    \item $\mathscr{X}(\gamma_1)$ is marginally softer than $\mathscr{X}(\gamma_2)$.
    \item There exists a path $P$ connecting $\gamma_1$ to $\gamma_2$, such that for any two elements (edges or vertices) $a, b\in P$ where $a$ is closer to $\gamma_1$ than $b$ is, we have that $\mathscr{X}(a)$ is softer than or equal to $\mathscr{X}(b)$.
\end{enumerate}
To illustrate this concept, recall the example in figure~\ref{mode_neighboring_allowed_config2}. There, either $SC_i$ component ($i=1,2$) is relevant to the $C_i^2$ and $C_3$ components. Notice that although an $SC_i$ component is not adjacent to the $C_3$ component, they are connected via an $S$ component, and along the path $SC_i \to S \to C_3$, the modes become progressively harder. The $S$ component, in contrast, is relevant to the $C_3$ component only.

\bigbreak\noindent\emph{Messengers.}

We now formulate the notion of messengers. Motivated by the examples in section~\ref{section-basic_idea_delivery_infrared_message}, it acts as a soft ``core'' mediating infrared scaling among three or more harder-mode components.

\begin{figure}[t]
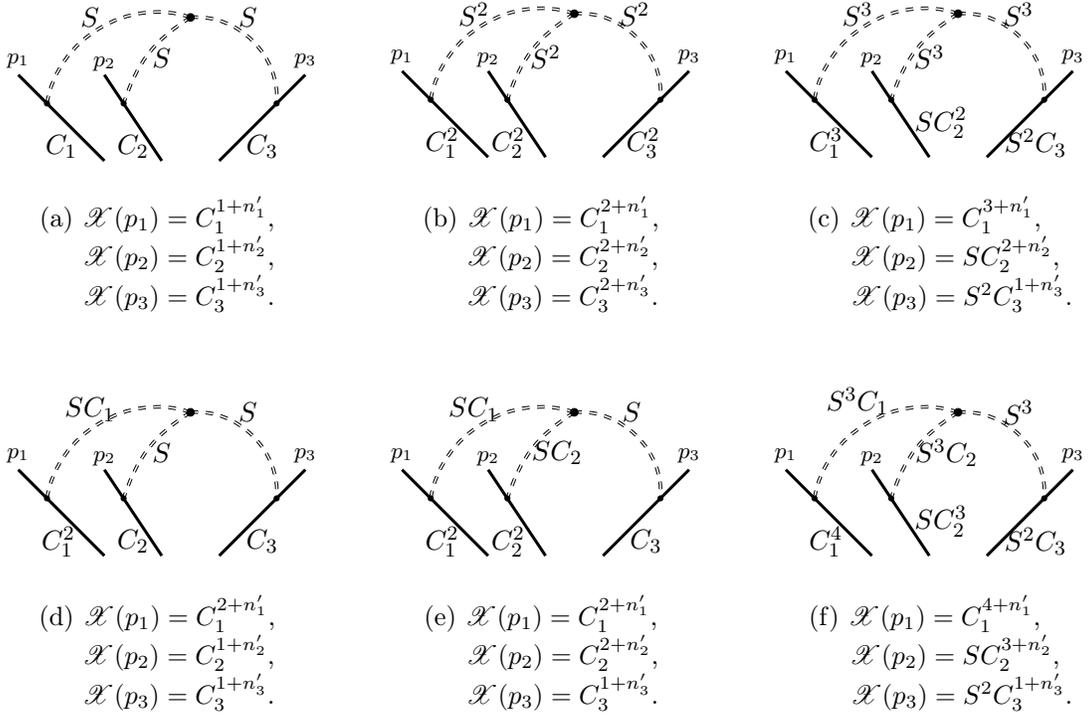

\centering
\begin{subfigure}[b]{0.32\textwidth}
\centering
\include{figs/messenger_example1}
\vspace{-1em}
\caption{\centering
    $\mathscr{X}(p_1)= C_1^{1+n'_1}$,\\
    $\phantom{\textup{(a) }}\mathscr{X}(p_2)= C_2^{1+n'_2}$,\\
    $\phantom{\textup{(a) }}\mathscr{X}(p_3)= C_3^{1+n'_3}$.
}
\label{messenger_example1}
\end{subfigure}
\begin{subfigure}[b]{0.32\textwidth}
\centering
\include{figs/messenger_example2}
\vspace{-1em}
\caption{\centering
    $\mathscr{X}(p_1)= C_1^{2+n'_1}$,\\
    $\phantom{\textup{(a) }}\mathscr{X}(p_2)= C_2^{2+n'_2}$,\\
    $\phantom{\textup{(a) }}\mathscr{X}(p_3)= C_3^{2+n'_3}$.
}
\label{messenger_example2}
\end{subfigure}
\begin{subfigure}[b]{0.32\textwidth}
\centering
\include{figs/messenger_example3}
\vspace{-1em}
\caption{\centering
    $\mathscr{X}(p_1)= C_1^{3+n'_1}$,\\
    $\phantom{\textup{(a)}\quad}\mathscr{X}(p_2)= SC_2^{2+n'_2}$,\\
    $\phantom{\textup{()}\qquad}\mathscr{X}(p_3)= S^2C_3^{1+n'_3}$.
}
\label{messenger_example3}
\end{subfigure}
\\
\centering\vspace{2em}
\begin{subfigure}[b]{0.32\textwidth}
\centering
\include{figs/messenger_example4}
\vspace{-1em}
\caption{\centering
    $\mathscr{X}(p_1)= C_1^{2+n'_1}$,\\
    $\phantom{\textup{(a) }}\mathscr{X}(p_2)= C_2^{1+n'_2}$,\\
    $\phantom{\textup{(a) }}\mathscr{X}(p_3)= C_3^{1+n'_3}$.
}
\label{messenger_example4}
\end{subfigure}
\begin{subfigure}[b]{0.32\textwidth}
\centering
\include{figs/messenger_example5}
\vspace{-1em}
\caption{\centering
    $\mathscr{X}(p_1)= C_1^{2+n'_1}$,\\
    $\phantom{\textup{(a) }}\mathscr{X}(p_2)= C_2^{2+n'_2}$,\\
    $\phantom{\textup{(a) }}\mathscr{X}(p_3)= C_3^{1+n'_3}$.
}
\label{messenger_example5}
\end{subfigure}
\begin{subfigure}[b]{0.32\textwidth}
\centering
\include{figs/messenger_example6}
\vspace{-1em}
\caption{\centering
    $\mathscr{X}(p_1)= C_1^{4+n'_1}$,\\
    $\phantom{\textup{(a)}\quad}\mathscr{X}(p_2)= SC_2^{3+n'_2}$,\\
    $\phantom{\textup{()}\qquad}\mathscr{X}(p_3)= S^2C_3^{1+n'_3}$.
}
\label{messenger_example6}
\end{subfigure}
\caption{Six examples of messengers (represented by dashed curves) and their relevant $S^{m-m_i}C_i^{m_i+n_i}$ components (solid lines). The modes of the edges are specified in each sub-figure, and those of the external momenta $p_1,p_2,p_3$ are specified in the caption, with $n'_i\in \{0,1,\dots,+\infty\}$.}
\label{figure-messenger_examples}
\end{figure}

Formally, a degree-$m$ messenger, denoted by $\Gamma^{[m]}$, is a connected subgraph consisting of a single $S^m$ component, called the kernel of $\Gamma^{[m]}$, and possibly some other mode components whose modes are of the form $S^mC^n$ ($n \in \mathbb{N}_+$). Furthermore, there exist \emph{three or more} mode components $\gamma_1,\dots,\gamma_I\subset \mathcal{G}\setminus \Gamma^{[m]}$ ($I\geqslant 3$), with $\mathscr{X}(\gamma_i) = S^{m-m_i}C_{j_i}^{m_i+n_i}$ ($m_i \in \{1,\dots,m\}$ and $n_i\in\mathbb{N}$), such that:
\begin{enumerate}
\item [1,] $j_1,\dots,j_I$, representing the lightlike directions, are all different;
\item [2,] for any $\gamma_i$ with $n_i=0$, the $S^m$ component of $\Gamma^{[m]}$ (the kernel) is relevant to it;\\
for any $\gamma_i$ with $n_i\neq 0$, $\Gamma^{[m]}$ contains an $S^mC_{j_i}^{n_i}$ component that is relevant to it;
\item [3,] each $\gamma_i$ is attached by an $S^{m-m_i}C_{j_i}^{m_i+n_i+n'_i}$-mode external momentum, with $n'_i \in \mathbb{N}^\infty$.
\end{enumerate}
To understand this, consider the examples in figure~\ref{figure-messenger_examples}, which show six degree-$m$ messengers $\Gamma^{[m]}$ (dashed curves, with $m=1$ in (a), (d), and (e), $m=2$ in (b), and $m=3$ in (c) and (f)), together with the $S^{m-m_i}C_i^{m_i+n_i}$ components $\Gamma^{[m]}$ is relevant to (solid lines).
For simplicity, we show cases where $\Gamma^{[m]}$ is relevant to exactly three such components; configurations with four or more are analogous. The modes of the edges are specified in the figure, and those of the external momenta $p_1,p_2,p_3$ are specified in the caption, with $n'_i\in \mathbb{N}^\infty$.

Let us go through some representatives in figure~\ref{figure-messenger_examples}. Figure~\ref{messenger_example1} generalizes the aforementioned example of figure~\ref{form_factor_like_region_example} (where we have $n'_1=0$ and $n'_2=n'_3=+\infty$). Figure~\ref{messenger_example3} shows that the $m_i$ in $S^{m-m_i}C_i^{m_i+n_i}$ components can have different values, and in this example, we have $m_1=3$, $m_2=2$, $m_3=1$, and $n_1=n_2=n_3=0$. Figure~\ref{messenger_example4} shows that a messenger can consist of multiple mode components, and here, it consists of an $SC_1$ and an $S$ components.

\begin{figure}[t]
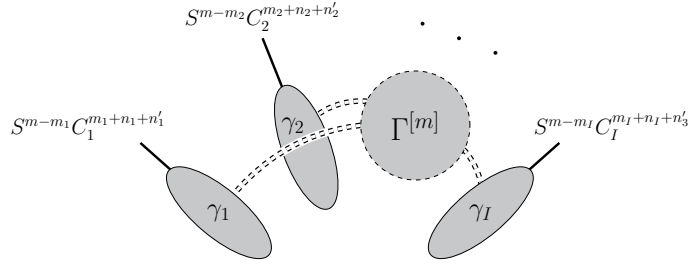

\centering
\include{figs/messenger_general_configuration}
\vspace{-1cm}\caption{The general configuration of a degree-$m$ messenger $\Gamma^{[m]}$ and its relevant components $\gamma_1,\dots,\gamma_I$ ($I\geqslant 3$). All modes inside $\Gamma^{[m]}$ are of the form $S^{m}C_i^{n_i}$ ($i\in\{1,\dots,I\}$). For each $i$, $\gamma_i$ has the mode $S^{m-m_i}C_i^{m_i+n_i}$, and its external momentum mode is specified in the figure.}
\label{figure-messenger_general_configuration}
\end{figure}

The general configuration of $\Gamma^{[m]}$, together with its relevant harder-mode components $\gamma_1,\dots,\gamma_I$ ($I\geqslant 3$), is depicted in figure~\ref{figure-messenger_general_configuration}.

\bigbreak\noindent\emph{Infrared Compatibility.}

With the preceding constructions in place, we now define the central concept of infrared compatibility. The definition is recursive: a mode component $\gamma$ is called \emph{infrared compatible}, if it satisfies at least one of the conditions below (where ``already confirmed'' components are those previously established as infrared compatible in the recursion):
\begin{enumerate}
    \item [\emph{1},] the partial sum over momenta entering $\gamma$, which are exclusively from the external kinematics and already confirmed infrared-compatible subgraphs, is of the mode $\mathscr{X}(\gamma)$;
    \item [\emph{2},] $\mathscr{X}(\gamma) = S^m$, and $\gamma$ is contained in a degree-$m$ messenger and relevant to some mode component $\gamma_0$ that is already confirmed infrared compatible;
    \item [\emph{3},] $\mathscr{X}(\gamma) = S^mC_i^n$ ($n\in\mathbb{N}$), and $\gamma$ is relevant to two mode components $\gamma_1$ and $\gamma_2$ that are already confirmed infrared compatible, with $\mathscr{X}(\gamma) = \mathscr{X}(\gamma_1)\wedge \mathscr{X}(\gamma_2)$.
\end{enumerate}
The recursive nature of this definition is essential: the infrared compatibility of a component may depend on prior confirmation of others (typically, starting from components directly tied to external kinematics and propagating inward).

The elegance of this concept lies in the fact that, once the connectivity requirements are satisfied, infrared compatibility provides a necessary and sufficient condition for the expanded integral to be scaleful.
\begin{theorem}
\label{theorem-infrared_compatibility_requirement}
    For any region, all mode components are infrared compatible.
\end{theorem}

This theorem will be proved in section~\ref{section-infrared_compatibility_requirements}. At this point, let us verify its content from the aforementioned examples. The soft components in figures~\ref{form_factor_like_nonregion_example1} and~\ref{form_factor_like_nonregion_example2} are not infrared compatible (failing all three conditions above, and recursion does not occur), whereas all mode components in figure~\ref{form_factor_like_region_example} are. This distinguishes the valid region in the latter case. As a more nontrivial illustration, all mode components in figure~\ref{infrared_compatibility_example1} become infrared compatible through the recursive process (with scaling propagating via the messengers $\gamma_a,\gamma_b,\gamma_c$). In contrast, figure~\ref{infrared_compatibility_example2} contains two non-infrared-compatible components: (1) the $S$ component $\gamma_d$, and (2) the $C_i$ component adjacent only to $\gamma_d$ (and not to $\gamma_a,\gamma_b,\gamma_c$). Accordingly, the former is a valid region while the latter is not. Some other examples are provided in sections~\ref{section-three_loop_example_2to2_scattering} and~\ref{section-six_loop_example_1to3_decay_plus_soft_emission}.

\subsection{Implications from the infrared-compatibility requirement}
\label{section-implication_infrared_compatibility_requirement}

The infrared‑compatibility requirement is strong: it constrains not only the connections among subgraphs but also the mode structure. Moreover, it automatically implies the Second Connectivity Theorem. We discuss these consequences in this subsection.

As a first corollary of theorem~\ref{theorem-infrared_compatibility_requirement}, the ``resolution'' of the set of possible modes is fixed by the external kinematics.
\begin{corollary}
\label{corollary-infrared_compatibility_requirement_mode_resolution}
    If all the external momentum modes are in the form $S^mC_i^n$ with $m,n\in k\mathbb{N}$ (for a fixed $k\in\{2,3,\dots\}$) and arbitrary $i$, then all the modes occurring in $\mathcal{G}$ must be of this form as well.
\end{corollary}
This explains why the momentum modes in the virtuality expansion (eq.~\eqref{eq:virtuality_expansion}) can always be expressed as $S^mC_i^n$ with $m,n\in\mathbb{N}$. For example, suppose that for some kinematics satisfying eq.~\eqref{eq:virtuality_expansion}, there existed a region containing the ``semihard'' mode $S^{1/2}\equiv (\lambda^{1/2},\lambda^{1/2},\lambda^{1/2})$. Rescaling $\lambda\to \lambda^2$ would produce a configuration in which all external momentum modes are of the form $S^mC_i^n$ with $m,n\in \{2,4,\dots\}$, while an $S$ mode (i.e., $S^1$) would appear internally. Such a configuration is forbidden by the corollary, so the $S^{1/2}$ mode cannot actually occur in the virtuality expansion. The same argument excludes any other modes $S^mC_i^n$ for which $(m,n)\notin \mathbb{N}^2$
\begin{proof}
    We prove the corollary by induction. In the process of verifying the infrared compatibility of any mode component $\gamma$, if all those previously confirmed infrared‑compatible mode subgraphs have modes of the form $S^mC_i^n$ with $m,n\in\mathbb{N}$, then the three infrared‑compatibility conditions force $\mathscr{X}(\gamma)$ to be of the same form as well.
\end{proof}

Another essential implication from the infrared-compatibility requirement is the \emph{absence of cascading modes} in the virtuality expansion. By \emph{cascading modes}, we mean the phenomenon in which, for fixed external kinematics, the number of distinct momentum modes required to describe the regions grows unbounded (e.g., increasingly softer modes like $S^k$ for arbitrarily large $k$) as the loop order increases. This has been observed in certain asymptotic expansions involving multiple on-shell massive momenta~\cite{Ma23,Fontes:2024yvw} or on-shell spacelike-collinear momenta~\cite{GrdHzgJnsMaprepare}. In contrast, within the scope of this work, i.e., the virtuality expansion in wide-angle kinematics (eq.~\eqref{eq:virtuality_expansion}), cascading modes never occur.
\begin{corollary}
\label{corollary-infrared_compatibility_requirement_no_cascade}
    For any fixed external kinematics satisfying eq.~\eqref{eq:virtuality_expansion}, the number of distinct momentum modes occurring at the amplitude level is bounded, independently of the loop order.
\end{corollary}
\begin{proof}
    Consider the modes of all partial sums (of one or more momenta) among the external momenta, which have nonzero virtuality:
    \begin{align}
    \label{eq:partial_sum_external_momenta_mode}
    S^{m_1}C_{i_1}^{n_1},\ S^{m_2}C_{i_2}^{n_2},\ \dots,\ S^{m_r}C_{i_r}^{n_r},
    \end{align}
    where $n_1,\dots,n_r \neq +\infty$. Define an index
    \begin{align}
       \kappa\equiv \max\{m_1+n_1,\dots,m_r+n_r\}.
    \end{align}
    Every mode in (\ref{eq:partial_sum_external_momenta_mode}), which corresponds to a partial sum of the external momenta, is harder than or equal to $S^\kappa$.
    
    The proof proceeds by induction on the recursive confirmation of infrared compatibility (theorem~\ref{theorem-infrared_compatibility_requirement}). Components confirmed first (those directly tied to external kinematics) have modes harder than or equal to $S^\kappa$. At any stage, if all previously confirmed components satisfy this bound, then any newly confirmed component (via any of the three infrared-compatibility conditions) is at most marginally softer and thus also satisfies the bound. Consequently, no mode component can be softer than $S^\kappa$, independent of loop order. This proves the absence of cascading modes.
\end{proof}

The proof above also shows the softest possible mode $S^\kappa$ associated with the given external kinematics. This information will prove useful for deriving the complete list of regions directly from our structural understanding, as discussed in section~\ref{section-summary_matching_results_known_expansions} and in later multiloop applications (section~\ref{section-multiloop_examples}). Finally, we note that theorem~\ref{theorem-infrared_compatibility_requirement} is strictly stronger than theorem~\ref{theorem-mode_subgraphs_connectivity2}.
\begin{corollary}
\label{corollary-infrared_compatibility_requirement_proving_connectivity2}
    If all mode components are infrared compatible, then the Second Connectivity Theorem (theorem~\ref{theorem-mode_subgraphs_connectivity2}) necessarily holds.
\end{corollary}
\begin{proof}
    Consider any mode component $\gamma$. Its infrared compatibility must be confirmed via one of the three conditions. If it is confirmed through condition \emph{1} or \emph{3}, the statement of the Second Connectivity Theorem is automatically satisfied. If it is confirmed through condition \emph{2}, then $\gamma$ has mode $S^m$ and is contained in a degree‑$m$ messenger $\Gamma^{[m]}$. By the definition of messengers, $\Gamma^{[m]}$ is relevant to three or more distinct mode components $\gamma_i$ (denoted $\gamma_1,\dots,\gamma_a$ with $a\geqslant3$ for simplicity), where $\mathscr{X}(\gamma_i) = S^{m-m_i}C_i^{m_i+n_i}$ and one of the following holds:
    \begin{itemize}
        \item at least two of the $n_1,\dots,n_a$ are zero;
        \item exactly one of the $n_1,\dots,n_a$ is zero.
    \end{itemize}
    In both cases, $\gamma$ fulfills the Second Connectivity Theorem.
    In the first case, there exist indices $i,j\in\{1,\dots,a\}$ with $n_i=n_j=0$. The corresponding components $\gamma_i$ and $\gamma_j$ then have modes $S^{m-m_i}C_i^{m_i}$ and $S^{m-m_j}C_j^{m_j}$, respectively. $\gamma$ is then relevant to both $\gamma_i$ and~$\gamma_j$, with $\mathscr{X}(\gamma) = \mathscr{X}(\gamma_i)\wedge\mathscr{X}(\gamma_j)$, which satisfies (the first part of) the Second Connectivity Theorem.
    In the second case, at least two indices $i,j\in\{1,\dots,a\}$ satisfy $n_i,n_j\neq0$. By definition, $\Gamma^{[m]}$ contains components $\gamma'_i$ and $\gamma'_j$ that are relevant to $\gamma_i$ and $\gamma_j$, respectively. We then have $\mathscr{X}(\gamma'_i)=S^{m}C_i^{n_i}$ and $\mathscr{X}(\gamma'_j)=S^{m}C_j^{n_j}$. Consequently, $\mathscr{X}(\gamma) = \mathscr{X}(\gamma'_i)\vee\mathscr{X}(\gamma'_j)$, which meets (the second part of) the Second Connectivity Theorem.
\end{proof}

\subsection{Summary and matching to known expansions}
\label{section-summary_matching_results_known_expansions}

For better readability, we collect the subgraphs requirements discussed in the preceding subsections and restate them below as the central conclusion of this work.

\begin{mdframed}[linewidth=1pt, backgroundcolor=Bittersweet!5, skipabove=10pt, skipbelow=10pt]
For any configuration consistent with the fundamental pattern (figure~\ref{figure-virtuality_expansion_fundamental_pattern}), it is a region \emph{if and only if} both of the following conditions hold:
\begin{enumerate}
    \item [1,] the graph $\bigcup_{\mathscr{V}(X)\leqslant n} \Gamma_X$ is connected for any $n\in \mathbb{N}$;
    \item [2,] all mode components are infrared compatible.
\end{enumerate}
\end{mdframed}
Note that the second and third statements of the First Connectivity Theorem (theorem~\ref{theorem-mode_subgraphs_connectivity1}) are not included above because they follow from the first statement, as shown in lemma~\ref{lemma-first_connectivity_theorem_1to23}. Likewise, the Second Connectivity Theorem (theorem~\ref{theorem-mode_subgraphs_connectivity2}) is omitted because it is a consequence of the infrared‑compatibility requirement, as established in corollary~\ref{corollary-infrared_compatibility_requirement_proving_connectivity2}.

We now compare the subgraph requirements derived for the general virtuality expansion with those known for certain special cases. One typical expansion is the \emph{on-shell expansion}, defined by the kinematics
\begin{align}
    p_i^2\sim \lambda \ \ (i=1,\dots,K),\quad q_j^2\sim 1\ \ (j=1,\dots,L),\quad p_{i_1}\cdot p_{i_2}\sim 1\ \ (i_1\neq i_2).
\label{eq:wideangle_onshell_kinematics}
\end{align}
This corresponds to assigning each external momentum $p_i$ in the virtuality expansion~\eqref{eq:virtuality_expansion} into the $C_i$ mode, while omitting all soft external momenta $l_k$.

Another similar expansion is the \emph{soft expansion}, defined by
\begin{align}
\label{eq:wideangle_soft_kinematics}
\begin{split}
    & p_i^2=0\ \ (i=1,\dots,K), \quad q_j^2\sim 1\ \ (j=1,\dots,L), \quad l_k^2=0\ \ (k=1,\dots,M),\\
    & p_{i_1}\cdot p_{i_2}\sim 1\ \ (i_1\neq i_2), \quad p_i\cdot l_k\sim q_j\cdot l_k\sim \lambda , \quad l_{k_1}\cdot l_{k_2}\sim \lambda^2 \ \ (k_1\neq k_2).
\end{split}
\end{align}
In comparison with the on-shell expansion above, here all $p_i$ are exactly on shell (i.e., in the $C_i^\infty$ mode), and we include soft external momenta $l_1,\dots,l_M$, each exactly on-shell (i.e., in the $SC_k^\infty$ mode).

All-order prescriptions for the region structures of these expansions have been derived in ref.~\cite{Ma23}. A rigorous proof exists for the on-shell expansion regions, while the subgraph requirements for the soft expansion regions have been derived assuming a specific set of momentum modes. These derivations, however, are lengthy and technical, and apply only to those special expansions. Below, we demonstrate that from the understanding of region structures developed in this work, those results can be obtained from a surprisingly concise proof.

A first observation is that the softest possible mode in these expansions is $S$, independent of loop order. This follows from the proof of no cascading modes (corollary~\ref{corollary-infrared_compatibility_requirement_no_cascade}). For generic on-shell-expansion kinematics, the modes of all possible partial sums of external momenta are $H$ and $C_i$ (any $i$); for generic soft-expansion kinematics, they are $H$, $C_i$, and $S$. In both cases, the index $\kappa$ from corollary~\ref{corollary-infrared_compatibility_requirement_no_cascade} equals 1, confirming that $S$ is the softest possible mode.

\begin{figure}[t]
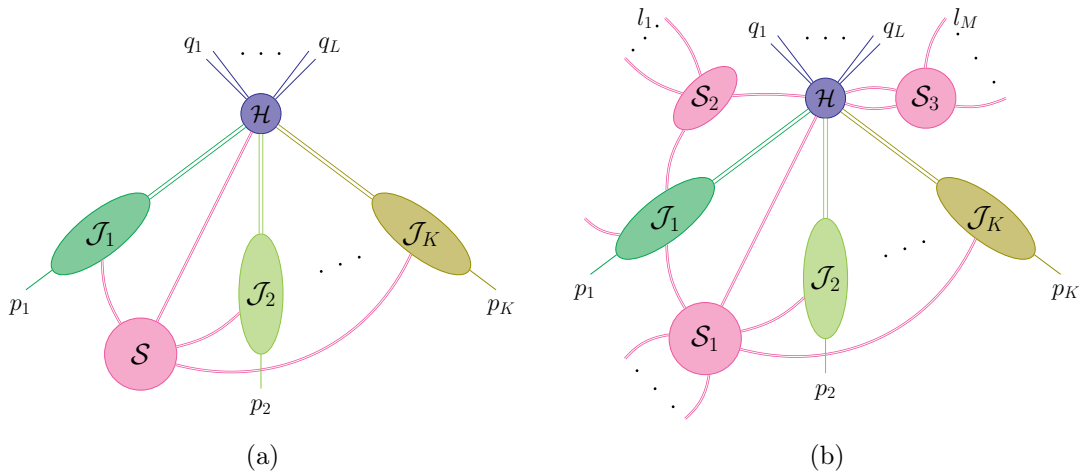

\centering
\begin{subfigure}[b]{0.45\textwidth}
\centering
\include{figs/onshell_generic_facet_region}
\vspace{-2em}\caption{}
\label{onshell_generic_facet_region}
\end{subfigure}
\quad
\begin{subfigure}[b]{0.45\textwidth}
\centering
\include{figs/soft_generic_facet_region}
\vspace{-2em}\caption{}
\label{soft_generic_facet_region}
\end{subfigure}
\caption{Generic facet regions in (a) the on-shell expansion and (b) the soft expansion.}
\label{figure-onshell_soft_generic_facet_regions}
\end{figure}

The fundamental pattern, figure~\ref{figure-virtuality_expansion_fundamental_pattern}, then reduces to figure~\ref{onshell_generic_facet_region} for the on-shell expansion, and figure~\ref{soft_generic_facet_region} for the soft expansion.
In these patterns, the jet $\mathcal{J}_i$ contains only $C_i$-mode components, and $\mathcal{S}$ contains only $S$-mode components. The First Connectivity Theorem is automatically satisfied.

Now consider additional constraints on the mode subgraphs. For both patterns in figure~\ref{figure-onshell_soft_generic_facet_regions}, the Second Connectivity Theorem requires:
\begin{enumerate}
    \item [(1)] the total momentum flowing into every 1VI component of $\mathcal{H}$ has the hard mode;
    \item [(2)] the total momentum flowing into every 1VI component of $\mathcal{J}_i$ has the $C_i$ mode;
    \item [(3)] if an $S$ component is attached by zero or one soft external momentum, then it must be adjacent to two or more jets.
\end{enumerate}
The infrared-compatibility requirement further imposes:
\begin{enumerate}
    \item [(4)] every mode component is infrared compatible.
\end{enumerate}
These requirements can be further simplified as follows. For the on-shell expansion, (1) and (2) are unchanged. Since there are no soft external momenta, (3) can be reduced to ``every $S$ component is adjacent to two or more distinct jets''. When these hold, all mode components are automatically infrared compatible, rendering (4) redundant. For the soft expansion, (1)--(3) are unchanged. When these hold, (4) reduces to ``all $C_i$ components are infrared compatible'', which suffices to ensure that all $S$ components are infrared compatible as well. These simplified requirements agree completely with those in previous works.

To end this section, we comment that our prescription can be directly applied to certain Feynman integrals with massive edges, where all the masses as large as the hard scale $Q$. In such cases, the hard subgraph $\mathcal{H}$ must contain all these massive edges, with the other requirements (regarding connectivity and infrared compatibility) unchanged.

\section{Proof of the subgraph requirements}
\label{section-proof_subgraph_requirements}

This section, as the most technical part of this paper, aims to show that the subgraph requirements listed in section~\ref{section-summary_matching_results_known_expansions} are necessary and sufficient for any configuration of figure~\ref{figure-virtuality_expansion_fundamental_pattern} to be a region. We begin by establishing some notation. For each term $s\!\cdot\!x_1^{r_1}x_2^{r_2}\dots x_N^{r_N}$ of $\mathcal{P}(\x;\s)$, we use $\x^{\r}$ to represent it, and $T(\r)$ to represent its corresponding spanning (2-)tree.

Our analysis will be based on two fundamental criteria for a given region $R$, which we term the \emph{facet criterion} and the \emph{minimum-weight criterion}.

\begin{mdframed}
\textbf{Facet criterion:} up to rescaling, the region vector $\boldsymbol{v}_R$ is the unique vector normal to the facet $f_R$.

\vspace{5pt}
\noindent\textbf{Minimum-weight criterion:} for any leading term $\x^{\r}$ and any other term $\x^{\r'}$ (including non-leading ones), we have $w(T(\r))\leqslant w(T(\r'))$.
\end{mdframed}

The weight for spanning (2-)trees is defined in eq.~(\ref{eq:definition_spanning_1and2_trees_weights}). These criteria follow directly from the defining properties of lower facets and leading polynomials and help justify the subgraph requirements by repeatedly applying proof by contradiction. That is, to establish a given graph-theoretical property, we assume it is violated and exhaustively consider the possible configurations. Contradictions with either fundamental criterion above then rules out the violation, thus proving the property. The following lemma exemplifies this methodology.
\begin{lemma}
\label{lemma-hard_mode_component_tree_structure}
    Suppose $\gamma$ is a mode component contained in $\mathcal{G}'\subseteq \mathcal{G}$ (where $\mathcal{G}'$ can be either the entire Feynman graph or any of its subgraphs), such that $\mathscr{X}(\gamma)$ is harder than any of its adjacent edges, then for any leading term where $\mathcal{G}'$ appears as a spanning tree $T^1$, the graph $\gamma\cap T^1$ is a tree subgraph.
\end{lemma}
\begin{proof}
    As $T^1$ is a tree, the graph $\gamma\cap T^1$ contains no loops. It then remains to show its connectedness. Assume, for contradiction, that $\gamma\cap T^1$ is disconnected. Because $\gamma$ itself is connected (by definition of a mode component), there must exist an edge $e \in \gamma\setminus T^1$ (i.e., in $\gamma$ but not in $T^1$) connecting two distinct components of $\gamma\cap T^1$.
    
    The graph $T^1\cup e$ then contains a unique loop. This loop must include at least one edge~$e'$ that is adjacent to $\gamma$. We can now construct a new spanning tree
    $$T'^1 \equiv T^1 \cup e \setminus e',$$
    whose weight is given by
    \begin{align}
        w(T'^1) = w(T^1) + w(e') - w(e).
    \end{align}
    Recall $w(e)$ denotes the weight associated with edge $e$, as defined in section~\ref{section-parametric_representations_spanning_trees}. By the lemma's hypothesis, $\mathscr{X}(\gamma)$ is harder than $\mathscr{X}(e')$, which implies $w(e) > w(e')$. Therefore, $w(T'^1) < w(T^1)$, contradicting the minimum-weight criterion. This contradiction forces us to conclude that $\gamma\cap T^1$ is indeed connected, and hence a tree.
\end{proof}
Using this same methodology, the following subsections will present more sophisticated analyses of the region structure. In section~\ref{section-proof_connectivity_requirement_theorem} we show the First Connectivity Theorem (theorem~\ref{theorem-mode_subgraphs_connectivity1}). Next we investigate the general structure of the leading terms in section~\ref{section-leading_terms}. We shall classify them into certain types and characterize each type by its mode component structures, which allows for an immediate proof of the Second Connectivity Theorem (theorem~\ref{theorem-mode_subgraphs_connectivity2}). Having established the connectivity theorems, we proceed to show in section~\ref{section-infrared_compatibility_requirements} that the infrared-compatibility requirement is necessary (theorem~\ref{theorem-infrared_compatibility_requirement}) and sufficient. To improve legibility, illustrative examples will accompany the key technical proofs.

\subsection{Proof of the First Connectivity Theorem}
\label{section-proof_connectivity_requirement_theorem}

Recall that in section~\ref{section-connectivity_requirement}, we have shown that the First Connectivity Theorem can be proved once those graphs of type \emph{1} in its statement are proved connected. In what follows, we shall see their connectedness by contradiction.

Specifically, consider a mode component $\gamma_*$ such that all the edges adjacent to $\gamma_*$ have larger virtualities than $\mathscr{V}(\mathscr{X}(\gamma_*))$. As adjacent subgraphs cannot be overlapping with each other (see section~\ref{section-mode_subgraphs}), this further implies that all the edges adjacent to $\gamma_*$ are softer than $\mathscr{X}(\gamma_*)$. By definition, the hard subgraph $\mathcal{H}$ satisfies this condition, and the First Connectivity Theorem is equivalent to the statement that no other such $\gamma_*$ would exist.

In principle, $\gamma_*$ can consist of only a single vertex, but we only need to consider the cases where $\gamma_*$ contains edges. The reason is, if one has managed to show that any such $\gamma_*$ leads to a scaleless integral, then contracting $\gamma_*$ to a single vertex amounts to removing its propagators from the integrand, and the integral would remain scaleless.

From now on, we assume the existence of such a nontrivial~$\gamma_*$ ($\neq \mathcal{H}$) in a given region $R$. We aim at proving the following statement:
\begin{align}
\label{eq:mode_subgraphs_connectivity_theorem_center_statement}
\boxed{\textit{ The graph } T(\r) \cap \gamma_* \textit{ is a spanning tree of } \gamma_* \textit{, for any leading term } \x^{\r}.\ }
\end{align}
This, in turn, will imply a homogeneity condition as an additional constraint on the dimension of $f_R$. Let us use $\mathfrak{n}_\gamma(\r)$ to denote the degree of the parameters associated with the edges of $\gamma$ in $\x^{\r}$. Equivalently, $\mathfrak{n}_\gamma(\r)$ equals the number of edges of $\gamma$ that are absent in the spanning (2-)tree $T(\r)$. The homogeneity condition reads:
\begin{eqnarray}
\label{eq:theorem_connectivity_proof_homogeneity}
    \mathfrak{n}_{\gamma_*}(\r) = L(\gamma_*), \textit{ for any leading term } \x^{\r}.
\end{eqnarray}
The resulting expanded integral will therefore be scaleless. This contradicts the fundamental premise that the expansion procedure yields scaleful contributions for the region $R$. Consequently, our assumption that a nontrivial mode component $\gamma_* \neq H$ with the stated property exists in $R$ must be false. This establishes the connectedness of the graphs of type \emph{1}, completing the proof.

\begin{proof}[Proof of statement (\ref{eq:mode_subgraphs_connectivity_theorem_center_statement})]
    To start with, note that $T(\r)\cap\gamma_*$ contains no loops. So the later analyses will focus on showing the connectedness of $T(\r)\cap\gamma_*$.
    
    Let us consider the possibility that the graph $T(\r)\cap\gamma_*$ contains multiple connected components. From lemma~\ref{lemma-hard_mode_component_tree_structure}, such a possibility is excluded if all the vertices of $T(\r)\cap\gamma_*$ lie in the same tree component. Therefore, $\x^{\r}$ can only be a leading $\mathcal{F}$ term, and $\gamma_*$ intersects both components of $T(\r)$, which we shall denote by $T^2(\r)$ from now on.
    
    Since $\gamma_*$ is itself connected, there exists an edge $e_0 \in \gamma_* \setminus T^2(\r)$ whose endpoints, $v_A$ and $v_B$, lie in the two distinct components of $T^2(\r)$. We denote these components by $t(\r;A)$ and $t(\r;B)$, respectively. We will now exhaust the possible configurations of $T^2(\r)$ and exclude them one by one.

    \begin{itemize}
    \item [\textbf{I.}]
    If the entire graph $\mathcal{G}$ has an off-shell external momentum $q_j^\mu$ with $q_j^2\sim 1$, then without loss of generality, we may assume it attached to $t(\r;A)$. The configuration of $T^2(\r)$ is then shown in figure~\ref{connectivity_theorem_proof_config_I}, where we have used a dashed line to emphasize that $e_0\not\in T^2(\r)$.
    \begin{figure}[t]
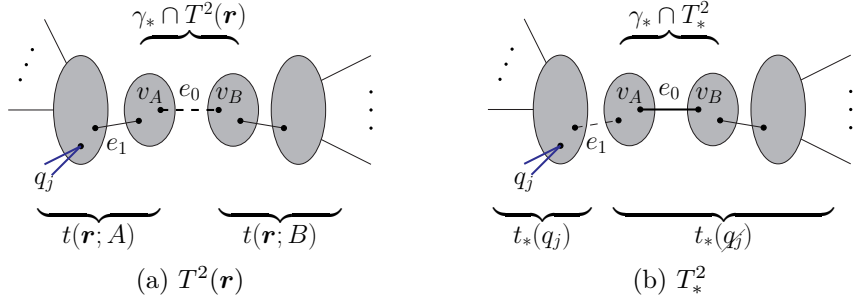

    \centering
    \hspace{-2em}
    \begin{subfigure}[b]{0.32\textwidth}
    \centering
    \include{figs/connectivity_theorem_proof_config_I}
    \vspace{-3em}
    \caption{$T^2(\r)$}
    \label{connectivity_theorem_proof_config_I}
    \end{subfigure}
    \hspace{3em}
    \begin{subfigure}[b]{0.32\textwidth}
    \centering
    \include{figs/connectivity_theorem_proof_config_I_comparison}
    \vspace{-3em}
    \caption{$T_*^2$}
    \label{connectivity_theorem_proof_config_I_comparison}
    \end{subfigure}
    \caption{The comparison of the two spanning 2-trees, $T^2(\r)$ and $T_*^2$, in case~\textbf{I}, where each blob represents a tree subgraph. The two components of each spanning 2-tree is specified. The two blobs in the middle, which $v_A$ and $v_B$ belongs to, are contained in $\gamma_*$.}
    \label{figure-connectivity_theorem_proof_I}
    \end{figure}
    From $v_A$ to $q_j^\mu$ there is a unique path $P_A\subset t(\r;A)$, and since the edges attaching to $\gamma_*$ are all softer than $\mathscr{X}(\gamma_*)$, there must be an edge $e_1\in P_A$ with $w(e_1)<w(e_0)$. We can now construct the following spanning 2-tree (see figure~\ref{connectivity_theorem_proof_config_I_comparison}):
    \begin{eqnarray}
    T_*^2\equiv T^2(\r)\cup e_0 \setminus e_1.
    \end{eqnarray}
    The weight of $T_*^2$ then satisfies
    \begin{eqnarray}
    \label{eq:connectivity_theorem_proof_config_I_weight}
    w(T_*^2) \leqslant w(T^2(\r)) +w(e_1) -w(e_0)< w(T^2(\r)).
    \end{eqnarray}
    The first inequality follows because $Q_{T_*^2}^2\gtrsim Q_{T^2(\r)}^2$ (recall that $Q_{T^2}$ denotes the total momentum flowing between the components of $T^2$); therefore, the kinematic contribution to $w(T_*^2)$---which is the power of $\lambda$ in $Q_{T_*^2}^2$, see eq.~(\ref{eq:definition_spanning_1and2_trees_weights})---cannot exceed the kinematic contribution to $w(T^2(\r))$. A detailed justification is provided in appendix~\ref{appendix-comparing_kinematic_factors}. The second inequality uses $w(e_1) < w(e_0)$, which holds by construction. Inequality (\ref{eq:connectivity_theorem_proof_config_I_weight}) thus contradicts the minimum‑weight criterion, ruling out the presence of any off‑shell external momenta.
    \end{itemize}
    
    If no external momenta is off shell, there must exist four or more $p_i$ (because three or fewer external lightlike momenta cannot sum to zero in the wide-angle kinematics), and one component of $T^2(\r)$ must be attached by at least two of them. Now we suppose $p_1$ and $p_2$ are attached to $t(\r;A)$, and define two specific paths $P_1,P_2\subset t(\r;A)$, which connect $v_A$ to the external momenta $p_1$ and $p_2$ respectively. By construction, both paths start from $v_A$, possibly sharing some initial edges and vertices (which constitute a sub-path $P_1 \cap P_2$), before diverging and ending at $p_1$ and $p_2$. We now inspect whether or not $P_1\cap P_2$ is contained in $\gamma_*$ (equivalently, whether $P_1$ and $P_2$ separate inside or outside $\gamma_*$).
    
    \begin{itemize}    
    \item [\textbf{II.}] For $P_1\cap P_2\not\subset\gamma_*$, which means that $P_1$ and $P_2$ separate \emph{outside} $\gamma_*$, there must exist an edge $e_1\in P_1\cap P_2$ that is adjacent to $\gamma_*$, as is shown in figure~\ref{connectivity_theorem_proof_config_II}. By construction, $w(e_1)<w(e_0)$. We can thus follow the analysis in case~\textbf{I} above to construct another spanning 2-tree, $T_*^2\equiv T^2(\r)\cup e_0 \setminus e_1$ (see figure~\ref{connectivity_theorem_proof_config_II_comparison}), with $w(T_*^2)< w(T^2(\r))$ due to the same reason as~(\ref{eq:connectivity_theorem_proof_config_I_weight}). (Note that the only difference between this case and case \textbf{I} is that $q_j$ is replaced by $p_1$ and $p_2$.) The same violation of the minimum-weight criterion then excludes this possibility as well.
    \begin{figure}[t]
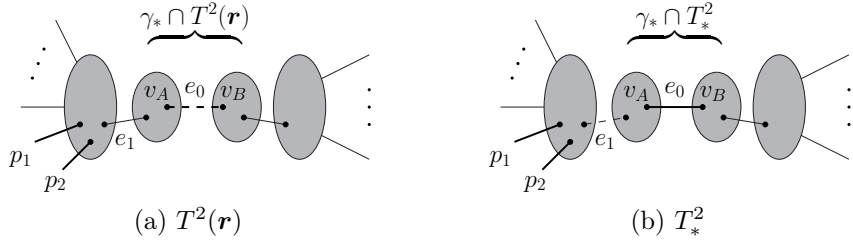

    \centering
    \hspace{-2em}
    \begin{subfigure}[b]{0.32\textwidth}
    \centering
    \include{figs/connectivity_theorem_proof_config_II}
    \vspace{-3em}
    \caption{$T^2(\r)$}
    \label{connectivity_theorem_proof_config_II}
    \end{subfigure}
    \hspace{3em}
    \begin{subfigure}[b]{0.32\textwidth}
    \centering
    \include{figs/connectivity_theorem_proof_config_II_comparison}
    \vspace{-3em}
    \caption{$T_*^2$}
    \label{connectivity_theorem_proof_config_II_comparison}
    \end{subfigure}
    \caption{The comparison of the two spanning 2-trees, $T^2(\r)$ and $T_*^2$, in case~\textbf{II}.}
    \label{figure-connectivity_theorem_proof_II}
    \end{figure}
    \end{itemize}
    
    From the preceding analyses, the remaining possible configurations of $T^2(\r)$ require that $P_1\cap P_2\subset \gamma_*$, i.e., the paths $P_1, P_2,\dots$ (possibly some other paths) separate at some vertex \emph{within} $\gamma_*$ before ending at $p_1,p_2,\dots$, respectively. To further rule out these configurations, we consider the location of the hard subgraph $\mathcal{H}$. First, we argue that $\mathcal{H}$ \emph{cannot} have nonzero intersections with both $t(\r;A)$ and $t(\r;B)$ simultaneously.
    
    \begin{itemize}
    \item [\textbf{III.}] If both $\mathcal{H}\cap t(\r;A)\neq \varnothing$ and $\mathcal{H}\cap t(\r;B)\neq \varnothing$, there is then an edge $e_H\in \mathcal{H}\setminus T^2(\r)$ connecting the two components of $T^2(\r)$. The spanning tree $T^1(\r_1)\equiv T^2(\r)\cup e_H$ has the weight
    \begin{eqnarray}
        w(T^1(\r_1)) = w(T^2(\r)) - w(e_H) - \mathscr{V}(Q_{T^2(\r)}) \leqslant w(T^2(\r)),
    \end{eqnarray}
    where we have used $w(e_H)=0$ and $\mathscr{V}(Q_{T^2(\r)})\in \mathbb{N}$ in the inequality. From the minimum-weight criterion, the equality must hold, immediately indicating: (1) $\mathscr{V}(Q_{T^2(\r)})=0$, i.e., the momentum flowing between the components of $T^2(\r)$ is $\mathcal{O}(1)$; (2) $\x^{\r_1}$ is a leading $\mathcal{U}$ term because $w(T^1(\r_1)) = w(T^2(\r))$.

    Let us then add $e_0$ to the spanning tree $T^1(\r_1)$ and obtain a graph $T^1(\r_1)\cup e_0$. From the defining property of spanning trees, there is a unique loop in this graph. This loop, by construction, simultaneously contains $e_H\in\mathcal{H}$ and $e_0\in \gamma_*$. Since $\mathcal{H}$ is not adjacent to $\gamma_*$ (recall that all the edges adjacent to $\gamma_*$ are softer than $\mathscr{X}(\gamma_*)$), this loop must also contain some edge $e_1$ that is adjacent to $\gamma_*$, with $w(e_1)<w(e_0)$. Then the following spanning tree $T^1(\r')\equiv T^1(\r_1)\cup e_0\setminus e_1$ has a smaller weight:
    \begin{eqnarray}
        w(T^1(\r')) = w(T^1(\r_1)) + w(e_1) - w(e_0) < w(T^1(\r_1)).
    \end{eqnarray}
    The minimum-weight criterion is again violated, ruling out this possibility.
    \end{itemize}
    Therefore, all the $\mathcal{H}$ vertices must simultaneously lie in either $t(\r;A)$ or $t(\r;B)$, and from now on, we assume they are in $t(\r;A)$.
    
    A more detailed depiction of $T^2(\r)$ is then shown in figure~\ref{connectivity_theorem_proof_config_IV}. Here, $m$ on-shell external momenta $p_1,\dots,p_m$ are attached to $t(\r;A)$, and $n$ others $p_{m+1},\dots, p_{m+n}$ to $t(\r;B)$, with $m+n\geqslant 4$. The $m$ paths in $t(\r;A)$ from $v_A$ to each $p_i$ ($1\leqslant i \leqslant m$) share no edges outside $\gamma_*$ for any two of them. Similarly, the $n$ paths in $t(\r;B)$ from $v_B$ to each $p_{m+i}$ ($1\leqslant i \leqslant n$) share no edges outside $\gamma_*$ for any two of them. The subgraph $\mathcal{H}\cap t(\r;A)$, which by lemma~\ref{lemma-hard_mode_component_tree_structure} is a tree (more precisely, a spanning tree of $\mathcal{H}$), forms part of $t(\r;A)$.\footnote{Here we have drawn $\mathcal{H}\cap t(\r;A)$ attached to $\gamma_*$. Actually, it could also be attached to one of the ``$p_i$ branches'' ($i=1,\dots,m$), but the subsequent analysis would be identical, so we do not treat this case separately.} We denote this scenario as the final case \textbf{IV} and analyze it below.
    \begin{figure}[t]
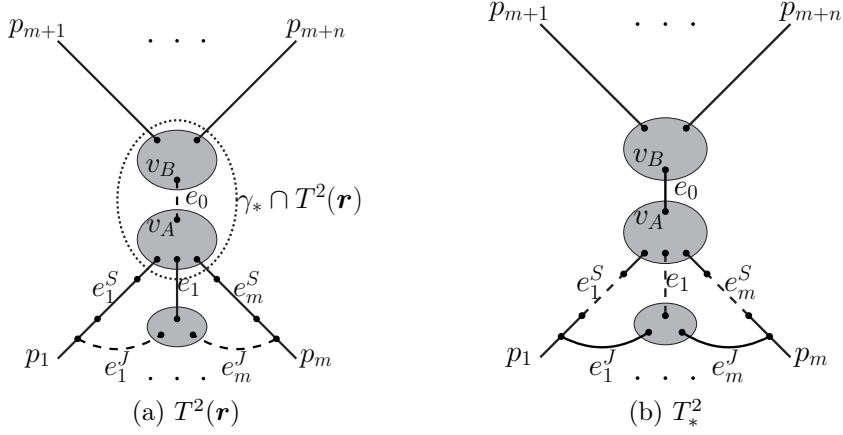

    \centering
    \hspace{-2em}
    \begin{subfigure}[b]{0.32\textwidth}
    \centering
    \include{figs/connectivity_theorem_proof_config_IV}
    \vspace{-3em}\caption{$T^2(\r)$}
    \label{connectivity_theorem_proof_config_IV}
    \end{subfigure}
    \hspace{3em}
    \begin{subfigure}[b]{0.32\textwidth}
    \centering
    \include{figs/connectivity_theorem_proof_config_IV_comparison}
    \vspace{-3em}\caption{$T_*^2$}
    \label{connectivity_theorem_proof_config_IV_comparison}
    \end{subfigure}
    \caption{The comparison of the two spanning 2-trees, $T^2(\r)$ and $T_*^2$, in case~\textbf{IV}. The subgraph $\gamma_* \cap T^2(\r)$ is enclosed by the dotted curve. The three gray blobs, from top to bottom, represent $\gamma_* \cap t(\r;B)$, $\gamma_* \cap t(\r;A)$, and $\mathcal{H} \cap t(\r;A)$, respectively.}
    \label{figure-connectivity_theorem_proof_IV}
    \end{figure}
    \begin{itemize}
    \item [\textbf{IV.}] For each external momentum $p_i$ ($i=1,\dots,m+n$), let $P_i$ be the path in $T^2(\r)$ that connects $p_i$ to a vertex $v_A\in\gamma_*$. We now demonstrate that at least two of these paths must contain edges from $\mathcal{S}$, by considering the possible locations of $\gamma_*$.
    \begin{itemize}
        \item $\gamma_* \subset \mathcal{S}$. Since any path in $\mathcal{G}$ connecting an external momentum $p_i$ to a vertex in $\gamma_*$ must include a soft edge adjacent to $\gamma_*$, in this case, all paths $P_1, \dots, P_{m+n}$ contain edges from $\mathcal{S}$.
        \item $\gamma_* \subset \mathcal{J}_{i_0}$ for a specific $i_0$. In this case, a path connecting $p_i$ (for any $i\neq i_0$) to $\gamma_*$ must either (1) pass through the hard subgraph $\mathcal{H}$, or (2) contain a soft edge. Since $\mathcal{H} \cap t(\r;A)$ is connected (as established previously), at most one of these paths can intersect $\mathcal{H}$. With three or more such paths in the set $\{P_i\}_{i \neq i_0}$, this means at least two of them must fulfill condition (2) and therefore contain soft edges.
    \end{itemize}
    We thus conclude from both scenarios that at least two paths $P_i$ contain soft edges.
    
    In figure~\ref{connectivity_theorem_proof_config_IV}, let us assume that $P_1$ and $P_m$ are two of these paths (the subsequent logic can apply identically to the other type of choices, i.e., $P_1$ and $P_{m+1}$), with $e_1^S$ and $e_m^S$ being the soft edges that are adjacent to $\mathcal{J}_1$ and $\mathcal{J}_m$ respectively. In the entire graph $\mathcal{G}$, consider the path in $\mathcal{J}_1$ which carries the soft momentum flow from $e_1^S$ to $\mathcal{H}$. On this path, there exists an edge $e_1^J\in \mathcal{J}_1\setminus T^2(\r)$, whose endpoints are in the $P_1$ and $\mathcal{H}\cap t(\r;A)$ respectively. Similarly, there exists an edge $e_m^J\in \mathcal{J}_m\setminus T^2(\r)$, whose endpoints are in $P_m$ and $\mathcal{H}\cap t(\r;A)$ respectively. Since $e_1^J$ ($e_m^J$) is a jet edge carrying the soft momentum in $e_1^S$ ($e_m^S$), we have two inequalities $w(e_1^S)<w(e_1^J)$ and $w(e_m^S)<w(e_m^J)$.
    
    Meanwhile, consider the path in $t(\r;A)$ which connects $\gamma_*\cap t(\r;A)$ and $\mathcal{H}\cap t(\r;A)$, which must contain an edge $e_1$ such that $w(e_1)<w(e_0)$. (This follows from the same reasoning as cases \textbf{I}-\textbf{III}.) We can then consider the following spanning 2-tree
    \begin{eqnarray}
        T_*^2\equiv T^2(\r)\cup (e_0\cup e_1^J\cup e_m^J) \setminus (e_1\cup e_1^S\cup e_m^S),
    \end{eqnarray}
    as is shown in figure~\ref{connectivity_theorem_proof_config_IV_comparison}, with the momentum between its two components being $p_1+p_m$, which is off shell. Its weight satisfies
    \begin{align}
        w(T_*^2) &\leqslant w(T^2(\r)) +w(e') +w(e_1^S) + w(e_2^S) - w(e_0) - w(e_1^J) - w(e_2^J) \nonumber\\
        &< w(T^2(\r)).
    \end{align}
    In the first inequality, we have used that $(p_1+p_m)^2\sim \mathcal{O}(1)$, implying that the kinematic contribution to $w(T_*^2)$ is zero; meanwhile, the kinematic contribution to $w(T^2(\r))$ is either zero or positive by definition. In the second inequality, we have employed the previously derived inequalities, $w(e')<w(e_0)$, $w(e_1^S)<w(e_1^J)$, and $w(e_2^S)<w(e_2^J)$. Therefore, the minimum-weight criterion is also violated in this case.
    \end{itemize}
    
    In conclusion, we have excluded all possible configurations of $T^2(\r)$ under the assumption that $T(\r)\cap\gamma_*$ contains multiple connected components, thus proving the statement~(\ref{eq:mode_subgraphs_connectivity_theorem_center_statement}).
\end{proof}
With $T(\r)\cap\gamma_*$ confirmed a spanning tree for any leading term $\x^{\r}$, a homogeneity condition from eq.~(\ref{eq:theorem_connectivity_proof_homogeneity}) arises, imposing as an additional reduction of the dimension of $f_R$ and making the expanded integral scaleless. Therefore, such a $\gamma_*$ cannot exist, and we have finally proved the First Connectivity Theorem.

Let us end this subsection by going through an example to see how our reasoning above works. Consider the aforementioned configuration in figure~\ref{configurations_violating_first_connectivity_theorem2}, in which the First Connectivity Theorem is violated due to the $S$ loop in the middle. By denoting this $S$ component as $\gamma_*$, one can verify through direct inspection that the graph $T(\r)\cap \gamma_*$ is a spanning tree of $\gamma_*$ for any leading term $\x^{\r}$; thus, a homogeneity condition arises and its corresponding expanded integral is scaleless. One can also see the tree structure of $T(\r)\cap \gamma_*$ without examining all the leading terms. Suppose there exists a leading $\mathcal{F}$ term $\x^{\r}$, for which the graph $T(\r)\cap \gamma_*$ is disconnected. One can then construct another spanning 2-tree with a smaller weight, thereby violating the minimal-weight criterion. For example,
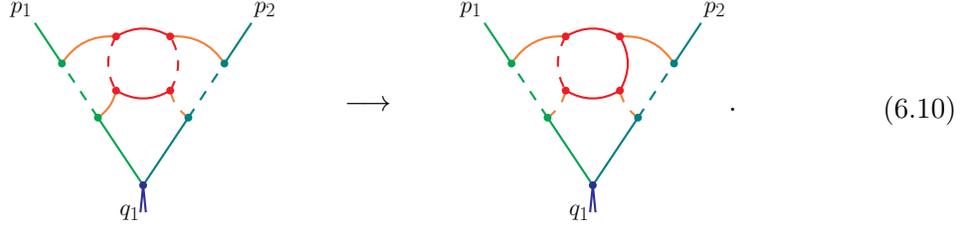
\begin{equation}
\begin{aligned}
    \begin{tikzpicture}[line width = 0.6, scale = 0.36, font=\large, mydot/.style={circle, fill, inner sep=.7pt}, transform shape, baseline=(current bounding box.center)]
    \draw (5,2) edge [thick, Green] (3.33,4.5) node [] {};
    \draw (3.33,4.5) edge [dash pattern=on 4pt off 4pt, thick, Green] (2,6.5) node [] {};
    \draw (2,6.5) edge [thick, Green] (1,8) node [] {};
    \draw (5,2) edge [thick, teal] (6.67,4.5) node [] {};
    \draw (6.67,4.5) edge [dash pattern=on 4pt off 4pt, thick, teal] (8,6.5) node [] {};
    \draw (8,6.5) edge [thick, teal] (9,8) node [] {};
    \draw (5,2) edge [thick, Blue] (4.9,1) node [] {};
    \draw (5,2) edge [thick,Blue] (5.1,1) node [] {};
    \draw (4,7.5) edge [dash pattern=on 4pt off 4pt, thick, Red, bend right = 30] (4,5.5) node [] {};
    \draw (4,5.5) edge [thick, Red, bend right = 30] (6,5.5) node [] {};
    \draw (6,5.5) edge [dash pattern=on 4pt off 4pt, thick, Red, bend right = 30] (6,7.5) node [] {};
    \draw (6,7.5) edge [thick, Red, bend right = 30] (4,7.5) node [] {};
    \draw (2,6.5) edge [thick, Orange, bend left = 30] (4,7.5) node [] {};
    \draw (3.33,4.5) edge [thick, Orange, bend right = 20] (4,5.5) node [] {};
    \draw (8,6.5) edge [thick, Orange, bend right = 30] (6,7.5) node [] {};
    \draw (6.67,4.5) edge [dash pattern=on 4pt off 4pt, thick, Orange, bend left = 20] (6,5.5) node [] {};    
    \node () at (0.5,8.5) {\Huge $p_1$};
    \node () at (9.5,8.5) {\Huge $p_2$};
    \node () at (4.5,1) {\Huge $q_1$};    
    \draw[fill, thick, Green] (2,6.5) circle (3pt);
    \draw[fill, thick, Green] (3.33,4.5) circle (3pt);
    \draw[fill, thick, teal] (8,6.5) circle (3pt);
    \draw[fill, thick, teal] (6.67,4.5) circle (3pt);
    \draw[fill, Blue, thick] (5,2) circle (3pt);
    \draw[fill, Red, thick] (4,5.5) circle (3pt);
    \draw[fill, Red, thick] (4,7.5) circle (3pt);
    \draw[fill, Red, thick] (6,5.5) circle (3pt);
    \draw[fill, Red, thick] (6,7.5) circle (3pt);
    \end{tikzpicture}
    \qquad\longrightarrow\qquad
    \begin{tikzpicture}[line width = 0.6, scale = 0.36, font=\large, mydot/.style={circle, fill, inner sep=.7pt}, transform shape, baseline=(current bounding box.center)]
    \draw (5,2) edge [thick, Green] (3.33,4.5) node [] {};
    \draw (3.33,4.5) edge [dash pattern=on 4pt off 4pt, thick, Green] (2,6.5) node [] {};
    \draw (2,6.5) edge [thick, Green] (1,8) node [] {};
    \draw (5,2) edge [thick, teal] (6.67,4.5) node [] {};
    \draw (6.67,4.5) edge [dash pattern=on 4pt off 4pt, thick, teal] (8,6.5) node [] {};
    \draw (8,6.5) edge [thick, teal] (9,8) node [] {};
    \draw (5,2) edge [thick, Blue] (4.9,1) node [] {};
    \draw (5,2) edge [thick,Blue] (5.1,1) node [] {};
    \draw (4,7.5) edge [dash pattern=on 4pt off 4pt, thick, Red, bend right = 30] (4,5.5) node [] {};
    \draw (4,5.5) edge [thick, Red, bend right = 30] (6,5.5) node [] {};
    \draw (6,5.5) edge [thick, Red, bend right = 30] (6,7.5) node [] {};
    \draw (6,7.5) edge [thick, Red, bend right = 30] (4,7.5) node [] {};
    \draw (2,6.5) edge [thick, Orange, bend left = 30] (4,7.5) node [] {};
    \draw (3.33,4.5) edge [dash pattern=on 4pt off 4pt, thick, Orange, bend right = 20] (4,5.5) node [] {};
    \draw (8,6.5) edge [thick, Orange, bend right = 30] (6,7.5) node [] {};
    \draw (6.67,4.5) edge [dash pattern=on 4pt off 4pt, thick, Orange, bend left = 20] (6,5.5) node [] {};    
    \node () at (0.5,8.5) {\Huge $p_1$};
    \node () at (9.5,8.5) {\Huge $p_2$};
    \node () at (4.5,1) {\Huge $q_1$};    
    \draw[fill, thick, Green] (2,6.5) circle (3pt);
    \draw[fill, thick, Green] (3.33,4.5) circle (3pt);
    \draw[fill, thick, teal] (8,6.5) circle (3pt);
    \draw[fill, thick, teal] (6.67,4.5) circle (3pt);
    \draw[fill, Blue, thick] (5,2) circle (3pt);
    \draw[fill, Red, thick] (4,5.5) circle (3pt);
    \draw[fill, Red, thick] (4,7.5) circle (3pt);
    \draw[fill, Red, thick] (6,5.5) circle (3pt);
    \draw[fill, Red, thick] (6,7.5) circle (3pt);
    \end{tikzpicture}.
\end{aligned}
\end{equation}
This modification is precisely from the analysis of case {\bf{I}} in our proof of statement~(\ref{eq:mode_subgraphs_connectivity_theorem_center_statement}). As a result, such an $\x^{\r}$ does not exist in the leading polynomial.

\subsection{Structures of leading terms}
\label{section-leading_terms}

To further investigate region structures, we need to understand the expressions of the leading terms in an arbitrary virtuality expansion. Generally, the number of leading terms with respect to a given region can be large, but as we will show, they can all be characterized graphically in terms of the spanning (2‑)tree structures of the mode components.

\begin{figure}[t]
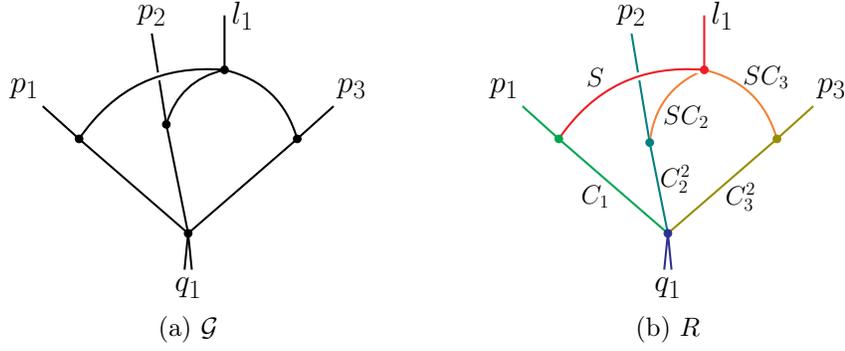

\centering
\hspace{-2em}
\begin{subfigure}[b]{0.32\textwidth}
\centering
\include{figs/leading_terms_motivating_example_graph}
\vspace{-3em}
\caption{$\mathcal{G}$}
\label{leading_terms_motivating_example_graph}
\end{subfigure}
\hspace{3em}
\begin{subfigure}[b]{0.32\textwidth}
\centering
\include{figs/leading_terms_motivating_example_region}
\vspace{-3em}
\caption{$R$}
\label{leading_terms_motivating_example_region}
\end{subfigure}
\caption{A two-loop graph $\mathcal{G}$ and one of its regions $R$.}
\label{figure-leading_terms_motivating_example}
\end{figure}

We begin with the two‑loop example in figure~\ref{figure-leading_terms_motivating_example}, where $\mathcal{G}$ is the Feynman graph and the external momenta $p_1$, $p_2$, $p_3$, $q_1$, $l_1$ are in the modes $C_1$, $C_2^2$, $C_3^2$, $H$, and $S$, respectively. For this kinematics, $R$ is one of the admissible regions (as can be verified by explicit computation). The modes of the internal edges are indicated by color: $C_1$ edges are {\color{Green}\bf green}, $C_2^2$ edges are {\color{teal}\bf teal}, $C_3^2$ edges are {\color{olive}\bf olive}, $S$ edges are {\color{Red}\bf red}, and $SC_1$ and $SC_2$ edges are {\color{Orange}\bf orange}.

From now on, we will use the superscript ``$(R)$'' to denote terms that are leading in the region $R$. That is, $\mathcal{U}^{(R)}$ and $\mathcal{F}^{(R)}$ represent the leading $\mathcal{U}$ and $\mathcal{F}$ polynomials, respectively.

We now consider the leading terms associated with $R$ and list their corresponding spanning (2-)trees. There is a unique $\mathcal{U}^{(R)}$ term,
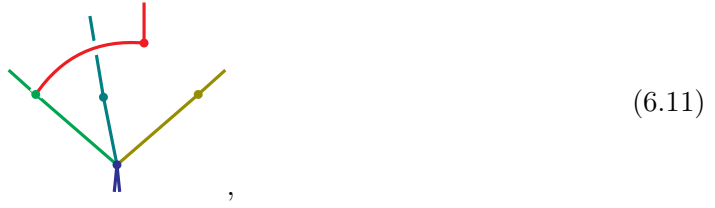
\begin{equation}
\label{eq:leading_terms_motivating_example_region_Uterm}
\begin{aligned}
    \begin{tikzpicture}[line width = 0.6, scale = 0.36, font=\large, mydot/.style={circle, fill, inner sep=.7pt}, transform shape]
    \draw (5,2.5) edge [very thick, Green] (2,5.1) node [] {};
    \draw (2,5.1) edge [very thick, Green] (1,6) node [] {};
    \draw (5,2.5) edge [very thick, teal] (4.5,5) node [] {};
    \draw (4.5,5) edge [very thick, teal] (4,8) node [] {};
    \draw (5,2.5) edge [very thick, olive] (8,5.1) node [] {};
    \draw (8,5.1) edge [very thick, olive] (9,6) node [] {};
    \draw (6,7) edge [very thick, draw=white, double=white, double distance=3pt, bend right = 30] (2,5.1) node [] {};\draw (6,7) edge [very thick, Red, bend right = 30] (2,5.1) node [] {};
    \draw (6,7) edge [very thick, Red] (6,8.5) node [] {};
    \draw (5.1,1.5) edge [very thick, Blue] (5,2.5) node [] {};
    \draw (4.9,1.5) edge [very thick, Blue] (5,2.5) node [] {};    
    \draw[fill, very thick, Green] (2,5.1) circle (3pt);
    \draw[fill, very thick, teal] (4.5,5) circle (3pt);
    \draw[fill, very thick, olive] (8,5.1) circle (3pt);
    \draw[fill, very thick, Red] (6,7) circle (3pt);
    \draw[fill, very thick, Blue] (5,2.5) circle (3pt);    
    \end{tikzpicture},
\end{aligned}
\end{equation}
together with nine $\mathcal{F}^{(R)}$ terms,
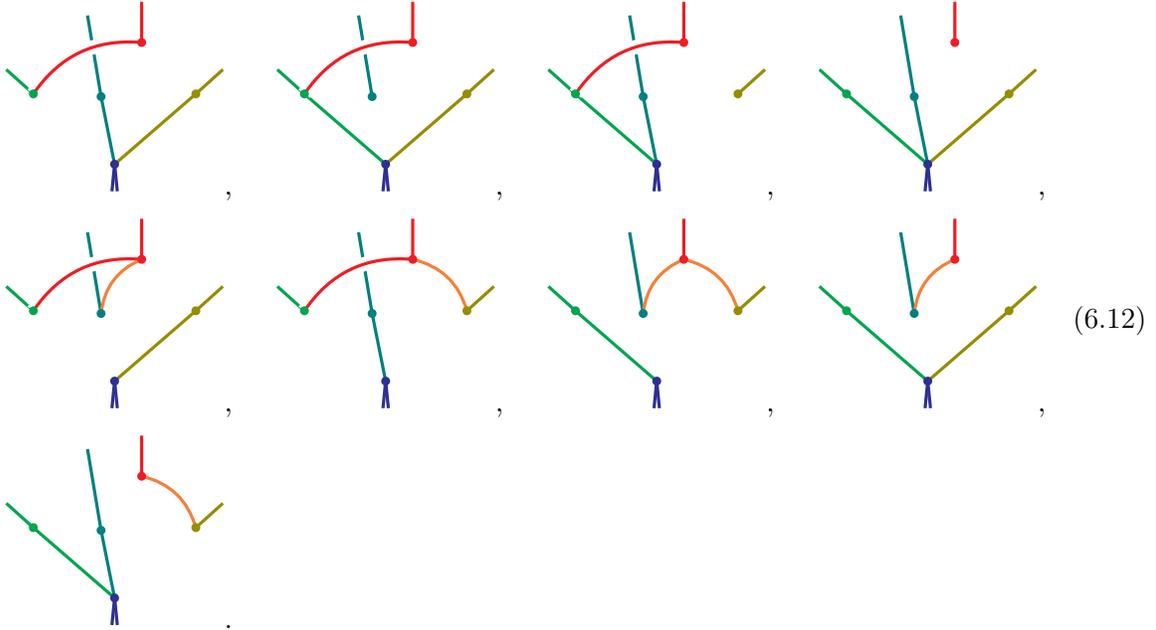
\begin{equation}
\label{eq:leading_terms_motivating_example_region_Fterms}
\begin{aligned}
    &\begin{tikzpicture}[line width = 0.6, scale = 0.36, font=\large, mydot/.style={circle, fill, inner sep=.7pt}, transform shape]
    \draw (2,5.1) edge [very thick, Green] (1,6) node [] {};
    \draw (5,2.5) edge [very thick, teal] (4.5,5) node [] {};
    \draw (4.5,5) edge [very thick, teal] (4,8) node [] {};
    \draw (5,2.5) edge [very thick, olive] (8,5.1) node [] {};
    \draw (8,5.1) edge [very thick, olive] (9,6) node [] {};
    \draw (6,7) edge [very thick, draw=white, double=white, double distance=3pt, bend right = 30] (2,5.1) node [] {};\draw (6,7) edge [very thick, Red, bend right = 30] (2,5.1) node [] {};
    \draw (6,7) edge [very thick, Red] (6,8.5) node [] {};
    \draw (5.1,1.5) edge [very thick, Blue] (5,2.5) node [] {};
    \draw (4.9,1.5) edge [very thick, Blue] (5,2.5) node [] {};    
    \draw[fill, very thick, Green] (2,5.1) circle (3pt);
    \draw[fill, very thick, teal] (4.5,5) circle (3pt);
    \draw[fill, very thick, olive] (8,5.1) circle (3pt);
    \draw[fill, very thick, Red] (6,7) circle (3pt);
    \draw[fill, very thick, Blue] (5,2.5) circle (3pt);    
    \end{tikzpicture},\quad\ 
    \begin{tikzpicture}[line width = 0.6, scale = 0.36, font=\large, mydot/.style={circle, fill, inner sep=.7pt}, transform shape]
    \draw (5,2.5) edge [very thick, Green] (2,5.1) node [] {};
    \draw (2,5.1) edge [very thick, Green] (1,6) node [] {};
    \draw (4.5,5) edge [very thick, teal] (4,8) node [] {};
    \draw (5,2.5) edge [very thick, olive] (8,5.1) node [] {};
    \draw (8,5.1) edge [very thick, olive] (9,6) node [] {};
    \draw (6,7) edge [very thick, draw=white, double=white, double distance=3pt, bend right = 30] (2,5.1) node [] {};\draw (6,7) edge [very thick, Red, bend right = 30] (2,5.1) node [] {};
    \draw (6,7) edge [very thick, Red] (6,8.5) node [] {};
    \draw (5.1,1.5) edge [very thick, Blue] (5,2.5) node [] {};
    \draw (4.9,1.5) edge [very thick, Blue] (5,2.5) node [] {};    
    \draw[fill, very thick, Green] (2,5.1) circle (3pt);
    \draw[fill, very thick, teal] (4.5,5) circle (3pt);
    \draw[fill, very thick, olive] (8,5.1) circle (3pt);
    \draw[fill, very thick, Red] (6,7) circle (3pt);
    \draw[fill, very thick, Blue] (5,2.5) circle (3pt);    
    \end{tikzpicture},\quad\ 
    \begin{tikzpicture}[line width = 0.6, scale = 0.36, font=\large, mydot/.style={circle, fill, inner sep=.7pt}, transform shape]
    \draw (5,2.5) edge [very thick, Green] (2,5.1) node [] {};
    \draw (2,5.1) edge [very thick, Green] (1,6) node [] {};
    \draw (5,2.5) edge [very thick, teal] (4.5,5) node [] {};
    \draw (4.5,5) edge [very thick, teal] (4,8) node [] {};
    \draw (8,5.1) edge [very thick, olive] (9,6) node [] {};
    \draw (6,7) edge [very thick, draw=white, double=white, double distance=3pt, bend right = 30] (2,5.1) node [] {};\draw (6,7) edge [very thick, Red, bend right = 30] (2,5.1) node [] {};
    \draw (6,7) edge [very thick, Red] (6,8.5) node [] {};
    \draw (5.1,1.5) edge [very thick, Blue] (5,2.5) node [] {};
    \draw (4.9,1.5) edge [very thick, Blue] (5,2.5) node [] {};    
    \draw[fill, very thick, Green] (2,5.1) circle (3pt);
    \draw[fill, very thick, teal] (4.5,5) circle (3pt);
    \draw[fill, very thick, olive] (8,5.1) circle (3pt);
    \draw[fill, very thick, Red] (6,7) circle (3pt);
    \draw[fill, very thick, Blue] (5,2.5) circle (3pt);    
    \end{tikzpicture},\quad\ 
    \begin{tikzpicture}[line width = 0.6, scale = 0.36, font=\large, mydot/.style={circle, fill, inner sep=.7pt}, transform shape]
    \draw (5,2.5) edge [very thick, Green] (2,5.1) node [] {};
    \draw (2,5.1) edge [very thick, Green] (1,6) node [] {};
    \draw (5,2.5) edge [very thick, teal] (4.5,5) node [] {};
    \draw (4.5,5) edge [very thick, teal] (4,8) node [] {};
    \draw (5,2.5) edge [very thick, olive] (8,5.1) node [] {};
    \draw (8,5.1) edge [very thick, olive] (9,6) node [] {};
    \draw (6,7) edge [very thick, Red] (6,8.5) node [] {};
    \draw (5.1,1.5) edge [very thick, Blue] (5,2.5) node [] {};
    \draw (4.9,1.5) edge [very thick, Blue] (5,2.5) node [] {};    
    \draw[fill, very thick, Green] (2,5.1) circle (3pt);
    \draw[fill, very thick, teal] (4.5,5) circle (3pt);
    \draw[fill, very thick, olive] (8,5.1) circle (3pt);
    \draw[fill, very thick, Red] (6,7) circle (3pt);
    \draw[fill, very thick, Blue] (5,2.5) circle (3pt);    
    \end{tikzpicture},\\
    &\begin{tikzpicture}[line width = 0.6, scale = 0.36, font=\large, mydot/.style={circle, fill, inner sep=.7pt}, transform shape]
    \draw (2,5.1) edge [very thick, Green] (1,6) node [] {};
    \draw (4.5,5) edge [very thick, teal] (4,8) node [] {};
    \draw (5,2.5) edge [very thick, olive] (8,5.1) node [] {};
    \draw (8,5.1) edge [very thick, olive] (9,6) node [] {};
    \draw (6,7) edge [very thick, draw=white, double=white, double distance=3pt, bend right = 30] (2,5.1) node [] {};\draw (6,7) edge [very thick, Red, bend right = 30] (2,5.1) node [] {};
    \draw (6,7) edge [very thick, Orange, bend right = 30] (4.5,5) node [] {};
    \draw (6,7) edge [very thick, Red] (6,8.5) node [] {};
    \draw (5.1,1.5) edge [very thick, Blue] (5,2.5) node [] {};
    \draw (4.9,1.5) edge [very thick, Blue] (5,2.5) node [] {};    
    \draw[fill, very thick, Green] (2,5.1) circle (3pt);
    \draw[fill, very thick, teal] (4.5,5) circle (3pt);
    \draw[fill, very thick, olive] (8,5.1) circle (3pt);
    \draw[fill, very thick, Red] (6,7) circle (3pt);
    \draw[fill, very thick, Blue] (5,2.5) circle (3pt);    
    \end{tikzpicture},\quad\ 
    \begin{tikzpicture}[line width = 0.6, scale = 0.36, font=\large, mydot/.style={circle, fill, inner sep=.7pt}, transform shape]
    \draw (2,5.1) edge [very thick, Green] (1,6) node [] {};
    \draw (5,2.5) edge [very thick, teal] (4.5,5) node [] {};
    \draw (4.5,5) edge [very thick, teal] (4,8) node [] {};
    \draw (8,5.1) edge [very thick, olive] (9,6) node [] {};
    \draw (6,7) edge [very thick, draw=white, double=white, double distance=3pt, bend right = 30] (2,5.1) node [] {};\draw (6,7) edge [very thick, Red, bend right = 30] (2,5.1) node [] {};
    \draw (6,7) edge [very thick, Orange, bend left = 30] (8,5.1) node [] {};
    \draw (6,7) edge [very thick, Red] (6,8.5) node [] {};
    \draw (5.1,1.5) edge [very thick, Blue] (5,2.5) node [] {};
    \draw (4.9,1.5) edge [very thick, Blue] (5,2.5) node [] {};    
    \draw[fill, very thick, Green] (2,5.1) circle (3pt);
    \draw[fill, very thick, teal] (4.5,5) circle (3pt);
    \draw[fill, very thick, olive] (8,5.1) circle (3pt);
    \draw[fill, very thick, Red] (6,7) circle (3pt);
    \draw[fill, very thick, Blue] (5,2.5) circle (3pt);    
    \end{tikzpicture},\quad\ 
    \begin{tikzpicture}[line width = 0.6, scale = 0.36, font=\large, mydot/.style={circle, fill, inner sep=.7pt}, transform shape]
    \draw (5,2.5) edge [very thick, Green] (2,5.1) node [] {};
    \draw (2,5.1) edge [very thick, Green] (1,6) node [] {};
    \draw (4.5,5) edge [very thick, teal] (4,8) node [] {};
    \draw (8,5.1) edge [very thick, olive] (9,6) node [] {};
    \draw (6,7) edge [very thick, Orange, bend right = 30] (4.5,5) node [] {};
    \draw (6,7) edge [very thick, Orange, bend left = 30] (8,5.1) node [] {};
    \draw (6,7) edge [very thick, Red] (6,8.5) node [] {};
    \draw (5.1,1.5) edge [very thick, Blue] (5,2.5) node [] {};
    \draw (4.9,1.5) edge [very thick, Blue] (5,2.5) node [] {};    
    \draw[fill, very thick, Green] (2,5.1) circle (3pt);
    \draw[fill, very thick, teal] (4.5,5) circle (3pt);
    \draw[fill, very thick, olive] (8,5.1) circle (3pt);
    \draw[fill, very thick, Red] (6,7) circle (3pt);
    \draw[fill, very thick, Blue] (5,2.5) circle (3pt);    
    \end{tikzpicture},\quad\ 
    \begin{tikzpicture}[line width = 0.6, scale = 0.36, font=\large, mydot/.style={circle, fill, inner sep=.7pt}, transform shape]
    \draw (5,2.5) edge [very thick, Green] (2,5.1) node [] {};
    \draw (2,5.1) edge [very thick, Green] (1,6) node [] {};
    \draw (4.5,5) edge [very thick, teal] (4,8) node [] {};
    \draw (5,2.5) edge [very thick, olive] (8,5.1) node [] {};
    \draw (8,5.1) edge [very thick, olive] (9,6) node [] {};
    \draw (6,7) edge [very thick, Orange, bend right = 30] (4.5,5) node [] {};
    \draw (6,7) edge [very thick, Red] (6,8.5) node [] {};
    \draw (5.1,1.5) edge [very thick, Blue] (5,2.5) node [] {};
    \draw (4.9,1.5) edge [very thick, Blue] (5,2.5) node [] {};    
    \draw[fill, very thick, Green] (2,5.1) circle (3pt);
    \draw[fill, very thick, teal] (4.5,5) circle (3pt);
    \draw[fill, very thick, olive] (8,5.1) circle (3pt);
    \draw[fill, very thick, Red] (6,7) circle (3pt);
    \draw[fill, very thick, Blue] (5,2.5) circle (3pt);    
    \end{tikzpicture},\\
    &\begin{tikzpicture}[line width = 0.6, scale = 0.36, font=\large, mydot/.style={circle, fill, inner sep=.7pt}, transform shape]
    \draw (5,2.5) edge [very thick, Green] (2,5.1) node [] {};
    \draw (2,5.1) edge [very thick, Green] (1,6) node [] {};
    \draw (5,2.5) edge [very thick, teal] (4.5,5) node [] {};
    \draw (4.5,5) edge [very thick, teal] (4,8) node [] {};
    \draw (8,5.1) edge [very thick, olive] (9,6) node [] {};
    \draw (6,7) edge [very thick, Orange, bend left = 30] (8,5.1) node [] {};
    \draw (6,7) edge [very thick, Red] (6,8.5) node [] {};
    \draw (5.1,1.5) edge [very thick, Blue] (5,2.5) node [] {};
    \draw (4.9,1.5) edge [very thick, Blue] (5,2.5) node [] {};    
    \draw[fill, very thick, Green] (2,5.1) circle (3pt);
    \draw[fill, very thick, teal] (4.5,5) circle (3pt);
    \draw[fill, very thick, olive] (8,5.1) circle (3pt);
    \draw[fill, very thick, Red] (6,7) circle (3pt);
    \draw[fill, very thick, Blue] (5,2.5) circle (3pt);    
    \end{tikzpicture}.
\end{aligned}
\end{equation}
This example provides a first indication that the $\mathcal{F}^{(R)}$ terms are intrinsically more intricate than the $\mathcal{U}^{(R)}$ terms. Later, we will further classify the $\mathcal{F}^{(R)}$ terms and show how each class can be generally characterized by the mode components of $R$.

\subsubsection{\texorpdfstring{$\mathcal{U}^{(R)}$ terms}{Leading U terms}}

We start by considering general $\mathcal{U}^{(R)}$ terms. As they correspond to the minimum spanning trees of $\mathcal{G}$ where the weight of each edge is given by $\boldsymbol{v}_R$, we can combine theorems~\ref{theorem-weight_hierarchical_partition_tree_structure} and~\ref{theorem-mode_subgraphs_connectivity1} (both proved at this point) to obtain the following:
\begin{lemma}
\label{lemma-leading_Uterms_subgraph_characterization}
    If $\x^{\r}$ is a $\mathcal{U}^{(R)}$ term, then each graph $\gamma\cap T^1(\r)$, where $\gamma$ denotes any mode component, is a spanning tree of $\widetilde{\gamma}$. That is,
    \begin{eqnarray}
    \mathcal{U}^{(R)}:\ \mathfrak{n}_{\gamma}^{}(\r) = L(\widetilde{\gamma}) \text{ for any mode component }\gamma\,.
    \label{eq:leading_Uterms_subgraph_characterization}
    \end{eqnarray}
\end{lemma}
Recall that $\mathfrak{n}_{\gamma}^{}(\r)$ denotes the degree of those parameters associated with the edges of $\gamma$ in $\x^{\r}$; or equivalently, the number of edges of $\gamma$ that are absent in the spanning (2-)tree $T(\r)$.

Conversely, by combining an arbitrary choice of spanning trees from all mode components, we obtain a minimum spanning tree of $\mathcal{G}$, which corresponds to a $\mathcal{U}^{(R)}$ term. To formalize this, let $\x^{[\gamma]}$ denote the parameters associated with the edges of $\gamma$. Then the leading polynomial $\mathcal{U}^{(R)}(\x)$ can be factorized as follows:
\begin{eqnarray}
    \mathcal{U}^{(R)}(\x) = \prod_\gamma \mathcal{U}_\gamma(\x^{[\gamma]})\,,
\label{eq:leading_Uterms_factorize}
\end{eqnarray}
where $\mathcal{U}_\gamma(\x^{[\gamma]})$ is the polynomial containing only $\x^{[\gamma]}$ and corresponding to a certain spanning tree of $\widetilde{\gamma}$. This factorization identity has been discussed previously in ref.~\cite{GrdHzgJnsMaSchlk22}.

\begin{corollary}
\label{lemma-leading_Uterms_subgraph_characterization_corollary1}
    If $\x^{\r}$ is a $\mathcal{U}^{(R)}$ term, then along the unique path in $T^1(\r)$ from any vertex $v_1\in \mathcal{G}$ to any hard vertex $v_2\in \mathcal{H}$, any two adjacent elements (edges and vertices) have either identical or harder modes.
\end{corollary}
\begin{proof}
    We prove it by contradiction. Suppose this is not the case, then from some point on this path, the corresponding mode will first get softer, possibly stay the same for the next few elements, and then becomes harder. This implies the existence of a mode component $\gamma\subset \mathcal{G}$, whose subgraph $\gamma\cap T^1(\r)$ is adjacent to two harder-mode vertices. Then by definition, $\widetilde{\gamma}\cap T^1(\r)$ would contain at least a loop, contradicting the statement of lemma~\ref{lemma-leading_Uterms_subgraph_characterization}.
\end{proof}
As we will see soon, this corollary is useful for the study of $\mathcal{F}^{(R)}$-term structures.

\subsubsection{\texorpdfstring{Classification of $\mathcal{F}^{(R)}$ terms}{Classification of leading F terms}}

In contrast to the previous cases, the structure of a generic spanning 2-tree $T^2$ corresponding to an $\mathcal{F}^{(R)}$ term is significantly more complex. To analyze it systematically, we first classify general $\mathcal{F}^{(R)}$ terms into two types, based on the following key property regarding the edges connecting the two components of $T^2$.\footnote{By ``edges connecting the two components of $T^2$'', we refer to those edges whose endpoints are in the two components of $T^2$, respectively. These edges themselves do not belong to $T^2$ by definition.}
\begin{lemma}
\label{lemma-leading_Fterms_connecting_edges_weight_property}
    For any edge $e\in \mathcal{G}\setminus T^2$ whose endpoints are in the two components of $T^2$ respectively, its weight must satisfy $w(e) \leqslant -\mathscr{V}(Q_{T^2})$, where $Q_{T^2}$ denotes the momentum flowing between the two components of $T^2$. If $w(e) = -\mathscr{V}(Q_{T^2})$, one can sharpen it into $\mathscr{X}(e) = \mathscr{X}(Q_{T^2})$.
\end{lemma}
\begin{proof}
    By definition, the graph $T^2\cup e$ is a spanning tree of $\mathcal{G}$, which must correspond to a $\mathcal{U}$ term. This $\mathcal{U}$ term has the weight
    \begin{eqnarray}
        w(T^2) - w(e) - \mathscr{V}(Q_{T^2}).
    \end{eqnarray}
    Based on the minimum-weight criterion, the expression above must be greater than or equal to $w(T^2)$. This directly leads to $w(e) \leqslant -\mathscr{V}(Q_{T^2})$.

    Now we derive $\mathscr{X}(e) = \mathscr{X}(Q_{T^2})$ from $w(e) = -\mathscr{V}(Q_{T^2})$. This is immediate if $Q_{T^2}$ is hard. To see this for other modes of $Q_{T^2}$, first notice that from above, there cannot be hard edges connecting the components of $T^2$; in other words, all the hard vertices are in the same component of $T^2$. For the component containing no hard vertices, the external momenta attached to it sum to $Q_{T^2}$ before entering the hard scattering, thus some vertex in this component, denoted by $v_1$, must have $\mathscr{X}(v_1) = \mathscr{X}(Q_{T^2})$. Now, consider the spanning tree $T^1 = T^2\cup e$, which corresponds to a $\mathcal{U}^{(R)}$ term since its weight achieves the minimum value. Then according to lemma~\ref{lemma-leading_Uterms_subgraph_characterization_corollary1}, on the path in $T^1$ joining $v_1$ and any hard vertex $v_2$ (which, by construction, contains $e$), as we go from $v_1$ to $v_2$, the modes of the successive vertices and edges are either identical or getting harder. Since $w(e) = -\mathscr{V}(Q_{T^2})$, the only allowed mode for $\mathscr{X}(e)$ is precisely $\mathscr{X}(Q_{T^2})$.
\end{proof}

Building on this observation, we classify general $\mathcal{F}^{(R)}$ terms into the following two types.
\begin{itemize}
    \item {\bf{Type-I (}$\boldsymbol{\mathcal{F}_\textup{I}^{(R)}}$\bf{)}}: there exists an edge $e \in \mathcal{G} \setminus T^2$ whose endpoints are in the two components of $T^2$ respectively, such that $\mathscr{X}(e) = \mathscr{X}(Q_{T^2})$ (and thus $w(e) = -\mathscr{V}(Q_{T^2})$).
    \item {\bf{Type-II (}$\boldsymbol{\mathcal{F}_\textup{II}^{(R)}}$\bf{)}}: for any edge $e \in \mathcal{G} \setminus T^2$ whose endpoints are in the two components of $T^2$ respectively, we have $w(e) < -\mathscr{V}(Q_{T^2})$.
\end{itemize}
Let us recall the previous example in (\ref{eq:leading_terms_motivating_example_region_Fterms}) to better understand these notions. There are four $\mathcal{F}_\textup{I}^{(R)}$ terms, which are:
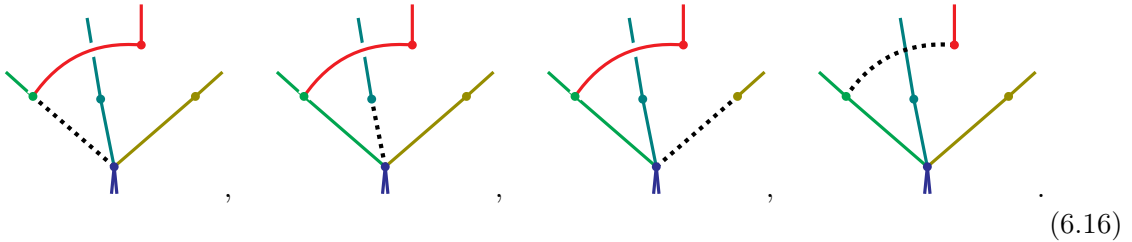
\begin{equation}
\label{eq:leading_terms_motivating_example_Fiterms}
\begin{aligned}
    &\begin{tikzpicture}[line width = 0.6, scale = 0.36, font=\large, mydot/.style={circle, fill, inner sep=.7pt}, transform shape]
    \draw (5,2.5) edge [ultra thick, dotted] (2,5.1) node [] {};
    \draw (2,5.1) edge [very thick, Green] (1,6) node [] {};
    \draw (5,2.5) edge [very thick, teal] (4.5,5) node [] {};
    \draw (4.5,5) edge [very thick, teal] (4,8) node [] {};
    \draw (5,2.5) edge [very thick, olive] (8,5.1) node [] {};
    \draw (8,5.1) edge [very thick, olive] (9,6) node [] {};
    \draw (6,7) edge [very thick, draw=white, double=white, double distance=3pt, bend right = 30] (2,5.1) node [] {};\draw (6,7) edge [very thick, Red, bend right = 30] (2,5.1) node [] {};
    \draw (6,7) edge [very thick, Red] (6,8.5) node [] {};
    \draw (5.1,1.5) edge [very thick, Blue] (5,2.5) node [] {};
    \draw (4.9,1.5) edge [very thick, Blue] (5,2.5) node [] {};    
    \draw[fill, very thick, Green] (2,5.1) circle (3pt);
    \draw[fill, very thick, teal] (4.5,5) circle (3pt);
    \draw[fill, very thick, olive] (8,5.1) circle (3pt);
    \draw[fill, very thick, Red] (6,7) circle (3pt);
    \draw[fill, very thick, Blue] (5,2.5) circle (3pt);    
    \end{tikzpicture},\quad\ 
    \begin{tikzpicture}[line width = 0.6, scale = 0.36, font=\large, mydot/.style={circle, fill, inner sep=.7pt}, transform shape]
    \draw (5,2.5) edge [very thick, Green] (2,5.1) node [] {};
    \draw (2,5.1) edge [very thick, Green] (1,6) node [] {};
    \draw (5,2.5) edge [ultra thick, dotted] (4.5,5) node [] {};
    \draw (4.5,5) edge [very thick, teal] (4,8) node [] {};
    \draw (5,2.5) edge [very thick, olive] (8,5.1) node [] {};
    \draw (8,5.1) edge [very thick, olive] (9,6) node [] {};
    \draw (6,7) edge [very thick, draw=white, double=white, double distance=3pt, bend right = 30] (2,5.1) node [] {};\draw (6,7) edge [very thick, Red, bend right = 30] (2,5.1) node [] {};
    \draw (6,7) edge [very thick, Red] (6,8.5) node [] {};
    \draw (5.1,1.5) edge [very thick, Blue] (5,2.5) node [] {};
    \draw (4.9,1.5) edge [very thick, Blue] (5,2.5) node [] {};    
    \draw[fill, very thick, Green] (2,5.1) circle (3pt);
    \draw[fill, very thick, teal] (4.5,5) circle (3pt);
    \draw[fill, very thick, olive] (8,5.1) circle (3pt);
    \draw[fill, very thick, Red] (6,7) circle (3pt);
    \draw[fill, very thick, Blue] (5,2.5) circle (3pt);    
    \end{tikzpicture},\quad\ 
    \begin{tikzpicture}[line width = 0.6, scale = 0.36, font=\large, mydot/.style={circle, fill, inner sep=.7pt}, transform shape]
    \draw (5,2.5) edge [very thick, Green] (2,5.1) node [] {};
    \draw (2,5.1) edge [very thick, Green] (1,6) node [] {};
    \draw (5,2.5) edge [very thick, teal] (4.5,5) node [] {};
    \draw (4.5,5) edge [very thick, teal] (4,8) node [] {};
    \draw (5,2.5) edge [ultra thick, dotted] (8,5.1) node [] {};
    \draw (8,5.1) edge [very thick, olive] (9,6) node [] {};
    \draw (6,7) edge [very thick, draw=white, double=white, double distance=3pt, bend right = 30] (2,5.1) node [] {};\draw (6,7) edge [very thick, Red, bend right = 30] (2,5.1) node [] {};
    \draw (6,7) edge [very thick, Red] (6,8.5) node [] {};
    \draw (5.1,1.5) edge [very thick, Blue] (5,2.5) node [] {};
    \draw (4.9,1.5) edge [very thick, Blue] (5,2.5) node [] {};    
    \draw[fill, very thick, Green] (2,5.1) circle (3pt);
    \draw[fill, very thick, teal] (4.5,5) circle (3pt);
    \draw[fill, very thick, olive] (8,5.1) circle (3pt);
    \draw[fill, very thick, Red] (6,7) circle (3pt);
    \draw[fill, very thick, Blue] (5,2.5) circle (3pt);    
    \end{tikzpicture},\quad\ 
    \begin{tikzpicture}[line width = 0.6, scale = 0.36, font=\large, mydot/.style={circle, fill, inner sep=.7pt}, transform shape]
    \draw (5,2.5) edge [very thick, Green] (2,5.1) node [] {};
    \draw (2,5.1) edge [very thick, Green] (1,6) node [] {};
    \draw (5,2.5) edge [very thick, teal] (4.5,5) node [] {};
    \draw (4.5,5) edge [very thick, teal] (4,8) node [] {};
    \draw (5,2.5) edge [very thick, olive] (8,5.1) node [] {};
    \draw (8,5.1) edge [very thick, olive] (9,6) node [] {};
    \draw (6,7) edge [ultra thick, dotted, bend right = 30] (2,5.1) node [] {};
    \draw (6,7) edge [very thick, Red] (6,8.5) node [] {};
    \draw (5.1,1.5) edge [very thick, Blue] (5,2.5) node [] {};
    \draw (4.9,1.5) edge [very thick, Blue] (5,2.5) node [] {};    
    \draw[fill, very thick, Green] (2,5.1) circle (3pt);
    \draw[fill, very thick, teal] (4.5,5) circle (3pt);
    \draw[fill, very thick, olive] (8,5.1) circle (3pt);
    \draw[fill, very thick, Red] (6,7) circle (3pt);
    \draw[fill, very thick, Blue] (5,2.5) circle (3pt);    
    \end{tikzpicture}.
\end{aligned}
\end{equation}
In each graph above, we have used black dotted lines to label the edge $e$ whose endpoints lie in the two components respectively and for which $w(e)=-\mathscr{V}(Q_{T^2})$. (In general, there can be multiple such $e$.) From left to right, their corresponding modes are $C_1$, $C_2^2$, $C_3^2$, and $S$.

The remaining five graphs in (\ref{eq:leading_terms_motivating_example_region_Fterms}) are $\mathcal{F}_\textup{II}^{(R)}$ terms:
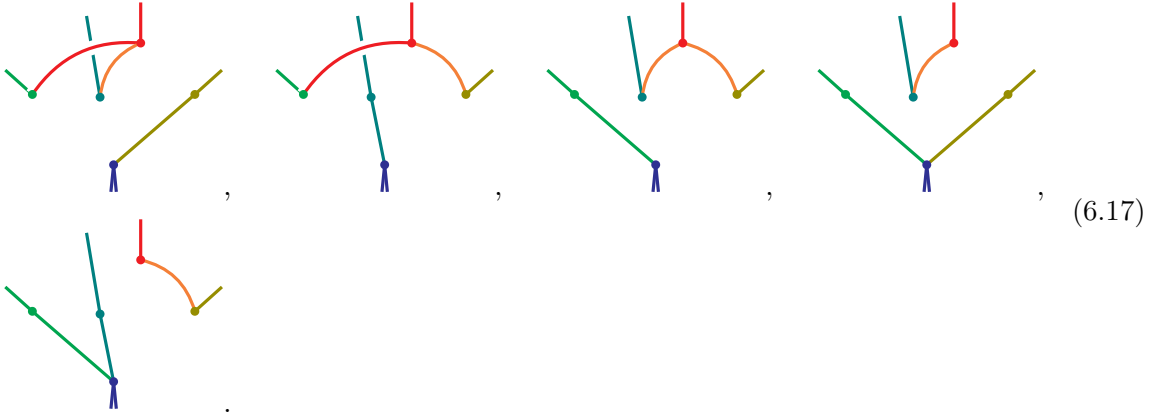
\begin{equation}
\label{eq:leading_terms_motivating_example_Fiiterms}
\begin{aligned}
    &\begin{tikzpicture}[line width = 0.6, scale = 0.36, font=\large, mydot/.style={circle, fill, inner sep=.7pt}, transform shape]
    \draw (2,5.1) edge [very thick, Green] (1,6) node [] {};
    \draw (4.5,5) edge [very thick, teal] (4,8) node [] {};
    \draw (5,2.5) edge [very thick, olive] (8,5.1) node [] {};
    \draw (8,5.1) edge [very thick, olive] (9,6) node [] {};
    \draw (6,7) edge [very thick, draw=white, double=white, double distance=3pt, bend right = 30] (2,5.1) node [] {};\draw (6,7) edge [very thick, Red, bend right = 30] (2,5.1) node [] {};
    \draw (6,7) edge [very thick, Orange, bend right = 30] (4.5,5) node [] {};
    \draw (6,7) edge [very thick, Red] (6,8.5) node [] {};
    \draw (5.1,1.5) edge [very thick, Blue] (5,2.5) node [] {};
    \draw (4.9,1.5) edge [very thick, Blue] (5,2.5) node [] {};    
    \draw[fill, very thick, Green] (2,5.1) circle (3pt);
    \draw[fill, very thick, teal] (4.5,5) circle (3pt);
    \draw[fill, very thick, olive] (8,5.1) circle (3pt);
    \draw[fill, very thick, Red] (6,7) circle (3pt);
    \draw[fill, very thick, Blue] (5,2.5) circle (3pt);    
    \end{tikzpicture},\quad\ 
    \begin{tikzpicture}[line width = 0.6, scale = 0.36, font=\large, mydot/.style={circle, fill, inner sep=.7pt}, transform shape]
    \draw (2,5.1) edge [very thick, Green] (1,6) node [] {};
    \draw (5,2.5) edge [very thick, teal] (4.5,5) node [] {};
    \draw (4.5,5) edge [very thick, teal] (4,8) node [] {};
    \draw (8,5.1) edge [very thick, olive] (9,6) node [] {};
    \draw (6,7) edge [very thick, draw=white, double=white, double distance=3pt, bend right = 30] (2,5.1) node [] {};\draw (6,7) edge [very thick, Red, bend right = 30] (2,5.1) node [] {};
    \draw (6,7) edge [very thick, Orange, bend left = 30] (8,5.1) node [] {};
    \draw (6,7) edge [very thick, Red] (6,8.5) node [] {};
    \draw (5.1,1.5) edge [very thick, Blue] (5,2.5) node [] {};
    \draw (4.9,1.5) edge [very thick, Blue] (5,2.5) node [] {};    
    \draw[fill, very thick, Green] (2,5.1) circle (3pt);
    \draw[fill, very thick, teal] (4.5,5) circle (3pt);
    \draw[fill, very thick, olive] (8,5.1) circle (3pt);
    \draw[fill, very thick, Red] (6,7) circle (3pt);
    \draw[fill, very thick, Blue] (5,2.5) circle (3pt);    
    \end{tikzpicture},\quad\ 
    \begin{tikzpicture}[line width = 0.6, scale = 0.36, font=\large, mydot/.style={circle, fill, inner sep=.7pt}, transform shape]
    \draw (5,2.5) edge [very thick, Green] (2,5.1) node [] {};
    \draw (2,5.1) edge [very thick, Green] (1,6) node [] {};
    \draw (4.5,5) edge [very thick, teal] (4,8) node [] {};
    \draw (8,5.1) edge [very thick, olive] (9,6) node [] {};
    \draw (6,7) edge [very thick, Orange, bend right = 30] (4.5,5) node [] {};
    \draw (6,7) edge [very thick, Orange, bend left = 30] (8,5.1) node [] {};
    \draw (6,7) edge [very thick, Red] (6,8.5) node [] {};
    \draw (5.1,1.5) edge [very thick, Blue] (5,2.5) node [] {};
    \draw (4.9,1.5) edge [very thick, Blue] (5,2.5) node [] {};    
    \draw[fill, very thick, Green] (2,5.1) circle (3pt);
    \draw[fill, very thick, teal] (4.5,5) circle (3pt);
    \draw[fill, very thick, olive] (8,5.1) circle (3pt);
    \draw[fill, very thick, Red] (6,7) circle (3pt);
    \draw[fill, very thick, Blue] (5,2.5) circle (3pt);    
    \end{tikzpicture},\quad\ 
    \begin{tikzpicture}[line width = 0.6, scale = 0.36, font=\large, mydot/.style={circle, fill, inner sep=.7pt}, transform shape]
    \draw (5,2.5) edge [very thick, Green] (2,5.1) node [] {};
    \draw (2,5.1) edge [very thick, Green] (1,6) node [] {};
    \draw (4.5,5) edge [very thick, teal] (4,8) node [] {};
    \draw (5,2.5) edge [very thick, olive] (8,5.1) node [] {};
    \draw (8,5.1) edge [very thick, olive] (9,6) node [] {};
    \draw (6,7) edge [very thick, Orange, bend right = 30] (4.5,5) node [] {};
    \draw (6,7) edge [very thick, Red] (6,8.5) node [] {};
    \draw (5.1,1.5) edge [very thick, Blue] (5,2.5) node [] {};
    \draw (4.9,1.5) edge [very thick, Blue] (5,2.5) node [] {};    
    \draw[fill, very thick, Green] (2,5.1) circle (3pt);
    \draw[fill, very thick, teal] (4.5,5) circle (3pt);
    \draw[fill, very thick, olive] (8,5.1) circle (3pt);
    \draw[fill, very thick, Red] (6,7) circle (3pt);
    \draw[fill, very thick, Blue] (5,2.5) circle (3pt);    
    \end{tikzpicture},\\
    &\begin{tikzpicture}[line width = 0.6, scale = 0.36, font=\large, mydot/.style={circle, fill, inner sep=.7pt}, transform shape]
    \draw (5,2.5) edge [very thick, Green] (2,5.1) node [] {};
    \draw (2,5.1) edge [very thick, Green] (1,6) node [] {};
    \draw (5,2.5) edge [very thick, teal] (4.5,5) node [] {};
    \draw (4.5,5) edge [very thick, teal] (4,8) node [] {};
    \draw (8,5.1) edge [very thick, olive] (9,6) node [] {};
    \draw (6,7) edge [very thick, Orange, bend left = 30] (8,5.1) node [] {};
    \draw (6,7) edge [very thick, Red] (6,8.5) node [] {};
    \draw (5.1,1.5) edge [very thick, Blue] (5,2.5) node [] {};
    \draw (4.9,1.5) edge [very thick, Blue] (5,2.5) node [] {};    
    \draw[fill, very thick, Green] (2,5.1) circle (3pt);
    \draw[fill, very thick, teal] (4.5,5) circle (3pt);
    \draw[fill, very thick, olive] (8,5.1) circle (3pt);
    \draw[fill, very thick, Red] (6,7) circle (3pt);
    \draw[fill, very thick, Blue] (5,2.5) circle (3pt);    
    \end{tikzpicture}.
\end{aligned}
\end{equation}
For each of these five spanning 2-trees, the total momentum flowing between its components is harder than the momentum of any edge connecting those components. For example, in the fourth graph, a $C_2^2$-mode external momentum $p_2$ and an $S$-mode external momentum $l_1$ attach to the same component, yielding a total momentum in the mode $C_2$ (because $C_2^2 \vee S = C_2$) that flows between the components of $T^2$. Meanwhile, there are three edges connecting the two components of the spanning 2-tree, whose associated modes are $S$, $C_2^2$, and $SC_3$ respectively, all softer than $C_2$.

In what follows, we analyze the general structure of each type of $\mathcal{F}^{(R)}$ term.

\subsubsection{\texorpdfstring{$\mathcal{F}_\textup{I}^{(R)}$ terms}{Leading type-i F terms}}

For any given $\mathcal{F}_\textup{I}^{(R)}$ term $\x^{\r}$, adding to $T^2(\r)$ the edge $e$---which connects the two components of $T^2(\r)$ and satisfies $\mathscr{X}(e) = \mathscr{X}(Q_{T^2})$---yields a spanning tree. One can verify that this spanning tree has the same weight as $T^2(\r)$ and therefore corresponds to a $\mathcal{U}^{(R)}$ term. For example, adding the dotted line in each graph of (\ref{eq:leading_terms_motivating_example_Fiterms}) produces the $\mathcal{U}^{(R)}$ term in (\ref{eq:leading_terms_motivating_example_region_Uterm}).

This correspondence between $\mathcal{F}_\textup{I}^{(R)}$ and $\mathcal{U}^{(R)}$ terms automatically allows us to characterize general $\mathcal{F}_\textup{I}^{(R)}$ terms via mode components, similar to lemma~\ref{lemma-leading_Uterms_subgraph_characterization}.
\begin{lemma}
\label{lemma-leading_Fterms_typeI_subgraph_characterization}
    If $\x^{\r}$ is an $\mathcal{F}_\textup{I}^{(R)}$ term, then there exists a corresponding mode component $\gamma'$ with $\mathscr{X}(\gamma') = \mathscr{X}(Q_{T^2(\r)})$, such that (1) $\gamma'\cap T^2(\r)$ is a spanning 2-tree of $\widetilde{\gamma}'$; (2) for all other mode components $\gamma \neq \gamma'$, $\gamma \cap T^2(\r)$ is a spanning tree of $\widetilde{\gamma}$. That is,
    \begin{align}
    \mathcal{F}_\textup{I}^{(R)}:\quad \mathfrak{n}_{\gamma'}^{} = L(\widetilde{\gamma}')+1,\quad \mathfrak{n}_{\gamma}^{} = L(\widetilde{\gamma})\ \ \ \forall\gamma\neq \gamma'\;.
    \label{eq:leading_Fterms_typeI_subgraph_characterization}
    \end{align}
\end{lemma}
From now on, we will use $\mathfrak{n}_{\gamma'}^{}$ to represent $\mathfrak{n}_{\gamma'}^{}(\r)$ unless $\x^{\r}$ refers to one specific term. Similar to eq.~(\ref{eq:leading_Uterms_factorize}), the sum of $\mathcal{F}_\textup{I}^{(R)}$ terms sharing the same kinematic factor $(-s)$ factorizes as follows:
\begin{eqnarray}
    \mathcal{F}_\textup{I}^{(R)}(\x) = (-s)\cdot \mathcal{F_{\gamma'}}(\x^{[\gamma']})\cdot\prod_{\gamma\neq \gamma'} \mathcal{U}_\gamma(\x^{[\gamma]})\,,
\label{eq:leading_Fterms_typeI_factorize}
\end{eqnarray}
where $\gamma'$ is determined by $(-s)$ such that the virtuality of the mode $\mathscr{X}(\gamma')$ is exactly $s$. The terms in the polynomial $\mathcal{F}_{\gamma'}(\x^{[\gamma']})$ correspond to certain spanning 2-trees of $\widetilde{\gamma}'$ for which the total momenta flowing between their components are of mode $\mathscr{X}(\gamma')$.

\subsubsection{\texorpdfstring{$\mathcal{F}_\textup{II}^{(R)}$ terms}{Leading type-ii F terms}}
\label{section-FIIR_terms}

Most complexities arise in characterizing the $\mathcal{F}_\textup{II}^{(R)}$ terms. To begin, for any given $\mathcal{F}_\textup{II}^{(R)}$ term $\x^{\r}$, we consider the positions of the hard vertices. A first observation is that they must all lie in the same component of $T^2(\r)$. To see this, consider the contrary, i.e., suppose hard vertices lie in both components of $T^2(\r)$. Since $\mathcal{H}$ is connected, there must be an edge $e^{[H]}\in \mathcal{H}$ with endpoints in the two components of $T^2(\r)$ respectively. We then have, from the definition of $\mathcal{F}_\textup{II}^{(R)}$ terms, that $0=w(e^{[H]})<-\mathscr{V}(Q(\r))$, which cannot hold since $\mathscr{V}(Q(\r))\in \mathbb{N}$.

\begin{figure}[t]
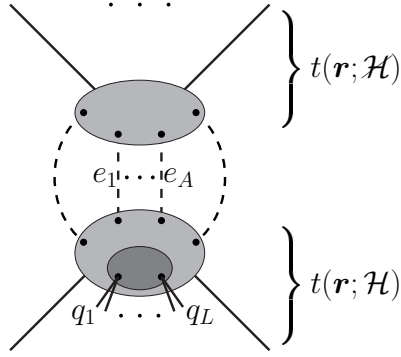

\centering
\include{figs/leading_Fterms_typeII_general_configuration}
\vspace{-3em}\caption{The configuration of $T^2(\r)$ where $x^{\r}$ is an $\mathcal{F}_\textup{II}^{(R)}$ term. The top and bottom gray blobs represent $t(\r;\cancel{\mathcal{H}})$ and $t(\r;\mathcal{H})$ respectively, and the dark gray blob contained in $t(\r;\mathcal{H})$ represents $\mathcal{H}\cap t(\r;\mathcal{H})$. Among the dashed edges that connect $t(\r;\cancel{\mathcal{H}})$ and $t(\r;\mathcal{H})$, a certain subset $\{e_1,\dots,e_A\}$ is selected accordingly (see the description above lemma~\ref{lemma-leading_FIIterms_components_mode_structure}).}
\label{figure-leading_Fterms_typeII_general_configuration}
\end{figure}

With this observation, let us denote the components of $T^2(\r)$ by $t(\r;\mathcal{H})$ (containing all the hard vertices) and $t(\r;\cancel{\mathcal{H}})$ (containing no hard vertices), as shown in figure~\ref{figure-leading_Fterms_typeII_general_configuration}. Among all the edges in $\mathcal{G}\setminus T^2(\r)$ that connect $t(\r;\mathcal{H})$ and $t(\r;\cancel{\mathcal{H}})$, we select those whose modes are \emph{not softer} than any others, and denote them as $e_1,\dots,e_A$. For example, suppose the set of edges connecting $t(\r;\mathcal{H})$ and $t(\r;\cancel{\mathcal{H}})$ are in the modes $C_2^2$, $S$, $SC_3$, and $C_4^3$ respectively; then the set $\{e_1,\dots,e_A\}$ includes those edges whose modes are $C_2^2$, $S$, and $C_4^3$ (those in the mode $SC_3$ are not included because $SC_3$ is softer than $S$).

In general, $e_1,\dots,e_A$ possess the following properties.
\begin{lemma}
\label{lemma-leading_FIIterms_components_mode_structure}
    The edges $e_1,\dots,e_A$, as constructed above, must satisfy:
    \begin{itemize}
        \item [\textup{(1)}] $A\geqslant 2$;
        \item [\textup{(2)}] $\Gamma_X\cap t(\r;\cancel{\mathcal{H}}) = \varnothing$, where $\Gamma_X$ is the $X$-mode subgraph, with $X$ being harder than $\mathscr{X}(e_a)$ for some $a\in \{1,\dots,A\}$.
    \end{itemize}
\end{lemma}
\begin{proof}
    To see (1), note that from the definition of $\mathcal{F}_\textup{II}^{(R)}$ terms, we have
    \begin{align}
        w(e_1),\dots,w(e_A)< -\mathscr{V}(Q(\r))\quad\textup{ and }\quad \bigvee_{a=1}^A \mathscr{X}(e_a) = \mathscr{X}(Q(\r)),
    \end{align}
    which cannot be simultaneously satisfied for $A=1$. So $A\geqslant2$.
    
    We prove (2) by contradiction. Suppose the mode $X$ is harder than a specific $\mathscr{X}(e_a)$, and $\Gamma_X\cap t(\r;\cancel{\mathcal{H}}) \neq \varnothing$. It then follows that the subgraph $\overline{\Gamma}_X$, which is defined as the union of $\Gamma_X$ and all the harder-mode subgraphs, overlaps with both $t(\r;\mathcal{H})$ and $t(\r;\cancel{\mathcal{H}})$. That is,
    \begin{align}
        \overline{\Gamma}_X\cap t(\r;\mathcal{H}) \neq \varnothing,\qquad \overline{\Gamma}_X\cap t(\r;\cancel{\mathcal{H}}) \neq \varnothing.
    \end{align}
    From theorem~\ref{theorem-mode_subgraphs_connectivity1}, $\overline{\Gamma}_X$ is connected, so there must be some edges $\overline{e}\in \overline{\Gamma}_X\setminus T^2(\r)$ connecting the two components of $T^2(\r)$. The mode $\mathscr{X}(e_a)$ is then softer than $\mathscr{X}(\overline{e})$, which contradicts the defining property of the set ${e_1,\dots,e_A}$: the edge $e_a$ should not have been an element of the set $\{e_1,\dots,e_A\}$ by construction. Therefore, we must have $\Gamma_X\cap t(\r;\cancel{\mathcal{H}}) = \varnothing$.
\end{proof}
For each $a=1,\dots,A$, we now define a corresponding tree subgraph $t_a\subset t(\r;\cancel{\mathcal{H}})$ such that: (1) it contains one endpoint of $e_a$, and (2) its edges and vertices are all of the same mode $\mathscr{X}(e_a)$. The tree graph $t(\r;\cancel{\mathcal{H}})$ thus contains $A$ distinct subtrees $t_1,\dots,t_A$\footnote{If multiple edges $e_a$ correspond to the same subtree $t_a$, the distinct subtrees are $t_1, \dots, t_{A'}$ with $A' \leqslant A$. Meanwhile, the following analysis applies equally in this case, so we shall not treat it separately.}, which are connected by paths within $t(\r;\cancel{\mathcal{H}})$. In fact, $t_1, \dots, t_A$ are not only connected but also \emph{aligned} by these paths---a property that we will specify as a corollary of lemma~\ref{lemma-leading_FIIterms_components_mode_structure}.
\begin{corollary}
\label{lemma-leading_FIIterms_components_mode_structure_corollary1}
    The subtrees $t_1,\dots,t_A$ are aligned by these paths. That is, $t(\r;\cancel{\mathcal{H}})$ can be represented by figure~\ref{figure-reduced_form_preliminary} where the nodes represent $t_1,\dots,t_A$ \textup{(}possibly with some attached branches\textup{)}, and each line represents a path connecting them.
    \begin{figure}[t]
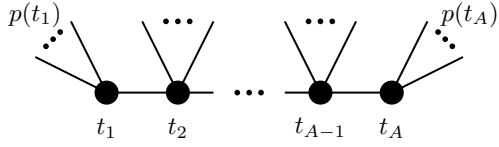

    \centering
    \include{figs/reduced_form_preliminary}
    \vspace{-2em}\caption{The schematic structure of the tree graph $t(\r;\cancel{\mathcal{H}})$, where the nodes represent the subtrees $t_1,\dots,t_A$ (possibly with some attached branches) and are aligned by the $A-1$ paths in $t(\r;\cancel{\mathcal{H}})$. The sum of external momenta attached to $t_a$ is $\mathfrak{p}(t_a)$, which can generally be zero for $a\neq 1,A$.}
    \label{figure-reduced_form_preliminary}
    \end{figure}
    Furthermore, denoting the sum of external momenta attached to $t_a$ by $\mathfrak{p}(t_a)$, we have
    \begin{align}
    \label{eq:lemma_leading_FIIterms_connection_structure}
        \mathscr{X}(\mathfrak{p}(t_1))\vee \mathscr{X}(\mathfrak{p}(t_A)) = \mathscr{X}(\mathfrak{p}(t_1))\vee\dots\vee \mathscr{X}(\mathfrak{p}(t_A)).
    \end{align}
\end{corollary}
\begin{proof}
    From theorem~\ref{theorem-mode_algebra_two_representatives}, there exist $a_1,a_2\in \{1,\dots,A\}$, such that
    \begin{align}
    \label{eq:lemma_leading_FIIterms_connection_structure_proof}
        \mathscr{X}(\mathfrak{p}(t_{a_1}))\vee \mathscr{X}(\mathfrak{p}(t_{a_2})) = \mathscr{X}(\mathfrak{p}(t_1))\vee\dots\vee \mathscr{X}(\mathfrak{p}(t_A)).
    \end{align}
    There is a unique path $P_{a_1a_2}\subset t(\r;\cancel{\mathcal{H}})$ that connects $t_{a_1}$ and $t_{a_2}$. Let us consider the possibility that there is a third subtree $t_b$ (with $b\neq a_1,a_2$) that does not intersect with $P_{a_1a_2}$, i.e., $P_{a_1a_2}\cap t_b = \varnothing$. In this case, one can construct another spanning 2-tree whose weight is lower than $w(T^2(\r))$ as follows. First, by construction there exists an edge $e_b$ whose endpoints are in $t(\r;\mathcal{H})$ and $t(\r;\cancel{\mathcal{H}})$ respectively, such that $\mathscr{X}(e_b)= \mathscr{X}(t_b)$. Second, there is a unique path $P'_b\subset t(\r;\cancel{\mathcal{H}})$ connecting $t_b$ to $P_{a_1a_2}$. The mode of one specific edge $e'_b\in P'_b$, which is adjacent to $t_b$, must be softer than $\mathscr{X}(t_b)$, otherwise the statement (2) in lemma~\ref{lemma-leading_FIIterms_components_mode_structure} would be violated. We thus have $w(e'_b)<w(e_b)$. Third, the spanning 2-tree $T'^2\equiv T^2(\r)\cup e_b\setminus e'_b$ does not disconnect $P_{a_1a_2}$, thus its kinematic factor is as large as that of $T^2(\r)$ according to eq.~(\ref{eq:lemma_leading_FIIterms_connection_structure_proof}). Combining these three observations, we obtain
    \begin{align}
        w(T'^2) = w(T^2(\r)) + w(e'_b) - w(e_b) < w(T^2(\r)),
    \end{align}
    which is a violation to the minimum-weight criterion.

    This analysis indicates that the path $P_{a_1a_2}$ must intersect \emph{all} the subtrees $t_1,\dots,t_A$. This is possible only if $t(\r;\cancel{\mathcal{H}})$ has the structure shown in figure~\ref{figure-reduced_form_preliminary}, with $t_{a_1}$ and $t_{a_2}$ at the endpoints (i.e., $t_1$ and $t_A$ as shown in figure~\ref{figure-reduced_form_preliminary}), which directly yields eq.~(\ref{eq:lemma_leading_FIIterms_connection_structure}).
\end{proof}

One can examine the examples in (\ref{eq:leading_terms_motivating_example_Fiiterms}) and see that they are consistent with corollary~\ref{lemma-leading_FIIterms_components_mode_structure_corollary1}. For each spanning 2-tree there, the component $t(\r;\cancel{\mathcal{H}})$ (the one not attached by the off-shell external momentum) conforms to the structure in figure~\ref{figure-reduced_form_preliminary}. For example, in the third term of (\ref{eq:leading_terms_motivating_example_Fiiterms}), the three subtrees $t_1$ ({\color{teal}\bf teal}), $t_2$ ({\color{Red}\bf red}), and $t_3$ ({\color{olive}\bf olive}) are aligned by two paths ({\color{Orange}\bf orange}). In contrast, the spanning 2-tree shown below does not correspond to any $\mathcal{F}_\textup{II}^{(R)}$ term and therefore does not appear in (\ref{eq:leading_terms_motivating_example_Fiiterms}), because the four subtrees of $t(\r;\cancel{\mathcal{H}})$ are not aligned by the paths:
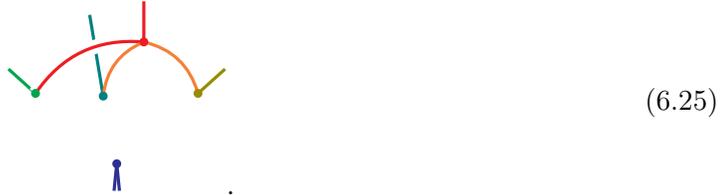
\begin{equation}
\begin{aligned}
    &\begin{tikzpicture}[line width = 0.6, scale = 0.36, font=\large, mydot/.style={circle, fill, inner sep=.7pt}, transform shape]
    \draw (2,5.1) edge [very thick, Green] (1,6) node [] {};
    \draw (4.5,5) edge [very thick, teal] (4,8) node [] {};
    \draw (8,5.1) edge [very thick, olive] (9,6) node [] {};
    \draw (6,7) edge [very thick, draw=white, double=white, double distance=3pt, bend right = 30] (2,5.1) node [] {};\draw (6,7) edge [very thick, Red, bend right = 30] (2,5.1) node [] {};
    \draw (6,7) edge [very thick, Orange, bend right = 30] (4.5,5) node [] {};
    \draw (6,7) edge [very thick, Orange, bend left = 30] (8,5.1) node [] {};
    \draw (6,7) edge [very thick, Red] (6,8.5) node [] {};
    \draw (5.1,1.5) edge [very thick, Blue] (5,2.5) node [] {};
    \draw (4.9,1.5) edge [very thick, Blue] (5,2.5) node [] {};    
    \draw[fill, very thick, Green] (2,5.1) circle (3pt);
    \draw[fill, very thick, teal] (4.5,5) circle (3pt);
    \draw[fill, very thick, olive] (8,5.1) circle (3pt);
    \draw[fill, very thick, Red] (6,7) circle (3pt);
    \draw[fill, very thick, Blue] (5,2.5) circle (3pt);    
    \end{tikzpicture}.
\end{aligned}
\end{equation}
Following the proof above, one can obtain three other spanning 2-trees (which are precisely the first three terms of (\ref{eq:leading_terms_motivating_example_Fiiterms})) whose weights are smaller than the spanning 2-tree above.

From the preceding conclusions, one can denote the paths described in corollary~\ref{lemma-leading_FIIterms_components_mode_structure_corollary1} by $P_{1,2},\dots,P_{A-1,A}$, where $P_{a,a+1}$ connects the subtrees $t_a$ and $t_{a+1}$. To further analyze the substructure of $P_{a,a+1}$, we label its edges in order as $e_1^a,\dots,e_i^a$, such that $e_1^a$ is adjacent to $t_a$, $e_2^a$ is adjacent to $e_1^a$, and so on, with $e_i^a$ adjacent to $t_{a+1}$. If there exist indices $i_1, i_2$ with $1\leqslant i_1<i_2\leqslant i$ such that the mode $\mathscr{X}_{i_1i_2}\equiv \mathscr{X}(e_{i_1}^a)=\mathscr{X}(e_{i_1+1}^a)=\dots=\mathscr{X}(e_{i_2-1}^a)$ is harder than both $\mathscr{X}(e_{i_1-1}^a)$ and $\mathscr{X}(e_{i_2}^a)$, we identify a corresponding subtree in $t(\r;\cancel{\mathcal{H}})$ that contains the edges $e_{i_1}^a,\dots,e_{i_2-1}^a$ and whose edges and vertices all have the same mode $\mathscr{X}_{i_1i_2}$.

We then include all such additional subtrees (if any) in the original set $\{t_1,\dots,t_A\}$ and obtain a larger set. By construction, these trees remain aligned by the paths in $t(\r;\cancel{\mathcal{H}})$. For simplicity, we shall still use $t_1,\dots,t_A$ to denote them with an enlarged $A$. Note that after this operation, statement (1) of lemma~\ref{lemma-leading_FIIterms_components_mode_structure} and corollary~\ref{lemma-leading_FIIterms_components_mode_structure_corollary1} still apply. Meanwhile, statement (2) of lemma~\ref{lemma-leading_FIIterms_components_mode_structure} no longer applies; nevertheless, we will not use it in the subsequent analysis.

With the set $\{t_1,\dots,t_A\}$ extended as above, from now on we call $t_1,\dots,t_A$ the \emph{characteristic subtrees} of the $\mathcal{F}_\textup{II}^{(R)}$ term $\x^{\r}$. The following corollary investigates the structure of their connecting paths, $P_{1,2},\dots,P_{A-1,A}$.
\begin{corollary}
\label{lemma-leading_FIIterms_components_mode_structure_corollary2}
    Each $P_{a,a+1}$ contains a single sub-path whose edges all have the same mode, which is equal to or softer than $\mathscr{X}(t_a)\wedge \mathscr{X}(t_{a+1})$; all the other edges in $P_{a,a+1}$ \textup{(}if any\textup{)} are of harder modes.
\end{corollary}
\begin{proof}
    For each $a\in\{1,\dots,A\}$, we denote the edges of $P_{a,a+1}$ in a certain order as $e_1^a,\dots,e_{i}^a$, such that $e_1^a$ is adjacent to $t_a$, $e_2^a$ is adjacent to $e_1^a$, etc., and $e_j^a$ is adjacent to $t_{a+1}$. From our construction above, there exist $i_1,i_2$ with $1\leqslant i_1< i_2\leqslant i$, such that
    \begin{itemize}
        \item $\mathscr{X}(e_b^a)$ is equal to or harder than $\mathscr{X}(e_{b+1}^a)$, for $b=1,\dots,i_1-1$;
        \item $\mathscr{X}(e_b^a) = \mathscr{X}(e_{b+1}^a)$, for $b=i_1,\dots,i_2-1$;
        \item $\mathscr{X}(e_b^a)$ is equal to or softer than $\mathscr{X}(e_{b+1}^a)$, for $b=i_2,\dots,i-1$;
        \end{itemize}
    In other words, the softest edges in $P_{a,a+1}$ are $e_{i_1}^a,\dots,e_{i_2-1}^a$, which form a sub-path of $P_{a,a+1}$. Their mode must be simultaneously softer than $\mathscr{X}(t_a)$ and $\mathscr{X}(t_{a+1})$, thus equal to or softer than $\mathscr{X}(t_a)\wedge \mathscr{X}(t_{a+1})$.
\end{proof}

Another crucial property is that edges adjacent to a characteristic subtree $t_a$ cannot be in any mode harder than $\mathscr{X}(t_a)$. A more precise description is given by the following lemma.
\begin{lemma}
\label{lemma-leading_FIIterms_subtree_adjacent_edges}
    For any edge $e \in \mathcal{G}$ adjacent to a characteristic subtree $t_a$, the following holds:
    \begin{enumerate}
        \item If both endpoints of $e$ lie in $t(\r;\cancel{\mathcal{H}})$, then $\mathscr{X}(e)$ is softer than $\mathscr{X}(t_a)$.
        \item If one endpoint of $e$ is in $t(\r;\cancel{\mathcal{H}})$ and the other is in $t(\r;\mathcal{H})$, then $\mathscr{X}(e)$ is softer than or equal to $\mathscr{X}(t_a)$.
        \item There exists at least one such edge $e$ with endpoints in $t(\r;\cancel{\mathcal{H}})$ and $t(\r;\mathcal{H})$, respectively, for which $\mathscr{X}(e) = \mathscr{X}(t_a)$.
    \end{enumerate}
\end{lemma}
Note that the statements above are already justified by lemma~\ref{lemma-leading_FIIterms_components_mode_structure} for the edges $e_1,\dots,e_A$ defined at the beginning of our $\mathcal{F}^{(R)}$-term analysis. Here, we argue that they continue to hold after $\{t_1, \dots, t_A\}$ has been extended to the set of characteristic subtrees as described.
\begin{proof}
    We first show that $\mathscr{X}(e)$ cannot be harder than $\mathscr{X}(t_a)$. Otherwise, by the First Connectivity Theorem (theorem~\ref{theorem-mode_subgraphs_connectivity1}), there would exist a path in $\mathcal{G}$ connecting $t_a$ and $\mathcal{H}$ whose edges all have modes harder than $\mathscr{X}(t_a)$. Starting from one endpoint $v_1$ that is adjacent to $t_a$, this path, denoted as $P_0$, must first reach a vertex $v_2$ that lies in either (1) another characteristic subtree $t_b$\footnote{We also include in this case the possibility that $v_2$ lies in one of the $A-1$ connecting paths of $t(\r;\cancel{\mathcal{H}})$, for which the same analysis would apply.}, or (2) the other component of $T^2(\r)$, which is $t(\r;\mathcal{H})$. Below we demonstrate that both possibilities would violate the minimum-weight criterion.

    In case (1), there is an edge $e_0\in P_0\setminus t(\r;\cancel{\mathcal{H}})$ such that $t(\r;\cancel{\mathcal{H}})\cup e_0$ contains a path through $e_0$, connecting $v_1$ and $v_2$. However, a unique path already exists within $t(\r;\cancel{\mathcal{H}})$ connecting $v_1$ and $v_2$, which must pass through either $P_{a-1,a}$ or $P_{a,a+1}$. (Figure~\ref{figure-leading_FIIterms_subtree_adjacent_edges_case1} illustrates the configuration where $e' \in P_{a-1,a}$.) Consequently, $t(\r;\cancel{\mathcal{H}})\cup e_0$ contains a loop. Within this loop, there is an edge $e' \in P_{a-1,a} \cup P_{a,a+1}$ for which $\mathscr{X}(e')$ is softer than $\mathscr{X}(t_a)$. We now construct another spanning 2-tree $T_0^2\equiv T^2(\r)\cup e_0\setminus e'$, whose weight is given by
    \begin{align}
        w(T_0^2) = w(T^2(\r)) + w(e') - w(e_0) < w(T^2(\r)),
    \end{align}
    where we have used the aforementioned fact that $\mathscr{X}(e')$ is softer than $\mathscr{X}(e_a)$ softer than $\mathscr{X}(e_0)$, from which $w(e')<w(e_0)$. The minimum-weight criterion is thus violated.
    \begin{figure}[t]
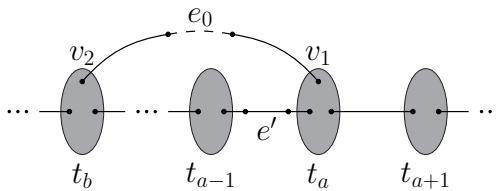

    \centering
    \include{figs/leading_FIIterms_subtree_adjacent_edges_case1}
    \vspace{-2em}\caption{The configuration of $t(\r;\cancel{\mathcal{H}})$ of case (1) in the proof of lemma~\ref{lemma-leading_FIIterms_subtree_adjacent_edges}. The gray blobs represent the subtrees $t_b$, $t_{a-1}$, $t_a$, and $t_{a+1}$. The endpoints of $e_0$ are directly connected to $t_a$ and $t_b$, respectively. The edge $e'$ belongs to the path $P_{a-1,a}$.}
    \label{figure-leading_FIIterms_subtree_adjacent_edges_case1}
    \end{figure}

    For case (2), there exists an edge $e_0\in P_0$ whose endpoints are in $t(\r;\cancel{\mathcal{H}})$ and $t(\r;\mathcal{H})$, respectively. Note that $\mathscr{X}(v_1)$ is harder than $\mathscr{X}(t_a)$, thus $v_1\notin t_a$, and $v_1$ must not disconnect $t_a$, as is shown in figure~\ref{figure-leading_FIIterms_subtree_adjacent_edges_case2}.
    \begin{figure}[t]
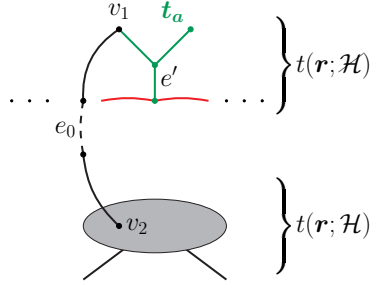

    \centering
    \include{figs/leading_FIIterms_subtree_adjacent_edges_case2}
    \vspace{-3em}\caption{The configuration of $T^2(\r)$ of case (2) in the proof of lemma~\ref{lemma-leading_FIIterms_subtree_adjacent_edges}. In the component $t(\r;\cancel{\mathcal{H}})$, the characteristic subtree $t_a$ is marked in green, while its adjacent paths are marked in red. The vertex $v_1$, which is one endpoint of the path $P_0$, does not belong to and cannot disconnect $t_a$.}
    \label{figure-leading_FIIterms_subtree_adjacent_edges_case2}
    \end{figure}
    This implies the existence of another edge $e'\in t_a$, such that the spanning 2-tree $T_0^2\equiv T^2(\r)\cup e_0\setminus e'$, with a weight given by
    \begin{align}
        w(T_0^2) = w(T^2(\r)) + w(e') - w(e_0) < w(T^2(\r)).
    \end{align}
    As in case (1), the minimum-weight criterion is violated.
    
    Therefore, for all $e$ adjacent to $t_a$, $\mathscr{X}(e)$ is either equal to or softer than $\mathscr{X}(t_a)$. Note that if both endpoints of $e$ are in $t(\r;\cancel{\mathcal{H}})$, $\mathscr{X}(e)$ can only be softer than $\mathscr{X}(t_a)$ otherwise it would be absorbed into $t_a$. The first two statements of the lemma are then proved.

    Finally, the First Connectivity Theorem indicates that the modes $\mathscr{X}(e)$ cannot be all softer than $\mathscr{X}(t_a)$. In other words, there exists at least one specific edge $e$ with $\mathscr{X}(e) = \mathscr{X}(t_a)$. From the first two statements, its endpoints must be in $t(\r;\cancel{\mathcal{H}})$ and $t(\r;\mathcal{H})$ respectively. The third statement is also proved.
\end{proof}

So far, the pattern of $T^2(\r)$ has been clarified as follows. It consists of two components, $t(\r;\mathcal{H})$ and $t(\r;\cancel{\mathcal{H}})$, where all the hard vertices are contained in $t(\r;\mathcal{H})$. In the component $t(\r;\cancel{\mathcal{H}})$, there are $A$ characteristic subtrees $t_1,\dots,t_A$ aligned by $A-1$ paths $P_{1,2},\dots,P_{A-1,A}$. For each $a\in \{1,\dots,A\}$, there exists some edge $e_a\in \mathcal{G}\setminus T^2(\r)$, whose endpoints are in $t_a$ and $t(\r;\mathcal{H})$, respectively, with $\mathscr{X}(e_a) = \mathscr{X}(t_a)$. For each $a\in \{1,\dots,A-1\}$, all the edges of the path $P_{a,a+1}$ are softer than either $\mathscr{X}(t_a)$ or $\mathscr{X}(t_{a+1})$.
From now on, we refer to the softest mode on the path $P_{a,a+1}$---that is, the mode of the edges in the sub-path identified in corollary~\ref{lemma-leading_FIIterms_components_mode_structure_corollary2}---as the \emph{characteristic mode} of $P_{a,a+1}$ and denote it as $\widehat{X}_a$. The following lemma relates $\widehat{X}_1,\dots,\widehat{X}_{A-1}$ and $\mathscr{X}(t_1),\dots,\mathscr{X}(t_A)$.

\begin{lemma}
\label{lemma-leading_FIIterms_paths_weight}
    The characteristic modes $\widehat{X}_1,\dots,\widehat{X}_{A-1}$ and the modes $\mathscr{X}(t_1),\dots,\mathscr{X}(t_A)$ satisfy the following relations:
    \begin{enumerate}
        \item $\widehat{X}_a = \mathscr{X}(t_a)\wedge \mathscr{X}(t_{a+1})$ for $a=1,\dots,A-1$;
        \item $\mathscr{X}(t_a) = \widehat{X}_{a-1}\vee \widehat{X}_a$ for $a=2,\dots,A-1$.
    \end{enumerate}
\end{lemma}
This lemma reveals a particularly elegant property of $t(\r;\cancel{\mathcal{H}})$. The characteristic mode of the path $P_{a,a+1}$ can be determined by its neighboring characteristic subtrees, $t_a$ and $t_{a+1}$, via the $\wedge$ operation. Conversely, the mode of each ``internal'' characteristic subtree $t_a$ (with $a\in\{2,\dots,A-1\}$) is determined by the characteristic modes of its two adjacent paths, $P_{a-1,a}$ and $P_{a,a+1}$, via the $\vee$ operation. This indicates that adjacent modes---$\mathscr{X}(t_a)$ and $\mathscr{X}(t_{a+1})$, or $\widehat{X}_{a-1}$ and $\widehat{X}_a$---must be overlapping.

\begin{proof}
    Before we justify the two relations, we first point out that the mode $\mathscr{X}(\mathfrak{p}(t_a))$ is softer than or equal to $\mathscr{X}(t_a)$ (recall that $\mathfrak{p}(t_a)$ refers to the sum of external momenta entering $t_a$). Otherwise, there would exist some edge adjacent to $t_a$ whose mode is harder than $\mathscr{X}(t_a)$, which contradicts lemma~\ref{lemma-leading_FIIterms_subtree_adjacent_edges}.

    To justify \emph{1}, we modify $T^2(\r)$ into a specific spanning tree as follows. For each $t_a$, we include to $T^2(\r)$ the edge $e_a$ (see the paragraph above this lemma, and note the relation $\mathscr{X}(e_a) = \mathscr{X}(t_a)$); for each path $P_{a,a+1}$, we delete any one edge $\widetilde{e}_a$ that is of its characteristic mode $\widehat{X}_a$. This operation includes $A$ more edges to $T^2(\r)$ and delete $A-1$ edges from $T^2(\r)$, turning it into a spanning tree of $\mathcal{G}$. Let us denote the corresponding $\mathcal{U}$ term as $\x^{\r'}$.

    On the one hand, from the minimum-weight criterion, $w(T^1(\r')) \geqslant w(T^2(\r)),$ where the equal sign holds if and only if $\x^{\r'}$ is a $\mathcal{U}^{(R)}$ term.
    
    On the other hand, by construction we have
    \begin{align}
    \label{ineq:lemma_leading_FIIterms_paths_weight}
        w(T^1(\r'))&= w(T^2(\r)) - \sum_{i=1}^A w(e_a) + \sum_{a=1}^{A-1} w(\widetilde{e}_a) -\mathscr{V}(\mathscr{X}(\mathfrak{p}(t_1))\vee\dots\vee \mathscr{X}(\mathfrak{p}(t_A))) \nonumber\\
        &\hspace{-1cm}= w(T^2(\r)) + \sum_{a=1}^A \mathscr{V}(\mathscr{X}(t_a)) - \sum_{a=1}^{A-1} \mathscr{V}(\widehat{X}_a) -\mathscr{V}(\mathscr{X}(\mathfrak{p}(t_1))\vee\dots\vee \mathscr{X}(\mathfrak{p}(t_A))) \nonumber\\
        &\hspace{-1cm}\leqslant w(T^2(\r)) + \sum_{i=1}^A \mathscr{V}(\mathscr{X}(t_a)) - \sum_{a=1}^{A-1} \mathscr{V}(\mathscr{X}(t_a)\wedge \mathscr{X}(t_{a+1})) -\mathscr{V}(\mathscr{X}(\mathfrak{p}(t_1))\vee\dots\vee \mathscr{X}(\mathfrak{p}(t_A))) \nonumber\\
        &\hspace{-1cm}\leqslant w(T^2(\r)) + \sum_{i=1}^A \mathscr{V}(\mathscr{X}(t_a)) - \sum_{a=1}^{A-1} \mathscr{V}(\mathscr{X}(t_a)\wedge \mathscr{X}(t_{a+1})) -\mathscr{V}(\mathscr{X}(t_1)\vee\dots\vee \mathscr{X}(t_A)) \nonumber\\
        &\hspace{-1cm}\leqslant w(T^2(\r)).
    \end{align}
    We now explain each line in (\ref{ineq:lemma_leading_FIIterms_paths_weight}). The first line expresses the defining property of $w(T^1(\r'))$: to obtain it from $w(T^2(\r))$, one subtracts the weights of the included edges $\{e_a\}_{a=1,\dots,A}$, adds the weights of the deleted edges $\{\widetilde{e}_a\}_{a=1,\dots,A-1}$, and subtracts the kinematic factor of $T^2(\r)$, which is $\mathscr{V}(\mathscr{X}(\mathfrak{p}(t_1))\vee\dots\vee \mathscr{X}(\mathfrak{p}(t_A)))$. In the second line, the edge weights are substituted with their virtuality degrees, i.e., $w(e) = -\mathscr{V}(\mathscr{X}(e))$; we have also employed $\mathscr{X}(e_a) = \mathscr{X}(t_a)$. In the third line, we have used the fact that the characteristic mode $\widehat{X}_a$ is either equal to or softer than $\mathscr{X}(t_a)\wedge \mathscr{X}(t_{a+1})$ (see corollary~\ref{lemma-leading_FIIterms_components_mode_structure_corollary2}), which implies $\mathscr{V}(\widehat{X}_a)\geqslant \mathscr{V}(\mathscr{X}(t_a)\wedge \mathscr{X}(t_{a+1}))$ for each $a=1,\dots,A-1$. In the fourth line, we have used the aforementioned property that $\mathscr{X}(\mathfrak{p}(t_a))$ is softer than or equal to $\mathscr{X}(t_a)$, from which $\mathscr{V}(\mathscr{X}(\mathfrak{p}(t_1))\vee\dots\vee \mathscr{X}(\mathfrak{p}(t_A)))\geqslant \mathscr{V}(\mathscr{X}(t_1)\vee\dots\vee \mathscr{X}(t_A))$. Finally, in the last line, we have employed corollary~\ref{theorem-virtualities_relation_corollary1}.

    In comparison with the previous conclusion that $w(T^1(\r')) \geqslant w(T^2(\r))$, we must have
    $$w(T^1(\r')) = w(T^2(\r)),$$
    which further indicates that every equality in (\ref{ineq:lemma_leading_FIIterms_paths_weight}) must hold. The mode $\widehat{X}_a$ is then precisely $\mathscr{X}(t_a)\wedge \mathscr{X}(t_{a+1})$. In other words, the softest edge in $P_{a,a+1}$ is of the mode $\mathscr{X}(t_a)\wedge \mathscr{X}(t_{a+1})$. We have thus justified the first statement.

    To see why the second statement is true, note that from above,
    \begin{align}
        \widehat{X}_{a-1}\vee \widehat{X}_a = (\mathscr{X}(t_{a-1})\wedge \mathscr{X}(t_a)) \vee (\mathscr{X}(t_a)\wedge \mathscr{X}(t_{a+1})).
    \end{align}
    The mode $\mathscr{X}(t_a)$ is harder than both brackets on the right-hand side above. Therefore, $\mathscr{X}(t_a)$ must be equal to or harder than $\widehat{X}_{a-1}\vee \widehat{X}_a$. Below we aim to rule out the possibility that $\mathscr{X}(t_{a_0})$ is harder than $\widehat{X}_{a_0-1}\vee \widehat{X}_{a_0}$ for some $a_0\in \{2,\dots,A-1\}$.

    As we have demonstrated above, all the equalities hold in (\ref{ineq:lemma_leading_FIIterms_paths_weight}), from which we have
    \begin{align}
    \label{eq:lemma_leading_FIIterms_paths_weight}
        \sum_{a=1}^A \mathscr{V}(\mathscr{X}(t_a)) = \sum_{a=1}^{A-1} \mathscr{V}(\mathscr{X}(t_a)\wedge \mathscr{X}(t_{a+1})) +\mathscr{V}(\mathscr{X}(t_1)\vee\dots\vee \mathscr{X}(t_A)).
    \end{align}
    Meanwhile, let us construct the following graph $\mathcal{G}'$, which can be regarded as a simplified version of $\mathcal{G}$, and consider a specific region:
    \begin{equation}
    \textup{Region of }\mathcal{G}':\quad
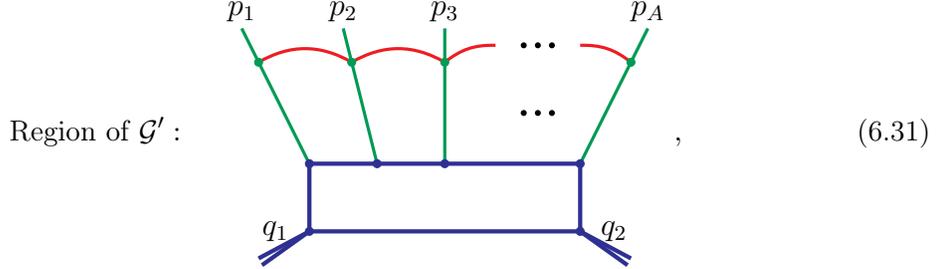
    \begin{tikzpicture}[baseline=10ex, scale=0.45, mydot/.style={circle, fill, inner sep=.8pt}]

    \draw [ultra thick, color=Blue] (2,3) -- (10,3);
    \draw [ultra thick, color=Blue] (2,1) -- (2,3);
    \draw [ultra thick, color=Blue] (10,1) -- (10,3);
    \draw [ultra thick, color=Blue] (2,1) -- (10,1);
    \draw [ultra thick, color=Blue] (0.5,0.2) -- (2,1);
    \draw [ultra thick, color=Blue] (0.6,0) -- (2,1);
    \draw [ultra thick, color=Blue] (11.5,0.2) -- (10,1);
    \draw [ultra thick, color=Blue] (11.4,0) -- (10,1);
    \draw [very thick, color=ForestGreen] (2,3) -- (0,7);
    \draw [very thick, color=ForestGreen] (4,3) -- (3,7);
    \draw [very thick, color=ForestGreen] (6,3) -- (6,7);
    \draw [very thick, color=ForestGreen] (10,3) -- (12,7);
    \draw [very thick, color=Red, bend left = 30] (0.5,6) to (3.25,6);
    \draw [very thick, color=Red, bend left = 30] (3.25,6) to (6,6);   
    \draw [very thick, color=Red, bend left = 20] (6,6) to (7.5,6.5);   
    \draw [very thick, color=Red, bend left = 20] (10,6.5) to (11.5,6);
    \node [draw=ForestGreen, circle, minimum size=3pt, fill=ForestGreen, inner sep=0pt, outer sep=0pt] () at (0.5,6) {};
    \node [draw=ForestGreen, circle, minimum size=3pt, fill=ForestGreen, inner sep=0pt, outer sep=0pt] () at (3.25,6) {};
    \node [draw=ForestGreen, circle, minimum size=3pt, fill=ForestGreen, inner sep=0pt, outer sep=0pt] () at (6,6) {};
    \node [draw=ForestGreen, circle, minimum size=3pt, fill=ForestGreen, inner sep=0pt, outer sep=0pt] () at (11.5,6) {};
    \node [draw=Blue, circle, minimum size=3pt, fill=Blue, inner sep=0pt, outer sep=0pt] () at (2,1) {};
    \node [draw=Blue, circle, minimum size=3pt, fill=Blue, inner sep=0pt, outer sep=0pt] () at (2,3) {};
    \node [draw=Blue, circle, minimum size=3pt, fill=Blue, inner sep=0pt, outer sep=0pt] () at (10,1) {};
    \node [draw=Blue, circle, minimum size=3pt, fill=Blue, inner sep=0pt, outer sep=0pt] () at (10,3) {};
    \node [draw=Blue, circle, minimum size=3pt, fill=Blue, inner sep=0pt, outer sep=0pt] () at (4,3) {};
    \node [draw=Blue, circle, minimum size=3pt, fill=Blue, inner sep=0pt, outer sep=0pt] () at (6,3) {};
    \path (7.5,6.5)-- node[mydot, pos=.333] {} node[mydot] {} node[mydot, pos=.666] {}(10,6.5);
    \path (7.5,4.5)-- node[mydot, pos=.333] {} node[mydot] {} node[mydot, pos=.666] {}(10,4.5);
    \node () at (1,1) {\large $q_1$};
    \node () at (11,1) {\large $q_2$};
    \node () at (0,7.5) {\large $p_1$};
    \node () at (3,7.5) {\large $p_2$};
    \node () at (6,7.5) {\large $p_3$};
    \node () at (12,7.5) {\large $p_A$};
    \end{tikzpicture},
    \end{equation}
    The external momenta are chosen with the following modes:
    $$\mathscr{X}(q_1) = \mathscr{X}(q_A) = H,\ \ \mathscr{X}(p_1) = \mathscr{X}(t_1),\ \ \mathscr{X}(p_A) = \mathscr{X}(t_A),\ \ \mathscr{X}(p_i) = \widehat{X}_{i-1}\vee \widehat{X}_i\ \ (i=2,\dots,A-1).$$
    In the region above, the blue edges are all hard, the green edges are in the same mode of their corresponding $p_i$, respectively, and the modes of the red edges (from left to right) are $\widehat{X}_1,\dots,\widehat{X}_{A-1}$, respectively. One can verify the existence of this region directly, by applying the weight-solving algorithm that will be introduced later in section~\ref{section-infrared_compatibility_requirements} (which is independent of the analysis in this lemma). For simplicity, we will not show it here.

    We then consider the following two terms in the Symanzik polynomials:
    \begin{equation}
    \x^{\r_1}=\ 
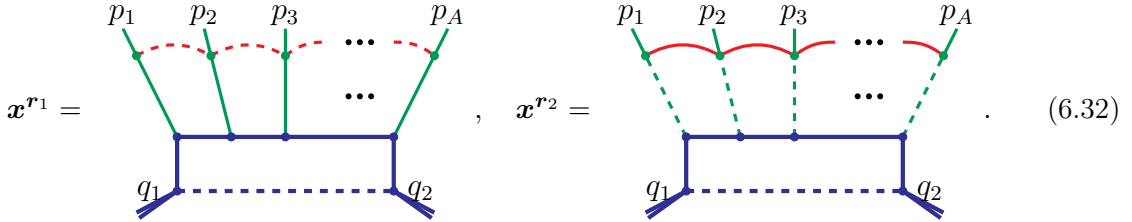
    \begin{tikzpicture}[baseline=8ex, scale=0.36, mydot/.style={circle, fill, inner sep=.8pt}]

    \draw [ultra thick, color=Blue] (2,3) -- (10,3);
    \draw [ultra thick, color=Blue] (2,1) -- (2,3);
    \draw [ultra thick, color=Blue] (10,1) -- (10,3);
    \draw [ultra thick, dashed, color=Blue] (2,1) -- (10,1);
    \draw [ultra thick, color=Blue] (0.5,0.2) -- (2,1);
    \draw [ultra thick, color=Blue] (0.6,0) -- (2,1);
    \draw [ultra thick, color=Blue] (11.5,0.2) -- (10,1);
    \draw [ultra thick, color=Blue] (11.4,0) -- (10,1);
    \draw [very thick, color=ForestGreen] (2,3) -- (0.5,6);
    \draw [very thick, color=ForestGreen] (0.5,6) -- (0,7);
    \draw [very thick, color=ForestGreen] (4,3) -- (3.25,6);
    \draw [very thick, color=ForestGreen] (3.25,6) -- (3,7);
    \draw [very thick, color=ForestGreen] (6,3) -- (6,6);
    \draw [very thick, color=ForestGreen] (6,6) -- (6,7);
    \draw [very thick, color=ForestGreen] (10,3) -- (11.5,6);
    \draw [very thick, color=ForestGreen] (11.5,6) -- (12,7);
    \draw [very thick, dashed, color=Red, bend left = 30] (0.5,6) to (3.25,6);
    \draw [very thick, dashed, color=Red, bend left = 30] (3.25,6) to (6,6);
    \draw [very thick, dashed, color=Red, bend left = 20] (6,6) to (7.5,6.5);
    \draw [very thick, dashed, color=Red, bend left = 20] (10,6.5) to (11.5,6);
    \node [draw=ForestGreen, circle, minimum size=3pt, fill=ForestGreen, inner sep=0pt, outer sep=0pt] () at (0.5,6) {};
    \node [draw=ForestGreen, circle, minimum size=3pt, fill=ForestGreen, inner sep=0pt, outer sep=0pt] () at (3.25,6) {};
    \node [draw=ForestGreen, circle, minimum size=3pt, fill=ForestGreen, inner sep=0pt, outer sep=0pt] () at (6,6) {};
    \node [draw=ForestGreen, circle, minimum size=3pt, fill=ForestGreen, inner sep=0pt, outer sep=0pt] () at (11.5,6) {};
    \node [draw=Blue, circle, minimum size=3pt, fill=Blue, inner sep=0pt, outer sep=0pt] () at (2,1) {};
    \node [draw=Blue, circle, minimum size=3pt, fill=Blue, inner sep=0pt, outer sep=0pt] () at (2,3) {};
    \node [draw=Blue, circle, minimum size=3pt, fill=Blue, inner sep=0pt, outer sep=0pt] () at (10,1) {};
    \node [draw=Blue, circle, minimum size=3pt, fill=Blue, inner sep=0pt, outer sep=0pt] () at (10,3) {};
    \node [draw=Blue, circle, minimum size=3pt, fill=Blue, inner sep=0pt, outer sep=0pt] () at (4,3) {};
    \node [draw=Blue, circle, minimum size=3pt, fill=Blue, inner sep=0pt, outer sep=0pt] () at (6,3) {};
    \path (7.5,6.5)-- node[mydot, pos=.333] {} node[mydot] {} node[mydot, pos=.666] {}(10,6.5);
    \path (7.5,4.5)-- node[mydot, pos=.333] {} node[mydot] {} node[mydot, pos=.666] {}(10,4.5);
    \node () at (1,1) {\large $q_1$};
    \node () at (11,1) {\large $q_2$};
    \node () at (0,7.5) {\large $p_1$};
    \node () at (3,7.5) {\large $p_2$};
    \node () at (6,7.5) {\large $p_3$};
    \node () at (12,7.5) {\large $p_A$};
    \end{tikzpicture},
    \quad \x^{\r_2}=\ 
    \begin{tikzpicture}[baseline=8ex, scale=0.36, mydot/.style={circle, fill, inner sep=.8pt}]

    \draw [ultra thick, color=Blue] (2,3) -- (10,3);
    \draw [ultra thick, color=Blue] (2,1) -- (2,3);
    \draw [ultra thick, color=Blue] (10,1) -- (10,3);
    \draw [ultra thick, dashed, color=Blue] (2,1) -- (10,1);
    \draw [ultra thick, color=Blue] (0.5,0.2) -- (2,1);
    \draw [ultra thick, color=Blue] (0.6,0) -- (2,1);
    \draw [ultra thick, color=Blue] (11.5,0.2) -- (10,1);
    \draw [ultra thick, color=Blue] (11.4,0) -- (10,1);
    \draw [very thick, dashed, color=ForestGreen] (2,3) -- (0.5,6);
    \draw [very thick, color=ForestGreen] (0.5,6) -- (0,7);
    \draw [very thick, dashed, color=ForestGreen] (4,3) -- (3.25,6);
    \draw [very thick, color=ForestGreen] (3.25,6) -- (3,7);
    \draw [very thick, dashed, color=ForestGreen] (6,3) -- (6,6);
    \draw [very thick, color=ForestGreen] (6,6) -- (6,7);
    \draw [very thick, dashed, color=ForestGreen] (10,3) -- (11.5,6);
    \draw [very thick, color=ForestGreen] (11.5,6) -- (12,7);
    \draw [very thick, color=Red, bend left = 30] (0.5,6) to (3.25,6);
    \draw [very thick, color=Red, bend left = 30] (3.25,6) to (6,6);   
    \draw [very thick, color=Red, bend left = 20] (6,6) to (7.5,6.5);   
    \draw [very thick, color=Red, bend left = 20] (10,6.5) to (11.5,6);   
    \node [draw=ForestGreen, circle, minimum size=3pt, fill=ForestGreen, inner sep=0pt, outer sep=0pt] () at (0.5,6) {};
    \node [draw=ForestGreen, circle, minimum size=3pt, fill=ForestGreen, inner sep=0pt, outer sep=0pt] () at (3.25,6) {};
    \node [draw=ForestGreen, circle, minimum size=3pt, fill=ForestGreen, inner sep=0pt, outer sep=0pt] () at (6,6) {};
    \node [draw=ForestGreen, circle, minimum size=3pt, fill=ForestGreen, inner sep=0pt, outer sep=0pt] () at (11.5,6) {};
    \node [draw=Blue, circle, minimum size=3pt, fill=Blue, inner sep=0pt, outer sep=0pt] () at (2,1) {};
    \node [draw=Blue, circle, minimum size=3pt, fill=Blue, inner sep=0pt, outer sep=0pt] () at (2,3) {};
    \node [draw=Blue, circle, minimum size=3pt, fill=Blue, inner sep=0pt, outer sep=0pt] () at (10,1) {};
    \node [draw=Blue, circle, minimum size=3pt, fill=Blue, inner sep=0pt, outer sep=0pt] () at (10,3) {};
    \node [draw=Blue, circle, minimum size=3pt, fill=Blue, inner sep=0pt, outer sep=0pt] () at (4,3) {};
    \node [draw=Blue, circle, minimum size=3pt, fill=Blue, inner sep=0pt, outer sep=0pt] () at (6,3) {};
    \path (7.5,6.5)-- node[mydot, pos=.333] {} node[mydot] {} node[mydot, pos=.666] {}(10,6.5);
    \path (7.5,4.5)-- node[mydot, pos=.333] {} node[mydot] {} node[mydot, pos=.666] {}(10,4.5);
    \node () at (1,1) {\large $q_1$};
    \node () at (11,1) {\large $q_2$};
    \node () at (0,7.5) {\large $p_1$};
    \node () at (3,7.5) {\large $p_2$};
    \node () at (6,7.5) {\large $p_3$};
    \node () at (12,7.5) {\large $p_A$};
    \end{tikzpicture}.
    \end{equation}
    The first term, $\x^{\r_1}$, corresponds to a spanning tree that is consistent with (\ref{eq:leading_Uterms_subgraph_characterization}). It is thus a $\mathcal{U}^{(R)}$ term. The second term, $\x^{\r_2}$, corresponds to a spanning 2-tree and is by definition an $\mathcal{F}_\textup{II}^{}$ term. (In fact, it is an $\mathcal{F}_\textup{II}^{(R)}$ term, as we will see later.) By comparing their weights, we have
    \begin{align}
    \label{eq:lemma_leading_FIItermsweight_comparison_constructed_terms}
        w(T^2(\r_2))&\geqslant w(T^1(\r_1))\nonumber\\
        &\hspace{-1em}= w(T^2(\r_2)) +\sum_{a=1}^A \mathscr{V}(\mathscr{X}(p_a)) - \sum_{a=1}^{A-1} \mathscr{V}(\widehat{X}_a) - \mathscr{V}(\mathscr{X}(t_1)\vee\dots\vee \mathscr{X}(t_A))\nonumber\\
        &\hspace{-1em}> w(T^2(\r_2)) +\sum_{a=1}^A \mathscr{V}(\mathscr{X}(t_a)) - \sum_{a=1}^{A-1} \mathscr{V}(\mathscr{X}(t_a)\wedge \mathscr{X}(t_{a+1})) - \mathscr{V}(\mathscr{X}(t_1)\vee\dots\vee \mathscr{X}(t_A)).
    \end{align}
    To see the last strict inequality, recall our assumption that there is an $a_0$ with $\mathscr{X}(t_{a_0})$ harder than $\widehat{X}_{a_0-1}\vee \widehat{X}_{a_0}$, from which $\mathscr{V}(\mathscr{X}(t_{a_0})) < \mathscr{V}(\widehat{X}_{a_0-1}\vee \widehat{X}_{a_0}) = \mathscr{V}(\mathscr{X}(p_a))$. However, (\ref{eq:lemma_leading_FIItermsweight_comparison_constructed_terms}) would then yield the following inequality
    \begin{align}
        \sum_{i=1}^A \mathscr{V}(\mathscr{X}(t_a)) <\sum_{a=1}^{A-1} \mathscr{V}(\mathscr{X}(t_a)\wedge \mathscr{X}(t_{a+1})) + \mathscr{V}(\mathscr{X}(t_1)\vee\dots\vee \mathscr{X}(t_A)),
    \end{align}
    contradicting eq.~(\ref{eq:lemma_leading_FIIterms_paths_weight}). As a result, $\mathscr{X}(t_a) = \widehat{X}_{a-1}\vee \widehat{X}_a$ must hold for all $a$. The second statement is thus proved as well.
\end{proof}
Let us now check the statement of lemma~\ref{lemma-leading_FIIterms_paths_weight} by inspecting the third term of (\ref{eq:leading_terms_motivating_example_Fiiterms}), i.e.,
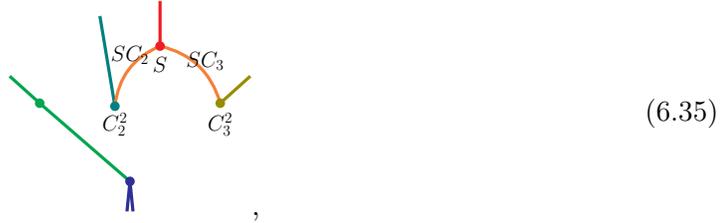
\begin{equation}
\begin{aligned}
    \begin{tikzpicture}[line width = 0.6, scale = 0.4, font=\large, mydot/.style={circle, fill, inner sep=.7pt}, transform shape]
    \draw (5,2.5) edge [very thick, Green] (2,5.1) node [] {};
    \draw (2,5.1) edge [very thick, Green] (1,6) node [] {};
    \draw (4.5,5) edge [very thick, teal] (4,8) node [] {};
    \draw (8,5.1) edge [very thick, olive] (9,6) node [] {};
    \draw (6,7) edge [very thick, Orange, bend right = 30] (4.5,5) node [] {};
    \draw (6,7) edge [very thick, Orange, bend left = 30] (8,5.1) node [] {};
    \draw (6,7) edge [very thick, Red] (6,8.5) node [] {};
    \draw (5.1,1.5) edge [very thick, Blue] (5,2.5) node [] {};
    \draw (4.9,1.5) edge [very thick, Blue] (5,2.5) node [] {};    
    \draw[fill, very thick, Green] (2,5.1) circle (3pt);
    \draw[fill, very thick, teal] (4.5,5) circle (3pt);
    \draw[fill, very thick, olive] (8,5.1) circle (3pt);
    \draw[fill, very thick, Red] (6,7) circle (3pt);
    \draw[fill, very thick, Blue] (5,2.5) circle (3pt);
    \node () at (4.5,4.4) {\huge $C_2^2$};
    \node () at (6,6.4) {\huge $S$};
    \node () at (8,4.4) {\huge $C_3^2$};
    \node () at (5,6.7) {\huge $SC_2$};
    \node () at (7.5,6.5) {\huge $SC_3$};
    \end{tikzpicture},
\end{aligned}
\end{equation}
where $A=3$, and the modes $\mathscr{X}(t_1)=C_2^2$, $\mathscr{X}(t_2)=S$, $\mathscr{X}(t_3)=C_3^2$, $\widehat{X}_1=SC_2$, and $\widehat{X}_2=SC_3$ have been specified. For this example, the relation $\widehat{X}_a = \mathscr{X}(t_a)\wedge \mathscr{X}(t_{a+1})$ becomes
\begin{align}
    SC_2=C_2^2\wedge S,\qquad SC_3=S\wedge C_3^2.
\end{align}
Meanwhile, the relation $\mathscr{X}(t_a) = \widehat{X}_{a-1}\vee \widehat{X}_a$ becomes
\begin{align}
    S= SC_2\vee SC_3.
\end{align}
These relations all hold by definition, serving as a check for lemma~\ref{lemma-leading_FIIterms_paths_weight}.

From the proof above, eq.~(\ref{eq:lemma_leading_FIIterms_paths_weight}) must hold for any $\mathcal{F}_\textup{II}^{(R)}$ term. That is, the equality in (\ref{ineq:virtualities_relation}) must hold if we take $\{X_1,\dots,X_n\} = \{\mathscr{X}(t_1),\dots,\mathscr{X}(t_A)\}$. A necessary and sufficient condition for this equality can be immediately obtained from corollary~\ref{theorem-virtualities_relation_corollary2}, which we summarize into the corollary below.
\begin{corollary}
\label{lemma-leading_FIIterms_paths_weight_corollary1}
Each $\mathcal{F}_\textup{II}^{(R)}$ term satisfies
\begin{align}
    \begin{split}
        &\left(\mathscr{X}(t_{a_1})\vee \mathscr{X}(t_{a_2})\right)\wedge \mathscr{X}(t_{a_3}) = \mathscr{X}(t_{*_3})\wedge \mathscr{X}(t_{a_3}),\\
        &\left(\mathscr{X}(t_{a_1})\vee \mathscr{X}(t_{a_2})\vee \mathscr{X}(t_{a_3})\right)\wedge \mathscr{X}(t_{a_4}) = \mathscr{X}(t_{*_4})\wedge \mathscr{X}(t_{a_4}),\\
        &\qquad\dots\\
        &\left(\mathscr{X}(t_{a_1})\vee \dots\vee \mathscr{X}(t_{a_{A-1}})\right)\wedge \mathscr{X}(t_{a_A}) = \mathscr{X}(t_{*_A})\wedge \mathscr{X}(t_{a_A}),
    \end{split}
\label{eq:virtuality_equation_condition_repeat}
\end{align}
with $\{a_1,\dots,a_A\}$ being any permutation of $\{1,\dots,A\}$, such that for each $k\in \{1,\dots,A\}$, the set $\{a_1,\dots,a_k\}$ consists of $k$ consecutive integers. The indices $*_3,\dots,*_A$ are uniquely determined such that each $*_j$ is from $\{a_1,\dots,a_{j-1}\}$ and adjacent to $a_j$.
\end{corollary}

\begin{figure}[t]
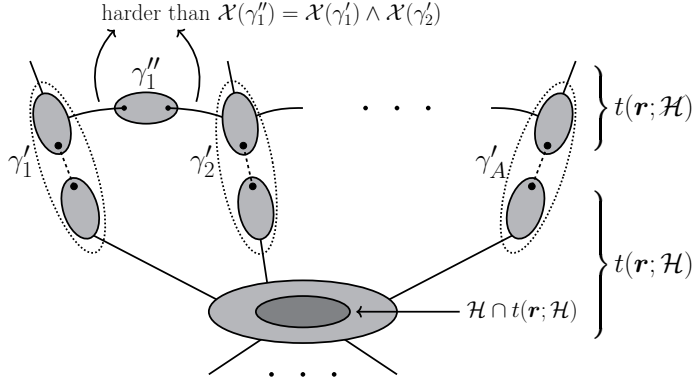

\centering
\include{figs/leading_Fterms_typeII_precise_configuration}
\vspace{-3em}\caption{General structure of an $\mathcal{F}_\textup{II}^{(R)}$ term $\x^{\r}$, where the spanning 2-tree $T^2(\r)$ consists of the components $t(\r;\mathcal{H})$ and $t(\r;\cancel{\mathcal{H}})$. Gray blobs represent tree subgraphs, with the dark gray blob being $\mathcal{H}\cap t(\r;\mathcal{H})$. The mode components $\gamma'_a$ ($a=1,\dots,A$), encircled by dotted curves, have their intersections $\gamma'_a\cap t(\r;\cancel{\mathcal{H}})$ connected by paths $P_{a,a+1}$. Each sub-path $\gamma''_a\cap P_{a,a+1}$ has the characteristic mode of $P_{a,a+1}$, which is $\mathscr{X}(\gamma'_a)\wedge \mathscr{X}(\gamma'_{a+1})$, with its endpoints belonging to harder-mode subgraphs.}
\label{figure-leading_Fterms_typeII_precise_configuration}
\end{figure}

In the entire graph $\mathcal{G}$, we now denote the mode component containing the subtree $t_a$ as $\gamma'_a$ (for $a=1,\dots,A$), and the mode component containing the softest (characteristic-mode) edges in $P_{a,a+1}$ as $\gamma''_{a}$ (for $a=1,\dots,A-1$). These subgraphs have certain structures in $T^2(\r)$, as described in the following corollary.
\begin{corollary}
\label{lemma-leading_FIIterms_paths_weight_corollary2}
    The subgraphs $\gamma'_a$ \textup{(}$a=1,\dots,A$\textup{)} and $\gamma''_a$ \textup{(}$a=1,\dots,A-1$\textup{)}, as defined above, satisfy:
    \begin{itemize}
        \item each $\widetilde{\gamma}'_a\cap T^2(\r)$ is a spanning 2-tree of $\widetilde{\gamma}'_a$;
        \item each $\widetilde{\gamma}''_a\cap T^2(\r)$ contains a single loop of $\widetilde{\gamma}''_a$.
    \end{itemize}
\end{corollary}
\begin{proof}
    We first show that the graph $\widetilde{\gamma}'_a\cap T^2(\r)$ has two connected components. By definition, $\widetilde{\gamma}'_a\cap T^2(\r)$ is obtained from $\gamma'_a\cap T^2(\r)$ by contracting to a single vertex all the harder-mode vertices which $\gamma'_a\cap T^2(\r)$ is adjacent to. From lemma~\ref{lemma-leading_FIIterms_subtree_adjacent_edges}, such vertices do not exist in $t(\r;\cancel{\mathcal{H}})$. Therefore, the contraction does not change the structure of $\gamma'_a\cap t(\r;\cancel{\mathcal{H}})$, which will remain as a tree subgraph, disconnected from $\widetilde{\gamma}'_a\cap t(\r;\mathcal{H})$.

    We next verify that $\widetilde{\gamma}'_a\cap t(\r;\mathcal{H})$ contains no loops in order to show the first statement. Since $t(\r;\mathcal{H})$ is a tree graph, a loop in $\widetilde{\gamma}'_a\cap t(\r;\mathcal{H})$ can only arise if $\gamma'_a$ is adjacent to two (or more) harder-mode vertices $v_1,v_2\in t(\r;\mathcal{H})$, and $\gamma'_a\cap t(\r;\mathcal{H})$ contains a (unique) path $P_1$ that connects $v_1$ and $v_2$. (After contracting $v_1$ and $v_2$ to the same vertex, the path $P_1$ turns into a closed loop.) To exclude this possibility, let us consider the union of mode subgraphs $\overline{\Gamma}_*\equiv \cup_X\Gamma_X$ with $X$ harder than $\mathscr{X}(\gamma'_a)$, which is connected according to theorem~\ref{theorem-mode_subgraphs_connectivity1}. It then follows that there is another path $P_2\subset \overline{\Gamma}_*$ joining $v_1$ and $v_2$, and furthermore, there is an edge $e_*\in P_2$ such that $t(\r;\mathcal{H})\cup e_*$ contains a loop that shares some edges with $P_1$. Let us denote any one of such edges by $e'$. The following spanning 2-tree
    $$T'^2 \equiv T^2(\r)\cup e_*\setminus e' =t(\r;\cancel{\mathcal{H}})\cup \left(t(\r;\mathcal{H})\cup e_*\setminus e'\right)$$
    has the same kinematic factor as $T^2(\r)$. Meanwhile, its weight is
    \begin{align}
        w(T'^2) = w(T^2(\r))+w(e')-w(e_*)<w(T^2(\r)),
    \end{align}
    where we have used the defining property that $\mathscr{X}(e')$ is softer than that $\mathscr{X}(e_*)$, from which $w(e')<w(e_*)$. The minimum-weight criterion is thus violated, indicating the absence of loops in $\widetilde{\gamma}'_a\cap t(\r;\mathcal{H})$. The first statement in the lemma is proved.
    
    Finally, let us consider the second statement. For each $a$, since $\gamma''_a\cap T^2(\r)$ contains no loops, any given loop in $\widetilde{\gamma}''_a\cap T^2(\r)$ must have the following origin: a path in $\gamma''_a\cap T^2(\r)$ has endpoints that are harder than $\mathscr{X}(\gamma''_a)$, thus a loop is formed after these endpoints are identified as the same vertex. Recall that $\gamma''_a$ is defined as the mode component containing the softest edges in $P_{a,a+1}$, which, according to lemma~\ref{lemma-leading_FIIterms_paths_weight}, form a \emph{single} sub-path of $P_{a,a+1}$. The two endpoints of this sub-path hence belongs to harder-mode subgraphs, which, in forming $\widetilde{\gamma}''_a$, are identified as a single vertex. A unique loop in $\widetilde{\gamma}''_a$ is then generated. The proof is then completed.
\end{proof}
With the results of corollary~\ref{lemma-leading_FIIterms_paths_weight_corollary2}, we immediately have:
\begin{subequations}
\label{eq:leading_FIIterms_2tree_loop_structure}
    \begin{align}
        & \mathfrak{n}_{\gamma'_a}^{} = L(\widetilde{\gamma}'_a)+1,\quad\textup{for }a=1,\dots,A;
        \label{eq:leading_FIIterms_spanning_2tree_structure}
        \\
        & \mathfrak{n}_{\gamma''_a}^{} = L(\widetilde{\gamma}''_a)-1,\quad\textup{for }a=1,\dots,A-1.
        \label{eq:leading_FIIterms_loop_structure}
    \end{align}
\end{subequations}
We have clarified the general structure of the $\mathcal{F}_\textup{II}^{(R)}$ terms $\x^{\r}$ as follows. To recapitulate, the spanning 2-tree $T^2(\r)$ can be represented by figure~\ref{figure-leading_Fterms_typeII_precise_configuration}, which consists of two components $t(\r;\mathcal{H})$ and $t(\r;\cancel{\mathcal{H}})$.
The component $t(\r;\mathcal{H})$ contains all the hard vertices. The other component, $t(\r;\cancel{\mathcal{H}})$, can be seen as $A$ distinct subgraphs $\gamma'_a\cap t(\r;\cancel{\mathcal{H}})$ (with $a=1,\dots,A$) aligned by $A-1$ paths $P_{a,a+1}$ (with $a=1,\dots,A-1$). Each gray blob in the figure represents a tree subgraph. The graph $\widetilde{\gamma}'_a\cap T^2(\r)$ is a spanning 2-tree of $\widetilde{\gamma}'_a$. The softest edges on each $P_{a,a+1}$ are of the mode $\mathscr{X}(\gamma'_a)\wedge \mathscr{X}(\gamma'_{a+1})$ (the characteristic mode of $P_{a,a+1}$), forming a sub-path of $P_{a,a+1}$. By denoting the mode component containing this sub-path as $\gamma''_a$ (with $a=1,\dots,A-1$), the graph $\widetilde{\gamma}''_a\cap T^2(\r)$ contains a single loop of $\widetilde{\gamma}''_a$.

\begin{figure}[t]
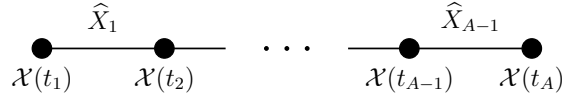

\centering
\include{figs/reduced_form}
\vspace{-2em}\caption{The reduced form of $t(\r;\cancel{\mathcal{H}})$, where the $A$ nodes represent the subtrees $t_1,\dots,t_A$, and the $A-1$ paths aligning them represent the paths $P_{1,2},\dots,P_{A-1,A}$. Here we have omitted any branches and external momenta attached to them. Below each $t_a$, we label its mode $\mathscr{X}(t_a)$, and above each $P_{a,a+1}$, we label its characteristic mode $\widehat{X}_a$.}
\label{figure-reduced_form}
\end{figure}

To characterize the mode structure of $t(\r;\cancel{\mathcal{H}})$, we define its \emph{reduced form} as a sequence of $A$ nodes aligned by $A-1$ paths (see figure~\ref{figure-reduced_form}). One can thus see the nodes as the characteristic subtrees $t_1,\dots,t_A$, and the paths as $P_{1,2},\dots,P_{A-1,A}$.
The characteristic subtree modes $\mathscr{X}(t_1),\dots,\mathscr{X}(t_A)$ are indicated on the corresponding nodes. Similarly, the characteristic modes $\widehat{X}_1,\dots,\widehat{X}_{A-1}$ are indicated on the corresponding paths. These modes satisfy the relations given in lemma~\ref{lemma-leading_FIIterms_paths_weight}.

We emphasize that the external momenta of $t(\r;\cancel{\mathcal{H}})$ are implicit in the reduced form, subject to certain constraints. In proving lemma~\ref{lemma-leading_FIIterms_paths_weight}, we have demonstrated that the mode of the total external momentum attached to $t_a$, which we denote by $\mathscr{X}(\mathfrak{p}(t_a))$, must be softer than or equal to $\mathscr{X}(t_a)$. It is also possible that some $t_a$ have no external momenta attached; however, according to corollary~\ref{lemma-leading_FIIterms_components_mode_structure_corollary1}, $t_1$ and $t_A$ must have at least one for each of them, satisfying eq.~(\ref{eq:lemma_leading_FIIterms_connection_structure}).

\begin{figure}[t]
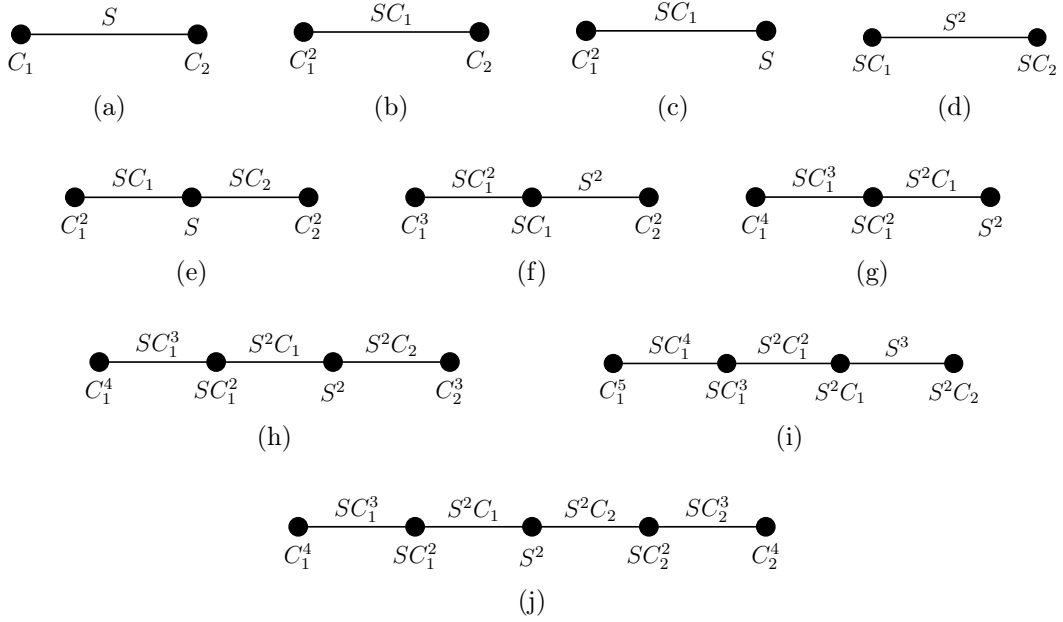

\centering
\begin{subfigure}[b]{0.2\textwidth}
\centering
\include{figs/leading_Fterms_typeII_2nodes_example1}
\vspace{-3em}\caption{}
\label{leading_Fterms_typeII_2nodes_example1}
\end{subfigure}
\hspace{1em}
\begin{subfigure}[b]{0.2\textwidth}
\centering
\include{figs/leading_Fterms_typeII_2nodes_example2}
\vspace{-3em}\caption{}
\label{leading_Fterms_typeII_2nodes_example2}
\end{subfigure}
\hspace{1em}
\begin{subfigure}[b]{0.2\textwidth}
\centering
\include{figs/leading_Fterms_typeII_2nodes_example3}
\vspace{-3em}\caption{}
\label{leading_Fterms_typeII_2nodes_example3}
\end{subfigure}
\hspace{1em}
\begin{subfigure}[b]{0.2\textwidth}
\centering
\include{figs/leading_Fterms_typeII_2nodes_example4}
\vspace{-3em}\caption{}
\label{leading_Fterms_typeII_2nodes_example4}
\end{subfigure}
\\
\centering\vspace{1em}
\begin{subfigure}[b]{0.25\textwidth}
\centering
\include{figs/leading_Fterms_typeII_3nodes_example1}
\vspace{-3em}\caption{}
\label{leading_Fterms_typeII_3nodes_example1}
\end{subfigure}
\hspace{1em}
\begin{subfigure}[b]{0.25\textwidth}
\centering
\include{figs/leading_Fterms_typeII_3nodes_example2}
\vspace{-3em}\caption{}
\label{leading_Fterms_typeII_3nodes_example2}
\end{subfigure}
\hspace{1em}
\begin{subfigure}[b]{0.25\textwidth}
\centering
\include{figs/leading_Fterms_typeII_3nodes_example3}
\vspace{-3em}\caption{}
\label{leading_Fterms_typeII_3nodes_example3}
\end{subfigure}
\\
\centering\vspace{1em}
\begin{subfigure}[b]{0.35\textwidth}
\centering
\include{figs/leading_Fterms_typeII_4nodes_example1}
\vspace{-3em}\caption{}
\label{leading_Fterms_typeII_4nodes_example1}
\end{subfigure}
\hspace{3em}
\begin{subfigure}[b]{0.35\textwidth}
\centering
\include{figs/leading_Fterms_typeII_4nodes_example2}
\vspace{-3em}\caption{}
\label{leading_Fterms_typeII_4nodes_example2}
\end{subfigure}
\\
\centering\vspace{1em}
\begin{subfigure}[b]{0.45\textwidth}
\centering
\include{figs/leading_Fterms_typeII_5nodes_example1}
\vspace{-3em}\caption{}
\label{leading_Fterms_typeII_5nodes_example1}
\end{subfigure}
\caption{Some examples of reduced forms of $t(\r;\cancel{\mathcal{H}})$, with the modes $\mathscr{X}(t_a)$ and $\widehat{X}_a$ specified.}
\label{figure-leading_Fterms_typeII_examples}
\end{figure}

So far, we have characterized $\mathcal{F}_\textup{II}^{(R)}$ terms by their subgraph structures via eq.~(\ref{eq:leading_FIIterms_2tree_loop_structure}), analogous to the previous characterizations of $\mathcal{U}^{(R)}$ and $\mathcal{F}_\textup{I}^{(R)}$ terms in eqs.~(\ref{eq:leading_Uterms_subgraph_characterization}) and (\ref{eq:leading_Fterms_typeI_subgraph_characterization}), respectively. Before proceeding to a summary of these results, we note that the modes $\mathscr{X}(\gamma'_a)$ and $\mathscr{X}(\gamma''_a)$ can be further constrained. To this end, we begin from the following lemma.
\begin{lemma}
\label{lemma-leading_FIIterms_specific_modes}
    The modes $\mathscr{X}(t_a)$ ($a=1,\dots,A$) are restricted to the following forms: there exist an index $b\in\{1,\dots,A\}$ and distinct directions $i\neq j$, such that
    \begin{itemize}
        \item[] $\mathscr{X}(t_a) = S^{m_a}C_i^{n_a}$ for $a\in \{1,\dots, b\}$, and $\mathscr{X}(t_a) = S^{m_a}C_j^{n_a}$ for $a\in \{b+1,\dots, A\}$,
    \end{itemize}
    with the exponents satisfying
    \begin{itemize}
        \item[] $m_1+n_1>m_2+n_2>\dots>m_{b}+n_{b}$, and $m_{b+1}^{}+n_{b+1}^{}<\dots<m_A+n_A$.
    \end{itemize}
\end{lemma}
Before going through a proof, we first discuss the meaning of this lemma by illustrating some concrete examples in figure~\ref{figure-leading_Fterms_typeII_examples}. Each figure represents the reduced form of a $t(\r;\cancel{\mathcal{H}})$, where (a)--(d) have $A=2$, (e)--(g) have $A=3$, (h) and (i) have $A=4$, and (j) has $A=5$.
One can verify the statement of lemma~\ref{lemma-leading_FIIterms_specific_modes} directly through these examples. Consider, for instance, case (h), where
\begin{align}
    \mathscr{X}(t_1) = C_1^4,\quad \mathscr{X}(t_2) = SC_1^2,\quad \mathscr{X}(t_3) = S^2,\quad \mathscr{X}(t_4) = C_2^3.
\end{align}
Each mode above is of the form $S^mC_1^n$ or $S^mC_2^n$ (for $\mathscr{X}(t_3) = S^2$, we may regard it as belonging to both forms by taking $n=0$), while the values $m+n$ satisfy $m_1+n_1 > m_2+n_2 > m_3+n_3$ and $m_3+n_3 < m_4+n_4$. The characteristic modes of the paths,
\begin{align}
    \widehat{X}_1 = SC_1^3,\quad \widehat{X}_2 = S^2C_1,\quad \widehat{X}_3 = S^2C_2,
\end{align}
are related to the subtree modes $\{\mathscr{X}(t_a)\}_{a=1,2,3,4}$ via the relations in lemma~\ref{lemma-leading_FIIterms_paths_weight}.

Below we prove lemma~\ref{lemma-leading_FIIterms_specific_modes}.
\begin{proof}
We first exclude an unwanted pattern. Suppose three consecutive characteristic subtrees $t_1,t_2,t_3$ all have modes of the same direction $i$:
$$\mathscr{X}(t_1)= S^{m_1}C_i^{n_1},\ \ \mathscr{X}(t_2) = S^{m_2}C_i^{n_2},\ \ \mathscr{X}(t_3) = S^{m_3}C_i^{n_3},$$
with $m_1+n_1<m_2+n_2$ and $m_3+n_3<m_2+n_2$. We show this cannot occur for a generic $\mathcal{F}_\text{II}^{(R)}$ term.

From corollary~\ref{lemma-leading_FIIterms_paths_weight_corollary1}, the following condition must hold
\begin{align}
    \left(S^{m_1}C_i^{n_1}\vee S^{m_2}C_i^{n_2}\right)\wedge S^{m_3}C_i^{n_3} = S^{m_2}C_i^{n_2}\wedge S^{m_3}C_i^{n_3}.
\end{align}
Moreover, as argued below lemma~\ref{lemma-leading_FIIterms_paths_weight}, adjacent modes ($S^{m_1}C_i^{n_1}$ and $S^{m_2}C_i^{n_2}$, or $S^{m_2}C_i^{n_2}$ and $S^{m_3}C_i^{n_3}$) are overlapping. We then apply the $\wedge$ and $\vee$ rules from theorem~\ref{theorem-overlapping_modes_intersection_union_rules} and obtain:
\begin{subequations}
\begin{align}
    \left(S^{m_1}C_i^{n_1}\vee S^{m_2}C_i^{n_2}\right)\wedge S^{m_3}C_i^{n_3} &= S^{m_3}C_i^{\max\{m_1+n_1,m_3+n_3\}-m_3},\\
    S^{m_2}C_i^{n_2}\wedge S^{m_3}C_i^{n_3} &= S^{m_3}C_i^{m_2+n_2-m_3}.
\end{align}
\end{subequations}
To derive the first equation above, note that $S^{m_1}C_i^{n_1}\vee S^{m_2}C_i^{n_2}$ can also be harder than or equal to $S^{m_3}C_i^{n_3}$; these cases match the right-hand side.

By construction, $\max\{m_1+n_1,m_3+n_3\} < m_2+n_2$, so the right-hand sides cannot be equal. Hence, we must have either $m_1+n_1 < m_2+n_2 < m_3+n_3$ or $m_1+n_1 > m_2+n_2 > m_3+n_3$. The same reasoning extends to any chain of consecutive subtrees $t_{a+1},t_{a+2},\dots,t_{a+b}$ ($b\geqslant 2$) with modes of the same direction $i$:
$$\mathscr{X}(t_{a+1}) = S^{m_{a+1}}C_i^{n_{a+1}},\ \ \mathscr{X}(t_{a+2}) = S^{m_{a+2}}C_i^{n_{a+2}},\ \ \dots,\ \ \mathscr{X}(t_{a+b}) = S^{m_{a+b}}C_i^{n_{a+b}}.$$
For such a chain one of the following two ordering patterns must hold:
\begin{align}
    \textup{(1) }\ &m_{a+1}+n_{a+1}<\dots<m_{a+b}+n_{a+b}\ \textup{ and }\ m_{a+1}>\dots>m_{a+b},\nonumber\\
    \textup{(2) }\ &m_{a+1}+n_{a+1}>\dots>m_{a+b}+n_{a+b}\ \textup{ and }\ m_{a+1}<\dots<m_{a+b}.\nonumber
\end{align}
The two cases are related by reversing the order of the subtrees; we may therefore assume, without loss of generality, that $m_{a+1}+n_{a+1} > \dots > m_{a+b}+n_{a+b}$ (case (2)). Theorem~\ref{theorem-overlapping_modes_intersection_union_rules} then yields
\begin{align}
\label{eq:lemma_leading_FIIterms_specific_modes_conclusion1}
    \mathscr{X}(t_{a+1})\vee\dots\vee \mathscr{X}(t_{a+b}) = S^{m_{a+1}}C_i^{m_{a+b}+n_{a+b}-m_{a+1}}.
\end{align}
Now let $t_{a+1},\dots,t_{a+b}$ be \emph{all} the characteristic subtrees whose modes can be written as $S^mC_i^n$. If any other characteristic subtrees remain, they must have a different direction $j\neq i$ and lie either before $t_{a+1}$ or after $t_{a+b}$. That leads to the following two possibilities.

\bigbreak\noindent\emph{Possibility 1:} $\mathscr{X}(t_a) = S^{m_a}C_j^{n_a}$ for some $j \neq i$. Corollary~\ref{lemma-leading_FIIterms_paths_weight_corollary1} requires
\begin{align}\label{eq:possibility1_condition}
\bigl(\mathscr{X}(t_{a+1})\vee\dots\vee\mathscr{X}(t_{a+b})\bigr)\wedge\mathscr{X}(t_a)
=\mathscr{X}(t_{a+1})\wedge\mathscr{X}(t_a).
\end{align}
Using \eqref{eq:lemma_leading_FIIterms_specific_modes_conclusion1} and theorem~\ref{theorem-overlapping_modes_intersection_union_rules}, both sides of \eqref{eq:possibility1_condition} can be evaluated. With the abbreviation $\sigma_\ell\equiv m_\ell+n_\ell$ we obtain
\begin{align}
\label{eq:lemma_leading_FIIterms_specific_modes_conclusion2_step1}
    \left(\mathscr{X}(t_{a+1})\vee\dots\vee \mathscr{X}(t_{a+b})\right)\wedge \mathscr{X}(t_a)&=S^{m_{a+1}}C_i^{m_{a+b}+n_{a+b}-m_{a+1}} \wedge S^{m_a}C_j^{n_a}\nonumber\\
    &\hspace{-5em}= \left\{\begin{array}{@{}l@{\quad}l}
    S^{\sigma_{a+b}}C_j^{\sigma_a-\sigma_{a+b}} & \textup{for } m_{a+b}+n_{a+b}\leqslant m_a+n_a, \\
    S^{\sigma_a}C_i^{\sigma_{a+b}-\sigma_a} & \textup{for } m_a+n_a\leqslant m_{a+b}+n_{a+b},
    \end{array}\right.
\end{align}
where we have defined $\sigma_a\equiv m_a+n_a$ throughout this proof for simplicity. Meanwhile,
\begin{align}
\label{eq:lemma_leading_FIIterms_specific_modes_conclusion2_step2}
    \mathscr{X}(t_{a+1})\wedge \mathscr{X}(t_a)&=S^{m_{a+1}}C_i^{n_{a+1}} \wedge S^{m_a}C_j^{n_a}\nonumber\\
    &= \left\{\begin{array}{@{}l@{\quad}l}
    S^{\sigma_{a+1}}C_j^{\sigma_a-\sigma_{a+1}} & \textup{for } m_{a+1}+n_{a+1}\leqslant m_a+n_a, \\
    S^{\sigma_a}C_i^{\sigma_{a+1}-\sigma_a} & \textup{for } m_a+n_a\leqslant m_{a+1}+n_{a+1}.
    \end{array}\right.
\end{align}
Because $i\neq j$ and $\sigma_{a+1}>\sigma_{a+b}$, the two expressions above would never coincide. Eq.~\eqref{eq:possibility1_condition} cannot be satisfied. Therefore, a subtree of direction $j$ cannot appear \emph{before} the aforementioned sequence of direction‑$i$ subtrees $t_{a+1},\dots,t_{a+b}$. In other words, $a=0$.

\bigbreak\noindent\emph{Possibility 2:} $\mathscr{X}(t_{a+b+1}) = S^{m_{a+b+1}}C_j^{n_{a+b+1}}$ for some $j \neq i$. Corollary~\ref{lemma-leading_FIIterms_paths_weight_corollary1} requires
\begin{align}
\label{eq:possibility2_condition}
    \left(\mathscr{X}(t_{a+1})\vee\dots\vee \mathscr{X}(t_{a+b})\right)\wedge \mathscr{X}(t_{a+b+1}) = \mathscr{X}(t_{a+b})\wedge \mathscr{X}(t_{a+b+1}).
\end{align}
A direct computation analogous to the previous one shows that \eqref{eq:possibility2_condition} holds identically, regardless of the ordering of $\sigma_{a+b}$ and $\sigma_{a+b+1}$:
\begin{align}
\label{eq:lemma_leading_FIIterms_specific_modes_conclusion2_step4}
    \left(\mathscr{X}(t_{a+1})\vee\dots\vee \mathscr{X}(t_{a+b})\right)\wedge \mathscr{X}(t_{a+b+1}) &=S^{m_{a+1}}C_i^{m_{a+b}+n_{a+b}-m_{a+1}} \wedge S^{m_{a+b+1}}C_j^{n_{a+b+1}}\nonumber\\
    &\hspace{-6em}= \left\{\begin{array}{@{}l@{\quad}l}
    S^{\sigma_{a+b}}C_j^{\sigma_{a+b+1}-\sigma_{a+b}} & \textup{for } m_{a+b}+n_{a+b}\leqslant m_{a+b+1}+n_{a+b+1} \\
    S^{\sigma_{a+b+1}}C_i^{\sigma_{a+b}-\sigma_{a+b+1}} & \textup{for } m_{a+b+1}+n_{a+b+1}\leqslant m_{a+b}+n_{a+b}
    \end{array}\right.\nonumber\\
    &\hspace{-6em}= \mathscr{X}(t_{a+b})\wedge \mathscr{X}(t_{a+b+1}),
\end{align}
That is, a subtree of a different direction $j$ may appear only \emph{after} the sequence of direction‑$i$ subtrees $t_{a+1},\dots,t_{a+b}$.

\bigbreak
We can repeat the same analysis for the new direction $j$ and obtain the following conclusion: all subtrees $t_1,\dots,t_A$ must be partitioned into at most two contiguous sequences, each containing subtrees of a single direction. Furthermore, within each sequence the values $\sigma_a=m_a+n_a$ are \emph{strictly monotonic}; the sequence containing $t_1$ has $\sigma_a$ strictly decreasing, while the other sequence (if it exists) has $\sigma_a$ strictly increasing. This is precisely the statement of the lemma.
\end{proof}

As a final remark, the external momenta attached to $t_a$ ($a=1,\dots,A$), which we have denoted as $\mathfrak{p}(t_a)$, must satisfy:
\begin{itemize}
    \item $\mathscr{X}(\mathfrak{p}(t_a))$ is equal to or softer than $\mathscr{X}(t_a)$;
    \item $\mathfrak{p}(t_1),\mathfrak{p}(t_A)\neq 0$, while other $\mathfrak{p}(t_a)$ ($a=2,\dots,A-1$) can be zero.;
    \item $\mathscr{X}(t_1)\vee \mathscr{X}(t_A) = \mathscr{X}(\mathfrak{p}(t_1))\vee \mathscr{X}(\mathfrak{p}(t_A))$.
\end{itemize}
These properties are automatic from the lemmas above.

\subsubsection{Summary}

Let us summarize all the main results for the leading-term structure, which we have obtained so far, and formulate them into the following theorem.
\begin{theorem}
\label{theorem-leading_terms_general_structures}
    For any given region $R$, the leading terms $\x^{\r}$ can be characterized as follows:
    \begin{itemize}
        \item $\mathcal{U}^{(R)}$ term: each graph $\gamma\cap T^1(\r)$, where $\gamma$ denotes any mode component, is a spanning tree of $\widetilde{\gamma}$. That is,
        \begin{eqnarray}
        \mathfrak{n}_{\gamma}^{} = L(\widetilde{\gamma}) \text{ for any mode component }\gamma\,.
        \end{eqnarray}
        \item $\mathcal{F}_\textup{I}^{(R)}$ term: there exists a corresponding mode component $\gamma'$ with $\mathscr{X}(\gamma') = \mathscr{X}(Q(\r))$, such that (1) $\gamma'\cap T^2(\r)$ is a spanning 2-tree of $\widetilde{\gamma}'$; (2) for any other mode component $\gamma \neq \gamma'$, the graph $\gamma \cap T^2(\r)$ is a spanning tree of $\widetilde{\gamma}$. That is,
        \begin{align}
        \label{eq:leading_Fterms_typeI_subgraph_characterization_repeat}
        \mathfrak{n}_{\gamma'}^{} = L(\widetilde{\gamma}')+1,\quad \mathfrak{n}_{\gamma}^{} = L(\widetilde{\gamma})\ \ \ \forall\gamma\neq \gamma'\;.
        \end{align}
        \item $\mathcal{F}_\textup{II}^{(R)}$ term: the spanning 2-tree $T^2(\r)$ can be represented by figure~\ref{figure-leading_Fterms_typeII_precise_configuration}, with mode components satisfying
        \begin{align}
        \begin{split}
        &\mathscr{X}(\gamma''_a) = \mathscr{X}(\gamma'_a) \wedge \mathscr{X}(\gamma'_{a+1})\ \ \textup{for }a=1,\dots,A-1,\\
        &\mathscr{X}(\gamma'_a) = \mathscr{X}(\gamma''_{a-1}) \vee \mathscr{X}(\gamma''_a)\ \ \textup{for }a=2,\dots,A-1,\\
        & \mathfrak{n}_{\gamma'_1}^{} = L(\widetilde{\gamma}'_1)+1,\dots,\ \mathfrak{n}_{\gamma'_{A}}^{} = L(\widetilde{\gamma}'_{A})+1,\\
        & \mathfrak{n}_{\gamma''_1}^{} = L(\widetilde{\gamma}''_1)-1,\dots,\ \mathfrak{n}_{\gamma''_{A-1}}^{} = L(\widetilde{\gamma}''_{A-1})-1,\\
        & \mathfrak{n}_{\gamma}^{} = L(\widetilde{\gamma}),\ \ \forall\gamma\notin \{\gamma'_1,\dots,\gamma'_A\}\cup \{ \gamma''_1,\dots,\gamma''_{A-1}\}.
        \end{split}
        \label{eq:leading_Fterms_typeII_subgraph_characterization}
        \end{align}
        Furthermore, the modes of $\{\gamma'_a\}_{a=1,\dots,A}$ and $\{\gamma''_a\}_{a=1,\dots,A-1}$ are restricted by lemma~\ref{lemma-leading_FIIterms_specific_modes}.
    \end{itemize}
\end{theorem}
This theorem will play an essential role in validating the infrared-compatibility requirement. Additionally, the Second Connectivity Theorem (theorem~\ref{theorem-mode_subgraphs_connectivity2}) admits a concise proof, which we present to conclude this subsection.

\subsubsection{Proof of the Second Connectivity Theorem}

Again, we proceed by contradiction. Suppose $\gamma$ is a mode component that satisfies neither condition in theorem~\ref{theorem-mode_subgraphs_connectivity2}. We will show that for any leading term $\x^{\r}$, the subgraph $\widetilde{\gamma} \cap T(\r)$ is always a spanning tree of $\widetilde{\gamma}$, thereby inducing a homogeneity condition that renders the expanded integral scaleless.

From theorem~\ref{theorem-leading_terms_general_structures}, $\widetilde{\gamma}\cap T(\r)$ is a spanning tree of $\widetilde{\gamma}$ as long as $\x^{\r}$ belongs to $\mathcal{U}^{(R)}(\x)$ or $\mathcal{F}_\textup{I}^{(R)}(\x;\s)$. For $\x^{\r}$ being an $\mathcal{F}_\textup{II}^{(R)}$ term, the only possibility for a different $\widetilde{\gamma}\cap T(\r)$ is that $\gamma\in \{ \gamma'_1,\dots,\gamma'_n,\gamma''_1,\dots,\gamma''_{n-1} \}$. However, each $\gamma''_i$ meets condition \emph{1} of theorem~\ref{theorem-mode_subgraphs_connectivity2}: it is relevant to two harder-mode subgraphs $\gamma'_i$ and $\gamma'_{i+1}$, with $\mathscr{X}(\gamma''_i) = \mathscr{X}(\gamma'_i)\wedge \mathscr{X}(\gamma'_{i+1})$. Meanwhile, each $\gamma'_i$ meets condition \emph{2} of theorem~\ref{theorem-mode_subgraphs_connectivity2}: there are two momenta $k_{i,1},k_{i,2}$---sourced from the external kinematics and/or softer-mode subgraphs relevant to $\gamma'_i$---entering $\gamma'_i$, such that $\mathscr{X}(\gamma'_i) = \mathscr{X}(k_{i,1})\vee \mathscr{X}(k_{i,2})$. Therefore, $\gamma$ can belong to neither $\{\gamma'_i\}$ nor $\{\gamma''_i\}$, and $\widetilde{\gamma}\cap T(\r)$ must be a spanning tree of $\widetilde{\gamma}$ as well. This completes the proof of the Second Connectivity Theorem.

\subsection{The infrared-compatibility requirement}
\label{section-infrared_compatibility_requirements}

This section demonstrates that for a configuration $R$ satisfying the connectivity requirements, the infrared-compatibility requirement is both necessary and sufficient for $R$ to be a region. This is equivalent to showing that the corresponding lower face $f_R$, which is spanned by all leading terms, is an $N$-dimensional space, where $N$ is the number of edges in $\mathcal{G}$.

There have been previous analyses in this direction for the on-shell expansion~\cite{GrdHzgJnsMaSchlk22} and soft expansion~\cite{Ma23}; they aim to determine the dimension of $f_R$ by identifying independent vectors within it: $R$ is a (facet) region if and only if one can find $N$ independent vectors. Here, in contrast, we shall employ an equivalent but formally more concise approach, namely, to investigate whether the space normal to $f_R$ is one dimensional.

This can be achieved by solving for the edge weights $\{w_e\}_{e\in\mathcal{G}}$ from the leading polynomial. Specifically, each term that is leading in the expansion with respect to $R$ yields a linear equation constraining these weights. Given the complete set of these constraints, the condition that $f_R$ is $N$ dimensional is then equivalent to the uniqueness of the solution for $\{w_e\}$.

Let us see this from the following two configurations of the one-loop form factor, with the external momenta in two distinct asymptotic limits:
\begin{equation}
    \begin{tikzpicture}[baseline=10ex, scale=0.45, mydot/.style={circle, fill, inner sep=.8pt}]

    \draw [ultra thick, color=Blue] (5.1,2) -- (5,3);
    \draw [ultra thick, color=Blue] (4.9,2) -- (5,3);
    \draw [very thick, color=ForestGreen] (2,7) -- (5,3);
    \draw [very thick, color=olive] (8,7) -- (5,3);
    \draw [very thick, color=Red, bend left = 30] (3,5.67) to (7,5.67);
    \node [draw=ForestGreen, circle, minimum size=3pt, fill=ForestGreen, inner sep=0pt, outer sep=0pt] () at (3,5.67) {};
    \node [draw=olive, circle, minimum size=3pt, fill=olive, inner sep=0pt, outer sep=0pt] () at (7,5.67) {};
    \node [draw=Blue, circle, minimum size=3pt, fill=Blue, inner sep=0pt, outer sep=0pt] () at (5,3) {};
    \node () at (5,1.5) {\large $q_1$};
    \node () at (2,7.5) {\large $p_1$};
    \node () at (8,7.5) {\large $p_2$};
    \node () at (3.5,4.2) {\large $1$};
    \node () at (6.5,4.2) {\large $2$};
    \node () at (5,6.8) {\large $3$};
    \end{tikzpicture}
    \xrightarrow[]{\textup{asymptotic limits}}
    \left\{\begin{matrix}
        p_1^2\sim\lambda,\ p_2^2\sim\lambda,\ q_1^2\sim 1;
        \\\\
        p_1^2\sim\lambda,\ p_2^2=0,\ q_1^2\sim 1.
    \end{matrix}\right.
    \label{eq:solving_weights_method_motivating_examples}
\end{equation}
The edges $e_1$, $e_2$, and $e_3$ are in the $C_1$, $C_2$, and $S$ modes respectively. This configuration, as has been explained in section~\ref{section-introduction}, is a region for the first asymptotic limit ($p_1^2\sim\lambda,\ p_2^2\sim\lambda,\ q_1^2\sim 1$), but not for the second one ($p_1^2\sim\lambda,\ p_2^2=0,\ q_1^2\sim 1$).

This conclusion can be derived by solving for the edge weights. Recall that we use $\x^{\r}$ to represent a term $s\!\cdot\!x_1^{r_1}\cdots x_N^{r_N}$ where $s\sim \lambda^{r_{N+1}}$. If $\x^{\r}$ is a leading term, we have:
\begin{align}
\label{eq:solving_weights_method_center_equation}
    \sum_{i=1}^N r_i\cdot w_{e_i} + r_{N+1} = w_\mathcal{G}^{},
\end{align}
where $w_\mathcal{G}^{}$ is called the leading weight, which is the same for all leading terms. From now on, we will refer to eq.~(\ref{eq:solving_weights_method_center_equation}) as the \emph{characteristic equation} of $\x^{\r}$. For the first asymptotic limit, the leading terms and their characteristic equations are:
\begin{align}
    \begin{aligned}
        x_3 \quad &\rightarrow\quad  w_{e_3} = w_\mathcal{G}^{}; \\
        (-p_1^2)x_1x_3 \quad &\rightarrow\quad  w_{e_1} + w_{e_3} + 1 = w_\mathcal{G}^{}; \\
        (-p_2^2)x_2x_3 \quad &\rightarrow\quad  w_{e_2} + w_{e_3} + 1 = w_\mathcal{G}^{}; \\
        (-q_1^2)x_1x_2 \quad &\rightarrow\quad  w_{e_1} + w_{e_2} = w_\mathcal{G}^{}.
    \end{aligned}
\end{align}
There is a unique solution to the equations above: $(w_{e_1},w_{e_2},w_{e_3}) = (-1,-1,-2)$, which confirms that the configuration is a region in this limit. In contrast, for the second limit, we only have three (instead of four) leading terms, corresponding to a subset of the equations above:
\begin{align}
    \begin{aligned}
        x_3 \quad &\rightarrow\quad  w_{e_3} = w_\mathcal{G}^{}; \\
        (-p_1^2)x_1x_3 \quad &\rightarrow\quad  w_{e_1} + w_{e_3} + 1 = w_\mathcal{G}^{}; \\
        (-q_1^2)x_1x_2 \quad &\rightarrow\quad  w_{e_1} + w_{e_2} = w_\mathcal{G}^{}.
    \end{aligned}
\end{align}
Here the solution is no longer unique: $(w_{e_1},w_{e_2},w_{e_3}) = (-1,-1+a,-2+a)$ contains a free parameter $a\in\mathbb{R}$. This means that the space normal to $f_R$ is at least two dimensional, so $f_R$ is at most $N-1$ dimensional, and the configuration is not a region in this limit.

In the following, we will apply this method to generic Feynman graphs in the wide-angle kinematics. In section~\ref{section-sufficiency_subgraph_requirements}, we show the sufficiency of the infrared-compatibility requirement: with this requirement fulfilled, one can specify the unique solution of $(w_{e_1},\dots,w_{e_N})$ based on the leading-term characteristic equations, using a weight-solving algorithm we shall provide. In section~\ref{section-necessity_subgraph_requirements}, we show the necessity of the infrared-compatibility requirement by contradiction: once it is not fulfilled, the solution always contains a free parameter $a\in \mathbb{R}$.

\subsubsection{Proof of sufficiency}
\label{section-sufficiency_subgraph_requirements}

We first introduce a lemma regarding the edge weights within the same mode component.
\begin{lemma}
\label{lemma-solving_weights_method_leading_Uterm}
    For any mode component $\gamma$, the characteristic equations with respect to the $\mathcal{U}^{(R)}$ terms constrain all edge weights in $\gamma$ to be equal. Consequently, solving for one of them automatically determines the others.
\end{lemma}
This lemma has been stated and proved (as theorem 1) in ref.~\cite{GrdHzgJnsMaSchlk22}; for brevity, we omit the proof here. From this, by setting all the edge weights in a mode component equal, all the characteristic equations for the $\mathcal{U}^{(R)}$ terms reduce to a single one. To solve for all edge weights in $\mathcal{G}$, it is therefore sufficient to solve for one arbitrary edge weight from each mode component. We now present a recursive procedure that accomplishes this, which we refer to as the \emph{weight-solving algorithm}.

\begin{mdframed}[linewidth=1pt, skipabove=10pt, skipbelow=10pt]
\noindent\textbf{Weight-solving algorithm}
\begin{enumerate}[leftmargin=*]
    \item []\emph{Step 1. }By comparing the characteristic equations from specific $\mathcal{U}^{(R)}$ and $\mathcal{F}_\textup{I}^{(R)}$ terms, solve the edge weight of any $\gamma$ for which the partial momentum—sourced exclusively from the external kinematics—is $\mathscr{X}(\gamma)$.
    \item []\emph{Step 2. }If edges with unsolved weights remain, compare specific $\mathcal{F}_\textup{II}^{(R)}$ terms to solve the weight of any 1VI component $\widetilde{\gamma}$ satisfying one of the following conditions:
    \begin{itemize}
        \item $\mathscr{X}(\gamma) = S^m$, and $\gamma$ is contained in a degree-$m$ messenger and relevant to some weight-solved mode component $\gamma_1$;
        \item $\mathscr{X}(\gamma) = S^mC_i^n$ ($n\in\mathbb{N}$), and $\gamma$ is relevant to two weight-solved mode components $\gamma_1$ and $\gamma_2$, with $\mathscr{X}(\gamma) = \mathscr{X}(\gamma_1)\wedge \mathscr{X}(\gamma_2)$.
    \end{itemize}
    \item []\emph{Step 3. }If edges with unsolved weights remain, attach to any vertex of those $\gamma$ (from the previous step) an $\mathscr{X}(\gamma)$-mode external momentum.
    \item []\emph{Step 4+. }Repeat Steps \emph{1}--\emph{3} until all edge weights are solved.
\end{enumerate}
\end{mdframed}
Recall the definition of infrared compatibility from section~\ref{section-formal_construction_statements}, which is closely related to the procedure described above. In fact, the weight-solving algorithm provides a concrete representation of how the condition of infrared compatibility ``propagates'' through the subgraphs. Specifically, if the process of solving for weights by comparing certain leading terms, as outlined in Steps \emph{1} and \emph{2}, proceeds successfully, then fulfilling the infrared-compatibility requirement (i.e., all the mode components are infrared compatible) guarantees that the procedure will terminate with all edge weights solved. This constitutes the core idea of our proof.

The purpose of Step \emph{3} is to induce the recursive procedure by enabling the repetition of Steps \emph{1} and \emph{2} on the modified graph. A crucial subtlety, however, arises here: attaching an additional external momentum to $\mathcal{G}$ can generate new leading terms, hence new characteristic equations constraining the weights $\{w_e\}$. We must therefore demonstrate that these new equations are (linearly) \emph{dependent} on the original ones; otherwise, the addition of such an external momentum would change the dimension of $f_R$, invalidating the whole proof.

Consequently, the remainder of this section will demonstrate:
\begin{enumerate}
\item [(1)] how the edge weights can be solved as indicated in Steps \emph{1} and \emph{2};
\item [(2)] why the additional characteristic equations from Step \emph{3} can be determined by the original ones.
\end{enumerate}

\begin{figure}[t]
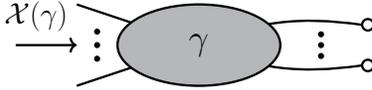

\centering
\include{figs/configuration_gamma_step1}
\vspace{-2em}\caption{The configuration of a mode component $\gamma$, whose total external momentum is in the mode $\mathscr{X}(\gamma)$. The circled vertices are those belonging to harder-mode subgraphs, hence not in $\gamma$.}
\label{figure-configuration_gamma_step1}
\end{figure}

Starting from Step \emph{1}, let us consider any given 1VI component $\widetilde{\gamma}$, whose total external momentum is of the mode $\mathscr{X}(\gamma)$. The configuration of $\gamma$ can then be described by figure~\ref{figure-configuration_gamma_step1}, where the circled vertices are those belonging to harder-mode subgraphs (hence not in $\gamma$), which are identified as a single vertex in $\widetilde{\gamma}$.

Among all the terms of $\mathcal{P}^{(R)}$, we choose two typical ones as follows. The first, $\x^{\r_1}$, is a $\mathcal{U}^{(R)}$ term. From lemma~\ref{lemma-leading_Uterms_subgraph_characterization}, $\widetilde{\gamma}\cap T^1(\r_1)$ is a spanning tree of $\widetilde{\gamma}$. Since the total external momentum flowing through $\widetilde{\gamma}$ is of the mode $\mathscr{X}(\gamma)$, we can always choose $\x^{\r_1}$ such that there is one edge $e\in \gamma\cap T^1(\r_1)$, with the momentum flowing through $e$ of the mode $\mathscr{X}(\gamma)$. Based on this, we can further construct an $\mathcal{F}$ term $\x^{\r_2}$, such that $T^2(\r_2) = T^1(\r_1)\setminus e$. It follows from eq.~(\ref{eq:leading_Fterms_typeI_factorize}) that $\x^{\r_2}$ is an $\mathcal{F}_{\textup{I}}^{(R)}$ term.

One can then examine the characteristic equations for $\x^{\r_1}$ and $\x^{\r_2}$. Their differences lie in a $w_e$ term (which the $\x^{\r_2}$ equation possesses while the $\x^{\r_1}$ equations does not) and their kinematic contributions (the $r_{N+1}$ term in eq.~(\ref{eq:solving_weights_method_center_equation}), which is $0$ for $\r=\r_1$ while $\mathscr{V}(\mathscr{X}(\gamma))$ for $\r=\r_2$). By subtracting these equations, we obtain
\begin{align}
\label{eq:solving_weights_method_step1_result}
    0 = w_e + \mathscr{V}(\mathscr{X}(\gamma))\quad\Leftrightarrow\quad w_e = -\mathscr{V}(\mathscr{X}(\gamma)).
\end{align}
The weight of any edge $e\in \gamma$ is then solved equally as $-\mathscr{V}(\mathscr{X}(\gamma))$, as the content of Step \emph{1}.

Now suppose some edges have unsolved weights at this point. We proceed to Step \emph{2}. Since all the mode components are infrared compatible, the definition in section~\ref{section-formal_construction_statements} guarantees that at least one component $\gamma$ satisfying the condition in Step \emph{2} must exist. We now explain how to solve for the weight of edges from such a $\gamma$, which we denote by $w_\gamma$ from now on.

\begin{figure}[t]
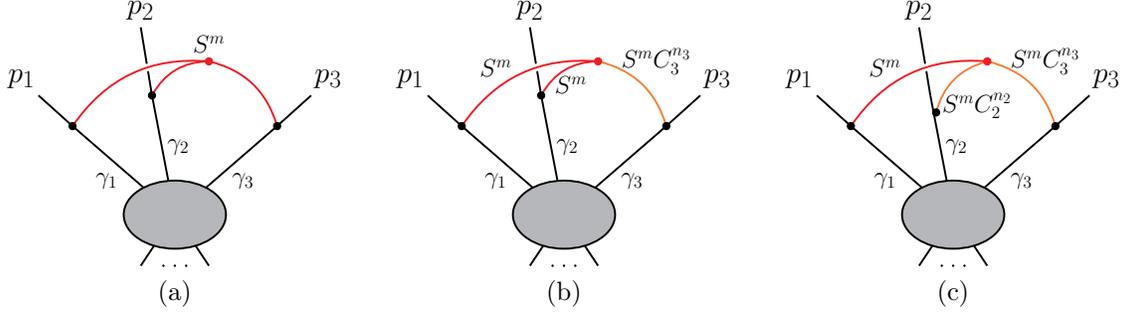

\centering
\begin{subfigure}[b]{0.3\textwidth}
\centering
\include{figs/Sm_component_messenger_config1}
\vspace{-3em}\subcaption{}
\label{Sm_component_messenger_config1}
\end{subfigure}
\quad
\begin{subfigure}[b]{0.3\textwidth}
\centering
\include{figs/Sm_component_messenger_config2}
\vspace{-3em}\caption{}
\label{Sm_component_messenger_config2}
\end{subfigure}
\quad
\begin{subfigure}[b]{0.3\textwidth}
\centering
\include{figs/Sm_component_messenger_config3}
\vspace{-3em}\caption{}
\label{Sm_component_messenger_config3}
\end{subfigure}
\caption{Three representative configurations for the $S^m$ component $\gamma$ within the messenger $\Gamma^{[m]}$, represented by red curves. The subgraphs $\gamma_i$ ($i=1,2,3$) are the mode components to which $\Gamma^{[m]}$ is relevant, and the $p_i$ ($i=1,2,3$) are their respective external momenta. Further loop structures within $\gamma_i$ ($i=1,2,3$) are omitted. The gray blob represents the remaining part of $\mathcal{G}$.}
\label{figure-Sm_component_messenger_configurations}
\end{figure}

Let us consider the first case of Step \emph{2}. As the degree-$m$ messenger $\Gamma^{[m]}$ is defined to be relevant to three or more mode components, one of which is $\gamma_1$, three possibilities occur:
\begin{enumerate}
    \item [(a)] $\gamma$ is relevant to $\geqslant 3$ mode components ($\gamma_1$ and another two or more);
    \item [(b)] $\gamma$ is relevant to $2$ mode components ($\gamma_1$ and another one);
    \item [(c)] $\gamma$ is relevant to $1$ mode component ($\gamma_1$).
\end{enumerate}
These three possibilities can be represented by figure~\ref{figure-Sm_component_messenger_configurations}, where for simplicity let us omit all additional loop structures. The analysis below would also apply in the existence of more complicated structures. For each possibility, $\Gamma^{[m]}$ is relevant to the mode components $\gamma_i$ ($i=1,2,3$), and $p_i$ are their external momenta respectively. In particular, $\gamma_1$ is the mode subgraph whose edge weights have been solved in Step \emph{1}. Below we shall analyze what modes $\gamma_i$ and $p_i$ can possibly have.

In figure~\ref{Sm_component_messenger_config1}, the $S^m$ component is relevant to $\gamma_1$, $\gamma_2$, and $\gamma_3$ by construction. Each mode $\mathscr{X}(\gamma_i)$ is of the form $S^{m-m_i}C_i^{m_i}$. Since the edge weights of $\gamma_1$ have been solved in Step \emph{1}, its external momentum $p_1$ must be of the mode $\mathscr{X}(\gamma_1)$. We therefore have
\begin{subequations}
\begin{alignat}{3}
    &\mathscr{X}(\gamma_1) = S^{m-m_1}C_1^{m_1}, &\quad&\mathscr{X}(\gamma_2) = S^{m-m_2}C_2^{m_2}, &\quad&\mathscr{X}(\gamma_3) = S^{m-m_3}C_3^{m_3};\\
    &\mathscr{X}(p_1) = S^{m-m_1}C_1^{m_1}, &\quad&\mathscr{X}(p_2) = S^{m-m_2}C_2^{m_2+n'_2}, &\quad&\mathscr{X}(p_3) = S^{m-m_3}C_3^{m_3+n'_3}.
\end{alignat}
\label{eq:Sm_component_messenger_config1_step2_general_modes}
\end{subequations}
From section~\ref{section-leading_terms}, the leading polynomial contains the following $\mathcal{U}^{(R)}$ terms
\begin{equation}
    \begin{tikzpicture}[line width = 0.6, scale=0.4, font=\large, mydot/.style={circle, fill, inner sep=.7pt}]
    \draw (5,2.5) edge [thick] (2,5.1) node [] {};
    \draw (2,5.1) edge [thick] (1,6) node [] {};
    \draw (5,2.5) edge [thick] (4.33,6) node [] {};
    \draw (4.33,6) edge [thick] (4,8) node [] {};
    \draw (5,2.5) edge [thick] (8,5.1) node [] {};
    \draw (8,5.1) edge [thick] (9,6) node [] {};
    \draw (6,7) edge [ thick, draw=white, double=white, double distance=3pt, bend right = 30] (2,5.1) node [] {};\draw (6,7) edge [ thick, Red, bend right = 30] (2,5.1) node [] {};
    \draw (5,2.5) edge [thick] (4,1) node [] {};
    \draw (5,2.5) edge [thick] (6,1) node [] {};
    \node () at (0.5,6.5) {$p_1$};
    \node () at (4,8.5) {$p_2$};
    \node () at (9.5,6.5) {$p_3$};
    \draw[fill, thick] (2,5.1) circle (3pt);
    \draw[fill, thick] (4.33,6) circle (3pt);
    \draw[fill, thick] (8,5.1) circle (3pt);
    \draw[fill, Red, thick] (6,7) circle (3pt);
    \node[ thick, draw, fill=Black!33, ellipse, minimum width=3em, minimum height=2em] () at (5,2.5){};
    \path (4,1)-- node[mydot, pos=.333] {} node[mydot] {} node[mydot, pos=.666] {}(6,1);
    \end{tikzpicture}
    \begin{tikzpicture}[line width = 0.6, scale=0.4, font=\large, mydot/.style={circle, fill, inner sep=.7pt}]
    \draw (5,2.5) edge [thick] (2,5.1) node [] {};
    \draw (2,5.1) edge [thick] (1,6) node [] {};
    \draw (5,2.5) edge [thick] (4.33,6) node [] {};
    \draw (4.33,6) edge [thick] (4,8) node [] {};
    \draw (5,2.5) edge [thick] (8,5.1) node [] {};
    \draw (8,5.1) edge [thick] (9,6) node [] {};
    \draw (6,7) edge [ thick, Red, bend right = 30] (4.33,6) node [] {};
    \draw (5,2.5) edge [thick] (4,1) node [] {};
    \draw (5,2.5) edge [thick] (6,1) node [] {};
    \node () at (0.5,6.5) {$p_1$};
    \node () at (4,8.5) {$p_2$};
    \node () at (9.5,6.5) {$p_3$};
    \draw[fill, thick] (2,5.1) circle (3pt);
    \draw[fill, thick] (4.33,6) circle (3pt);
    \draw[fill, thick] (8,5.1) circle (3pt);
    \draw[fill, Red, thick] (6,7) circle (3pt);
    \node[ thick, draw, fill=Black!33, ellipse, minimum width=3em, minimum height=2em] () at (5,2.5){};
    \path (4,1)-- node[mydot, pos=.333] {} node[mydot] {} node[mydot, pos=.666] {}(6,1);
    \end{tikzpicture}
    \begin{tikzpicture}[line width = 0.6, scale=0.4, font=\large, mydot/.style={circle, fill, inner sep=.7pt}]
    \draw (5,2.5) edge [thick] (2,5.1) node [] {};
    \draw (2,5.1) edge [thick] (1,6) node [] {};
    \draw (5,2.5) edge [thick] (4.33,6) node [] {};
    \draw (4.33,6) edge [thick] (4,8) node [] {};
    \draw (5,2.5) edge [thick] (8,5.1) node [] {};
    \draw (8,5.1) edge [thick] (9,6) node [] {};
    \draw (6,7) edge [ thick, Red, bend left = 30] (8,5.1) node [] {};
    \draw (5,2.5) edge [thick] (4,1) node [] {};
    \draw (5,2.5) edge [thick] (6,1) node [] {};
    \node () at (0.5,6.5) {$p_1$};
    \node () at (4,8.5) {$p_2$};
    \node () at (9.5,6.5) {$p_3$};
    \draw[fill, thick] (2,5.1) circle (3pt);
    \draw[fill, thick] (4.33,6) circle (3pt);
    \draw[fill, thick] (8,5.1) circle (3pt);
    \draw[fill, Red, thick] (6,7) circle (3pt);
    \node[ thick, draw, fill=Black!33, ellipse, minimum width=3em, minimum height=2em] () at (5,2.5){};
    \path (4,1)-- node[mydot, pos=.333] {} node[mydot] {} node[mydot, pos=.666] {}(6,1);
    \end{tikzpicture},
    \label{eq:Sm_component_messenger_config1_leading_Uterms}
\end{equation}
$\mathcal{F}_\textup{I}^{(R)}$ terms
\begin{equation}
    \begin{tikzpicture}[line width = 0.6, scale=0.4, font=\large, mydot/.style={circle, fill, inner sep=.7pt}]
    \draw (2,5.1) edge [thick] (1,6) node [] {};
    \draw (5,2.5) edge [thick] (4.33,6) node [] {};
    \draw (4.33,6) edge [thick] (4,8) node [] {};
    \draw (5,2.5) edge [thick] (8,5.1) node [] {};
    \draw (8,5.1) edge [thick] (9,6) node [] {};
    \draw (6,7) edge [ thick, draw=white, double=white, double distance=3pt, bend right = 30] (2,5.1) node [] {};\draw (6,7) edge [ thick, Red, bend right = 30] (2,5.1) node [] {};
    \draw (5,2.5) edge [thick] (4,1) node [] {};
    \draw (5,2.5) edge [thick] (6,1) node [] {};
    \node () at (0.5,6.5) {$p_1$};
    \node () at (4,8.5) {$p_2$};
    \node () at (9.5,6.5) {$p_3$};
    \draw[fill, thick] (2,5.1) circle (3pt);
    \draw[fill, thick] (4.33,6) circle (3pt);
    \draw[fill, thick] (8,5.1) circle (3pt);
    \draw[fill, Red, thick] (6,7) circle (3pt);
    \node[ thick, draw, fill=Black!33, ellipse, minimum width=3em, minimum height=2em] () at (5,2.5){};
    \path (4,1)-- node[mydot, pos=.333] {} node[mydot] {} node[mydot, pos=.666] {}(6,1);
    \end{tikzpicture}
    \begin{tikzpicture}[line width = 0.6, scale=0.4, font=\large, mydot/.style={circle, fill, inner sep=.7pt}]
    \draw (2,5.1) edge [thick] (1,6) node [] {};
    \draw (5,2.5) edge [thick] (4.33,6) node [] {};
    \draw (4.33,6) edge [thick] (4,8) node [] {};
    \draw (5,2.5) edge [thick] (8,5.1) node [] {};
    \draw (8,5.1) edge [thick] (9,6) node [] {};
    \draw (6,7) edge [ thick, Red, bend right = 30] (4.33,6) node [] {};
    \draw (5,2.5) edge [thick] (4,1) node [] {};
    \draw (5,2.5) edge [thick] (6,1) node [] {};
    \node () at (0.5,6.5) {$p_1$};
    \node () at (4,8.5) {$p_2$};
    \node () at (9.5,6.5) {$p_3$};
    \draw[fill, thick] (2,5.1) circle (3pt);
    \draw[fill, thick] (4.33,6) circle (3pt);
    \draw[fill, thick] (8,5.1) circle (3pt);
    \draw[fill, Red, thick] (6,7) circle (3pt);
    \node[ thick, draw, fill=Black!33, ellipse, minimum width=3em, minimum height=2em] () at (5,2.5){};
    \path (4,1)-- node[mydot, pos=.333] {} node[mydot] {} node[mydot, pos=.666] {}(6,1);
    \end{tikzpicture}
    \begin{tikzpicture}[line width = 0.6, scale=0.4, font=\large, mydot/.style={circle, fill, inner sep=.7pt}]
    \draw (2,5.1) edge [thick] (1,6) node [] {};
    \draw (5,2.5) edge [thick] (4.33,6) node [] {};
    \draw (4.33,6) edge [thick] (4,8) node [] {};
    \draw (5,2.5) edge [thick] (8,5.1) node [] {};
    \draw (8,5.1) edge [thick] (9,6) node [] {};
    \draw (6,7) edge [ thick, Red, bend left = 30] (8,5.1) node [] {};
    \draw (5,2.5) edge [thick] (4,1) node [] {};
    \draw (5,2.5) edge [thick] (6,1) node [] {};
    \node () at (0.5,6.5) {$p_1$};
    \node () at (4,8.5) {$p_2$};
    \node () at (9.5,6.5) {$p_3$};
    \draw[fill, thick] (2,5.1) circle (3pt);
    \draw[fill, thick] (4.33,6) circle (3pt);
    \draw[fill, thick] (8,5.1) circle (3pt);
    \draw[fill, Red, thick] (6,7) circle (3pt);
    \node[ thick, draw, fill=Black!33, ellipse, minimum width=3em, minimum height=2em] () at (5,2.5){};
    \path (4,1)-- node[mydot, pos=.333] {} node[mydot] {} node[mydot, pos=.666] {}(6,1);
    \end{tikzpicture},
    \label{eq:Sm_component_messenger_config1_leading_Fiterms}
\end{equation}
and $\mathcal{F}_\textup{II}^{(R)}$ terms
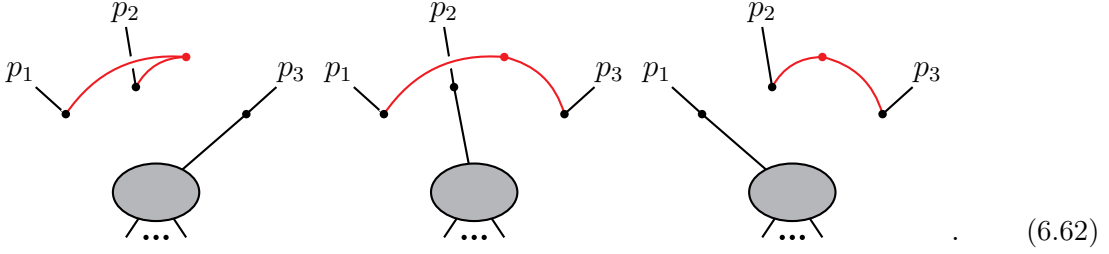
\begin{equation}
    \begin{tikzpicture}[line width = 0.6, scale=0.4, font=\large, mydot/.style={circle, fill, inner sep=.7pt}]
    \draw (2,5.1) edge [thick] (1,6) node [] {};
    \draw (4.33,6) edge [thick] (4,8) node [] {};
    \draw (5,2.5) edge [thick] (8,5.1) node [] {};
    \draw (8,5.1) edge [thick] (9,6) node [] {};
    \draw (6,7) edge [ thick, draw=white, double=white, double distance=3pt, bend right = 30] (2,5.1) node [] {};\draw (6,7) edge [ thick, Red, bend right = 30] (2,5.1) node [] {};
    \draw (6,7) edge [ thick, Red, bend right = 30] (4.33,6) node [] {};
    \draw (5,2.5) edge [thick] (4,1) node [] {};
    \draw (5,2.5) edge [thick] (6,1) node [] {};
    \node () at (0.5,6.5) {$p_1$};
    \node () at (4,8.5) {$p_2$};
    \node () at (9.5,6.5) {$p_3$};
    \draw[fill, thick] (2,5.1) circle (3pt);
    \draw[fill, thick] (4.33,6) circle (3pt);
    \draw[fill, thick] (8,5.1) circle (3pt);
    \draw[fill, Red, thick] (6,7) circle (3pt);
    \node[ thick, draw, fill=Black!33, ellipse, minimum width=3em, minimum height=2em] () at (5,2.5){};
    \path (4,1)-- node[mydot, pos=.333] {} node[mydot] {} node[mydot, pos=.666] {}(6,1);
    \end{tikzpicture}
    \begin{tikzpicture}[line width = 0.6, scale=0.4, font=\large, mydot/.style={circle, fill, inner sep=.7pt}]
    \draw (2,5.1) edge [thick] (1,6) node [] {};
    \draw (5,2.5) edge [thick] (4.33,6) node [] {};
    \draw (4.33,6) edge [thick] (4,8) node [] {};
    \draw (8,5.1) edge [thick] (9,6) node [] {};
    \draw (6,7) edge [ thick, draw=white, double=white, double distance=3pt, bend right = 30] (2,5.1) node [] {};\draw (6,7) edge [ thick, Red, bend right = 30] (2,5.1) node [] {};
    \draw (6,7) edge [ thick, Red, bend left = 30] (8,5.1) node [] {};
    \draw (5,2.5) edge [thick] (4,1) node [] {};
    \draw (5,2.5) edge [thick] (6,1) node [] {};
    \node () at (0.5,6.5) {$p_1$};
    \node () at (4,8.5) {$p_2$};
    \node () at (9.5,6.5) {$p_3$};
    \draw[fill, thick] (2,5.1) circle (3pt);
    \draw[fill, thick] (4.33,6) circle (3pt);
    \draw[fill, thick] (8,5.1) circle (3pt);
    \draw[fill, Red, thick] (6,7) circle (3pt);
    \node[ thick, draw, fill=Black!33, ellipse, minimum width=3em, minimum height=2em] () at (5,2.5){};
    \path (4,1)-- node[mydot, pos=.333] {} node[mydot] {} node[mydot, pos=.666] {}(6,1);
    \end{tikzpicture}
    \begin{tikzpicture}[line width = 0.6, scale=0.4, font=\large, mydot/.style={circle, fill, inner sep=.7pt}]
    \draw (5,2.5) edge [thick] (2,5.1) node [] {};
    \draw (2,5.1) edge [thick] (1,6) node [] {};
    \draw (4.33,6) edge [thick] (4,8) node [] {};
    \draw (8,5.1) edge [thick] (9,6) node [] {};
    \draw (6,7) edge [ thick, Red, bend right = 30] (4.33,6) node [] {};
    \draw (6,7) edge [ thick, Red, bend left = 30] (8,5.1) node [] {};
    \draw (5,2.5) edge [thick] (4,1) node [] {};
    \draw (5,2.5) edge [thick] (6,1) node [] {};
    \node () at (0.5,6.5) {$p_1$};
    \node () at (4,8.5) {$p_2$};
    \node () at (9.5,6.5) {$p_3$};
    \draw[fill, thick] (2,5.1) circle (3pt);
    \draw[fill, thick] (4.33,6) circle (3pt);
    \draw[fill, thick] (8,5.1) circle (3pt);
    \draw[fill, Red, thick] (6,7) circle (3pt);
    \node[ thick, draw, fill=Black!33, ellipse, minimum width=3em, minimum height=2em] () at (5,2.5){};
    \path (4,1)-- node[mydot, pos=.333] {} node[mydot] {} node[mydot, pos=.666] {}(6,1);
    \end{tikzpicture}.
    \label{eq:Sm_component_messenger_config1_leading_Fiiterms}
\end{equation}
Note that the gray blobs in (\ref{eq:Sm_component_messenger_config1_leading_Uterms})-(\ref{eq:Sm_component_messenger_config1_leading_Fiiterms}) represent certain tree subgraphs. By comparing the characteristic equations of the $\mathcal{U}^{(R)}$ and $\mathcal{F}_\textup{I}^{(R)}$ terms, we have set $w_{e_1} = -\mathscr{V}(\mathscr{X}(\gamma_1))$ as in eq.~(\ref{eq:solving_weights_method_step1_result}), as the content of Step \emph{1}.

The kinematic coefficient for the $\mathcal{F}_\textup{II}^{(R)}$ terms, from eq.~(\ref{eq:Sm_component_messenger_config1_step2_general_modes}), are of the following scales:
\begin{align}
    (p_1+p_2)^2\sim \lambda^{2m-m_1-m_2},\quad (p_1+p_3)^2\sim \lambda^{2m-m_1-m_3},\quad (p_2+p_3)^2\sim \lambda^{2m-m_2-m_3},
\end{align}
which can be derived from the $\vee$-operation rule (see theorem~\ref{theorem-overlapping_modes_intersection_union_rules}).
We then consider the characteristic equations of the $\mathcal{F}_\textup{II}^{(R)}$ terms and compare them with those of the $\mathcal{U}^{(R)}$ terms. In detail, the differences between the first $\mathcal{U}^{(R)}$ and $\mathcal{F}_\textup{II}^{(R)}$ terms lie in the terms $w_{\gamma_1}$ and $w_{\gamma_2}$ (present in the $\mathcal{U}^{(R)}$-term equation but absent from the $\mathcal{F}_\textup{II}^{(R)}$-term equation), a $w_{\gamma}$ term (present in the $\mathcal{F}_\textup{II}^{(R)}$-term equation but absent from the $\mathcal{U}^{(R)}$-term equation), and the kinematic contribution (which is $0$ for the $\mathcal{U}^{(R)}$ term and $2m-m_1-m_2$ for the $\mathcal{F}_\textup{II}^{(R)}$ term). The sum over these contributions must be zero, which yields one equation.

By comparing the $\mathcal{U}^{(R)}$ term and the other two $\mathcal{F}_\textup{II}^{(R)}$ terms, we obtain another two equations. We then have:
\begin{align}
\label{eq:Sm_component_messenger_config1_comparing_equations}
\begin{split}
    & \left.\begin{matrix}
    w_{\gamma_1}+w_{\gamma_2}-w_{\gamma}+ (2m-m_1-m_2) = 0\\
    w_{\gamma_1}+w_{\gamma_3}-w_{\gamma}+ (2m-m_1-m_3) = 0\\
    w_{\gamma_2}+w_{\gamma_3}-w_{\gamma}+ (2m-m_2-m_3) = 0
    \end{matrix}\right\}
    \quad\Rightarrow\quad w_{\gamma} = -2m.
\end{split}
\end{align}
Furthermore, from the above we obtain $w_{\gamma_2} = -2m + m_2$ and $w_{\gamma_3} = -2m + m_3$. These weights are determined by repeatedly applying Step \emph{1} in subsequent recursions. In this manner, we have solved for $w_\gamma$, completing Step \emph{2}. For the other two cases shown in figures~\ref{Sm_component_messenger_config2} and \ref{Sm_component_messenger_config3}, nearly identical analyses yield $w_\gamma$, which are detailed in appendix~\ref{appendix-analyses_other_two_cases}.

\begin{figure}[t]
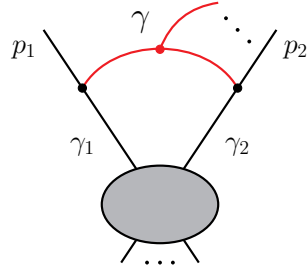

\centering
\include{figs/solving_weights_method_step2_case2}
\vspace{-3em}\caption{The second case of $\gamma$ in Step \emph{2}, represented by the red curves. Further loop structures within $\gamma_1$, $\gamma_2$, and $\gamma$ are omitted.}
\label{figure-solving_weights_method_step2_case2}
\end{figure}

Now we consider the second case of Step \emph{2}, where $\gamma$ is relevant to two weight-solved mode components $\gamma_1$ and $\gamma_2$ (and possibly other mode components), with $\mathscr{X}(\gamma) = \mathscr{X}(\gamma_1)\wedge \mathscr{X}(\gamma_2)$. This case can be represented by figure~\ref{figure-solving_weights_method_step2_case2}. We denote $\mathscr{X}(\gamma)=S^mC_i^{n}$ (note that $n=0$ is possible), $\mathscr{X}(\gamma_1)=S^m_1C_1^{n_1}$, and $\mathscr{X}(\gamma_2)=S^m_2C_2^{n_2}$.
The following $\mathcal{F}_\textup{II}^{(R)}$ term exists:
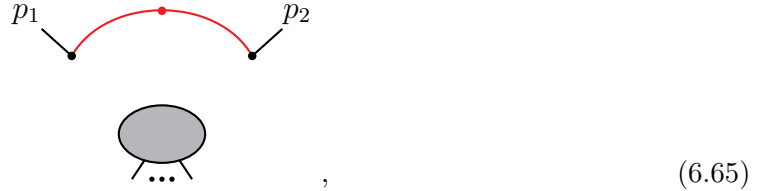
\begin{equation}
    \begin{tikzpicture}[line width = 0.6, scale =0.4, font=\large, mydot/.style={circle, fill, inner sep=.7pt}]
    \draw (2,5.1) edge [thick] (1,6) node [] {};
    \draw (8,5.1) edge [thick] (9,6) node [] {};
    \draw (8,5.1) edge [thick, Red, bend right = 60] (2,5.1) node [] {};
    \draw (5,2.5) edge [thick] (4,1) node [] {};
    \draw (5,2.5) edge [thick] (6,1) node [] {};
    \draw[fill, thick] (2,5.1) circle (3pt);
    \draw[fill, thick] (8,5.1) circle (3pt);
    \draw[fill, thick, Red] (5,6.6) circle (3pt);
    \node[thick, draw, fill=Black!33, ellipse, minimum width=3em, minimum height=2em] () at (5,2.5){};
    \path (4,1)-- node[mydot, pos=.333] {} node[mydot] {} node[mydot, pos=.666] {}(6,1);
    \node () at (0.5,6.5) {$p_1$};
    \node () at (9.5,6.5) {$p_2$};
    \end{tikzpicture},
    \label{eq:solving_weights_method_step2_case2_leading_terms}
\end{equation}
where the gray blob represents certain tree structures. Note that the momentum flowing between the two components of this spanning 2-tree is of the mode $\mathscr{X}(\gamma_1)\vee\mathscr{X}(\gamma_2)$.\footnote{By definition, the mode is $\mathscr{X}(p_1)\vee\mathscr{X}(p_2)$. Since the edge weights of $\gamma_1$ and $\gamma_2$ have been solved in Step~\emph{1}, we must also have $\mathscr{X}(p_1)=\mathscr{X}(\gamma_1)$ and $\mathscr{X}(p_2)=\mathscr{X}(\gamma_2)$ here.} Comparing the characteristic equations of this $\mathcal{F}_\textup{II}^{(R)}$ term and the $\mathcal{U}^{(R)}$ terms yields
\begin{align}
    0&=w_{\gamma_1}+w_{\gamma_2}-w_\gamma+\mathscr{V}(\mathscr{X}(\gamma_1)\vee\mathscr{X}(\gamma_2)) \nonumber\\
    &= -\mathscr{V}(\mathscr{X}(\gamma_1)) -\mathscr{V}(\mathscr{X}(\gamma_2)) -w_\gamma+\mathscr{V}(\mathscr{X}(\gamma_1)\vee\mathscr{X}(\gamma_2))\quad
    \Rightarrow\quad w_\gamma = \mathscr{V}(\mathscr{X}(\gamma_1)\wedge \mathscr{X}(\gamma_2)),
\end{align}
where lemma~\ref{theorem-virtualities_relation} has been applied. We have thus justified Step \emph{2} completely.

We proceed to Step \emph{3} of the weight-solving algorithm, where the operation is to attach some additional external momenta to the vertices of those $\gamma$ in Step \emph{2}. For example, this operation modifies the three configurations in figure~\ref{figure-Sm_component_messenger_configurations} into the following (at this moment let us ignore the dotted curves, which are for later use):
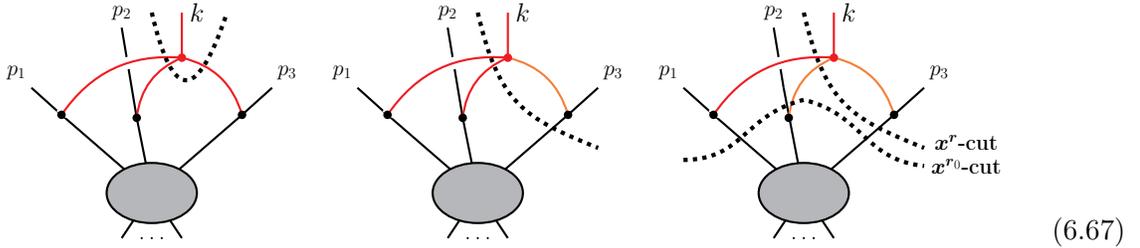
\begin{equation}
\begin{tikzpicture}[line width = 0.6, scale=0.4, font=\large, mydot/.style={circle, fill, inner sep=.7pt}, transform shape]
\draw (5,2.5) edge [thick] (2,5.1) node [] {};
\draw (2,5.1) edge [thick] (1,6) node [] {};
\draw (5,2.5) edge [thick] (4.5,5) node [] {};
\draw (4.5,5) edge [thick] (4,8) node [] {};
\draw (5,2.5) edge [thick] (8,5.1) node [] {};
\draw (8,5.1) edge [thick] (9,6) node [] {};
\draw (6,7) edge [thick, draw=white, double=white, double distance=3pt, bend right = 30] (2,5.1) node [] {};\draw (6,7) edge [thick, Red, bend right = 30] (2,5.1) node [] {};
\draw (6,7) edge [thick, Red, bend right = 30] (4.5,5) node [] {};
\draw (6,7) edge [thick, Red, bend left = 30] (8,5.1) node [] {};
\draw (6,7) edge [thick, Red] (6,8.5) node [] {};
\draw (5,2.5) edge [thick] (4,1) node [] {};
\draw (5,2.5) edge [thick] (6,1) node [] {};
\node () at (0.5,6.5) {\huge $p_1$};
\node () at (4,8.5) {\huge $p_2$};
\node () at (9.5,6.5) {\huge $p_3$};
\node () at (6.5,8.5) {\Huge $k$};
\draw[fill, thick] (2,5.1) circle (3pt);
\draw[fill, thick] (4.5,5) circle (3pt);
\draw[fill, thick] (8,5.1) circle (3pt);
\draw[fill, Red, thick] (6,7) circle (3pt);
\node[thick, draw, fill=Black!33, ellipse, minimum width=3cm, minimum height=2cm] () at (5,2.5){};
\path (4,1)-- node[mydot, pos=.333] {} node[mydot] {} node[mydot, pos=.666] {}(6,1);
\draw (5,8.5) .. controls (5.7,5.5) and (6.5,5.5) .. (7.5,8.5) [ultra thick, dotted];
\end{tikzpicture}
\quad
\begin{tikzpicture}[line width = 0.6, scale=0.4, font=\large, mydot/.style={circle, fill, inner sep=.7pt}, transform shape]
\draw (5,2.5) edge [thick] (2,5.1) node [] {};
\draw (2,5.1) edge [thick] (1,6) node [] {};
\draw (5,2.5) edge [thick] (4.5,5) node [] {};
\draw (4.5,5) edge [thick] (4,8) node [] {};
\draw (5,2.5) edge [thick] (8,5.1) node [] {};
\draw (8,5.1) edge [thick] (9,6) node [] {};
\draw (6,7) edge [thick, draw=white, double=white, double distance=3pt, bend right = 30] (2,5.1) node [] {};\draw (6,7) edge [thick, Red, bend right = 30] (2,5.1) node [] {};
\draw (6,7) edge [thick, Red, bend right = 30] (4.5,5) node [] {};
\draw (6,7) edge [thick, Orange, bend left = 30] (8,5.1) node [] {};
\draw (6,7) edge [thick, Red] (6,8.5) node [] {};
\draw (5,2.5) edge [thick] (4,1) node [] {};
\draw (5,2.5) edge [thick] (6,1) node [] {};
\node () at (0.5,6.5) {\huge $p_1$};
\node () at (4,8.5) {\huge $p_2$};
\node () at (9.5,6.5) {\huge $p_3$};
\node () at (6.5,8.5) {\Huge $k$};
\draw[fill, thick] (2,5.1) circle (3pt);
\draw[fill, thick] (4.5,5) circle (3pt);
\draw[fill, thick] (8,5.1) circle (3pt);
\draw[fill, Red, thick] (6,7) circle (3pt);
\node[thick, draw, fill=Black!33, ellipse, minimum width=3cm, minimum height=2cm] () at (5,2.5){};
\path (4,1)-- node[mydot, pos=.333] {} node[mydot] {} node[mydot, pos=.666] {}(6,1);
\draw (5,8.5) .. controls (5.7,5.5) and (6.5,5) .. (9,4) [ultra thick, dotted];
\end{tikzpicture}
\quad
\begin{tikzpicture}[line width = 0.6, scale=0.4, font=\large, mydot/.style={circle, fill, inner sep=.7pt}, transform shape]
\draw (5,2.5) edge [thick] (2,5.1) node [] {};
\draw (2,5.1) edge [thick] (1,6) node [] {};
\draw (5,2.5) edge [thick] (4.33,6) node [] {};
\draw (4.33,6) edge [thick] (4,8) node [] {};
\draw (5,2.5) edge [thick] (8,5.1) node [] {};
\draw (8,5.1) edge [thick] (9,6) node [] {};
\draw (6,7) edge [thick, draw=white, double=white, double distance=3pt, bend right = 30] (2,5.1) node [] {};\draw (6,7) edge [thick, Red, bend right = 30] (2,5.1) node [] {};
\draw (6,7) edge [thick, Orange, bend right = 30] (4.5,5) node [] {};
\draw (6,7) edge [thick, Orange, bend left = 30] (8,5.1) node [] {};
\draw (6,7) edge [thick, Red] (6,8.5) node [] {};
\draw (5,2.5) edge [thick] (4,1) node [] {};
\draw (5,2.5) edge [thick] (6,1) node [] {};
\node () at (0.5,6.5) {\huge $p_1$};
\node () at (4,8.5) {\huge $p_2$};
\node () at (9.5,6.5) {\huge $p_3$};
\node () at (6.5,8.5) {\Huge $k$};
\node () at (10.4,4.2) {\huge\bf $\x^{\r}$-cut};
\node () at (10.4,3.4) {\huge\bf $\x^{\r_0}$-cut};
\draw[fill, thick] (2,5.1) circle (3pt);
\draw[fill, thick] (4.5,5) circle (3pt);
\draw[fill, thick] (8,5.1) circle (3pt);
\draw[fill, Red, thick] (6,7) circle (3pt);
\node[thick, draw, fill=Black!33, ellipse, minimum width=3cm, minimum height=2cm] () at (5,2.5){};
\path (4,1)-- node[mydot, pos=.333] {} node[mydot] {} node[mydot, pos=.666] {}(6,1);
\draw (5,8.5) .. controls (5.7,5.5) and (6.5,5) .. (9,4) [ultra thick, dotted];
\draw (1,3.6) .. controls (3,3.6) and (3,5.1) .. (5,5.6) .. controls (7,5.1) and (7,3.6) .. (9,3.4) [ultra thick, dotted];
\end{tikzpicture}
\label{eq:Sm_component_messenger_extra_external}
\end{equation}
As we have explained, additional terms can be introduced to the original leading polynomial, and we must confirm that their corresponding characteristic equations can be obtained from the original ones.

To this end, let us inspect the structure of these extra leading terms. A first observation is that they can only be $\mathcal{F}^{(R)}$ terms, because $\mathcal{U}^{(R)}$ terms correspond to minimum spanning trees of $\mathcal{G}$, the set of which does not depend on the external momenta. Let us then take any of these terms, denote it as $\x^{\r}$, and suppose the total momentum flowing between the components of $T^2(\r)$ to be in the mode $X_*$. Since $k$ is one of the external momenta, $\mathscr{X}(k)$ must be either equal to or softer than $X_*$. We now consider these possibilities.

For $\mathscr{X}(k)$ equal to $X_*$, the external momentum $k$ is the hardest one among all attached to the same component of $T^2(\r)$. The extra $\mathcal{F}^{(R)}$ term, which does not exist in the original Symanzik polynomial, must then be an $\mathcal{F}_\textup{I}^{(R)}$ term. The first graph in (\ref{eq:Sm_component_messenger_extra_external}) is an example, where the dotted curve represents the unitarity cut separating the two components of $T^2(\r)$. Subtracting the characteristic equation of this $\mathcal{F}_\textup{I}^{(R)}$ term from those of the $\mathcal{U}^{(R)}$ terms yields the trivial identity $w_\gamma - w_\gamma =0$. Hence, these additional $\mathcal{F}^{(R)}$ terms do not produce independent characteristic equations.
    
For $\mathscr{X}(k)$ softer than $X_*$, we can in fact sharpen the statement: $\mathscr{X}(k)$ must be \emph{marginally softer than} $X_*$ (a concept defined in section~\ref{section-formal_construction_statements}). If it were not marginally softer, another external momentum attached to the same component would dominate the total momentum transfer, causing the term to already appear in the original $\mathcal{F}^{(R)}$ polynomial. One example is the second graph in (\ref{eq:Sm_component_messenger_extra_external}), where $\mathscr{X}(k)=\mathscr{X}(\gamma)=S^m$, which is marginally softer than the total momentum $p_3 + k$ flowing across the unitarity cut, whose mode is $X_*=S^{m-m_3}C_3^{m_3}$ (can be obtained from eq.~(\ref{eq:Sm_component_messenger_config1_step2_general_modes})).

In general, let the total momentum flowing between the components of $T^2(\r)$ be $k + p'$, where the external momenta $k$ and $p'$ are both attached to $t(\r;\cancel{\mathcal{H}})$ (defined in section~\ref{section-FIIR_terms}), with $k$ added in Step~\emph{3} and $p'$ from the original kinematics. In our previous example, the second graph in (\ref{eq:Sm_component_messenger_extra_external}), $p'=p_3$. As argued above, the mode $\mathscr{X}(k)$ is marginally softer than $X_* = \mathscr{X}(k) \vee \mathscr{X}(p')$, and therefore, $\mathscr{X}(k)$ is overlapping with $\mathscr{X}(p')$.\footnote{Here we do not need to consider the possibility that $\mathscr{X}(k)$ is marginally softer than $\mathscr{X}(p')$, because in this case, having an additional $k$ does not affect any original characteristic equations.} To demonstrate that the characteristic equation of this $\mathcal{F}^{(R)}$ term is linearly dependent on the original equations, we now state a key property concerning the virtuality degrees of general modes, captured by the following lemma.
\begin{lemma}
\label{lemma-virtuality_property}
    Let $X_1=S^{m_1}C_i^{n_1}$ and $X_2=S^{m_2}C_i^{n_2}$ with $m_1+n_1 = m_2+n_2$, then for any $X_3=S^{m_3}C_j^{n_3}$ (with $j\neq i$) such that both $X_1$ and $X_2$ are overlapping with $X_3$, we have
    \begin{align}
        \mathscr{V}(X_1\vee X_3) - \mathscr{V}(X_1) = \mathscr{V}(X_2\vee X_3) -\mathscr{V}(X_2).
    \end{align}
\end{lemma}
\begin{proof}
    By employing theorem~\ref{theorem-overlapping_modes_intersection_union_rules}, we have
    \begin{align}
        \mathscr{V}(X_i)=2m_i+n_i,\quad \mathscr{V}(X_i\vee X_3) = m_i+m_3\quad \textup{for }i=1,2.
    \end{align}
    Therefore, $\mathscr{V}(X_1\vee X_3) - \mathscr{V}(X_1) = m_3-m_1-n_1= m_3-m_2-n_2 = \mathscr{V}(X_2\vee X_3) -\mathscr{V}(X_2)$.
\end{proof}
We aim to employ lemma~\ref{lemma-virtuality_property} to illustrate that any new $\mathcal{F}_\textup{II}^{(R)}$ term $\x^{\r}$ arising from $k$ corresponds to an original $\mathcal{F}_\textup{II}^{(R)}$ term $\x^{\r_0}$, such that they have the same characteristic equation.

As noted earlier, the external momenta attached to $t(\r;\cancel{\mathcal{H}})$ are $k$ and $p'$, where $p'$ originates from the original external kinematics and $\mathscr{X}(p')$ overlaps with $\mathscr{X}(k)$. A direct inspection of figures~\ref{figure-Sm_component_messenger_configurations} and \ref{figure-solving_weights_method_step2_case2}, which exhaust all possible configurations of $\gamma$ in Step~\emph{2}, reveals that there always exists another external momentum $p_0$ attached to a mode component $\gamma_0$ whose weight $w_{\gamma_0}$ was solved as $-\mathscr{V}(\mathscr{X}(p_0))$ in Step~\emph{1}, such that:
\begin{enumerate}
    \item [(1)] $\mathscr{X}(p_0)$ is overlapping with $\mathscr{X}(p')$;
    \item [(2)] There is another $\mathcal{F}^{(R)}$ term $\x^{\r_0}$ with both $p'$ and $p_0$ attached to the $t(\r_0;\cancel{\mathcal{H}})$.
\end{enumerate}
For example, in the third graph of (\ref{eq:Sm_component_messenger_extra_external}), we have $p'=p_3$, $p_0=p_1$, $\gamma_0=\gamma_1$, and the unitarity cuts corresponding to $\x^{\r}$ and $\x^{\r_0}$ are labeled ``$\x^{\r}$-cut'' and ``$\x^{\r_0}$-cut'', respectively. The characteristic equations of $\x^{\r}$ and $\x^{\r_0}$ read:
\begin{align}
\label{eq:weight_solving_algorithm_step3_equivalence}
\begin{split}
    \x^{\r}:&\quad w_{\gamma}+\mathscr{V}(\mathscr{X}(k)\vee\mathscr{X}(p')) + \dots =0;\\
    \x^{\r_0}:&\quad w_{\gamma_0^{}}+\mathscr{V}(\mathscr{X}(p_0)\vee\mathscr{X}(p')) + \dots =0.
\end{split}
\end{align}
Here the terms represented by ``$\dots$'' are identical. Recall that the weights $w_{\gamma_0} = -\mathscr{V}(\mathscr{X}(p_0))$ and $w_\gamma = -\mathscr{V}(\mathscr{X}(k))$ have already been determined in Steps \emph{1} and \emph{2}. Applying lemma~\ref{lemma-virtuality_property} with $X_1 = \mathscr{X}(p_0)$, $X_2 = \mathscr{X}(k)$, and $X_3 = \mathscr{X}(p')$ shows that the two equations in (\ref{eq:weight_solving_algorithm_step3_equivalence}) are indeed equivalent. In other words, the characteristic equation for $\x^{\r}$ can always be derived from that of an original leading term $\x^{\r_0}$. This confirms that Step \emph{3} introduces \emph{no} new independent characteristic equations beyond those obtained in previous steps.

With this established, one can iterate Steps \emph{1}--\emph{3} until all edge weights are solved—a process guaranteed to terminate provided all mode components are infrared compatible, as demonstrated earlier. This completes the proof of the sufficiency of the infrared-compatibility requirement.

\subsubsection{Proof of necessity}
\label{section-necessity_subgraph_requirements}

We now show the necessity of the infrared-compatibility requirement. We will see that once it is not fulfilled in a configuration, the solution of the edge weights from the set of leading-term characteristic equations, (\ref{eq:solving_weights_method_center_equation}), will always contain a free parameter $a\in \mathbb{R}$. Such freedom would render the expanded integral scaleless and exclude the configuration from the region list.

To proceed, we consider a specific subgraph $\Gamma_\Delta$ composed exclusively of non-infrared-compatible mode components, with the following properties after every component in $\mathcal{G} \setminus \Gamma_\Delta$ has been set to infrared compatible:
\begin{enumerate}
\item [1,] \emph{none} of the components in $\Gamma_\Delta$ becomes infrared compatible;
\item [2,] setting \emph{any} single component in $\Gamma_\Delta$ to infrared compatible causes \emph{all} the others to become infrared compatible as a consequence.
\end{enumerate}
In other words, $\Gamma_\Delta$ is a ``minimal piece'' of the complete set of non-infrared-compatible components. The first condition means that the infrared compatibility of $\Gamma_\Delta$ cannot be induced from outside. The second condition means that no proper subset of $\Gamma_\Delta$ has this property.

\begin{figure}[t]
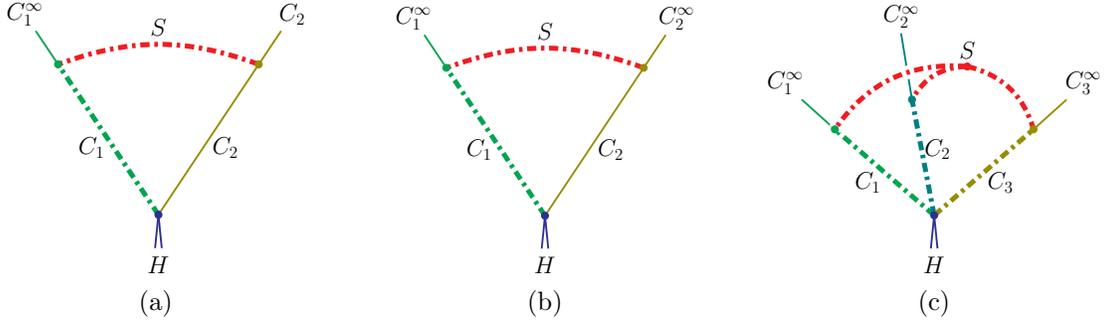

\centering
\begin{subfigure}[b]{0.3\textwidth}
\centering
\include{figs/GammaDelta_example1}
\vspace{-3em}\subcaption{}
\label{GammaDelta_example1}
\end{subfigure}
\quad
\begin{subfigure}[b]{0.3\textwidth}
\centering
\include{figs/GammaDelta_example2}
\vspace{-3em}\caption{}
\label{GammaDelta_example2}
\end{subfigure}
\quad
\begin{subfigure}[b]{0.3\textwidth}
\centering
\include{figs/GammaDelta_example3}
\vspace{-3em}\caption{}
\label{GammaDelta_example3}
\end{subfigure}
\caption{Three examples of $\Gamma_\Delta$; its edges are drawn bold and dash‑dotted. In each configuration the whole integral is scaleless because the infrared‑compatibility requirement is not satisfied.}
\label{figure-GammaDelta_examples}
\end{figure}

Three examples of $\Gamma_\Delta$ are shown in figure~\ref{figure-GammaDelta_examples}, where the edges belonging to $\Gamma_\Delta$ are drawn bold and dash‑dotted.
Each configuration in figure~\ref{figure-GammaDelta_examples} yields a scaleless integral because certain mode components are not infrared compatible; $\Gamma_\Delta$ is a particular subset of those components. In figure~\ref{GammaDelta_example1}, the non‑infrared‑compatible components are the $S$ and $C_1$ edges, and the only possible $\Gamma_\Delta$ is their union (together with their shared vertex), because setting either edge infrared compatible automatically makes the other infrared compatible.
In figure~\ref{GammaDelta_example2}, none of the edges are infrared compatible, but $\Gamma_\Delta$ cannot be the union of all three edges because it can be reduced: one can set the $C_2$ edge infrared compatible without forcing the other two to become infrared compatible. Consequently, in this case $\Gamma_\Delta$ consists of the $S$ and $C_1$ edges (and their shared vertex).
In figure~\ref{GammaDelta_example3}, none of the edges are infrared compatible, and a degree‑$1$ messenger (defined in section~\ref{section-formal_construction_statements}) is present. Hence, setting any single mode component infrared compatible forces all the others to become infrared compatible. As a result, $\Gamma_\Delta$ must be the union of all these edges.

Note that in a given scaleless configuration, the choice of $\Gamma_\Delta$ may not be unique. For example, in figure~\ref{GammaDelta_example2}, $\Gamma_\Delta$ was chosen as the union of the $S$ and $C_1$ edges, and their shared vertex. Due to symmetry, it could equally be chosen in the other direction, consisting of the $S$ and $C_2$ edges, and their shared vertex.

Two fundamental structures within $\Gamma_\Delta$ are already visible from these examples. First, if a component $\gamma \subset \Gamma_\Delta$ is an $S^M$ component that also serves as a degree-$M$ messenger (like the $S$ component in figure~\ref{GammaDelta_example3}), then all $S^mC_i^{M-m}$ components to which $\gamma$ is relevant (such as the $C_i$ components in figure~\ref{GammaDelta_example3}, where $(m,M)=(0,1)$) must also lie inside $\Gamma_\Delta$. Second, for each such $S^mC_i^{M-m}$ component, if another $S^M$ component is relevant to it but not part of a degree-$M$ messenger (as are the $S$ components in figures~\ref{GammaDelta_example1} and~\ref{GammaDelta_example2}), it must also belong to $\Gamma_\Delta$. In the following, we will put these statements on a rigorous footing to construct a general $\Gamma_\Delta$, and demonstrate its consistency with the definition.

To begin, pick any $S^M$ component $\gamma_A \subset \Gamma_\Delta$ that is contained in a degree-$M$ messenger. (If no such $\gamma_A$ exists, we skip the steps that involve it.) By definition, $\gamma_A$ must be relevant to some $S^mC_i^n$ components with $m+n = M$. All of these $S^mC_i^n$ components must belong to $\Gamma_\Delta$; otherwise, declaring all the components in $\mathcal{G}\setminus\Gamma_\Delta$ infrared compatible would make some of them infrared compatible, further making $\gamma_A$ infrared compatible (as we have seen in analyzing the three possibilities in figure~\ref{figure-Sm_component_messenger_configurations}), which contradicts the definition of $\Gamma_\Delta$. We denote the union of all such $\gamma_A$ by $\Gamma_{[\mathcal{S}_A]}$, with ``$\mathcal{S}_A$'' indicating specific subgraphs of $\mathcal{S}$. By construction, the edges and vertices of $\Gamma_{[\mathcal{S}_A]}$ are all in the same mode $S^M$.

Now consider the $S^mC_i^n$ components $\Gamma_{[\mathcal{S}_A]}$ is relevant to. Suppose they involve directions $i=1,\dots,I$ ($I \geqslant 1$), and for each $i$, the hardest mode among them is $S^{m_i}C_i^{n_i}$ ($m_i+n_i = M$). We then group the components into the following subgraphs:
\begin{align}
\begin{split}
    \Gamma_{[\mathcal{SJ}_1]}&:=\gamma_{S^{m_1}C_1^{n_1}}\cup \gamma_{S^{m_1+1}C_1^{n_1-1}}\cup \dots\cup \gamma_{S^{m_1+n_1-1}C_1},\\
    \Gamma_{[\mathcal{SJ}_2]}&:=\gamma_{S^{m_2}C_2^{n_2}}\cup \gamma_{S^{m_2+1}C_2^{n_2-1}}\cup \dots\cup \gamma_{S^{m_2+n_2-1}C_2},\\
    \dots&\\
    \Gamma_{[\mathcal{SJ}_I]}&:=\gamma_{S^{m_I}C_I^{n_I}}\cup \gamma_{S^{m_I+1}C_I^{n_I-1}}\cup \dots\cup \gamma_{S^{m_I+n_I-1}C_I}.
\end{split}
\end{align}
The subscript ``$[\mathcal{SJ}_i]$'' indicates that the edges and vertices of $\Gamma_{[\mathcal{SJ}_i]}$ are in $\mathcal{S}\cup \mathcal{J}_i$. A key observation about the external momentum entering $\Gamma_{[\mathcal{SJ}_i]}$ is that its mode must be of the form $S^{m_i}C_i^{n_i+n'_i}$, with $n'_i\in\mathbb{N}_+\cup\{\infty\}$. This is the only form that ensures: (1) no mode component inside $\Gamma_{[\mathcal{SJ}_i]}$ is infrared compatible, and (2) momentum conservation is consistent with the wide-angle kinematics in (\ref{eq:virtuality_expansion}). The examples below for $M=2$ illustrate this point, showing three different configurations of $\Gamma_{[\mathcal{SJ}_1]}$ and the momenta attached to them. In the figures, thick solid lines represent $\Gamma_{[\mathcal{SJ}_1]}$, the red dashed line labeled ``$S^2$'' denotes the momentum from $\Gamma_{[\mathcal{S}_A]}$ entering $\Gamma_{[\mathcal{SJ}_1]}$, and the remaining dashed line represents the external momentum of $\Gamma_{[\mathcal{SJ}_1]}$, whose mode is $C_1^2$ in the first example, $SC_1$ in the second, and $C_1^3$ in the third.
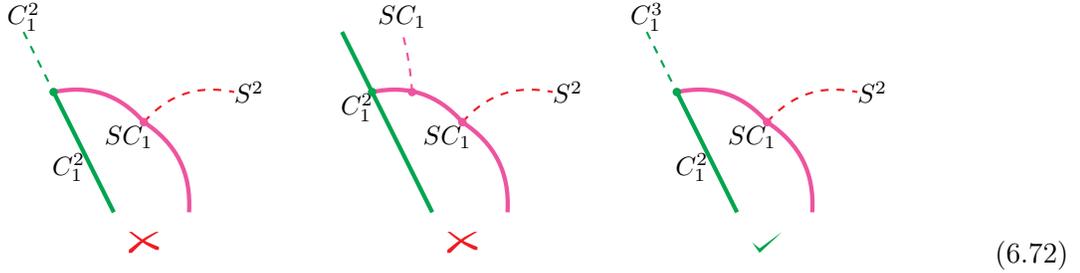
\begin{equation}
    \begin{tikzpicture}[line width = 0.6, scale=0.4, font=\small, mydot/.style={circle, fill, inner sep=.7pt}]
    \draw (1,7) edge [thick, dashed, Green] (2,5) node [] {};
    \draw (2,5) edge [ultra thick, Green] (4,1) node [] {};
    \draw (5,4) edge [ultra thick, Rhodamine, bend right = 30] (2,5) node [] {};
    \draw (5,4) edge [ultra thick, Rhodamine, bend left = 30] (6.5,1) node [] {};
    \draw (5,4) edge [thick, dashed, Red, bend left = 30] (8,5) node [] {};
    \node () at (1,7.5) {$C_1^2$};
    \node () at (2.5,2.5) {$C_1^2$};
    \node () at (4.5,3.5) {$SC_1$};
    \node () at (8.5,5) {$S^2$};
    \node () at (5,0) {$\crossmark[Red]$};
    \draw[fill, thick, Green] (2,5) circle (3pt);
    \draw[fill, Rhodamine, thick] (5,4) circle (3pt);
    \end{tikzpicture}
    \qquad
    \begin{tikzpicture}[line width = 0.6, scale=0.4, font=\small, mydot/.style={circle, fill, inner sep=.7pt}]
    \draw (1,7) edge [ultra thick, Green] (2,5) node [] {};
    \draw (2,5) edge [ultra thick, Green] (4,1) node [] {};
    \draw (3.33,5) edge [thick, dashed, Rhodamine, bend right = 5] (3,7) node [] {};
    \draw (5,4) edge [ultra thick, Rhodamine, bend right = 30] (2,5) node [] {};
    \draw (5,4) edge [ultra thick, Rhodamine, bend left = 30] (6.5,1) node [] {};
    \draw (5,4) edge [thick, dashed, Red, bend left = 30] (8,5) node [] {};
    \node () at (1.5,4.5) {$C_1^2$};
    \node () at (3,7.5) {$SC_1$};
    \node () at (4.5,3.5) {$SC_1$};
    \node () at (8.5,5) {$S^2$};
    \node () at (5,0) {$\crossmark[Red]$};
    \draw[fill, thick, Green] (2,5) circle (3pt);
    \draw[fill, Rhodamine, thick] (5,4) circle (3pt);
    \draw[fill, Rhodamine, thick] (3.33,5) circle (3pt);
    \end{tikzpicture}
    \quad
    \begin{tikzpicture}[line width = 0.6, scale=0.4, font=\small, mydot/.style={circle, fill, inner sep=.7pt}]
    \draw (1,7) edge [thick, dashed, Green] (2,5) node [] {};
    \draw (2,5) edge [ultra thick, Green] (4,1) node [] {};
    \draw (5,4) edge [ultra thick, Rhodamine, bend right = 30] (2,5) node [] {};
    \draw (5,4) edge [ultra thick, Rhodamine, bend left = 30] (6.5,1) node [] {};
    \draw (5,4) edge [thick, dashed, Red, bend left = 30] (8,5) node [] {};
    \node () at (1,7.5) {$C_1^3$};
    \node () at (2.5,2.5) {$C_1^2$};
    \node () at (4.5,3.5) {$SC_1$};
    \node () at (8.5,5) {$S^2$};
    \node () at (5,0) {$\greencheckmark[Green]$};
    \draw[fill, thick, Green] (2,5) circle (3pt);
    \draw[fill, Rhodamine, thick] (5,4) circle (3pt);
    \end{tikzpicture}
\end{equation}
The first configuration is not possible because the external momentum entering $\Gamma_{[\mathcal{SJ}_1]}$ is of the mode $C_1$. This would render the $C_1$ component of $\Gamma_{[\mathcal{SJ}_1]}$ infrared compatible, violating the defining property of $\Gamma_\Delta$. The second configuration is also impossible: by momentum conservation, the $C_1$ component would require a $p_1$-type external momentum input (see~(\ref{eq:virtuality_expansion_p_scaling})), which cannot coexist with the $SC_1^2$ external momentum in our wide‑angle kinematics, (\ref{eq:virtuality_expansion_wide_angle}). The third configuration, in contrast, is possible: the mode components in $\Gamma_{[\mathcal{SJ}_1]}$ could become infrared compatible only after $\Gamma_{[\mathcal{S}_A]}$ is set infrared compatible.

Similarly, there can be external momenta attached directly to a given $\gamma_A$. Denote the total momentum entering $\gamma_A$ as $\mathfrak{p}(\gamma_A)$.\footnote{More generally, $\mathfrak{p}(\gamma_A)$ can also include those line momenta of softer infrared‑compatible subgraphs entering $\gamma_A$. The analysis here applies equally to that case.} By momentum conservation, the mode $\mathscr{X}(\mathfrak{p}(\gamma_A))$ cannot be harder than or overlapping with $S^M$, and in fact, we can sharpen this requirement: $\mathscr{X}(\mathfrak{p}(\gamma_A))$ must be equal to or softer than $S^{M'}$ for some $M'>M$. If this were not the case, $\mathscr{X}(\mathfrak{p}(\gamma_A))$ would be of the form $S^MC_j^{n_j}$ with $j\notin \{1,\dots,I\}$, making $\mathfrak{p}(\gamma_A)$ relevant to every $\Gamma_{[\mathcal{SJ}_i]}$. The $S^{m_i}C_i^{n_i}$ components inside $\Gamma_{[\mathcal{SJ}_i]}$ would become infrared compatible due to its external momentum flow, again violating the definition of $\Gamma_\Delta$.

So far we have identified the following subgraphs within $\Gamma_\Delta$: $\Gamma_{[\mathcal{S}_A]},\Gamma_{[\mathcal{SJ}_1]},\dots,\Gamma_{[\mathcal{SJ}_I]}$. In addition, let us define a further subgraph $\Gamma_{[\mathcal{S}_B]}$ consisting of the following mode components:
\begin{itemize}
    \item the $S^M$ components, each relevant to precisely one $\Gamma_{[\mathcal{SJ}_i]}$;
    \item the $S^MC_j^{n'_j}$ components ($j\notin\{1,\dots,I\}$ and $n'_j\in \mathbb{N}_+$), each relevant to one or more of $\Gamma_{[\mathcal{S}_A]},\Gamma_{[\mathcal{SJ}_1]},\dots,\Gamma_{[\mathcal{SJ}_I]}$.
\end{itemize}
As in the analysis above, any external momentum entering such a component (if present) must be equal to or softer than $S^{M'}$, for some $M'>M$. For simplicity, we do not repeat the analysis here.

Crucially, $\Gamma_{[\mathcal{S}_B]}$ must itself be contained in $\Gamma_\Delta$. Suppose, for contradiction, that an $S^MC_j^{n'_j}$ component relevant to $\Gamma_{[\mathcal{SJ}_1]}$ lies in $\mathcal{G}\setminus \Gamma_\Delta$ instead. After we set the whole of $\mathcal{G}\setminus \Gamma_\Delta$ infrared compatible, this component becomes infrared compatible, and the total momentum flowing into $\Gamma_{[\mathcal{SJ}_1]}$ would acquire mode $S^{m_1}C_1^{n_1}$, thereby rendering $\Gamma_{[\mathcal{SJ}_1]}$ infrared compatible as well. This contradicts the definition of $\Gamma_\Delta$. Hence $\Gamma_{[\mathcal{S}_B]}\subset \Gamma_\Delta$.

\begin{figure}[t]
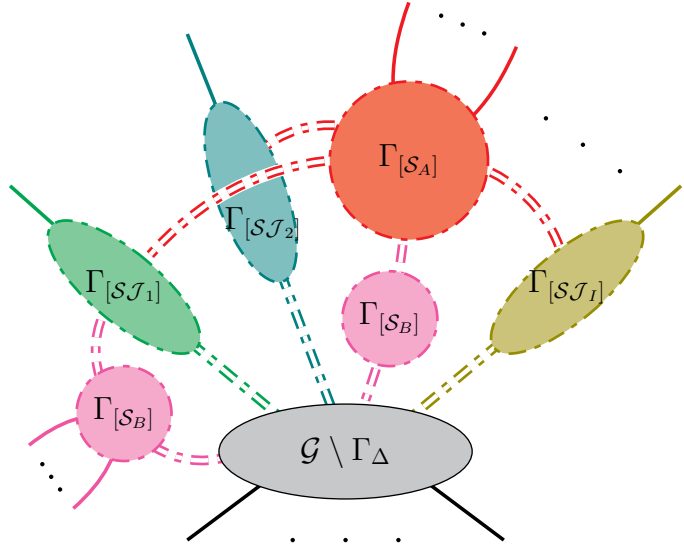

\centering
\include{figs/GammaDelta_config_onshell_external}
\vspace{-3em}\caption{Illustration of the general structure of $\Gamma_\Delta$ (the union of colored subgraphs, whose edges are dash-dotted) and $\mathcal{G}\setminus \Gamma_\Delta$ (the gray blob). Double lines indicate multiple connections. Here $\Gamma_{[\mathcal{S}_A]}$ is shown as a single component, and $\Gamma_{[\mathcal{S}_B]}$ consists of two components: one adjacent to $\Gamma_{[\mathcal{SJ}_1]}$ and $\mathcal{G}\setminus \Gamma_\Delta$, and the other adjacent to $\Gamma_{[\mathcal{S}_A]}$ and $\mathcal{G}\setminus \Gamma_\Delta$. The analysis regarding this pattern applies equally to more general configurations of $\Gamma_{[\mathcal{S}_A]}$ and $\Gamma_{[\mathcal{S}_B]}$.}
\label{figure-GammaDelta_config_onshell_external}
\end{figure}

We finalize our construction for $\Gamma_\Delta$ at this point, and obtain its pattern in figure~\ref{figure-GammaDelta_config_onshell_external}. We color the subgraphs as follows: $\Gamma_{[\mathcal{SJ}_i]}$ are drawn in different shades of {\color{Green}\bf green}, $\Gamma_{[\mathcal{S}_A]}$ in {\color{Red}\bf red}, and $\Gamma_{[\mathcal{S}_B]}$ in {\color{Rhodamine}\bf rhodamine}. The edges of $\Gamma_\Delta$ are drawn dash-dotted, and its complement $\mathcal{G}\setminus\Gamma_\Delta$ in {\color{darkgray}\bf gray}.
It remains to verify that the constructed subgraph of $\mathcal{G}$ satisfies the definition of $\Gamma_\Delta$. In other words, the collection $\{\Gamma_{[\mathcal{SJ}_1]},\dots,\Gamma_{[\mathcal{SJ}_I]},\Gamma_{[\mathcal{S}_A]},\Gamma_{[\mathcal{S}_B]}\}$ should comprise all subgraphs of $\Gamma_\Delta$ precisely. This is established by the following lemma.
\begin{lemma}
    The graph $\Gamma_\Delta$ constructed above satisfies the following conditions.
    \begin{enumerate}
        \item For every mode component $\gamma_B \subset \Gamma_{[\mathcal{S}_B]}$, there exist two mode components $\gamma_1 \subset \Gamma_\Delta$ and $\gamma_2 \subset \mathcal{G}\setminus\Gamma_\Delta$, such that $\gamma_B$ is relevant to them both, and $\mathscr{X}(\gamma_B) = \mathscr{X}(\gamma_1) \wedge \mathscr{X}(\gamma_2)$.
        \item No mode component in $\Gamma_\Delta$ would become infrared compatible even if all the mode components in $\mathcal{G}\setminus \Gamma_\Delta$ are infrared compatible.
        \item For every mode component $\gamma'\subset \Gamma_{[\mathcal{SJ}_i]}$ that is softer than $S^{m_i}C_i^{n_i}$, there exists a mode component $\gamma'' \subset \mathcal{G}\setminus\Gamma_\Delta$, such that $\gamma'$ is relevant to $\gamma''$, and $\mathscr{X}(\gamma') = S^{m_i}C_i^{n_i}\wedge \mathscr{X}(\gamma'')$.
        \item Suppose all mode components in $\mathcal{G}\setminus \Gamma_\Delta$ are infrared compatible. Then setting any single mode component in $\Gamma_\Delta$ infrared compatible forces all other mode components in $\Gamma_\Delta$ to become infrared compatible.
    \end{enumerate}
\end{lemma}
\begin{proof}
    We have shown that the total momentum entering $\gamma_B$, sourced exclusively from the external kinematics or from softer infrared‑compatible subgraphs entering $\gamma_B$, must be equal to or softer than $S^{M'}$ for some $M'>M$. Consequently, condition \emph{2} of the Second Connectivity Theorem (theorem~\ref{theorem-mode_subgraphs_connectivity2}) applies to $\gamma_B$. That is, $\gamma_B$ must be adjacent to two harder‑mode subgraphs $\gamma_1$ and $\gamma_2$ such that $\mathscr{X}(\gamma_B) = \mathscr{X}(\gamma_1) \wedge \mathscr{X}(\gamma_2)$.
    
    Given the known mode structure of $\Gamma_\Delta$, one can directly verify that $\gamma_1$ and $\gamma_2$ cannot both lie inside $\Gamma_\Delta$. Nor can they both belong to $\mathcal{G}\setminus \Gamma_\Delta$; otherwise, setting $\mathcal{G}\setminus \Gamma_\Delta$ infrared compatible would also make $\gamma_B$ infrared compatible. Hence, one of them must be contained in $\Gamma_\Delta$ with the other in $\mathcal{G}\setminus\Gamma_\Delta$. This completes the proof of the first statement.

    We next investigate when an $S^mC_i^n$ component $\gamma_{S^mC_i^n}^{}\subset \Gamma_{[\mathcal{SJ}_i]}$ can become infrared compatible. According to the definition of infrared compatibility, two scenarios are possible:
    \begin{enumerate}
        \item [(1)] There exist two softer‑mode subgraphs $\gamma_a$ and $\gamma_b$, both relevant to $\gamma_{S^mC_i^n}^{}$ and already confirmed infrared compatible, with $S^mC_i^n = \mathscr{X}(\gamma_a)\vee \mathscr{X}(\gamma_b)$.
        \item [(2)] $\gamma$ is relevant to two harder‑mode subgraphs $\gamma_a$ and $\gamma_b$ that are both confirmed infrared compatible, with $S^mC_i^n = \mathscr{X}(\gamma_a)\wedge \mathscr{X}(\gamma_b)$.
    \end{enumerate}    
    For the first case, a key observation is that the relation $S^mC_i^n = \mathscr{X}(\gamma_a) \vee \mathscr{X}(\gamma_b)$ forces $\mathscr{X}(\gamma_a)$ and $\mathscr{X}(\gamma_b)$ to take the following forms (up to an exchange):
    \begin{align}
    \begin{split}
        &\mathscr{X}(\gamma_a) = S^{m+n'}C_i^{n-n'}\ \ (n'\leqslant n)\quad\textup{or}\quad S^{m+n}C_j^{n'_j}\ \ (j\neq i,\ n'_j\in\mathbb{N}_+),\\
        &\mathscr{X}(\gamma_b) = S^m C_i^{n''}\ \ (n''\geqslant n+1).
    \end{split}
    \label{eq:harder_XaXb_possible_modes}
    \end{align}
    Similarly, for the second case, $S^mC_i^n = \mathscr{X}(\gamma_a) \wedge \mathscr{X}(\gamma_b)$ implies the following form:
    \begin{align}
    \begin{split}
        &\mathscr{X}(\gamma_a) = S^{m-m'}C_i^{n+m'}\ \ (m'\leqslant m),\\
        &\mathscr{X}(\gamma_b) = S^m C_i^{n''}\ \ (n''\leqslant n-1).
    \end{split}
    \label{eq:softer_XaXb_possible_modes}
    \end{align}
    Both eqs.~(\ref{eq:harder_XaXb_possible_modes}) and (\ref{eq:softer_XaXb_possible_modes}) follow from theorem~\ref{theorem-overlapping_modes_intersection_union_rules}, as we will detail in appendix~\ref{appendix-XaXb_possible_modes}.

    From our construction of $\Gamma_\Delta$, no matter which case above we take, $\gamma_a$ have been included in $\Gamma_\Delta$ already. This indicates that $\gamma_{S^mC_i^n}^{}$ can become infrared compatible only if some specific mode component in $\Gamma_\Delta$ has been set infrared compatible. In other words, the infrared compatibility of $\mathcal{G}\setminus \Gamma_\Delta$ alone \emph{cannot} induce that of $\gamma_{S^mC_i^n}^{}$. The same analyses can be made for the other mode components in $\Gamma_\Delta$. We have thus proved the second statement.

    The proof of the third statement follows a line of reasoning very similar to that of the first one. From the external momentum structure of $\gamma'$ and the Second Connectivity Theorem, $\gamma'$ must be adjacent to two harder‑mode subgraphs $\gamma_1$ and $\gamma''$ satisfying $\mathscr{X}(\gamma') = \mathscr{X}(\gamma_1) \wedge \mathscr{X}(\gamma'')$. Since $\mathscr{X}(\gamma') = S^{m_i+m'_i}C_i^{n_i-m'_i}$ with $0 \leqslant m'_i \leqslant n_i-1$ here, eq.~(\ref{eq:softer_XaXb_possible_modes}) can be employed to indicate that either $\mathscr{X}(\gamma_1)$ or $\mathscr{X}(\gamma'')$ is equal to $S^{m_i}C_i^{n_i}$. Choosing $\mathscr{X}(\gamma_1) = S^{m_i}C_i^{n_i}$ and noting the mode structure of $\Gamma_\Delta$, we conclude that $\gamma''$ must lie inside $\mathcal{G} \setminus \Gamma_\Delta$. This establishes the third statement.

    To show the fourth statement, assume that all mode components in $\mathcal{G}\setminus \Gamma_\Delta$ are already infrared compatible. We will show that once any single additional component in $\Gamma_\Delta$ is made infrared compatible, all the remaining components in $\Gamma_\Delta$ become infrared compatible as well.

    First, we demonstrate that making any component in a given $\Gamma_{[\mathcal{SJ}_i]}$ infrared compatible forces every other component in that $\Gamma_{[\mathcal{SJ}_i]}$ to become infrared compatible. On the one hand, by the third statement (already proved), setting the $S^{m_i}C_i^{n_i}$ component infrared compatible renders all other components of $\Gamma_{[\mathcal{SJ}_i]}$---those with modes $S^{m_i+1}C_i^{n_i-1}$, $S^{m_i+2}C_i^{n_i-2}$, \dots, $S^{m_i+n_i-1}C_i$---infrared compatible. On the other hand, if we instead set any $S^{m_i+k}C_i^{n_i-k}$ component (with $1\leqslant k\leqslant n_i-1$) infrared compatible, the sum of its momentum with the external momenta entering the $S^{m_i}C_i^{n_i}$ component acquires mode $S^{m_i}C_i^{n_i}$, which in turn forces the $S^{m_i}C_i^{n_i}$ component to become infrared compatible.

    Then, we show that infrared compatibility propagates among $\Gamma_{[\mathcal{SJ}_i]}$, $\Gamma_{[\mathcal{S}_A]}$, and $\Gamma_{[\mathcal{S}_B]}$. From the defining properties of messengers and infrared compatibility, setting any mode component in $\Gamma_* \equiv \Gamma_{[\mathcal{S}_A]} \cup \Gamma_{[\mathcal{SJ}_1]} \cup \dots \cup \Gamma_{[\mathcal{SJ}_I]}$ makes all the others in $\Gamma_*$ infrared compatible. By the first statement (already proved), this also forces $\Gamma_{[\mathcal{S}_B]}$ to become infrared compatible. Conversely, if we set any mode component in $\Gamma_{[\mathcal{S}_B]}$ infrared compatible, then some components of $\Gamma_*$ would acquire external momentum of precisely its mode, thus becoming infrared compatible. After this step, the previous argument can be iterated.

    Collectively, these observations establish the fourth statement.
\end{proof}

\begin{figure}[t]
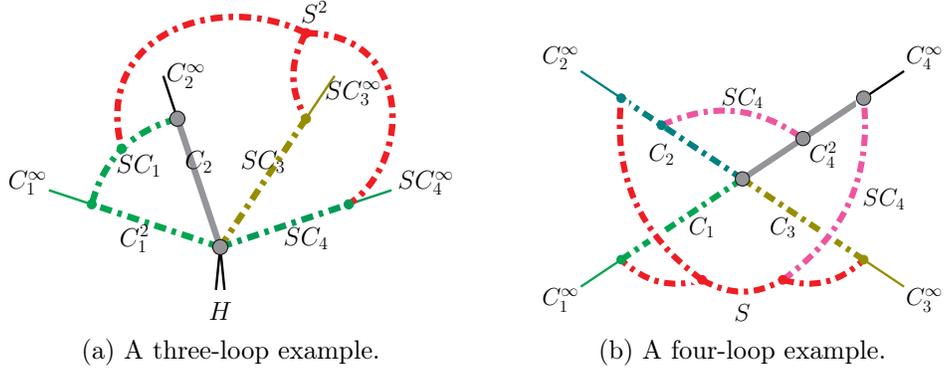

\centering
\begin{subfigure}[b]{0.4\textwidth}
\centering
\include{figs/GammaDelta_nontrivial_example1}
\vspace{-3em}\caption{A three-loop example.}
\label{GammaDelta_nontrivial_example1}
\end{subfigure}
\qquad
\begin{subfigure}[b]{0.36\textwidth}
\centering
\include{figs/GammaDelta_nontrivial_example2}
\vspace{-3em}\caption{A four-loop example.}
\label{GammaDelta_nontrivial_example2}
\end{subfigure}
\caption{Two examples of $\Gamma_\Delta$ whose edges are drawn bold and dash-dotted, and colored according to figure~\ref{figure-GammaDelta_config_onshell_external}.}
\label{figure-GammaDelta_nontrivial_examples}
\end{figure}

Before proceeding, we illustrate two more nontrivial examples of $\Gamma_\Delta$ in figure~\ref{figure-GammaDelta_nontrivial_examples}, where every external momentum and edge is labeled with its mode.
In each example, every edge belongs to some non-infrared-compatible mode component, and those belonging to $\Gamma_\Delta$ are drawn bold and dash-dotted.

We have color‑coded the edges and vertices according to the general pattern shown in figure~\ref{figure-GammaDelta_config_onshell_external}. In figure~\ref{GammaDelta_nontrivial_example1}, we have $M=2$, and the degree-$2$ messenger, which is precisely $\Gamma_{[\mathcal{S}_A]}$ here, is relevant to $\Gamma_{[\mathcal{SJ}_i]}$ with $i=1,3,4$. The graph $\Gamma_{[\mathcal{SJ}_1]}$ consists of a $C_1$ component and an $SC_1$ component, $\Gamma_{[\mathcal{SJ}_3]}$ consists of an $SC_3$ component, and $\Gamma_{[\mathcal{SJ}_4]}$ consists of an $SC_4$ component. The subgraph $\Gamma_{[\mathcal{S}_B]}$ is empty in this configuration.

In figure~\ref{GammaDelta_nontrivial_example2}, we have $M=1$, and the degree-1 messenger $\Gamma_{[\mathcal{S}_A]}$ is relevant to $\Gamma_{[\mathcal{SJ}_i]}$ with $i=1,2,3$. Each $\Gamma_{[\mathcal{SJ}_i]}$ consists of $C_i$ components only. The subgraph $\Gamma_{[\mathcal{S}_B]}$ consists of two $SC_4$ components in this configuration.

Now that the structure of $\Gamma_\Delta$ has been clarified, we show that if $\Gamma_\Delta\neq \varnothing$ in a configuration $R$, there exists a free parameter $a \in \mathbb{R}$ among the edge weights. This is sufficient to prove the necessity of the infrared‑compatibility requirement. Concretely, while $w_e = -\mathscr{V}(\mathscr{X}(e))$ for each $e \in \mathcal{G}$ is one solution to the characteristic equations, we may also deform the weights as follows:
\begin{align}
\label{eq:GammaDelta_edge_weight_modification}
\begin{split}
    &w_e\to -\mathscr{V}(\mathscr{X}(e))+a,\textup{ for each }e\in \Gamma_{[\mathcal{SJ}_1]}\cup\dots\cup\Gamma_{[\mathcal{SJ}_I]}\cup \Gamma_{[\mathcal{S}_B]}; \\
    &w_e\to -\mathscr{V}(\mathscr{X}(e))+2a,\textup{ for each }e\in \Gamma_{[\mathcal{S}_A]}.
\end{split}
\end{align}
The space normal to the lower face $f_R$ is thus at least one dimensional.

\begin{figure}[t]
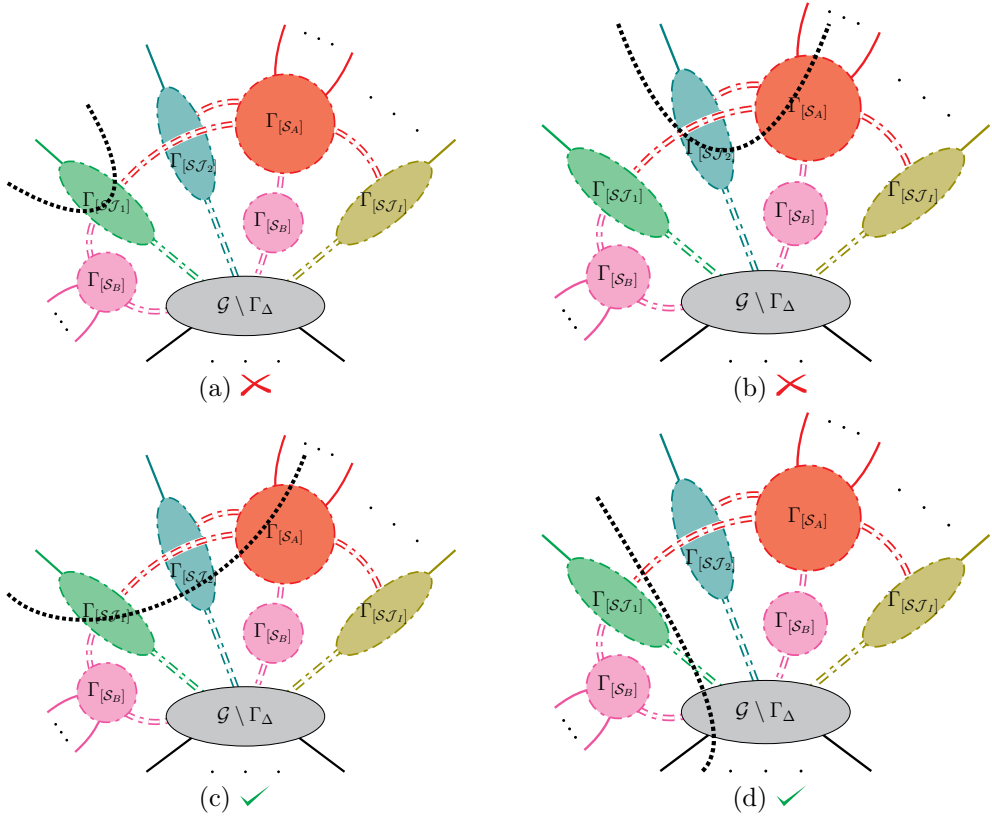

\centering
\begin{subfigure}[b]{0.4\textwidth}
\centering
\include{figs/GammaDelta_Fiiterm_unitairity_cut1}
\vspace{-3em}\caption{$\crossmark[Red]$}
\label{GammaDelta_Fiiterm_unitairity_cut1}
\end{subfigure}
\qquad
\begin{subfigure}[b]{0.4\textwidth}
\centering
\include{figs/GammaDelta_Fiiterm_unitairity_cut2}
\vspace{-3em}\caption{$\crossmark[Red]$}
\label{GammaDelta_Fiiterm_unitairity_cut2}
\end{subfigure}
\\
\centering
\begin{subfigure}[b]{0.4\textwidth}
\centering
\include{figs/GammaDelta_Fiiterm_unitairity_cut3}
\vspace{-3em}\caption{$\greencheckmark[Green]$}
\label{GammaDelta_Fiiterm_unitairity_cut3}
\end{subfigure}
\qquad
\begin{subfigure}[b]{0.4\textwidth}
\centering
\include{figs/GammaDelta_Fiiterm_unitairity_cut4}
\vspace{-3em}\caption{$\greencheckmark[Green]$}
\label{GammaDelta_Fiiterm_unitairity_cut4}
\end{subfigure}
\caption{Some typical $\mathcal{F}$-term unitarity cuts passing through $\Gamma_\Delta$. We argue that (a) and (b) cannot correspond to any $\mathcal{F}^{(R)}$ term (marked with ``$\crossmark[Red]$'') while (c) and (d) can (marked with ``$\greencheckmark[Green]$'').}
\label{figure-GammaDelta_Fiiterm_unitairity_cuts}
\end{figure}

To see this, we must understand the structure of the leading‑term characteristic equations. As remarked below lemma~\ref{lemma-solving_weights_method_leading_Uterm}, setting all edge weights in a mode component equal reduces all characteristic equations for the $\mathcal{U}^{(R)}$ terms to a single equation, which determines the overall weight of the entire graph. The characteristic equations from the $\mathcal{F}^{(R)}$ terms are then needed to determine individual edge weights. Specifically, to solve for the weight $w_e$ of an edge, we require those $\mathcal{F}^{(R)}$ terms in which $e\in \gamma'$ in eq.~(\ref{eq:leading_Fterms_typeI_subgraph_characterization_repeat}), or $e\in \{\gamma'_1,\dots,\gamma'_A\}\cup \{ \gamma''_1,\dots,\gamma''_{A-1}\}$ in eq.~(\ref{eq:leading_Fterms_typeII_subgraph_characterization}). Since each $\mathcal{F}^{(R)}$ term can be expressed as a unitarity cut of the entire graph, to solve for the weights $w_e$ with $e\in \Gamma_\Delta$ we must consider those unitarity cuts passing through $\Gamma_\Delta$. Let us discuss all the possible cases, some of which are illustrated in figure~\ref{figure-GammaDelta_Fiiterm_unitairity_cuts}.

First, there cannot be any $\mathcal{F}^{(R)}$-term unitarity cut that isolates only the external momentum of a single $\Gamma_{[\mathcal{SJ}_i]}$ (see figure~\ref{GammaDelta_Fiiterm_unitairity_cut1}). Otherwise, it would correspond to an $\mathcal{F}_\textup{I}^{(R)}$ term, and the difference between its characteristic equation and the $\mathcal{U}^{(R)}$-term characteristic equation would determine some edge weights in $\Gamma_{[\mathcal{SJ}_i]}$, contradicting to the fact that no mode component in $\Gamma_\Delta$ is infrared compatible. Out of the same reason, no $\mathcal{F}^{(R)}$-term unitarity cut can isolate only the external momenta of $\Gamma_{[\mathcal{S}_A]}$ or $\Gamma_{[\mathcal{S}_B]}$.

This reasoning can be extended to rule out those $\mathcal{F}^{(R)}$-term unitarity cuts, which isolate the external momentum of a single $\Gamma_{[\mathcal{SJ}_i]}$ together with some external momenta from the graph $\Gamma_{[\mathcal{S}_A]}\cup \Gamma_{[\mathcal{S}_B]}$ (see figure~\ref{GammaDelta_Fiiterm_unitairity_cut2}). This is because those external momenta of $\Gamma_{[\mathcal{S}_A]}\cup \Gamma_{[\mathcal{S}_B]}$ are not relevant to $\Gamma_{[\mathcal{SJ}_i]}$ (recall that they must be equal to or softer than $S^{M'}$ with some $M'>M$, as demonstrated earlier). Therefore, if such a unitarity cut did correspond to an $\mathcal{F}^{(R)}$-term, removing all the external momenta of $\Gamma_{[\mathcal{S}_A]}\cup \Gamma_{[\mathcal{S}_B]}$ would also lead to an $\mathcal{F}^{(R)}$-term, leading back to the configuration in figure~\ref{GammaDelta_Fiiterm_unitairity_cut1}, which has already been ruled out.

From the observations above, each possible unitarity cut must correspond to an $\mathcal{F}_\textup{II}^{(R)}$ term and satisfy one of the following:
\begin{itemize}
    \item [(1)] it isolates the external momenta of multiple $\Gamma_{[\mathcal{SJ}_i]}$ (see figure~\ref{GammaDelta_Fiiterm_unitairity_cut3});
    \item [(2)] it isolates the external momenta of a single $\Gamma_{[\mathcal{SJ}_i]}$ together with some other external momenta in $\mathcal{G}\setminus \Gamma_\Delta$ (see figure~\ref{GammaDelta_Fiiterm_unitairity_cut4}).
\end{itemize}
We now analyze these two cases separately. For case (1), from lemma~\ref{lemma-leading_FIIterms_specific_modes}, which constrains the mode structure of generic $\mathcal{F}_\textup{II}^{(R)}$ terms, implies that the unitarity cut must pass through precisely two of the $\Gamma_{[\mathcal{SJ}_i]}$ graphs. Furthermore, a path in $\Gamma_{[\mathcal{S}_A]}$ must connect their external momenta in the spanning 2-tree. Figure~\ref{GammaDelta_Fiiterm_unitairity_cut3} illustrates the pattern where these are $\Gamma_{[\mathcal{SJ}_1]}$ and $\Gamma_{[\mathcal{SJ}_2]}$. Theorem~\ref{theorem-leading_terms_general_structures} then gives
\begin{align}
    \mathfrak{n}_{\gamma_{[\mathcal{SJ}_1]}} = L(\widetilde{\Gamma}_{[\mathcal{SJ}_1]})+1,\quad \mathfrak{n}_{\gamma_{[\mathcal{SJ}_2]}} = L(\widetilde{\Gamma}_{[\mathcal{SJ}_2]})+1,\quad \mathfrak{n}_{\gamma_{[\mathcal{S}_A]}} = L(\widetilde{\Gamma}_{[\mathcal{S}_A]})-1.
\end{align}
In other words, compared with the $\mathcal{U}^{(R)}$-term characteristic equation, the characteristic equation for this $\mathcal{F}_\textup{II}^{(R)}$ term contains two additional terms $w_{e_1}$ and $w_{e_2}$ (with $e_1\in \Gamma_{[\mathcal{SJ}_1]}$ and $e_2\in \Gamma_{[\mathcal{SJ}_2]}$) and one fewer term $w_{e_A}$ (with $e_A\in \Gamma_{[\mathcal{S}_A]}$). Under the edge‑weight deformation in (\ref{eq:GammaDelta_edge_weight_modification}), $w_{e_1}\to w_{e_1}+a$, $w_{e_2}\to w_{e_2}+a$, and $w_{e_A}\to w_{e_A}+2a$. This characteristic equation remains to hold.

For case (2), similarly, the unitarity cut passes through one $\Gamma_{[\mathcal{SJ}_i]}$, and the external momentum of this $\Gamma_{[\mathcal{SJ}_i]}$ is connected to another external momentum attached to $\mathcal{G}\setminus\Gamma_\Delta$ by a path in $\Gamma_{[\mathcal{S}_B]}$. Figure~\ref{GammaDelta_Fiiterm_unitairity_cut4} illustrates the pattern for $i=1$. From theorem~\ref{theorem-leading_terms_general_structures}, we then have
\begin{align}
    \mathfrak{n}_{\gamma_{[\mathcal{SJ}_1]}} = L(\widetilde{\Gamma}_{[\mathcal{SJ}_1]})+1,\quad \mathfrak{n}_{\gamma_{[\mathcal{S}_A]}} = L(\widetilde{\Gamma}_{[\mathcal{S}_B]})-1.
\end{align}
Again, compared with the $\mathcal{U}^{(R)}$-term characteristic equation, the characteristic equation for this $\mathcal{F}_\textup{II}^{(R)}$ term contains two additional terms $w_{e_1}$ and $w_{e_2}$ (with $e_1\in \Gamma_{[\mathcal{SJ}_1]}$ and $e_2\in \mathcal{G}\setminus \Gamma_\Delta$) and one fewer term $w_{e_B}$ (with $e_B\in \Gamma_{[\mathcal{S}_B]}$). It remains to hold under the edge‑weight deformation in (\ref{eq:GammaDelta_edge_weight_modification}), where $w_{e_1}\to w_{e_1}+a$ and $w_{e_B}\to w_{e_B}+a$.

\bigbreak
Let us summarize our analysis and results above. We have shown that if the infrared-compatibility requirement is not satisfied in a configuration, a subgraph $\Gamma_\Delta\subset \mathcal{G}$ can always be identified, such that the leading-term characteristic equations do not have a unique solution, because they still hold under the deformation shown in (\ref{eq:GammaDelta_edge_weight_modification}). Thus, the expanded integral would be scaleless, and this configuration is not a region. We have then proved the necessity of the infrared-compatibility requirement.

\section{Demonstration with more multiloop examples}
\label{section-multiloop_examples}

Based on the knowledge in previous sections, we now demonstrate the region structures of two nontrivial multiloop examples. We begin with a three-loop two-to-two scattering graph in section~\ref{section-three_loop_example_2to2_scattering}. The associated expansion contains 81 regions; we explain how these arise from the fundamental pattern in section~\ref{section-fundamental_facet_region_structure} while satisfying all subgraph requirements in section~\ref{section-further_requirements_mode_subgraphs}. Then, in section~\ref{section-six_loop_example_1to3_decay_plus_soft_emission}, we consider a more complicated six-loop graph that contributes to the one-to-three decay process with an additional soft emission, and we examine all its 199 regions.

Finally, we note that although the present work is formulated for massless Feynman integrals, the extension to massive cases could be straightforward. In particular, with only minor adaptations, our analysis can be applied to a recent study of $gg \to HH$~\cite{Jaskiewicz:2024xkd}, which involves massive integrals, and it automatically reproduces the results obtained there. We demonstrate this extension in section~\ref{section-massive_extension_revisiting_recent_work_ggHH}.

\subsection{A three-loop example: two-to-two scattering}
\label{section-three_loop_example_2to2_scattering}

\begin{figure}[t]
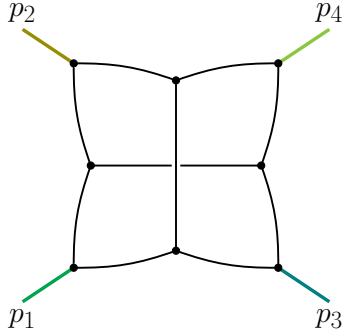

\centering
\include{figs/4pt3loop/three_loop_2to2_scattering}
\vspace{-3em}\caption{A three-loop graph contributing to the two-to-two scattering $p_1+p_2\to p_3+p_4$.}
\label{figure-three_loop_2to2_scattering}
\end{figure}

In this subsection, we analyze the region structure of the three-loop graph shown in figure~\ref{figure-three_loop_2to2_scattering}.
It represents a two-to-two scattering process. For the asymptotic expansion, we assign the following modes to the external momenta, which are color coded accordingly in the figure:
\begin{align}
\label{eq:three_loop_example_external_modes}
    \mathscr{X}(p_1) = C_1,\quad \mathscr{X}(p_2) = C_2^2,\quad \mathscr{X}(p_3) = C_3^\infty,\quad \mathscr{X}(p_4) = C_4^\infty. 
\end{align}
That is, the momenta $p_i$ ($i=1,2,3,4$) lie along distinct lightcones, with $p_1^2\sim \lambda$, $p_2^2\sim \lambda^2$, and $p_3^2=p_4^2=0$.
In this kinematic setup, a straightforward verification by computer codes yields \emph{81} distinct (facet) regions, each expressed in parameter space (see section~\ref{section-parametric_representations_spanning_trees} for the parametric representation of regions). Below, we shall derive the momentum-space interpretations of these regions directly from our understanding of region structures, and confirm that they match the output of the computer codes exactly.

Before proceeding, we note that figure~\ref{figure-three_loop_2to2_scattering} exhibits \emph{a unique hidden region} in which hard scattering occurs in two disconnected subgraphs: the horizontal and vertical central rungs~\cite{GrdHzgJnsMa24,Ma25}. This Landshoff-type configuration does not correspond to any facet region and therefore cannot be derived directly from the geometric method, but it gives an essential contribution to the asymptotic expansion, as verified by recent analytic results~\cite{Bargiela:2021wuy}. Nevertheless, in this work, we restrict our analysis solely to facet regions.

First, let us derive the regions containing the softest mode(s) in the expansion. From section~\ref{section-implication_infrared_compatibility_requirement}, the possibly softest mode is $S^{\max_i\{m_i+n_i\}}$, where the index $i$ goes through those external modes $S^{m_i}C_i^{n_i}$ with $n_i\neq \infty$. From eq.~(\ref{eq:three_loop_example_external_modes}), we have
\begin{align}
    i\in\{1,2\},\quad m_1=m_2=0,\quad n_1=1,\quad n_2=2.
\end{align}
Thus $\max_i\{m_i+n_i\} = 2$, and the possibly softest mode is $S^2$.

\begin{figure}[t]
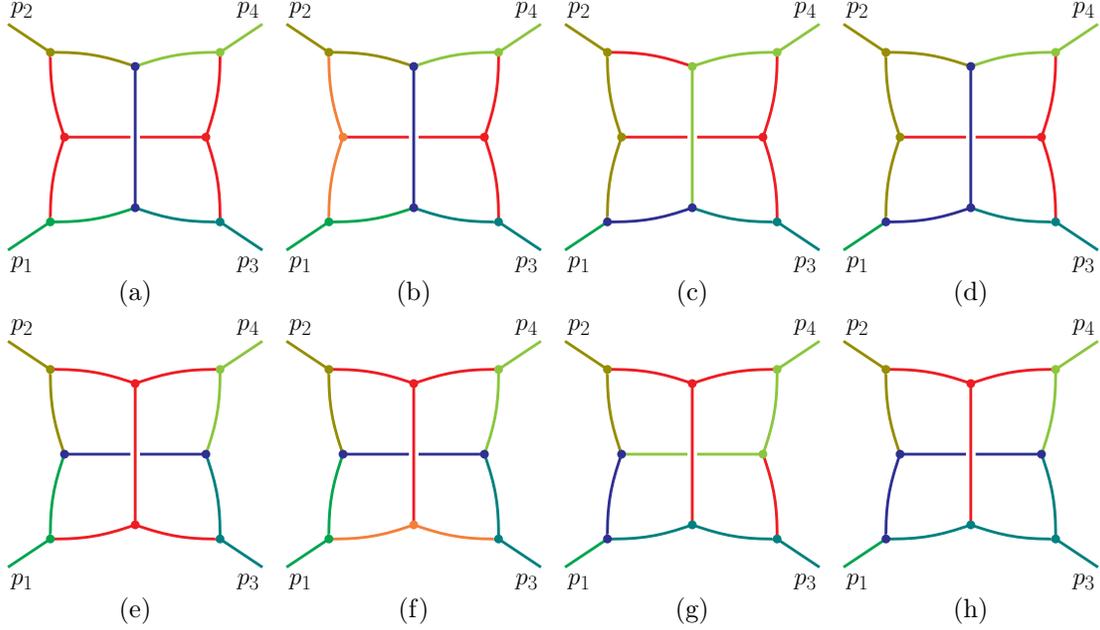

\centering
\begin{subfigure}[b]{0.23\textwidth}
\centering
\include{figs/4pt3loop/S2region1}
\vspace{-3.3em}\caption{}
\label{4pt3loop_S2region1}
\end{subfigure}
\begin{subfigure}[b]{0.23\textwidth}
\centering
\include{figs/4pt3loop/S2region2}
\vspace{-3.3em}\caption{}
\label{4pt3loop_S2region2}
\end{subfigure}
\begin{subfigure}[b]{0.23\textwidth}
\centering
\include{figs/4pt3loop/S2region3}
\vspace{-3.3em}\caption{}
\label{4pt3loop_S2region3}
\end{subfigure}
\begin{subfigure}[b]{0.23\textwidth}
\centering
\include{figs/4pt3loop/S2region4}
\vspace{-3.3em}\caption{}
\label{4pt3loop_S2region4}
\end{subfigure}
\\
\centering
\begin{subfigure}[b]{0.23\textwidth}
\centering
\include{figs/4pt3loop/S2region5}
\vspace{-3.3em}\caption{}
\label{4pt3loop_S2region5}
\end{subfigure}
\begin{subfigure}[b]{0.23\textwidth}
\centering
\include{figs/4pt3loop/S2region6}
\vspace{-3.3em}\caption{}
\label{4pt3loop_S2region6}
\end{subfigure}
\begin{subfigure}[b]{0.23\textwidth}
\centering
\include{figs/4pt3loop/S2region7}
\vspace{-3.3em}\caption{}
\label{4pt3loop_S2region7}
\end{subfigure}
\begin{subfigure}[b]{0.23\textwidth}
\centering
\include{figs/4pt3loop/S2region8}
\vspace{-3.3em}\caption{}
\label{4pt3loop_S2region8}
\end{subfigure}
\caption{The 8 regions of figure~\ref{figure-three_loop_2to2_scattering} characterized by the {\color{Red}$\boldsymbol{S^2}$} mode.}
\label{figure-4pt3loop_S2regions}
\end{figure}

We now exhaust all configurations of figure~\ref{figure-three_loop_2to2_scattering} that satisfy (1) the fundamental pattern (figure~\ref{figure-virtuality_expansion_fundamental_pattern}), (2) the subgraph requirements, and (3) the condition that the $S^2$ component (denoted by $\gamma_{S^2}^{}$) is nonempty.\footnote{In practice, the most efficient procedure is first to enumerate all fundamental patterns containing a $\gamma_{S^2}$ and satisfying the connectivity requirements, and then to discard those that fail the infrared‑compatibility requirement.} There are \emph{8} such configurations, shown in figure~\ref{figure-4pt3loop_S2regions}.
In accordance with the external modes, we adopt the following color scheme throughout this subsection: $H$ mode ({\color{Blue}\bf blue}), $C_1$ mode ({\color{Green}\bf green}), $C_2^n$ mode ({\color{olive}\bf olive}) with $n\in\{1,2\}$, $C_3^n$ mode ({\color{teal}\bf teal}) with $n\in\mathbb{N}_+$, $C_4^n$ mode ({\color{LimeGreen}\bf lime}) with $n\in\mathbb{N}_+$, $SC_i$ modes with $i\in\{2,3,4\}$ ({\color{Orange}\bf orange}), and $S^2$ mode ({\color{Red}\bf red}).

One can directly verify that all the subgraph requirements are satisfied in each region. For example, in figure~\ref{4pt3loop_S2region1}, the First Connectivity Theorem holds because all subgraphs of the form $\bigcup_{\mathscr{V}(X)\leqslant n}\Gamma_{X}$ (with $n=0,1,2,3,4$) are connected. Moreover, each mode component is infrared compatible, as can be checked in the following sequence:
\begin{itemize}
    \item [(1)] $\gamma_{C_1}^{}$ and $\gamma_{C_2^2}^{}$ (because the external momenta $p_1$ and $p_2$, which carry precisely the modes $C_1$ and $C_2^2$ respectively, enter them respectively);
    \item [(2)] $\gamma_{S^2}^{}$ (since it is a degree‑$2$ messenger relevant to $\gamma_{C_2^2},\gamma_{C_3^2},\gamma_{C_4^2}$, and $\gamma_{C_2^2}$ was already confirmed infrared compatible in the previous step);
    \item [(3)] $\gamma_{C_3^2}^{}$ and $\gamma_{C_4^2}^{}$ (because the total momentum entering each of them---summed from the external kinematics and previously confirmed infrared‑compatible components---belongs to mode $C_3^2$ or $C_4^2$, respectively).
\end{itemize}

\bigbreak
Next, we determine the regions where the softest mode(s) are of the form $SC_i$ for some~$i$. A key observation is that the only possible such mode is $SC_2$. To see why, note that no external momentum has mode $SC_i$. Then by the Second Connectivity Theorem (theorem~\ref{theorem-mode_subgraphs_connectivity2}), any $SC_i$ component must be relevant to two harder‑mode components $\gamma_{X_1}^{}$ and $\gamma_{X_2}^{}$ with $SC_i = X_1\wedge X_2$. From the $\wedge$ operation rule (theorem~\ref{theorem-overlapping_modes_intersection_union_rules}), one of $X_1$ or $X_2$ must be $C_i^2$. Because $\mathscr{X}(p_1)=C_1$, which is already harder than $C_1^2$, there cannot be a $C_1^2$ mode component, ruling out $i=1$. If $i=3$, then the region would contain both an $SC_3$ component $\gamma_{SC_3}^{}$ and a $C_3^2$ component $\gamma_{C_3^2}^{}$, both of which must be infrared compatible (theorem~\ref{theorem-infrared_compatibility_requirement}). This, however, leads to a circular dependency:
\begin{itemize}
\item For $\gamma_{C_3^2}^{}$ to be infrared compatible, the component $\gamma_{SC_3}^{}$ (whose momentum enters $\gamma_{C_3^2}^{}$) must be confirmed infrared compatible first.
\item For $\gamma_{SC_3}^{}$ to be infrared compatible, it must be relevant to two already confirmed infrared‑compatible components; since one of them must be $\gamma_{C_3^2}^{}$, this in turn requires $\gamma_{C_3^2}^{}$ to be infrared compatible first.
\end{itemize}
Hence the infrared‑compatibility requirement cannot be satisfied when an $SC_3$ component is present. The same reasoning excludes $SC_4$ components. We therefore conclude that in this scenario the unique softest mode is $SC_2$, which substantially simplifies the enumeration of regions.

\begin{figure}[t]
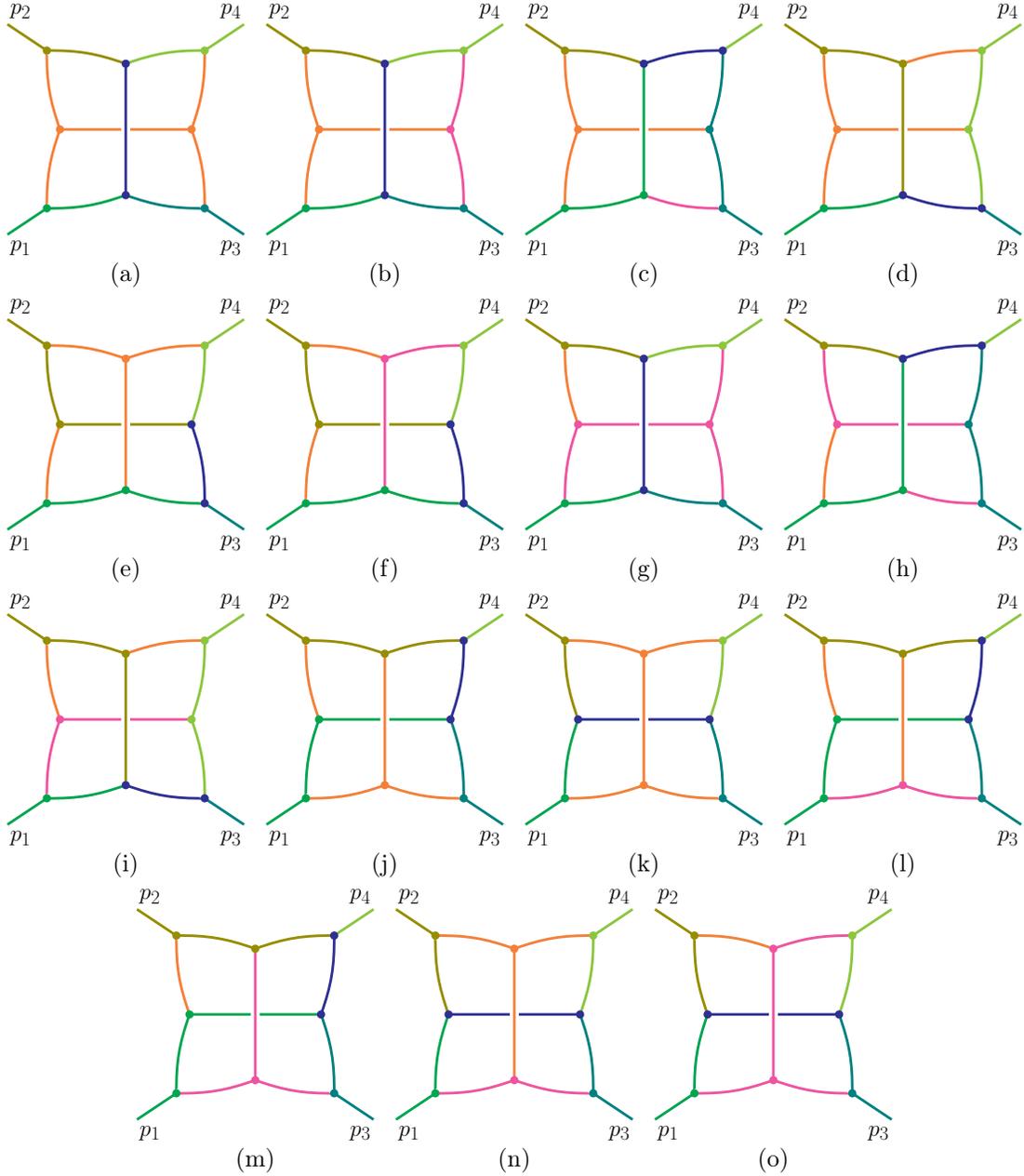

\centering
\begin{subfigure}[b]{0.23\textwidth}
\centering
\include{figs/4pt3loop/SCiregion1}
\vspace{-3.3em}\caption{}
\label{4pt3loop_SCiregion1}
\end{subfigure}
\begin{subfigure}[b]{0.23\textwidth}
\centering
\include{figs/4pt3loop/SCiregion2}
\vspace{-3.3em}\caption{}
\label{4pt3loop_SCiregion2}
\end{subfigure}
\begin{subfigure}[b]{0.23\textwidth}
\centering
\include{figs/4pt3loop/SCiregion3}
\vspace{-3.3em}\caption{}
\label{4pt3loop_SCiregion3}
\end{subfigure}
\begin{subfigure}[b]{0.23\textwidth}
\centering
\include{figs/4pt3loop/SCiregion4}
\vspace{-3.3em}\caption{}
\label{4pt3loop_SCiregion4}
\end{subfigure}
\\
\centering
\begin{subfigure}[b]{0.23\textwidth}
\centering
\include{figs/4pt3loop/SCiregion5}
\vspace{-3.3em}\caption{}
\label{4pt3loop_SCiregion5}
\end{subfigure}
\begin{subfigure}[b]{0.23\textwidth}
\centering
\include{figs/4pt3loop/SCiregion6}
\vspace{-3.3em}\caption{}
\label{4pt3loop_SCiregion6}
\end{subfigure}
\begin{subfigure}[b]{0.23\textwidth}
\centering
\include{figs/4pt3loop/SCiregion7}
\vspace{-3.3em}\caption{}
\label{4pt3loop_SCiregion7}
\end{subfigure}
\begin{subfigure}[b]{0.23\textwidth}
\centering
\include{figs/4pt3loop/SCiregion8}
\vspace{-3.3em}\caption{}
\label{4pt3loop_SCiregion8}
\end{subfigure}
\\
\centering
\begin{subfigure}[b]{0.23\textwidth}
\centering
\include{figs/4pt3loop/SCiregion9}
\vspace{-3.3em}\caption{}
\label{4pt3loop_SCiregion9}
\end{subfigure}
\begin{subfigure}[b]{0.23\textwidth}
\centering
\include{figs/4pt3loop/SCiregion10}
\vspace{-3.3em}\caption{}
\label{4pt3loop_SCiregion10}
\end{subfigure}
\begin{subfigure}[b]{0.23\textwidth}
\centering
\include{figs/4pt3loop/SCiregion11}
\vspace{-3.3em}\caption{}
\label{4pt3loop_SCiregion11}
\end{subfigure}
\begin{subfigure}[b]{0.23\textwidth}
\centering
\include{figs/4pt3loop/SCiregion12}
\vspace{-3.3em}\caption{}
\label{4pt3loop_SCiregion12}
\end{subfigure}
\\
\centering
\begin{subfigure}[b]{0.23\textwidth}
\centering
\include{figs/4pt3loop/SCiregion13}
\vspace{-3.3em}\caption{}
\label{4pt3loop_SCiregion13}
\end{subfigure}
\begin{subfigure}[b]{0.23\textwidth}
\centering
\include{figs/4pt3loop/SCiregion14}
\vspace{-3.3em}\caption{}
\label{4pt3loop_SCiregion14}
\end{subfigure}
\begin{subfigure}[b]{0.23\textwidth}
\centering
\include{figs/4pt3loop/SCiregion15}
\vspace{-3.3em}\caption{}
\label{4pt3loop_SCiregion15}
\end{subfigure}
\caption{The 15 regions of figure~\ref{figure-three_loop_2to2_scattering} characterized by the {\color{Orange}$\boldsymbol{SC_2}$} mode with a single edge in $\mathcal{H}$.}
\label{figure-4pt3loop_SCiregions_partI}
\end{figure}

For configurations with a single edge in $\mathcal{H}$, there are \emph{15} regions, shown in figure~\ref{figure-4pt3loop_SCiregions_partI}, where we have colored the $S$ mode in {\color{Rhodamine}\bf rhodamine}.
One can verify that in each case the orange‑colored mode is indeed $SC_2$. Moreover, all subgraph requirements are satisfied. For example, the mode components in figure~\ref{4pt3loop_SCiregion2} are infrared compatible, as confirmed by the following sequence:
\begin{itemize}
    \item [(1)] $\gamma_{C_1}^{}$ and $\gamma_{C_2^2}^{}$ (from the external momenta $p_1$ and $p_2$);
    \item [(2)] $\gamma_{SC_2}^{}$ (since it is relevant to $\gamma_{C_1}^{}$ and $\gamma_{C_2^2}^{}$, with $SC_2=C_1\wedge C_2^2$);
    \item [(3)] $\gamma_{C_3}^{}$ and $\gamma_{C_4}^{}$ (via the momentum flow from $p_3$, $p_4$, and $\gamma_{SC_2}^{}$);
    \item [(4)] $\gamma_{S}^{}$ (since it is relevant to $\gamma_{C_3}^{}$ and $\gamma_{C_4}^{}$, with $S=C_3\wedge C_4$);
    \item [(5)] $\gamma_H^{}$ (via momentum flow from $\gamma_{C_1}^{}$ and $\gamma_{C_3}^{}$, or from $\gamma_{C_2^2}^{}$ and $\gamma_{C_4}^{}$).
\end{itemize}
This pattern of establishing infrared compatibility applies to most of the other regions in figure~\ref{figure-4pt3loop_SCiregions_partI}. When a particular mode component is absent (for example, $\gamma_{C_4^2}=\varnothing$ in figure~\ref{4pt3loop_SCiregion3}), we simply omit it.

An exception is figure~\ref{4pt3loop_SCiregion13}. There, steps (1) and (2) above still hold. Next, observe that the $\gamma_{S}^{}$ component is relevant to three harder‑mode components: $\gamma_{C_1}^{}$, $\gamma_{C_2}^{}$, and $\gamma_{C_3}^{}$, whose external momentum modes are $C_1$, $C_2^2$, and $C_3^\infty$, respectively. Hence $\gamma_{S}^{}$ satisfies the conditions of a degree‑$1$ messenger, and it is infrared compatible due to the already‑established infrared compatibility of $\gamma_{C_1}^{}$. The remaining components ($\gamma_{C_2}^{}$, $\gamma_{C_3}^{}$, and $\gamma_H^{}$) can then be shown to be infrared compatible via their momentum flows.

\begin{figure}[t]
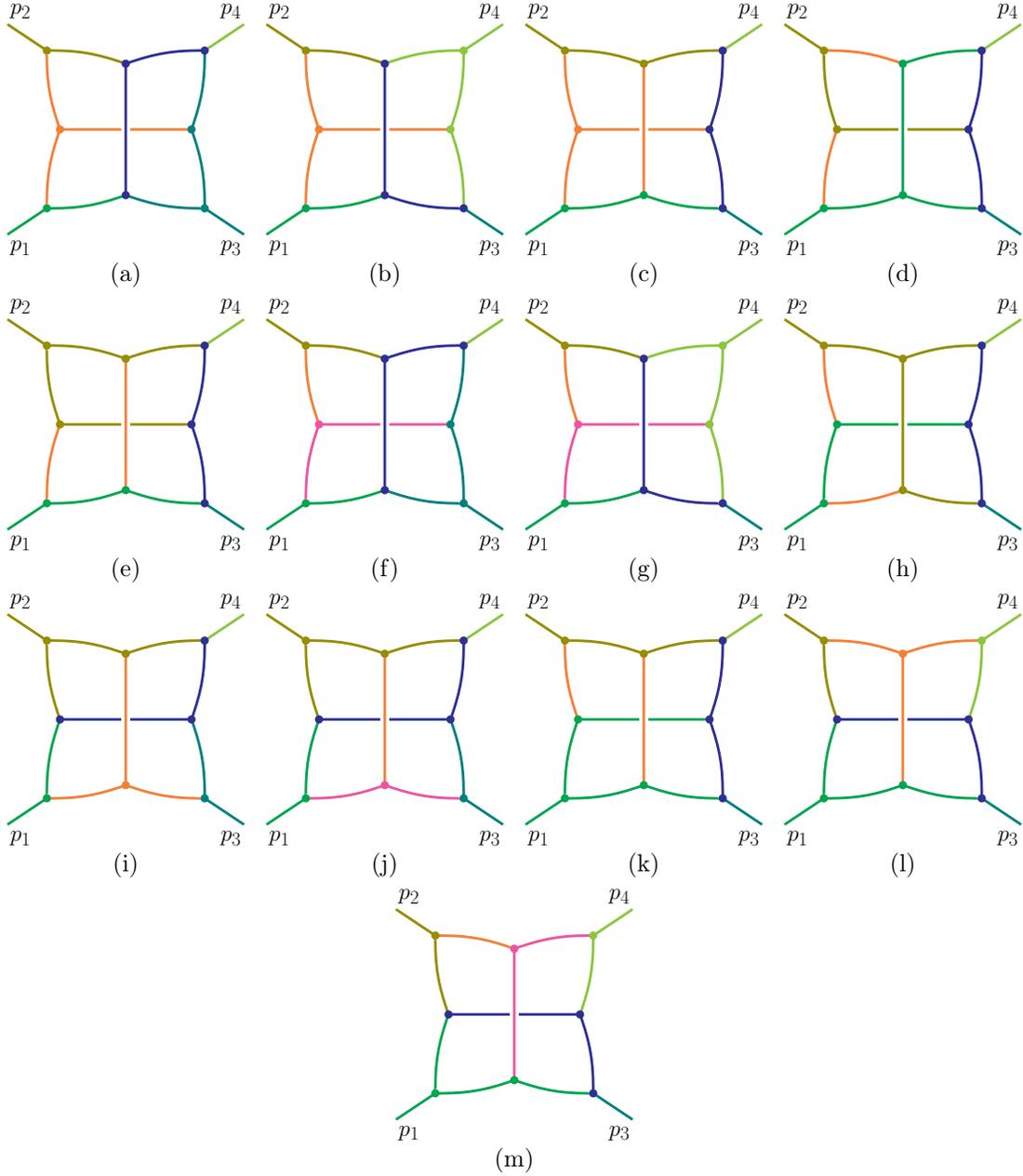

\centering
\begin{subfigure}[b]{0.23\textwidth}
\centering
\include{figs/4pt3loop/SCiregion16}
\vspace{-3.3em}\caption{}
\label{4pt3loop_SCiregion16}
\end{subfigure}
\begin{subfigure}[b]{0.23\textwidth}
\centering
\include{figs/4pt3loop/SCiregion17}
\vspace{-3.3em}\caption{}
\label{4pt3loop_SCiregion17}
\end{subfigure}
\begin{subfigure}[b]{0.23\textwidth}
\centering
\include{figs/4pt3loop/SCiregion18}
\vspace{-3.3em}\caption{}
\label{4pt3loop_SCiregion18}
\end{subfigure}
\begin{subfigure}[b]{0.23\textwidth}
\centering
\include{figs/4pt3loop/SCiregion19}
\vspace{-3.3em}\caption{}
\label{4pt3loop_SCiregion19}
\end{subfigure}
\\
\centering
\begin{subfigure}[b]{0.23\textwidth}
\centering
\include{figs/4pt3loop/SCiregion20}
\vspace{-3.3em}\caption{}
\label{4pt3loop_SCiregion20}
\end{subfigure}
\begin{subfigure}[b]{0.23\textwidth}
\centering
\include{figs/4pt3loop/SCiregion21}
\vspace{-3.3em}\caption{}
\label{4pt3loop_SCiregion21}
\end{subfigure}
\begin{subfigure}[b]{0.23\textwidth}
\centering
\include{figs/4pt3loop/SCiregion22}
\vspace{-3.3em}\caption{}
\label{4pt3loop_SCiregion22}
\end{subfigure}
\begin{subfigure}[b]{0.23\textwidth}
\centering
\include{figs/4pt3loop/SCiregion23}
\vspace{-3.3em}\caption{}
\label{4pt3loop_SCiregion23}
\end{subfigure}
\\
\centering
\begin{subfigure}[b]{0.23\textwidth}
\centering
\include{figs/4pt3loop/SCiregion24}
\vspace{-3.3em}\caption{}
\label{4pt3loop_SCiregion24}
\end{subfigure}
\begin{subfigure}[b]{0.23\textwidth}
\centering
\include{figs/4pt3loop/SCiregion25}
\vspace{-3.3em}\caption{}
\label{4pt3loop_SCiregion25}
\end{subfigure}
\begin{subfigure}[b]{0.23\textwidth}
\centering
\include{figs/4pt3loop/SCiregion26}
\vspace{-3.3em}\caption{}
\label{4pt3loop_SCiregion26}
\end{subfigure}
\begin{subfigure}[b]{0.23\textwidth}
\centering
\include{figs/4pt3loop/SCiregion27}
\vspace{-3.3em}\caption{}
\label{4pt3loop_SCiregion27}
\end{subfigure}
\\
\centering
\begin{subfigure}[b]{0.23\textwidth}
\centering
\include{figs/4pt3loop/SCiregion28}
\vspace{-3.3em}\caption{}
\label{4pt3loop_SCiregion28}
\end{subfigure}
\caption{The 13 regions of figure~\ref{figure-three_loop_2to2_scattering} characterized by the {\color{Orange}$\boldsymbol{SC_2}$} mode with two edges in $\mathcal{H}$.}
\label{figure-4pt3loop_SCiregions_partII}
\end{figure}

For the remaining regions characterized by the $SC_2$ mode, \emph{13} of them have exactly two edges in $\mathcal{H}$ (see figure~\ref{figure-4pt3loop_SCiregions_partII}); \emph{14} of them have three or more edges in $\mathcal{H}$ (see figure~\ref{figure-4pt3loop_SCiregions_partIII}).
\begin{figure}[t]
\centering
\begin{subfigure}[b]{0.23\textwidth}
\centering
\include{figs/4pt3loop/SCiregion29}
\vspace{-3.3em}\caption{}
\label{4pt3loop_SCiregion29}
\end{subfigure}
\begin{subfigure}[b]{0.23\textwidth}
\centering
\include{figs/4pt3loop/SCiregion30}
\vspace{-3.3em}\caption{}
\label{4pt3loop_SCiregion30}
\end{subfigure}
\begin{subfigure}[b]{0.23\textwidth}
\centering
\include{figs/4pt3loop/SCiregion31}
\vspace{-3.3em}\caption{}
\label{4pt3loop_SCiregion31}
\end{subfigure}
\begin{subfigure}[b]{0.23\textwidth}
\centering
\include{figs/4pt3loop/SCiregion32}
\vspace{-3.3em}\caption{}
\label{4pt3loop_SCiregion32}
\end{subfigure}
\\
\centering
\begin{subfigure}[b]{0.23\textwidth}
\centering
\include{figs/4pt3loop/SCiregion33}
\vspace{-3.3em}\caption{}
\label{4pt3loop_SCiregion33}
\end{subfigure}
\begin{subfigure}[b]{0.23\textwidth}
\centering
\include{figs/4pt3loop/SCiregion34}
\vspace{-3.3em}\caption{}
\label{4pt3loop_SCiregion34}
\end{subfigure}
\begin{subfigure}[b]{0.23\textwidth}
\centering
\include{figs/4pt3loop/SCiregion35}
\vspace{-3.3em}\caption{}
\label{4pt3loop_SCiregion35}
\end{subfigure}
\begin{subfigure}[b]{0.23\textwidth}
\centering
\include{figs/4pt3loop/SCiregion36}
\vspace{-3.3em}\caption{}
\label{4pt3loop_SCiregion36}
\end{subfigure}
\\
\centering
\begin{subfigure}[b]{0.23\textwidth}
\centering
\include{figs/4pt3loop/SCiregion37}
\vspace{-3.3em}\caption{}
\label{4pt3loop_SCiregion37}
\end{subfigure}
\begin{subfigure}[b]{0.23\textwidth}
\centering
\include{figs/4pt3loop/SCiregion38}
\vspace{-3.3em}\caption{}
\label{4pt3loop_SCiregion38}
\end{subfigure}
\begin{subfigure}[b]{0.23\textwidth}
\centering
\include{figs/4pt3loop/SCiregion39}
\vspace{-3.3em}\caption{}
\label{4pt3loop_SCiregion39}
\end{subfigure}
\begin{subfigure}[b]{0.23\textwidth}
\centering
\include{figs/4pt3loop/SCiregion40}
\vspace{-3.3em}\caption{}
\label{4pt3loop_SCiregion40}
\end{subfigure}
\\
\centering
\begin{subfigure}[b]{0.23\textwidth}
\centering
\include{figs/4pt3loop/SCiregion41}
\vspace{-3.3em}\caption{}
\label{4pt3loop_SCiregion41}
\end{subfigure}
\begin{subfigure}[b]{0.23\textwidth}
\centering
\include{figs/4pt3loop/SCiregion42}
\vspace{-3.3em}\caption{}
\label{4pt3loop_SCiregion42}
\end{subfigure}
\caption{The 14 regions of figure~\ref{figure-three_loop_2to2_scattering} characterized by the {\color{Orange}$\boldsymbol{SC_2}$} mode with $\geqslant3$ edges in $\mathcal{H}$.}
\label{figure-4pt3loop_SCiregions_partIII}
\end{figure}
For each of these regions, the verification of the subgraph requirements follows similarly as before, which we do not elaborate here for simplicity. So far we have come up with all those regions featuring the modes $S^2$ and/or $SC_i$.

\begin{figure}[t]
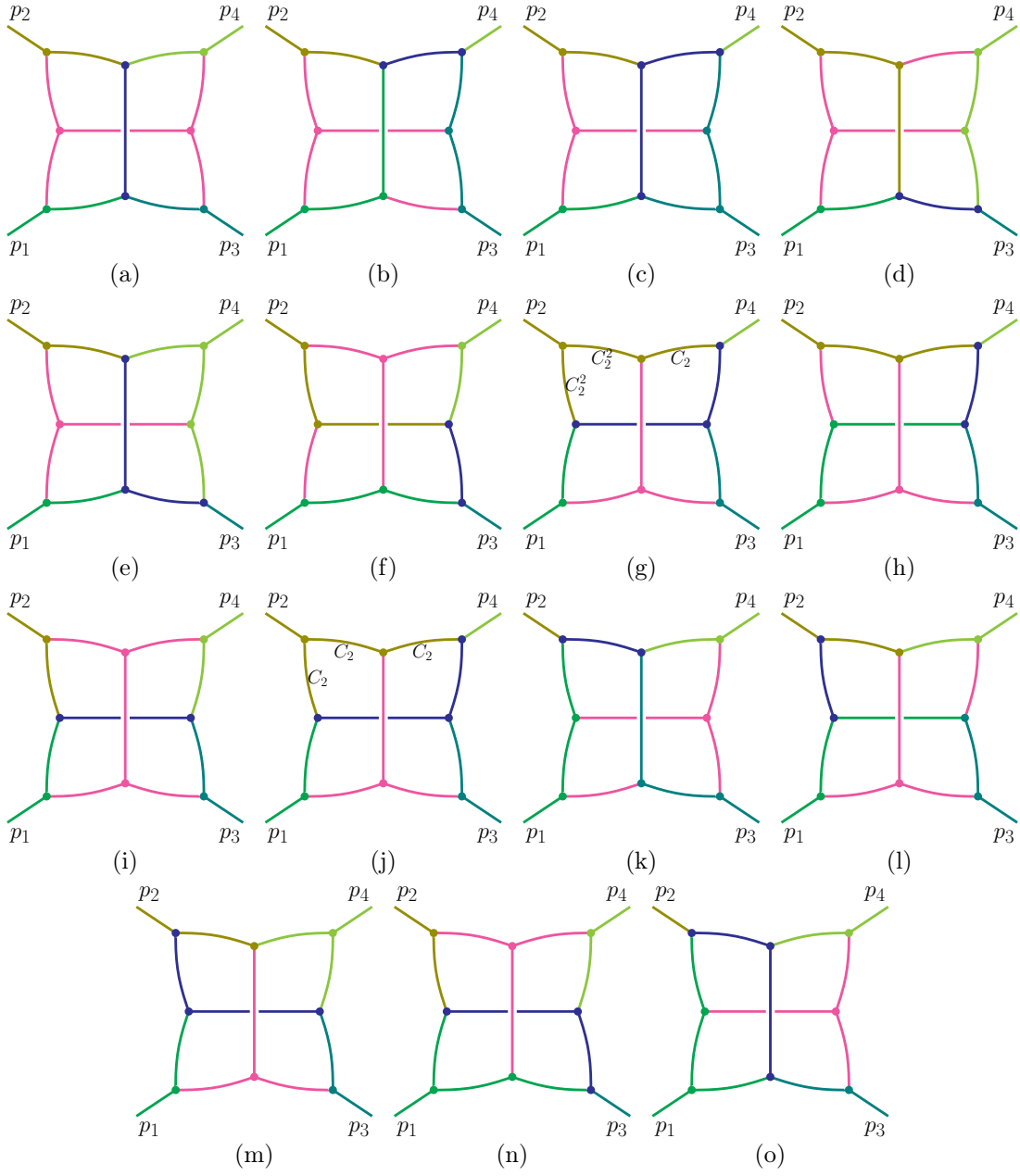

\centering
\begin{subfigure}[b]{0.23\textwidth}
\centering
\include{figs/4pt3loop/Sregion1}
\vspace{-3.3em}\caption{}
\label{4pt3loop_Sregion1}
\end{subfigure}
\begin{subfigure}[b]{0.23\textwidth}
\centering
\include{figs/4pt3loop/Sregion2}
\vspace{-3.3em}\caption{}
\label{4pt3loop_Sregion2}
\end{subfigure}
\begin{subfigure}[b]{0.23\textwidth}
\centering
\include{figs/4pt3loop/Sregion3}
\vspace{-3.3em}\caption{}
\label{4pt3loop_Sregion3}
\end{subfigure}
\begin{subfigure}[b]{0.23\textwidth}
\centering
\include{figs/4pt3loop/Sregion4}
\vspace{-3.3em}\caption{}
\label{4pt3loop_Sregion4}
\end{subfigure}
\\
\centering
\begin{subfigure}[b]{0.23\textwidth}
\centering
\include{figs/4pt3loop/Sregion5}
\vspace{-3.3em}\caption{}
\label{4pt3loop_Sregion5}
\end{subfigure}
\begin{subfigure}[b]{0.23\textwidth}
\centering
\include{figs/4pt3loop/Sregion6}
\vspace{-3.3em}\caption{}
\label{4pt3loop_Sregion6}
\end{subfigure}
\begin{subfigure}[b]{0.23\textwidth}
\centering
\include{figs/4pt3loop/Sregion7}
\vspace{-3.3em}\caption{}
\label{4pt3loop_Sregion7}
\end{subfigure}
\begin{subfigure}[b]{0.23\textwidth}
\centering
\include{figs/4pt3loop/Sregion8}
\vspace{-3.3em}\caption{}
\label{4pt3loop_Sregion8}
\end{subfigure}
\\
\centering
\begin{subfigure}[b]{0.23\textwidth}
\centering
\include{figs/4pt3loop/Sregion9}
\vspace{-3.3em}\caption{}
\label{4pt3loop_Sregion9}
\end{subfigure}
\begin{subfigure}[b]{0.23\textwidth}
\centering
\include{figs/4pt3loop/Sregion10}
\vspace{-3.3em}\caption{}
\label{4pt3loop_Sregion10}
\end{subfigure}
\begin{subfigure}[b]{0.23\textwidth}
\centering
\include{figs/4pt3loop/Sregion11}
\vspace{-3.3em}\caption{}
\label{4pt3loop_Sregion11}
\end{subfigure}
\begin{subfigure}[b]{0.23\textwidth}
\centering
\include{figs/4pt3loop/Sregion12}
\vspace{-3.3em}\caption{}
\label{4pt3loop_Sregion12}
\end{subfigure}
\\
\centering
\begin{subfigure}[b]{0.23\textwidth}
\centering
\include{figs/4pt3loop/Sregion13}
\vspace{-3.3em}\caption{}
\label{4pt3loop_Sregion13}
\end{subfigure}
\begin{subfigure}[b]{0.23\textwidth}
\centering
\include{figs/4pt3loop/Sregion14}
\vspace{-3.3em}\caption{}
\label{4pt3loop_Sregion14}
\end{subfigure}
\begin{subfigure}[b]{0.23\textwidth}
\centering
\include{figs/4pt3loop/Sregion15}
\vspace{-3.3em}\caption{}
\label{4pt3loop_Sregion15}
\end{subfigure}
\caption{The 15 regions of figure~\ref{figure-three_loop_2to2_scattering} characterized by the {\color{Rhodamine}$\boldsymbol{S}$} mode.}
\label{figure-4pt3loop_Sregions}
\end{figure}

\bigbreak
Then, we determine the regions characterized by the mode $S$. That is, in each such region, there is a nonempty $S$ component $\gamma_S^{}$, and no modes are softer than $S$. There are \emph{15} such regions in total, shown in figure~\ref{figure-4pt3loop_Sregions}.

Special attention is required for figures~\ref{4pt3loop_Sregion7} and \ref{4pt3loop_Sregion10}. While they appear identical due to our shared use of {\color{olive}\bf olive} for both $C_2$ and $C_2^2$ modes, they represent distinct configurations. In figure~\ref{4pt3loop_Sregion7}, the two edges adjacent to $p_2$ are in the $C_2^2$ mode, whereas in figure~\ref{4pt3loop_Sregion10} they are in the $C_2$ mode. Both configurations satisfy all subgraph requirements and are therefore valid regions in the expansion. For example, in figure~\ref{4pt3loop_Sregion7}, $\gamma_{C_2^2}^{}$ is infrared compatible due to the external momentum $p_2$ (which is in the $C_2$ mode), while in figure~\ref{4pt3loop_Sregion10}, $\gamma_{C_2}^{}$ becomes infrared compatible via the momentum flow from $p_2$ and the $S$ component (which is confirmed infrared compatible in advance).

\begin{figure}[t]
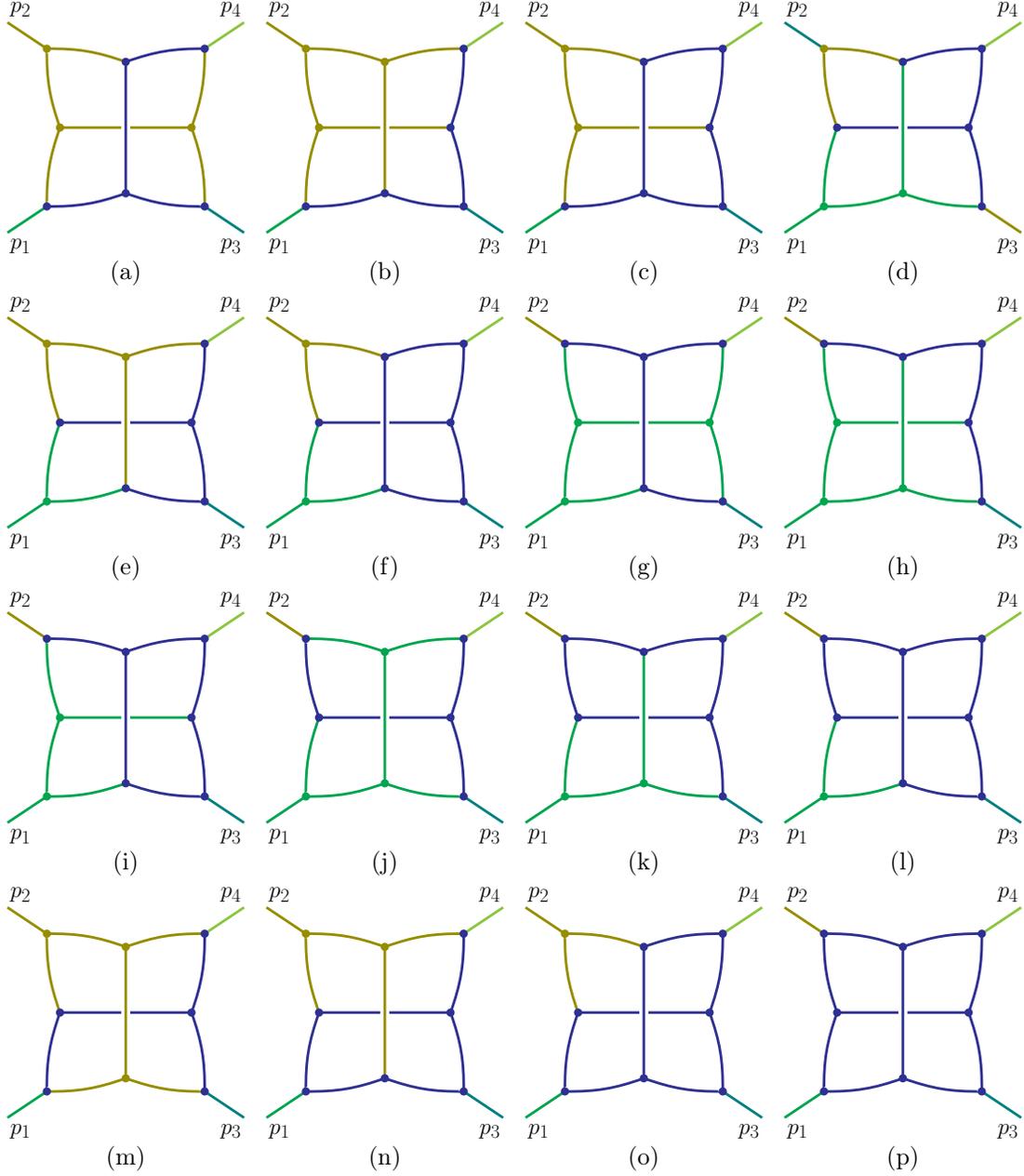

\centering
\begin{subfigure}[b]{0.23\textwidth}
\centering
\include{figs/4pt3loop/HCregion1}
\vspace{-3.3em}\caption{}
\label{4pt3loop_HCregion1}
\end{subfigure}
\begin{subfigure}[b]{0.23\textwidth}
\centering
\include{figs/4pt3loop/HCregion2}
\vspace{-3.3em}\caption{}
\label{4pt3loop_HCregion2}
\end{subfigure}
\begin{subfigure}[b]{0.23\textwidth}
\centering
\include{figs/4pt3loop/HCregion3}
\vspace{-3.3em}\caption{}
\label{4pt3loop_HCregion3}
\end{subfigure}
\begin{subfigure}[b]{0.23\textwidth}
\centering
\include{figs/4pt3loop/HCregion4}
\vspace{-3.3em}\caption{}
\label{4pt3loop_HCregion4}
\end{subfigure}
\\
\centering
\begin{subfigure}[b]{0.23\textwidth}
\centering
\include{figs/4pt3loop/HCregion5}
\vspace{-3.3em}\caption{}
\label{4pt3loop_HCregion5}
\end{subfigure}
\begin{subfigure}[b]{0.23\textwidth}
\centering
\include{figs/4pt3loop/HCregion6}
\vspace{-3.3em}\caption{}
\label{4pt3loop_HCregion6}
\end{subfigure}
\begin{subfigure}[b]{0.23\textwidth}
\centering
\include{figs/4pt3loop/HCregion7}
\vspace{-3.3em}\caption{}
\label{4pt3loop_HCregion7}
\end{subfigure}
\begin{subfigure}[b]{0.23\textwidth}
\centering
\include{figs/4pt3loop/HCregion8}
\vspace{-3.3em}\caption{}
\label{4pt3loop_HCregion8}
\end{subfigure}
\\
\centering
\begin{subfigure}[b]{0.23\textwidth}
\centering
\include{figs/4pt3loop/HCregion9}
\vspace{-3.3em}\caption{}
\label{4pt3loop_HCregion9}
\end{subfigure}
\begin{subfigure}[b]{0.23\textwidth}
\centering
\include{figs/4pt3loop/HCregion10}
\vspace{-3.3em}\caption{}
\label{4pt3loop_HCregion10}
\end{subfigure}
\begin{subfigure}[b]{0.23\textwidth}
\centering
\include{figs/4pt3loop/HCregion11}
\vspace{-3.3em}\caption{}
\label{4pt3loop_HCregion11}
\end{subfigure}
\begin{subfigure}[b]{0.23\textwidth}
\centering
\include{figs/4pt3loop/HCregion12}
\vspace{-3.3em}\caption{}
\label{4pt3loop_HCregion12}
\end{subfigure}
\\
\centering
\begin{subfigure}[b]{0.23\textwidth}
\centering
\include{figs/4pt3loop/HCregion13}
\vspace{-3.3em}\caption{}
\label{4pt3loop_HCregion13}
\end{subfigure}
\begin{subfigure}[b]{0.23\textwidth}
\centering
\include{figs/4pt3loop/HCregion14}
\vspace{-3.3em}\caption{}
\label{4pt3loop_HCregion14}
\end{subfigure}
\begin{subfigure}[b]{0.23\textwidth}
\centering
\include{figs/4pt3loop/HCregion15}
\vspace{-3.3em}\caption{}
\label{4pt3loop_HCregion15}
\end{subfigure}
\begin{subfigure}[b]{0.23\textwidth}
\centering
\include{figs/4pt3loop/HCregion16}
\vspace{-3.3em}\caption{}
\label{4pt3loop_HCregion16}
\end{subfigure}
\caption{The 16 regions of figure~\ref{figure-three_loop_2to2_scattering} with only the hard and collinear (including $C_2^2$) modes.}
\label{figure-4pt3loop_HCregions}
\end{figure}

\bigbreak
Finally, we derive the regions with only the hard and collinear (including $C_2^2$) modes. Note that in this situation, there cannot be any $C_3$ or $C_4$ component, otherwise the infrared-compatibility requirement would be violated (recall that $\mathscr{X}(p_3)=C_3^\infty$ and $\mathscr{X}(p_4)=C_4^\infty$, which do not carry the modes $C_3$ or $C_4$).
There are \emph{16} such regions, depicted in figure~\ref{figure-4pt3loop_HCregions}. The last one, figure~\ref{4pt3loop_HCregion16}, is the hard region.

We have thus derived all the regions of figure~\ref{figure-three_loop_2to2_scattering} in the associated virtuality expansion. The total number of regions is
\begin{align}
    8+13+15+14+15+16=81,
\end{align}
which matches the count obtained from \texttt{pySecDec}. One can further translate these regions from momentum space to parameter space using the relation in eq.~(\ref{eq:region_scaling_relation_momentum_LPparameter}), again finding perfect agreement.

\subsection{A six-loop example: one-to-three decay plus a soft emission}
\label{section-six_loop_example_1to3_decay_plus_soft_emission}

\begin{figure}[t]
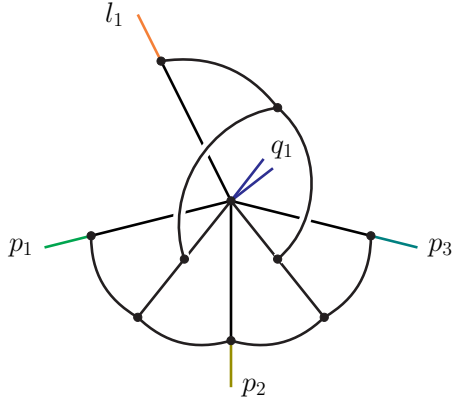

\centering
\include{figs/5pt6loop/six_loop_example_1to3_decay_plus_soft_emission}
\vspace{-3.5em}\caption{A six-loop graph contributing to the one-to-three decay process with a soft emission, i.e., $q_1\to p_1+p_2+p_3+l_1$.}
\label{figure-six_loop_example_1to3_decay_plus_soft_emission}
\end{figure}

In this subsection, we consider the six-loop graph in figure~\ref{figure-six_loop_example_1to3_decay_plus_soft_emission}. It represents a one-to-three decay process with an additional soft emission ($q_1 \to p_1 + p_2 + p_3 + l_1$).
For the asymptotic expansion, we assign the following modes to the external momenta and color them accordingly in the figure:
\begin{align}
\label{eq:six_loop_example_external_modes}
    \mathscr{X}(p_1) = C_1^\infty,\quad \mathscr{X}(p_2) = C_2,\quad \mathscr{X}(p_3) = C_3^\infty,\quad \mathscr{X}(l_1) = SC_4,\quad \mathscr{X}(q_1) = H.
\end{align}
In this kinematic setup, a straightforward verification by codes yields 199 distinct regions (expressed in parameter space). Below, we derive their momentum-space interpretations based on our understanding.

We start from those regions with the softest mode(s). Following the same reasoning as in section~\ref{section-three_loop_example_2to2_scattering}, the possibly softest mode is $S^{\textup{max}_i\{m_i+n_i\}}=S^2$. In order that a nonempty $S^2$ component (i.e., $\gamma_{S^2}^{}\neq \varnothing$) satisfies the infrared-compatibility requirement (theorem~\ref{theorem-infrared_compatibility_requirement}), it must acquire its infrared scaling from its harder-mode neighbors that have already been confirmed infrared compatible. Among all possible harder-mode neighbors of $\gamma_{S^2}^{}$, only one of them, $\gamma_{SC_4}^{}$, can become infrared compatible in advance due to the $SC_4$-mode external momentum $l_1$. That is, the infrared compatibility of $\gamma_{S^2}^{}$ is assured only if it exist as a degree-$2$ messenger, simultaneously relevant to nonempty mode components $\gamma_{C_1^2}^{}$, $\gamma_{C_3^2}^{}$, and $\gamma_{SC_4}^{}$. It then follows that the jet components $\gamma_{C_1}^{}, \gamma_{C_3}^{} \neq \varnothing$, containing $C_1^2$ and $C_3^2$ modes respectively.

\begin{figure}[t]
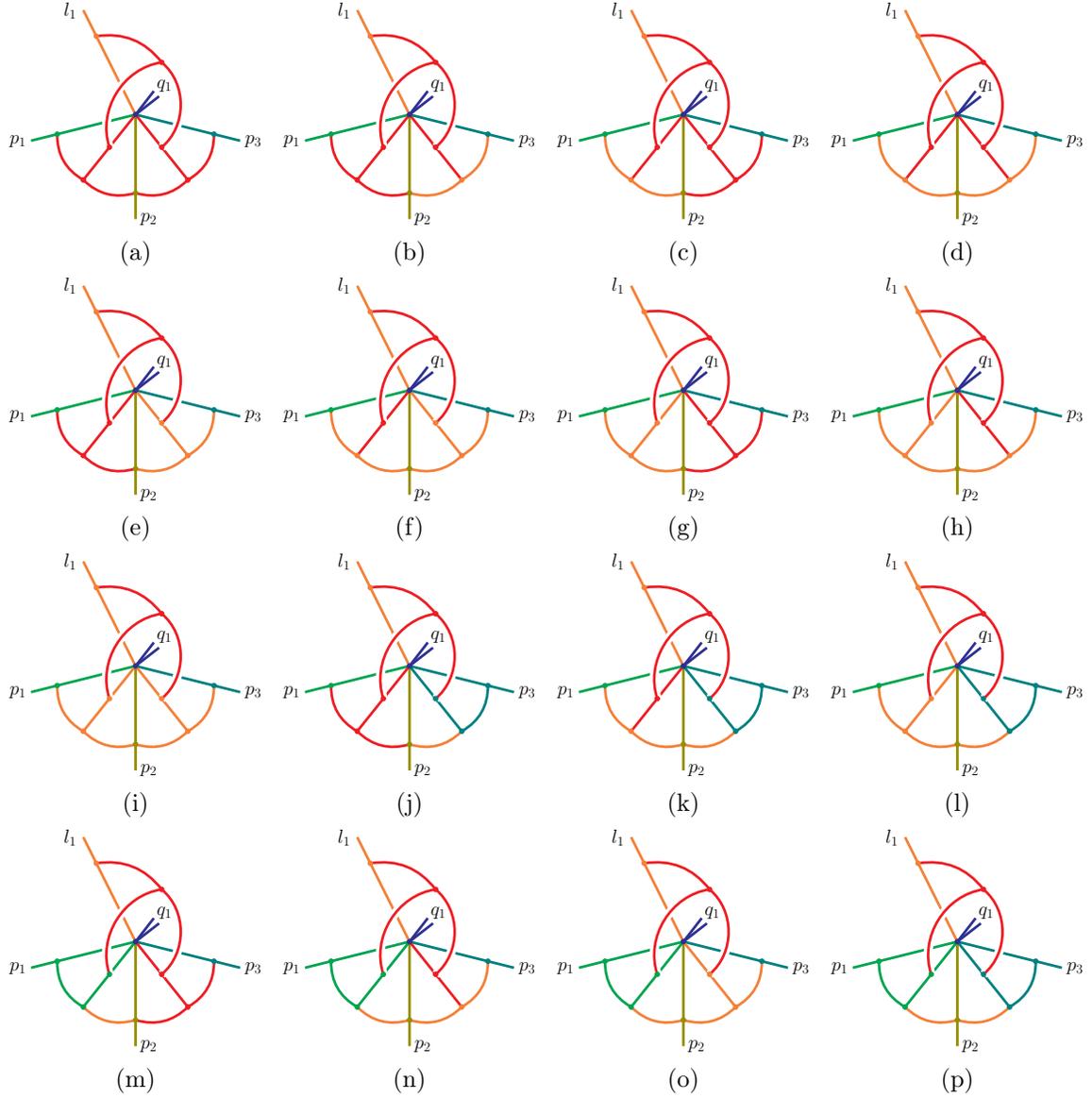

\centering
\begin{subfigure}[b]{0.24\textwidth}
\centering
\include{figs/5pt6loop/S2region1}
\vspace{-3em}\caption{}
\label{5pt6loop_S2region1}
\end{subfigure}
\begin{subfigure}[b]{0.24\textwidth}
\centering
\include{figs/5pt6loop/S2region2}
\vspace{-3em}
\caption{}
\label{5pt6loop_S2region2}
\end{subfigure}
\begin{subfigure}[b]{0.24\textwidth}
\centering
\include{figs/5pt6loop/S2region3}
\vspace{-3em}
\caption{}
\label{5pt6loop_S2region3}
\end{subfigure}
\begin{subfigure}[b]{0.24\textwidth}
\centering
\include{figs/5pt6loop/S2region4}
\vspace{-3em}
\caption{}
\label{5pt6loop_S2region4}
\end{subfigure}
\\
\centering
\begin{subfigure}[b]{0.24\textwidth}
\centering
\include{figs/5pt6loop/S2region5}
\vspace{-3em}\caption{}
\label{5pt6loop_S2region5}
\end{subfigure}
\begin{subfigure}[b]{0.24\textwidth}
\centering
\include{figs/5pt6loop/S2region6}
\vspace{-3em}
\caption{}
\label{5pt6loop_S2region6}
\end{subfigure}
\begin{subfigure}[b]{0.24\textwidth}
\centering
\include{figs/5pt6loop/S2region7}
\vspace{-3em}
\caption{}
\label{5pt6loop_S2region7}
\end{subfigure}
\begin{subfigure}[b]{0.24\textwidth}
\centering
\include{figs/5pt6loop/S2region8}
\vspace{-3em}
\caption{}
\label{5pt6loop_S2region8}
\end{subfigure}
\\
\centering
\begin{subfigure}[b]{0.24\textwidth}
\centering
\include{figs/5pt6loop/S2region9}
\vspace{-3em}\caption{}
\label{5pt6loop_S2region9}
\end{subfigure}
\begin{subfigure}[b]{0.24\textwidth}
\centering
\include{figs/5pt6loop/S2region10}
\vspace{-3em}
\caption{}
\label{5pt6loop_S2region10}
\end{subfigure}
\begin{subfigure}[b]{0.24\textwidth}
\centering
\include{figs/5pt6loop/S2region11}
\vspace{-3em}
\caption{}
\label{5pt6loop_S2region11}
\end{subfigure}
\begin{subfigure}[b]{0.24\textwidth}
\centering
\include{figs/5pt6loop/S2region12}
\vspace{-3em}
\caption{}
\label{5pt6loop_S2region12}
\end{subfigure}
\\
\centering
\begin{subfigure}[b]{0.24\textwidth}
\centering
\include{figs/5pt6loop/S2region13}
\vspace{-3em}\caption{}
\label{5pt6loop_S2region13}
\end{subfigure}
\begin{subfigure}[b]{0.24\textwidth}
\centering
\include{figs/5pt6loop/S2region14}
\vspace{-3em}
\caption{}
\label{5pt6loop_S2region14}
\end{subfigure}
\begin{subfigure}[b]{0.24\textwidth}
\centering
\include{figs/5pt6loop/S2region15}
\vspace{-3em}
\caption{}
\label{5pt6loop_S2region15}
\end{subfigure}
\begin{subfigure}[b]{0.24\textwidth}
\centering
\include{figs/5pt6loop/S2region16}
\vspace{-3em}
\caption{}
\label{5pt6loop_S2region16}
\end{subfigure}
\caption{The 16 regions of figure~\ref{figure-six_loop_example_1to3_decay_plus_soft_emission}, which are characterized by the {\color{Red}$\boldsymbol{S^2}$} mode with all three jets $\mathcal{J}_1,\mathcal{J}_2,\mathcal{J}_3$ nonempty ($\gamma_{C_1}^{},\gamma_{C_2}^{},\gamma_{C_3}^{}\neq \varnothing$).}
\label{figure-5pt6loop_S2regions_partI}
\end{figure}

In this situation, let us further assume that $\gamma_{C_2}^{} \neq \varnothing$, for which \emph{16} regions are allowed, as depicted in figure~\ref{figure-5pt6loop_S2regions_partI}.
In accordance with the external modes, here (and throughout this subsection) we color the mode subgraphs as follows: $H$ mode ({\color{Blue}\bf blue}), $C_1^2$ mode ({\color{Green}\bf green}), $C_2$ mode ({\color{olive}\bf olive}), $C_3^2$ mode ({\color{teal}\bf teal}), $SC_i$ modes with $i=1,3,4$ ({\color{Orange}\bf orange}), and $S^2$ mode ({\color{Red}\bf red}). One can verify that for each of these regions, all subgraph requirements are satisfied. In particular, to satisfy the infrared-compatibility requirement, mode components are confirmed to be infrared compatible in the following sequence:
\begin{itemize}
    \item [(1)] $\gamma_{SC_4}^{}$ and $\gamma_{C_2}^{}$ (from the external momenta $l_1$ and $p_2$);
    \item [(2)] $\gamma_{S^2}^{}$ (as it is a messenger relevant to $\gamma_{SC_4}^{}$ component);
    \item [(3)] $\gamma_{C_1^2}^{}$ and $\gamma_{C_3^2}^{}$ (via the momentum flow from $p_1$, $p_3$, and $\gamma_{S^2}^{}$);
    \item [(4)] $\gamma_{SC_1}^{}$ and $\gamma_{SC_3}^{}$ (since they are relevant to either $\gamma_{C_1^2}^{}$ or $\gamma_{C_3^2}^{}$, and to $\gamma_{C_2}^{}$).
\end{itemize}
Conversely, any other configuration with $\gamma_{S^2}^{}\neq \varnothing$ and $\gamma_{C_1}^{},\gamma_{C_2}^{},\gamma_{C_3}^{}\neq \varnothing$ would violate at least one subgraph requirement, thus not appearing in figure~\ref{figure-5pt6loop_S2regions_partI}.

For configurations with $\gamma_{S^2}^{}\neq \varnothing$ and $\gamma_{C_2}^{}= \varnothing$ (in other words, $p_2$ is attached to $\mathcal{H}$ directly), there are \emph{8} regions, as shown in figure~\ref{figure-5pt6loop_S2regions_partII}.
\begin{figure}[t]
\centering
\begin{subfigure}[b]{0.24\textwidth}
\centering
\include{figs/5pt6loop/S2region17}
\vspace{-3.3em}\caption{}
\label{5pt6loop_S2region17}
\end{subfigure}
\begin{subfigure}[b]{0.24\textwidth}
\centering
\include{figs/5pt6loop/S2region18}
\vspace{-3.3em}
\caption{}
\label{5pt6loop_S2region18}
\end{subfigure}
\begin{subfigure}[b]{0.24\textwidth}
\centering
\include{figs/5pt6loop/S2region19}
\vspace{-3.3em}
\caption{}
\label{5pt6loop_S2region19}
\end{subfigure}
\begin{subfigure}[b]{0.24\textwidth}
\centering
\include{figs/5pt6loop/S2region20}
\vspace{-3.3em}
\caption{}
\label{5pt6loop_S2region20}
\end{subfigure}
\\
\centering
\begin{subfigure}[b]{0.24\textwidth}
\centering
\include{figs/5pt6loop/S2region21}
\vspace{-3.3em}\caption{}
\label{5pt6loop_S2region21}
\end{subfigure}
\begin{subfigure}[b]{0.24\textwidth}
\centering
\include{figs/5pt6loop/S2region22}
\vspace{-3.3em}
\caption{}
\label{5pt6loop_S2region22}
\end{subfigure}
\begin{subfigure}[b]{0.24\textwidth}
\centering
\include{figs/5pt6loop/S2region23}
\vspace{-3.3em}
\caption{}
\label{5pt6loop_S2region23}
\end{subfigure}
\begin{subfigure}[b]{0.24\textwidth}
\centering
\include{figs/5pt6loop/S2region24}
\vspace{-3.3em}
\caption{}
\label{5pt6loop_S2region24}
\end{subfigure}
\caption{The 8 regions of figure~\ref{figure-six_loop_example_1to3_decay_plus_soft_emission}, which are characterized by the {\color{Red}$\boldsymbol{S^2}$} mode with $\mathcal{J}_1,\mathcal{J}_3\neq \varnothing$ and $\mathcal{J}_2= \varnothing$.}
\label{figure-5pt6loop_S2regions_partII}
\end{figure}
Identical to the region structure in figure~\ref{figure-5pt6loop_S2regions_partI}, here $\gamma_{S^2}^{}$ still acts as a degree-$2$ messenger, which is relevant to $\gamma_{SC_4}^{}$, $\gamma_{C_1}^{}$, and $\gamma_{C_3}^{}$. One difference lies in the absence of $SC_1$ and $SC_3$ components, i.e., $\gamma_{SC_1}^{}=\gamma_{SC_3}^{}=\varnothing$. This is due to the infrared-compatibility requirement: since $\gamma_{C_2}^{}=\varnothing$ here, any nonempty $\gamma_{SC_1}^{}$ or $\gamma_{SC_3}^{}$ would not be infrared compatible.

\bigbreak
Next, we consider regions characterized by $SC_i$ modes; namely, regions whose softest modes are $SC_1$, $SC_3$, and/or $SC_4$. A key observation is that only the $SC_4$ mode can appear in such regions. To see why, suppose for example that $\gamma_{SC_1}^{}\neq \varnothing$. Since there are no external momenta of mode $SC_1$, the infrared‑compatibility requirement would force $\gamma_{SC_1}^{}$ to be relevant to two already infrared‑compatible components, which must include a $\gamma_{C_1^2}^{}$. However, the infrared compatibility of $\gamma_{C_1^2}^{}$ requires the confirmation of infrared compatibility of $\gamma_{SC_1}^{}$ in advance, which leads to a similar circular dependency discussed in section~\ref{section-three_loop_example_2to2_scattering}. In other words, $\gamma_{SC_1}^{}$ cannot satisfy the infrared‑compatibility requirement. Therefore, all regions of this type contain the $SC_4$ mode, while $SC_1$ modes for $i=1,2,3$ are absent.

Following the analysis above, we first examine configurations where $\mathcal{J}_1$, $\mathcal{J}_2$, and $\mathcal{J}_3$ are all nonempty (from now on, we use $\mathcal{J}_i$ to represent the mode component $\gamma_{C_i}^{}$).
\begin{figure}[t]
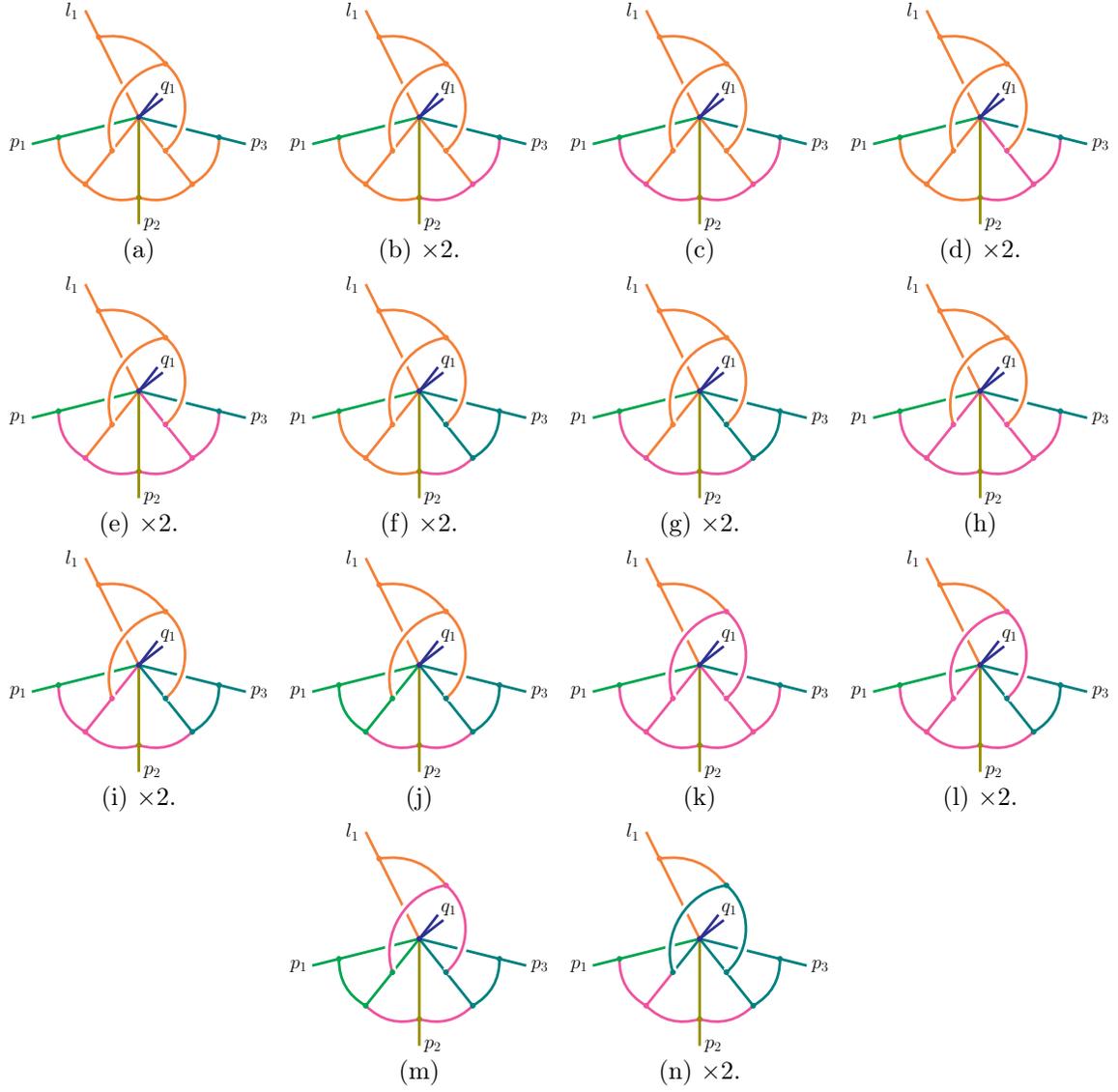

\centering
\begin{subfigure}[b]{0.24\textwidth}
\centering
\include{figs/5pt6loop/SCiregion1}
\vspace{-3.3em}\caption{}
\label{5pt6loop_SCiregion1}
\end{subfigure}
\begin{subfigure}[b]{0.24\textwidth}
\centering
\include{figs/5pt6loop/SCiregion2}
\vspace{-3.3em}\caption{$\times2$.}
\label{5pt6loop_SCiregion2}
\end{subfigure}
\begin{subfigure}[b]{0.24\textwidth}
\centering
\include{figs/5pt6loop/SCiregion3}
\vspace{-3.3em}\caption{}
\label{5pt6loop_SCiregion3}
\end{subfigure}
\begin{subfigure}[b]{0.24\textwidth}
\centering
\include{figs/5pt6loop/SCiregion4}
\vspace{-3.3em}\caption{$\times2$.}
\label{5pt6loop_SCiregion4}
\end{subfigure}
\\
\begin{subfigure}[b]{0.24\textwidth}
\centering
\include{figs/5pt6loop/SCiregion5}
\vspace{-3.3em}\caption{$\times2$.}
\label{5pt6loop_SCiregion5}
\end{subfigure}
\begin{subfigure}[b]{0.24\textwidth}
\centering
\include{figs/5pt6loop/SCiregion6}
\vspace{-3.3em}\caption{$\times2$.}
\label{5pt6loop_SCiregion6}
\end{subfigure}
\begin{subfigure}[b]{0.24\textwidth}
\centering
\include{figs/5pt6loop/SCiregion7}
\vspace{-3.3em}\caption{$\times2$.}
\label{5pt6loop_SCiregion7}
\end{subfigure}
\begin{subfigure}[b]{0.24\textwidth}
\centering
\include{figs/5pt6loop/SCiregion8}
\vspace{-3.3em}\caption{}
\label{5pt6loop_SCiregion8}
\end{subfigure}
\\
\begin{subfigure}[b]{0.24\textwidth}
\centering
\include{figs/5pt6loop/SCiregion9}
\vspace{-3.3em}\caption{$\times2$.}
\label{5pt6loop_SCiregion9}
\end{subfigure}
\begin{subfigure}[b]{0.24\textwidth}
\centering
\include{figs/5pt6loop/SCiregion10}
\vspace{-3.3em}\caption{}
\label{5pt6loop_SCiregion10}
\end{subfigure}
\begin{subfigure}[b]{0.24\textwidth}
\centering
\include{figs/5pt6loop/SCiregion11}
\vspace{-3.3em}\caption{}
\label{5pt6loop_SCiregion11}
\end{subfigure}
\begin{subfigure}[b]{0.24\textwidth}
\centering
\include{figs/5pt6loop/SCiregion12}
\vspace{-3.3em}\caption{$\times2$.}
\label{5pt6loop_SCiregion12}
\end{subfigure}
\\
\begin{subfigure}[b]{0.24\textwidth}
\centering
\include{figs/5pt6loop/SCiregion13}
\vspace{-3.3em}\caption{}
\label{5pt6loop_SCiregion13}
\end{subfigure}
\begin{subfigure}[b]{0.24\textwidth}
\centering
\include{figs/5pt6loop/SCiregion14}
\vspace{-3.3em}\caption{$\times2$.}
\label{5pt6loop_SCiregion14}
\end{subfigure}
\caption{The 22 regions of figure~\ref{figure-six_loop_example_1to3_decay_plus_soft_emission} characterized by the {\color{Orange}$\boldsymbol{SC_4}$} mode and with all three jets $\mathcal{J}_1,\mathcal{J}_2,\mathcal{J}_3$ nonempty. The color {\color{Rhodamine}\bf rhodamine} represents the $S$ mode. Only 14 distinct diagrams are shown explicitly; the remaining 8 are obtained by applying the reflection $1\leftrightarrow 3$ to those with ``$\times2$''.}
\label{figure-5pt6loop_SCiregions_partI}
\end{figure}
There are \emph{22} such regions, shown in figure~\ref{figure-5pt6loop_SCiregions_partI}, where we have colored the $S$ mode in {\color{Rhodamine}\bf rhodamine}.
For each of them, the external momentum $l_1$ must be relevant to both $\mathcal{J}_1$ and $\mathcal{J}_3$. Since
\begin{align}
    \mathscr{X}(l_1)\vee \mathscr{X}(p_1) = SC_4\vee C_1^\infty =  C_1,\quad \mathscr{X}(l_1)\vee \mathscr{X}(p_3) = SC_4\vee C_3^\infty =  C_3,
\end{align}
only $C_1$ and $C_3$ modes are allowed in $\mathcal{J}_1$ and $\mathcal{J}_3$, respectively. Note that $l_1$ can enter some $S$ component before reaching $\mathcal{J}_1$ or $\mathcal{J}_3$.

Only 14 of the 22 regions are explicitly shown in figure~\ref{figure-5pt6loop_SCiregions_partI}. Due to symmetry, the remaining 8 regions can be obtained by applying the reflection $1\leftrightarrow 3$ (i.e., $p_1\leftrightarrow p_3$ and $C_1\leftrightarrow C_3$) to those marked ``$\times2$''.

\begin{figure}[t]
\centering
\begin{subfigure}[b]{0.24\textwidth}
\centering
\include{figs/5pt6loop/SCiregion15}
\vspace{-3.3em}\caption{}
\label{5pt6loop_SCiregion15}
\end{subfigure}
\begin{subfigure}[b]{0.24\textwidth}
\centering
\include{figs/5pt6loop/SCiregion16}
\vspace{-3.3em}\caption{}
\label{5pt6loop_SCiregion16}
\end{subfigure}
\begin{subfigure}[b]{0.24\textwidth}
\centering
\include{figs/5pt6loop/SCiregion17}
\vspace{-3.3em}\caption{}
\label{5pt6loop_SCiregion17}
\end{subfigure}
\begin{subfigure}[b]{0.24\textwidth}
\centering
\include{figs/5pt6loop/SCiregion18}
\vspace{-3.3em}\caption{}
\label{5pt6loop_SCiregion18}
\end{subfigure}
\\
\begin{subfigure}[b]{0.24\textwidth}
\centering
\include{figs/5pt6loop/SCiregion19}
\vspace{-3.3em}\caption{}
\label{5pt6loop_SCiregion19}
\end{subfigure}
\begin{subfigure}[b]{0.24\textwidth}
\centering
\include{figs/5pt6loop/SCiregion20}
\vspace{-3.3em}\caption{}
\label{5pt6loop_SCiregion20}
\end{subfigure}
\begin{subfigure}[b]{0.24\textwidth}
\centering
\include{figs/5pt6loop/SCiregion21}
\vspace{-3.3em}\caption{}
\label{5pt6loop_SCiregion21}
\end{subfigure}
\begin{subfigure}[b]{0.24\textwidth}
\centering
\include{figs/5pt6loop/SCiregion22}
\vspace{-3.3em}\caption{}
\label{5pt6loop_SCiregion22}
\end{subfigure}
\\
\begin{subfigure}[b]{0.24\textwidth}
\centering
\include{figs/5pt6loop/SCiregion23}
\vspace{-3.3em}\caption{}
\label{5pt6loop_SCiregion23}
\end{subfigure}
\begin{subfigure}[b]{0.24\textwidth}
\centering
\include{figs/5pt6loop/SCiregion24}
\vspace{-3.3em}\caption{}
\label{5pt6loop_SCiregion24}
\end{subfigure}
\begin{subfigure}[b]{0.24\textwidth}
\centering
\include{figs/5pt6loop/SCiregion25}
\vspace{-3.3em}\caption{}
\label{5pt6loop_SCiregion25}
\end{subfigure}
\begin{subfigure}[b]{0.24\textwidth}
\centering
\include{figs/5pt6loop/SCiregion26}
\vspace{-3.3em}\caption{}
\label{5pt6loop_SCiregion26}
\end{subfigure}
\\
\begin{subfigure}[b]{0.24\textwidth}
\centering
\include{figs/5pt6loop/SCiregion27}
\vspace{-3.3em}\caption{}
\label{5pt6loop_SCiregion27}
\end{subfigure}
\begin{subfigure}[b]{0.24\textwidth}
\centering
\include{figs/5pt6loop/SCiregion28}
\vspace{-3.3em}\caption{}
\label{5pt6loop_SCiregion28}
\end{subfigure}
\begin{subfigure}[b]{0.24\textwidth}
\centering
\include{figs/5pt6loop/SCiregion29}
\vspace{-3.3em}\caption{}
\label{5pt6loop_SCiregion29}
\end{subfigure}
\begin{subfigure}[b]{0.24\textwidth}
\centering
\include{figs/5pt6loop/SCiregion30}
\vspace{-3.3em}\caption{}
\label{5pt6loop_SCiregion30}
\end{subfigure}
\caption{16 of the 22 regions of figure~\ref{figure-six_loop_example_1to3_decay_plus_soft_emission}, which are characterized by the {\color{Orange}$\boldsymbol{SC_4}$} mode with $\mathcal{J}_1,\mathcal{J}_2\neq \varnothing$ and $\mathcal{J}_3=\varnothing$.}
\label{figure-5pt6loop_SCiregions_partII}
\end{figure}
For configurations with $\mathcal{J}_1,\mathcal{J}_2\neq \varnothing$ and $\mathcal{J}_3=\varnothing$, there are \emph{22} regions, 16 of which are presented in figure~\ref{figure-5pt6loop_SCiregions_partII} due to page limit. Nevertheless, the remaining 6 follow very similar structures.
In these regions, since $\mathcal{J}_3$ is empty, $l_1$ only needs to be relevant to $\mathcal{J}_1$, and like above, it can enter some $S$ component before reaching $\mathcal{J}_1$.

By applying the reflection $1\leftrightarrow3$ to the regions in figure~\ref{figure-5pt6loop_SCiregions_partII}, we obtain another \emph{22} regions in which $\mathcal{J}_1=\varnothing$ and $\mathcal{J}_2,\mathcal{J}_3\neq \varnothing$.

For the configurations with $\mathcal{J}_1,\mathcal{J}_3\neq \varnothing$ and $\mathcal{J}_2=\varnothing$, there are \emph{16} regions. 10 of them are explicitly shown in figures~\ref{5pt6loop_SCiregion31}--\ref{5pt6loop_SCiregion40}, while the remaining 6 can be obtained by applying the reflection $1\leftrightarrow3$ to the graphs marked ``$\times2$''.
\begin{figure}[t]
\centering
\begin{subfigure}[b]{0.24\textwidth}
\centering
\include{figs/5pt6loop/SCiregion31}
\vspace{-3.5em}\caption{$\times2$.}
\label{5pt6loop_SCiregion31}
\end{subfigure}
\begin{subfigure}[b]{0.24\textwidth}
\centering
\include{figs/5pt6loop/SCiregion32}
\vspace{-3.5em}\caption{}
\label{5pt6loop_SCiregion32}
\end{subfigure}
\begin{subfigure}[b]{0.24\textwidth}
\centering
\include{figs/5pt6loop/SCiregion33}
\vspace{-3.5em}\caption{$\times2$.}
\label{5pt6loop_SCiregion33}
\end{subfigure}
\begin{subfigure}[b]{0.24\textwidth}
\centering
\include{figs/5pt6loop/SCiregion34}
\vspace{-3.5em}\caption{$\times2$.}
\label{5pt6loop_SCiregion34}
\end{subfigure}
\\
\begin{subfigure}[b]{0.24\textwidth}
\centering
\include{figs/5pt6loop/SCiregion35}
\vspace{-3.5em}\caption{}
\label{5pt6loop_SCiregion35}
\end{subfigure}
\begin{subfigure}[b]{0.24\textwidth}
\centering
\include{figs/5pt6loop/SCiregion36}
\vspace{-3.5em}\caption{$\times2$.}
\label{5pt6loop_SCiregion36}
\end{subfigure}
\begin{subfigure}[b]{0.24\textwidth}
\centering
\include{figs/5pt6loop/SCiregion37}
\vspace{-3.5em}\caption{}
\label{5pt6loop_SCiregion37}
\end{subfigure}
\begin{subfigure}[b]{0.24\textwidth}
\centering
\include{figs/5pt6loop/SCiregion38}
\vspace{-3.5em}\caption{$\times2$.}
\label{5pt6loop_SCiregion38}
\end{subfigure}
\\
\begin{subfigure}[b]{0.24\textwidth}
\centering
\include{figs/5pt6loop/SCiregion39}
\vspace{-3.5em}\caption{$\times2$.}
\label{5pt6loop_SCiregion39}
\end{subfigure}
\begin{subfigure}[b]{0.24\textwidth}
\centering
\include{figs/5pt6loop/SCiregion40}
\vspace{-3.5em}\caption{}
\label{5pt6loop_SCiregion40}
\end{subfigure}
\begin{subfigure}[b]{0.24\textwidth}
\centering
\include{figs/5pt6loop/SCiregion41}
\vspace{-3.5em}\caption{}
\label{5pt6loop_SCiregion41}
\end{subfigure}
\begin{subfigure}[b]{0.24\textwidth}
\centering
\include{figs/5pt6loop/SCiregion42}
\vspace{-3.5em}\caption{}
\label{5pt6loop_SCiregion42}
\end{subfigure}
\\
\begin{subfigure}[b]{0.24\textwidth}
\centering
\include{figs/5pt6loop/SCiregion43}
\vspace{-3.5em}\caption{}
\label{5pt6loop_SCiregion43}
\end{subfigure}
\begin{subfigure}[b]{0.24\textwidth}
\centering
\include{figs/5pt6loop/SCiregion44}
\vspace{-3.5em}\caption{}
\label{5pt6loop_SCiregion44}
\end{subfigure}
\begin{subfigure}[b]{0.24\textwidth}
\centering
\include{figs/5pt6loop/SCiregion45}
\vspace{-3.5em}\caption{}
\label{5pt6loop_SCiregion45}
\end{subfigure}
\begin{subfigure}[b]{0.24\textwidth}
\centering
\include{figs/5pt6loop/SCiregion46}
\vspace{-3.5em}\caption{}
\label{5pt6loop_SCiregion46}
\end{subfigure}
\\
\begin{subfigure}[b]{0.24\textwidth}
\centering
\include{figs/5pt6loop/SCiregion47}
\vspace{-3.5em}\caption{}
\label{5pt6loop_SCiregion47}
\end{subfigure}
\begin{subfigure}[b]{0.24\textwidth}
\centering
\include{figs/5pt6loop/SCiregion48}
\vspace{-3.5em}\caption{}
\label{5pt6loop_SCiregion48}
\end{subfigure}
\begin{subfigure}[b]{0.24\textwidth}
\centering
\include{figs/5pt6loop/SCiregion49}
\vspace{-3.5em}\caption{}
\label{5pt6loop_SCiregion49}
\end{subfigure}
\caption{Some regions of figure~\ref{figure-six_loop_example_1to3_decay_plus_soft_emission} characterized by the {\color{Orange}$\boldsymbol{SC_4}$} mode. Graphs (a)--(j): 10 representatives of the 16 regions with $\mathcal{J}_1,\mathcal{J}_3\neq \varnothing$ and $\mathcal{J}_2=\varnothing$, while the remaining 6 can be obtained by applying the reflection $1\leftrightarrow3$ to those marked ``$\times2$''. Graphs (k)--(s): 9 regions with $\mathcal{J}_1\neq \varnothing$ and $\mathcal{J}_2,\mathcal{J}_3=\varnothing$.}
\label{figure-5pt6loop_SCiregions_partIII}
\end{figure}

For the configurations with $\mathcal{J}_1\neq \varnothing$ and $\mathcal{J}_2,\mathcal{J}_3=\varnothing$, there are \emph{9} regions, shown in figures~\ref{5pt6loop_SCiregion41}--\ref{5pt6loop_SCiregion49}. By applying the reflection $1\leftrightarrow3$ to these graphs, we obtain another \emph{9} regions corresponding to $\mathcal{J}_3\neq \varnothing$ and $\mathcal{J}_1,\mathcal{J}_2=\varnothing$.

\begin{figure}[t]
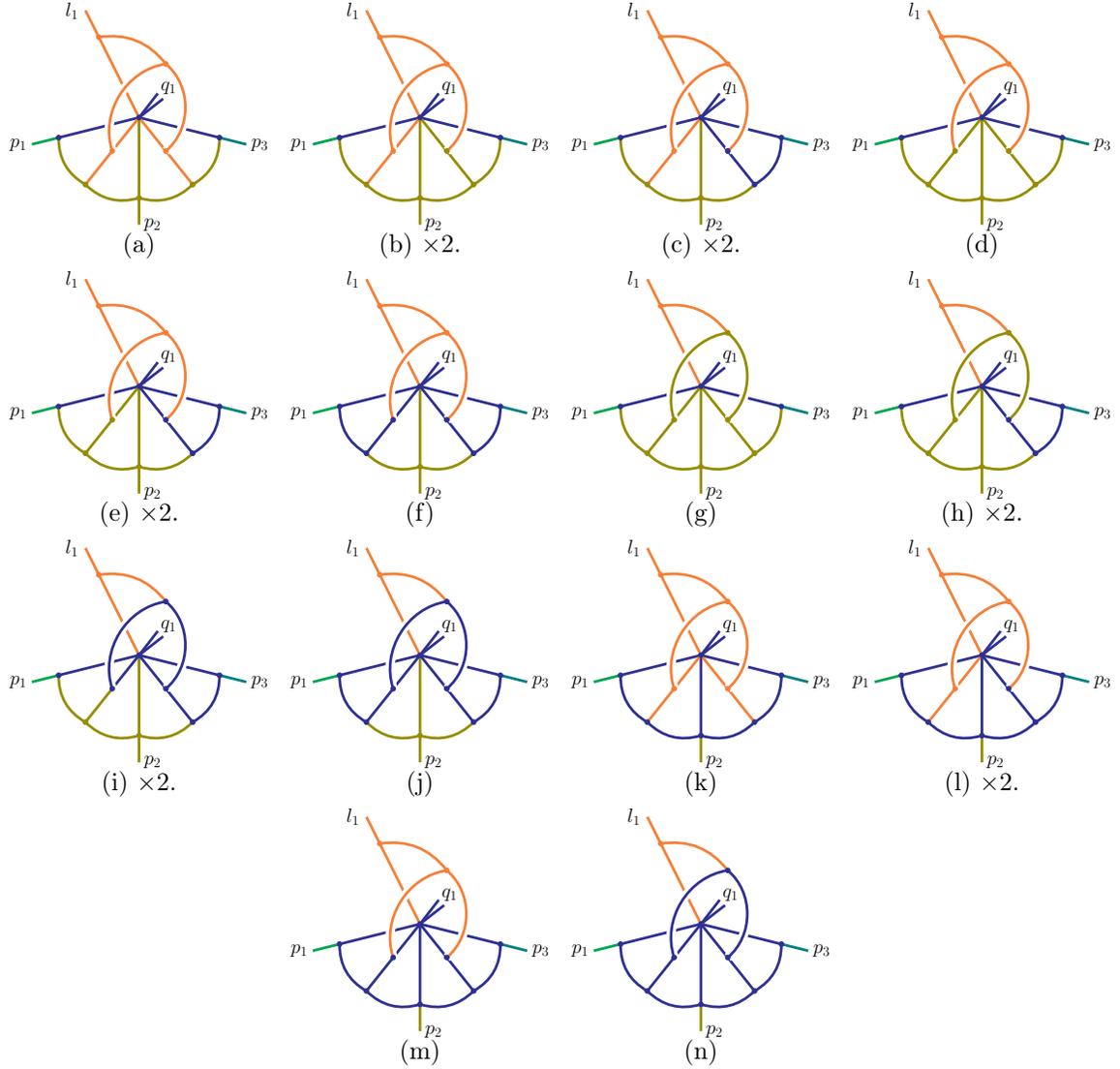

\centering
\begin{subfigure}[b]{0.24\textwidth}
\centering
\include{figs/5pt6loop/SCiregion50}
\vspace{-3.5em}\caption{}
\label{5pt6loop_SCiregion50}
\end{subfigure}
\begin{subfigure}[b]{0.24\textwidth}
\centering
\include{figs/5pt6loop/SCiregion51}
\vspace{-3.5em}\caption{$\times2$.}
\label{5pt6loop_SCiregion51}
\end{subfigure}
\begin{subfigure}[b]{0.24\textwidth}
\centering
\include{figs/5pt6loop/SCiregion52}
\vspace{-3.5em}\caption{$\times2$.}
\label{5pt6loop_SCiregion52}
\end{subfigure}
\begin{subfigure}[b]{0.24\textwidth}
\centering
\include{figs/5pt6loop/SCiregion53}
\vspace{-3.5em}\caption{}
\label{5pt6loop_SCiregion53}
\end{subfigure}
\\
\begin{subfigure}[b]{0.24\textwidth}
\centering
\include{figs/5pt6loop/SCiregion54}
\vspace{-3.5em}\caption{$\times2$.}
\label{5pt6loop_SCiregion54}
\end{subfigure}
\begin{subfigure}[b]{0.24\textwidth}
\centering
\include{figs/5pt6loop/SCiregion55}
\vspace{-3.5em}\caption{}
\label{5pt6loop_SCiregion55}
\end{subfigure}
\begin{subfigure}[b]{0.24\textwidth}
\centering
\include{figs/5pt6loop/SCiregion56}
\vspace{-3.5em}\caption{}
\label{5pt6loop_SCiregion56}
\end{subfigure}
\begin{subfigure}[b]{0.24\textwidth}
\centering
\include{figs/5pt6loop/SCiregion57}
\vspace{-3.5em}\caption{$\times2$.}
\label{5pt6loop_SCiregion57}
\end{subfigure}
\\
\begin{subfigure}[b]{0.24\textwidth}
\centering
\include{figs/5pt6loop/SCiregion58}
\vspace{-3.5em}\caption{$\times2$.}
\label{5pt6loop_SCiregion58}
\end{subfigure}
\begin{subfigure}[b]{0.24\textwidth}
\centering
\include{figs/5pt6loop/SCiregion59}
\vspace{-3.5em}\caption{}
\label{5pt6loop_SCiregion59}
\end{subfigure}
\begin{subfigure}[b]{0.24\textwidth}
\centering
\include{figs/5pt6loop/SCiregion60}
\vspace{-3.5em}\caption{}
\label{5pt6loop_SCiregion60}
\end{subfigure}
\begin{subfigure}[b]{0.24\textwidth}
\centering
\include{figs/5pt6loop/SCiregion61}
\vspace{-3.5em}\caption{$\times2$.}
\label{5pt6loop_SCiregion61}
\end{subfigure}
\\
\begin{subfigure}[b]{0.24\textwidth}
\centering
\include{figs/5pt6loop/SCiregion62}
\vspace{-3.5em}\caption{}
\label{5pt6loop_SCiregion62}
\end{subfigure}
\begin{subfigure}[b]{0.24\textwidth}
\centering
\include{figs/5pt6loop/SCiregion63}
\vspace{-3.5em}\caption{}
\label{5pt6loop_SCiregion63}
\end{subfigure}
\caption{Some regions of figure~\ref{figure-six_loop_example_1to3_decay_plus_soft_emission} characterized by the {\color{Orange}$\boldsymbol{SC_4}$} mode. Graphs (a)--(j): 10 representatives of the 15 regions with $\mathcal{J}_1,\mathcal{J}_3= \varnothing$ and $\mathcal{J}_2\neq \varnothing$, while the remaining 5 can be obtained by applying the reflection $1\leftrightarrow3$ to those marked ``$\times2$''. Graphs (k)--(s): 4 representatives of the 5 regions with $\mathcal{J}_1,\mathcal{J}_2,\mathcal{J}_3=\varnothing$, with the remaining one obtained from (l) by reflecting $1\leftrightarrow3$.}
\label{figure-5pt6loop_SCiregions_partIV}
\end{figure}

For the configurations with $\mathcal{J}_2\neq \varnothing$ and $\mathcal{J}_1,\mathcal{J}_3=\varnothing$, there are \emph{15} regions, where 10 are explicitly shown in figures~\ref{5pt6loop_SCiregion50}--\ref{5pt6loop_SCiregion59}, with the remaining obtainable by applying the reflection $1\leftrightarrow3$ to those marked ``$\times2$''.

For the configurations with $\mathcal{J}_1,\mathcal{J}_2,\mathcal{J}_3=\varnothing$, there are \emph{5} regions, where 4 are shown in figures~\ref{5pt6loop_SCiregion60}--\ref{5pt6loop_SCiregion63}, with the remaining one obtained from figure~\ref{5pt6loop_SCiregion61} by reflecting $1\leftrightarrow3$.

So far we have examined all the regions containing $S^2$ and/or $SC_i$ modes. An interesting feature is that in those regions characterized by the $SC_4$ mode, some $S$ components can also exist, only if two or more jets ($\mathcal{J}_1,\mathcal{J}_2,\mathcal{J}_3$) are nonempty. This is consistent with the Second Connectivity Theorem (theorem~\ref{theorem-mode_subgraphs_connectivity2}), because there is only one mode softer than $S$, thus condition \emph{2} of the theorem cannot apply to any given $S$ component $\gamma_S^{}$. In other words, $\gamma_S^{}$ must be relevant to harder-mode components, which can only be from nonempty jets.

\begin{figure}[t]
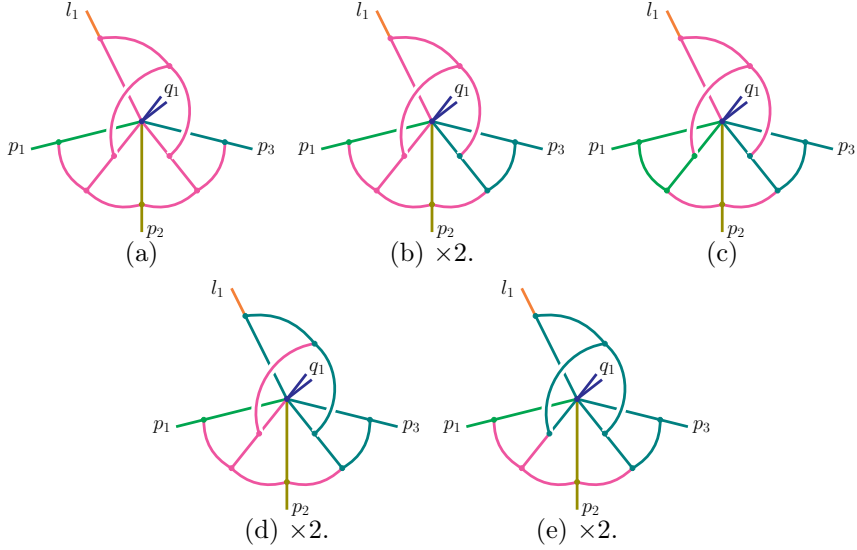

\centering
\begin{subfigure}[b]{0.24\textwidth}
\centering
\include{figs/5pt6loop/SCiregion64}
\vspace{-3.5em}\caption{}
\label{5pt6loop_SCiregion64}
\end{subfigure}
\begin{subfigure}[b]{0.24\textwidth}
\centering
\include{figs/5pt6loop/SCiregion65}
\vspace{-3.5em}\caption{$\times2$.}
\label{5pt6loop_SCiregion65}
\end{subfigure}
\begin{subfigure}[b]{0.24\textwidth}
\centering
\include{figs/5pt6loop/SCiregion66}
\vspace{-3.5em}\caption{}
\label{5pt6loop_SCiregion66}
\end{subfigure}
\\
\begin{subfigure}[b]{0.24\textwidth}
\centering
\include{figs/5pt6loop/SCiregion67}
\vspace{-3.5em}\caption{$\times2$.}
\label{5pt6loop_SCiregion67}
\end{subfigure}
\begin{subfigure}[b]{0.24\textwidth}
\centering
\include{figs/5pt6loop/SCiregion68}
\vspace{-3.5em}\caption{$\times2$.}
\label{5pt6loop_SCiregion68}
\end{subfigure}
\caption{The 8 regions of figure~\ref{figure-six_loop_example_1to3_decay_plus_soft_emission} characterized by the {\color{Rhodamine}$\boldsymbol{S}$} mode and with all three jets $\mathcal{J}_1,\mathcal{J}_2,\mathcal{J}_3$ nonempty. Here we explicitly show 5 of them, with the remaining 3 obtainable from reflecting $1\leftrightarrow3$ for those with ``$\times2$''.}
\label{figure-5pt6loop_SCiregions_partV}
\end{figure}

\bigbreak
Then, we consider regions characterized by the $S$ mode; namely, regions in which the softest mode is $S$. In this case, the remaining modes can only be $H$ or $C_i$ (with $i=1,2,3$ if $\mathcal{J}_i\neq \varnothing$). Let us first consider those configurations with $\mathcal{J}_1,\mathcal{J}_2,\mathcal{J}_3\neq \varnothing$. Since $\mathscr{X}(p_1)=C_1^\infty$ and $\mathscr{X}(p_3)=C_3^\infty$, in order that $\mathcal{J}_1$ and $\mathcal{J}_3$ satisfy the infrared-compatibility requirement, either of the following must be satisfied.
\begin{itemize}
    \item [(1)] The external momentum $l_1$ is relevant to both $\mathcal{J}_1$ and $\mathcal{J}_3$, simultaneously making them infrared compatible.
    \item [(2)] The external momentum $l_1$ is relevant to $\mathcal{J}_1$ ($\mathcal{J}_3$), making it infrared compatible, while there is an $S$ component acting as a degree-$1$ messenger that is relevant to both jets, thus making the other infrared compatible.
\end{itemize}
Following this idea, we come up with \emph{8} such regions, 5 of which are explicitly shown in figure~\ref{figure-5pt6loop_SCiregions_partV}. The first row, figures~\ref{5pt6loop_SCiregion64}--\ref{5pt6loop_SCiregion66}, follow from (1) above, while the second row, figure~\ref{5pt6loop_SCiregion67} and \ref{5pt6loop_SCiregion68}, follow from (2).

\begin{figure}[t]
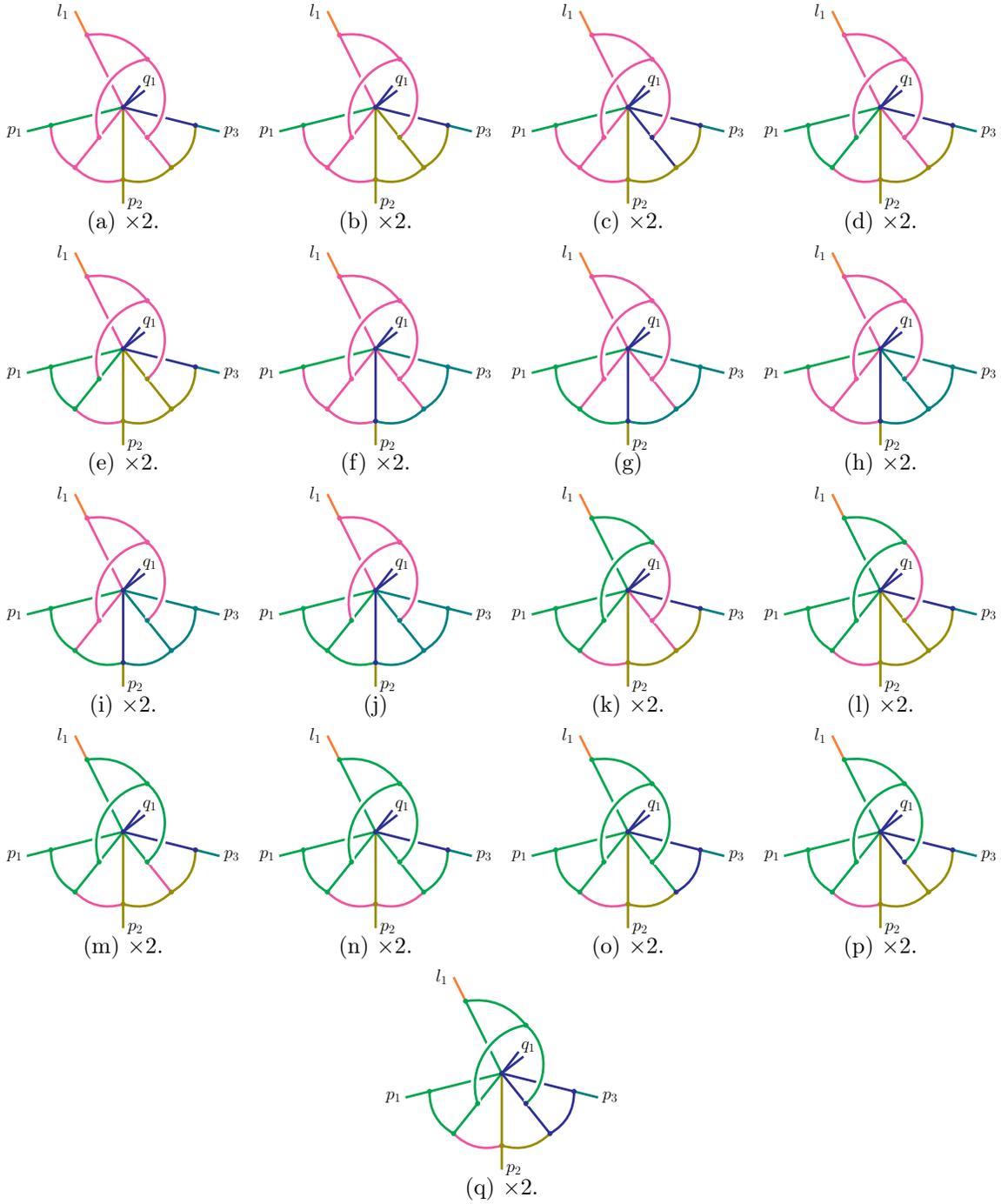

\centering
\begin{subfigure}[b]{0.24\textwidth}
\centering
\include{figs/5pt6loop/SCiregion69}
\vspace{-3.5em}\caption{$\times2$.}
\label{5pt6loop_SCiregion69}
\end{subfigure}
\begin{subfigure}[b]{0.24\textwidth}
\centering
\include{figs/5pt6loop/SCiregion70}
\vspace{-3.5em}\caption{$\times2$.}
\label{5pt6loop_SCiregion70}
\end{subfigure}
\begin{subfigure}[b]{0.24\textwidth}
\centering
\include{figs/5pt6loop/SCiregion71}
\vspace{-3.5em}\caption{$\times2$.}
\label{5pt6loop_SCiregion71}
\end{subfigure}
\begin{subfigure}[b]{0.24\textwidth}
\centering
\include{figs/5pt6loop/SCiregion72}
\vspace{-3.5em}\caption{$\times2$.}
\label{5pt6loop_SCiregion72}
\end{subfigure}
\\
\begin{subfigure}[b]{0.24\textwidth}
\centering
\include{figs/5pt6loop/SCiregion73}
\vspace{-3.5em}\caption{$\times2$.}
\label{5pt6loop_SCiregion73}
\end{subfigure}
\begin{subfigure}[b]{0.24\textwidth}
\centering
\include{figs/5pt6loop/SCiregion74}
\vspace{-3.5em}\caption{$\times2$.}
\label{5pt6loop_SCiregion74}
\end{subfigure}
\begin{subfigure}[b]{0.24\textwidth}
\centering
\include{figs/5pt6loop/SCiregion75}
\vspace{-3.5em}\caption{}
\label{5pt6loop_SCiregion75}
\end{subfigure}
\begin{subfigure}[b]{0.24\textwidth}
\centering
\include{figs/5pt6loop/SCiregion76}
\vspace{-3.5em}\caption{$\times2$.}
\label{5pt6loop_SCiregion76}
\end{subfigure}
\\
\begin{subfigure}[b]{0.24\textwidth}
\centering
\include{figs/5pt6loop/SCiregion77}
\vspace{-3.5em}\caption{$\times2$.}
\label{5pt6loop_SCiregion77}
\end{subfigure}
\begin{subfigure}[b]{0.24\textwidth}
\centering
\include{figs/5pt6loop/SCiregion78}
\vspace{-3.5em}\caption{}
\label{5pt6loop_SCiregion78}
\end{subfigure}
\begin{subfigure}[b]{0.24\textwidth}
\centering
\include{figs/5pt6loop/SCiregion79}
\vspace{-3.5em}\caption{$\times2$.}
\label{5pt6loop_SCiregion79}
\end{subfigure}
\begin{subfigure}[b]{0.24\textwidth}
\centering
\include{figs/5pt6loop/SCiregion80}
\vspace{-3.5em}\caption{$\times2$.}
\label{5pt6loop_SCiregion80}
\end{subfigure}
\\
\begin{subfigure}[b]{0.24\textwidth}
\centering
\include{figs/5pt6loop/SCiregion81}
\vspace{-3.5em}\caption{$\times2$.}
\label{5pt6loop_SCiregion81}
\end{subfigure}
\begin{subfigure}[b]{0.24\textwidth}
\centering
\include{figs/5pt6loop/SCiregion82}
\vspace{-3.5em}\caption{$\times2$.}
\label{5pt6loop_SCiregion82}
\end{subfigure}
\begin{subfigure}[b]{0.24\textwidth}
\centering
\include{figs/5pt6loop/SCiregion83}
\vspace{-3.5em}\caption{$\times2$.}
\label{5pt6loop_SCiregion83}
\end{subfigure}
\begin{subfigure}[b]{0.24\textwidth}
\centering
\include{figs/5pt6loop/SCiregion84}
\vspace{-3.5em}\caption{$\times2$.}
\label{5pt6loop_SCiregion84}
\end{subfigure}
\\
\begin{subfigure}[b]{0.24\textwidth}
\centering
\include{figs/5pt6loop/SCiregion85}
\vspace{-3.5em}\caption{$\times2$.}
\label{5pt6loop_SCiregion85}
\end{subfigure}
\caption{32 regions of figure~\ref{figure-six_loop_example_1to3_decay_plus_soft_emission} characterized by the {\color{Rhodamine}$\boldsymbol{S}$} mode with exactly two jets. Here we explicitly show 17 of them, with the remaining 15 obtainable from reflecting $1\leftrightarrow3$ for those with ``$\times2$''.}
\label{figure-5pt6loop_SCiregions_partVI}
\end{figure}

For the regions without three jets, we observe that for each of them, there must be precisely two jets. This is due to the Second Connectivity Theorem: any $S$ component must be relevant to two harder-mode subgraphs, which are jets in this situation. We can then consider all such possibilities and come up with \emph{32} regions, depicted in figure~\ref{figure-5pt6loop_SCiregions_partVI}.

\begin{figure}[t]
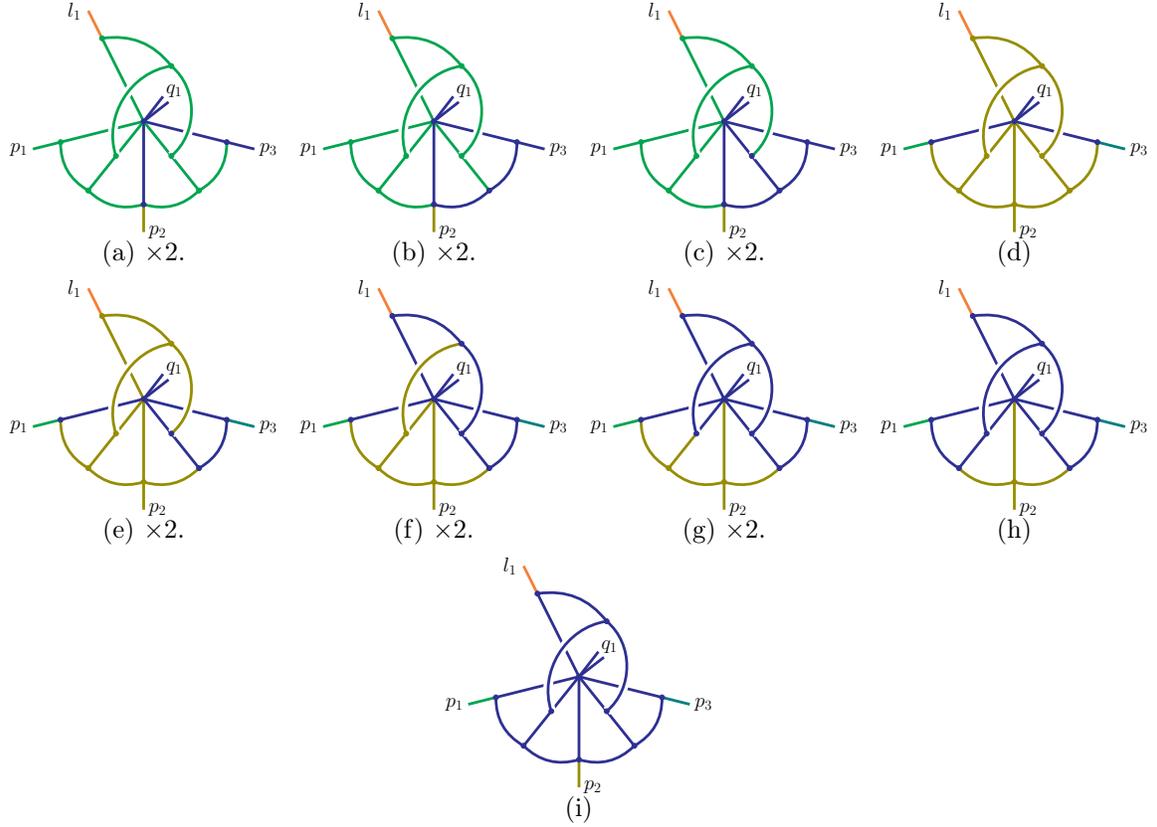

\centering
\begin{subfigure}[b]{0.24\textwidth}
\centering
\include{figs/5pt6loop/Ciregion1}
\vspace{-3.5em}\caption{$\times2$.}
\label{5pt6loop_Ciregion1}
\end{subfigure}
\begin{subfigure}[b]{0.24\textwidth}
\centering
\include{figs/5pt6loop/Ciregion2}
\vspace{-3.5em}\caption{$\times2$.}
\label{5pt6loop_Ciregion2}
\end{subfigure}
\begin{subfigure}[b]{0.24\textwidth}
\centering
\include{figs/5pt6loop/Ciregion3}
\vspace{-3.5em}\caption{$\times2$.}
\label{5pt6loop_Ciregion3}
\end{subfigure}
\begin{subfigure}[b]{0.24\textwidth}
\centering
\include{figs/5pt6loop/Ciregion4}
\vspace{-3.5em}\caption{}
\label{5pt6loop_Ciregion4}
\end{subfigure}
\\
\begin{subfigure}[b]{0.24\textwidth}
\centering
\include{figs/5pt6loop/Ciregion5}
\vspace{-3.5em}\caption{$\times2$.}
\label{5pt6loop_Ciregion5}
\end{subfigure}
\begin{subfigure}[b]{0.24\textwidth}
\centering
\include{figs/5pt6loop/Ciregion6}
\vspace{-3.5em}\caption{$\times2$.}
\label{5pt6loop_Ciregion6}
\end{subfigure}
\begin{subfigure}[b]{0.24\textwidth}
\centering
\include{figs/5pt6loop/Ciregion7}
\vspace{-3.5em}\caption{$\times2$.}
\label{5pt6loop_Ciregion7}
\end{subfigure}
\begin{subfigure}[b]{0.24\textwidth}
\centering
\include{figs/5pt6loop/Ciregion8}
\vspace{-3.5em}\caption{}
\label{5pt6loop_Ciregion8}
\end{subfigure}
\\
\begin{subfigure}[b]{0.24\textwidth}
\centering
\include{figs/5pt6loop/Ciregion9}
\vspace{-3.5em}\caption{}
\label{5pt6loop_Ciregion9}
\end{subfigure}
\caption{15 regions of figure~\ref{figure-six_loop_example_1to3_decay_plus_soft_emission} characterized by the $C_i$ and $H$ modes only. Here we explicitly show 9 of them, with the remaining 6 obtainable from reflecting $1\leftrightarrow3$ for those with ``$\times2$''. The last graph represents the hard region.}
\label{figure-5pt6loop_Ciregions}
\end{figure}

\bigbreak
Finally, we construct regions with only $H$ and $C_i$ modes and depict them in figure~\ref{figure-5pt6loop_Ciregions}.
There are \emph{15} regions in total, with 14 featuring $C_i$ modes and one hard region.

We have thus obtained all the regions directly from the knowledge of region structures. The total number is:
\begin{align}
    16+8+22+22+22+16+9+9+15+5+8+32+15 = 199,
\end{align}
coinciding with the count in the output of \texttt{pySecDec}. One can further check their one-to-one correspondence via the relation (\ref{eq:region_scaling_relation_momentum_LPparameter}).

\subsection{\texorpdfstring{Extension to massive cases: revisiting a recent work on $gg\to HH$}{Extension to massive cases: revisiting a recent work on gg-to-HH}}
\label{section-massive_extension_revisiting_recent_work_ggHH}

As emphasized earlier, the present work concentrates on massless Feynman integrals, although the methodology could be extended to massive cases. We now take a first step in this direction by examining the regions in a recent paper~\cite{Jaskiewicz:2024xkd}. That study computed top‑mass corrections to the $gg\to HH$ amplitude in the high‑energy limit, i.e.,
\begin{align}
    g(p_1)+g(p_2)\to H(p_3)+H(p_4),\qquad s,|t|,|u|\gg m_t^2\gg m_H^2.
\end{align}
The asymptotic expansion takes the following form:
\begin{align}
\label{eq:asymptotic_expansion_JJSU}
    \frac{m_t^2}{s}\sim \lambda^2\ll 1,\quad p_1^2=p_2^2=p_3^2=p_4^2=0,\quad m_H=0.
\end{align}
The limit $m_H=0$ is permissible because the dependence of the $gg\to HH$ amplitude on the Higgs mass is analytic~\cite{Davies:2018qvx}. The external momenta lie along distinct lightlike directions.

Unlike the virtuality expansion in eq.~(\ref{eq:virtuality_expansion}), here each external momentum is exactly on a lightcone, and the small scale is provided by the top‑quark mass $m_t$. This difference could profoundly alter the structure of the Symanzik polynomials in two ways:
\begin{enumerate}
    \item The presence of massive propagators introduces quadratic terms in the second Symanzik polynomial $\mathcal{F}$, in contrast to the linear form in eq.~(\ref{eq:UFterm_massless_general_expression}).
    \item Because every external momentum is strictly lightlike, no $\mathcal{F}$ term could correspond to a unitarity cut that isolates a single external momentum.
\end{enumerate}
Consequently, a fully general all‑order analysis would require substantial modification. Below, however, we observe that the fundamental pattern and the subgraph requirements remain applicable, with only a slight modification on the concept of infrared compatibility. We shall provide neither an all-order statement nor a rigorous proof of this observation, which will be deferred to future research.

\noindent\textbf{Observation. }Each facet region in the expansion (\ref{eq:asymptotic_expansion_JJSU}) corresponds to a configuration of the fundamental pattern shown in figure~\ref{figure-virtuality_expansion_fundamental_pattern}. Moreover, a configuration of the fundamental pattern is a region if and only if the following two conditions hold:
\begin{enumerate}
    \item [1,] the graph $\bigcup_{\mathscr{V}(X)\leqslant n} \Gamma_X$ is connected for any $n\in \mathbb{N}$.
    \item [2,] all mode components are infrared compatible.
\end{enumerate}
Here, a mode component $\gamma$ is called \emph{infrared compatible} if it satisfies at least one of the following:
\begin{itemize}
    \item [(1)] the partial sum over external momenta flowing into $\gamma$ has mode $\mathscr{X}(\gamma)$;
    \item [(2)] $\gamma$ contains an edge $e$ whose mass $m_e$ scales as $m_e^2\sim \lambda^{\mathscr{V}(\mathscr{X}(\gamma))}$.
    \item [(3)] $\gamma$ is among the set of three or more mode components, which an already confirmed infrared‑compatible mode component $\gamma'$ is simultaneously relevant to.
    \item [(4)] \dots.
\end{itemize}
These conditions parallel the ``message‑delivery pattern'' introduced for massless wide‑angle scattering in section~\ref{section-formal_construction_statements}, where the ``message'' refers to the infrared scaling of $\gamma$. The first condition corresponds to obtaining the scaling directly from external kinematics, the second to obtaining it from its internal mass (which is $m_t$ in our case), and the third to receiving it via a ``messenger'' $\gamma'$. The ellipsis ``$\dots$'' stands for further possible conditions not occurring in the graphs studied here, which are therefore omitted.

\begin{figure}[t]
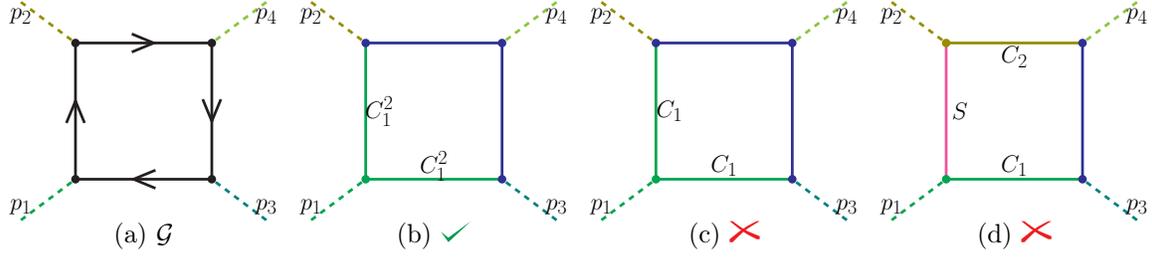

\centering
\begin{subfigure}[b]{0.24\textwidth}
\centering
\include{figs/JJSU/1loop}
\vspace{-3.5em}\caption{$\mathcal{G}$}
\label{JJSU_1loop}
\end{subfigure}
\begin{subfigure}[b]{0.24\textwidth}
\centering
\include{figs/JJSU/1loop_region}
\vspace{-3.5em}\caption{$\greencheckmark[ForestGreen]$}
\label{JJSU_1loop_region}
\end{subfigure}
\begin{subfigure}[b]{0.24\textwidth}
\centering
\include{figs/JJSU/1loop_nonregion1}
\vspace{-3.5em}\caption{$\crossmark[Red]$}
\label{JJSU_1loop_nonregion1}
\end{subfigure}
\begin{subfigure}[b]{0.24\textwidth}
\centering
\include{figs/JJSU/1loop_nonregion2}
\vspace{-3.5em}\caption{$\crossmark[Red]$}
\label{JJSU_1loop_nonregion2}
\end{subfigure}
\caption{The one-loop graph (a) and one of its four collinear$^2$ regions (b), together with two non-region configurations (c) and (d).}
\label{figure-JJSU_1loop_region_nonregions}
\end{figure}

Now let us collate the observation above with some known region examples. At one‑loop level, the external gluons and Higgs bosons attach to a top‑quark loop (see figure~\ref{JJSU_1loop}). Here massive ($m_t$) edges are solid, and massless edges are dashed. The expansion contains one hard region and four infrared regions, each involving a loop momentum of $C_i^2$ mode collinear to one of the external momenta.
Figure~\ref{JJSU_1loop_region} shows one example, where the loop momentum is in the $C_1^2$ mode---note that the mode is $C_1^2$ rather than $C_1$ because $m_t^2\sim \lambda^{\mathscr{V}(C_1^2)}$. Figures~\ref{JJSU_1loop_nonregion1} and~\ref{JJSU_1loop_nonregion2} illustrate two non‑region configurations (marked with ``$\crossmark[Red]$''), each containing mode components that are not infrared compatible. Specifically, the $C_1$ component in figure~\ref{JJSU_1loop_nonregion1} fails to satisfy any of the conditions (1)--(3) above. In figure~\ref{JJSU_1loop_nonregion2}, the $S$ component is infrared compatible due to condition (2), whereas the $C_1$ and $C_2$ components are not.

\begin{figure}[t]
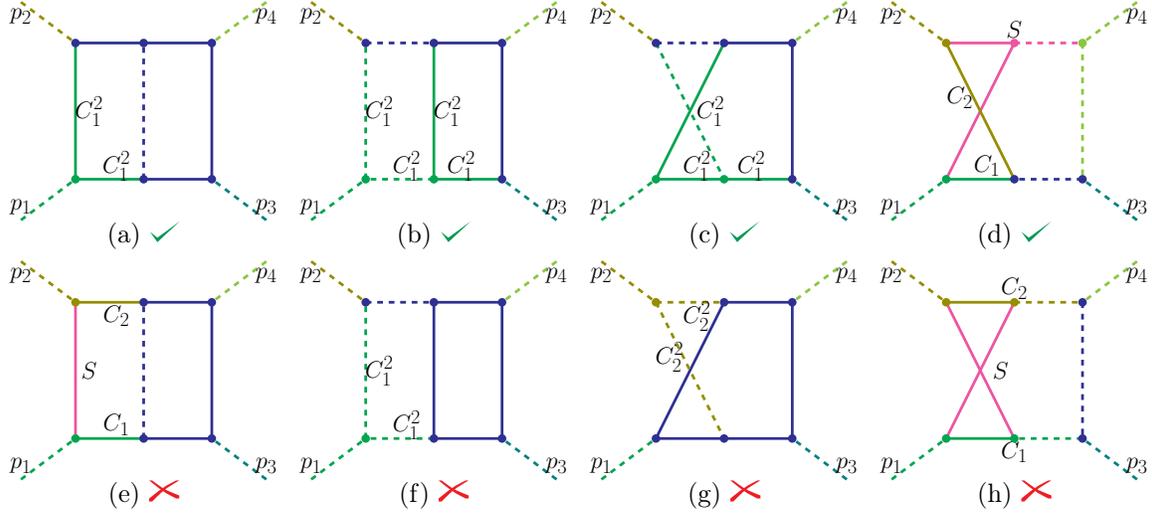

\centering
\begin{subfigure}[b]{0.24\textwidth}
\centering
\include{figs/JJSU/2loop_region1}
\vspace{-3.5em}\caption{$\greencheckmark[ForestGreen]$}
\label{JJSU_2loop_region1}
\end{subfigure}
\begin{subfigure}[b]{0.24\textwidth}
\centering
\include{figs/JJSU/2loop_region2}
\vspace{-3.5em}\caption{$\greencheckmark[ForestGreen]$}
\label{JJSU_2loop_region2}
\end{subfigure}
\begin{subfigure}[b]{0.24\textwidth}
\centering
\include{figs/JJSU/2loop_region3}
\vspace{-3.5em}\caption{$\greencheckmark[ForestGreen]$}
\label{JJSU_2loop_region3}
\end{subfigure}
\begin{subfigure}[b]{0.24\textwidth}
\centering
\include{figs/JJSU/2loop_region4}
\vspace{-3.5em}\caption{$\greencheckmark[ForestGreen]$}
\label{JJSU_2loop_region4}
\end{subfigure}
\\
\begin{subfigure}[b]{0.24\textwidth}
\centering
\include{figs/JJSU/2loop_nonregion1}
\vspace{-3.5em}\caption{$\crossmark[Red]$}
\label{JJSU_2loop_nonregion1}
\end{subfigure}
\begin{subfigure}[b]{0.24\textwidth}
\centering
\include{figs/JJSU/2loop_nonregion2}
\vspace{-3.5em}\caption{$\crossmark[Red]$}
\label{JJSU_2loop_nonregion2}
\end{subfigure}
\begin{subfigure}[b]{0.24\textwidth}
\centering
\include{figs/JJSU/2loop_nonregion3}
\vspace{-3.5em}\caption{$\crossmark[Red]$}
\label{JJSU_2loop_nonregion3}
\end{subfigure}
\begin{subfigure}[b]{0.24\textwidth}
\centering
\include{figs/JJSU/2loop_nonregion4}
\vspace{-3.5em}\caption{$\crossmark[Red]$}
\label{JJSU_2loop_nonregion4}
\end{subfigure}
\caption{Some examples of regions and non-region configurations at two-loop level, where (a)--(d) are regions, while (e)--(h) are not due to the violation of the infrared-compatibility requirement.}
\label{figure-JJSU_2loop_region_nonregions}
\end{figure}

At two‑loop level, the graphs can be divided into planar and non‑planar families. Some examples of regions and non‑region configurations are displayed in figure~\ref{figure-JJSU_2loop_region_nonregions}. Again, massive ($m_t$) edges are solid, and massless edges are dashed.
One can verify directly that each region (figures~\ref{JJSU_2loop_region1}–\ref{JJSU_2loop_region4}) satisfies both subgraph requirements stated in our observation above. In particular, we note that the $S$ component in figure~\ref{JJSU_2loop_region4} acts as a messenger, rendering the $C_1$, $C_2$, and $C_4$ components infrared compatible via condition (3). By contrast, each non‑region configuration violates the infrared‑compatibility requirement because it contains at least one mode component that does not meet any of conditions (1)--(3).

\section{Validation through codes: ``Facet Region Interpreter''}
\label{section-validation_through_codes}

The examples in sections~\ref{section-three_loop_example_2to2_scattering} and \ref{section-six_loop_example_1to3_decay_plus_soft_emission} provide strong supports for the theory. However, because the virtuality expansion~(\ref{eq:virtuality_expansion}), covers infinitely many asymptotic expansions, and because we claim a prescription valid at all loop orders, demonstrating only two examples may still not be convincing. To provide stronger validation, we have implemented this procedure as a computer program, which we call the ``\texttt{Facet Region Interpreter}'' (FRI).

The program is available from \href{https://github.com/yaomaphysics/FRI-public}{https://github.com/yaomaphysics/FRI-public}. Given a graph topology and external kinematics specified by the user, the program enumerates all facet regions expressed in terms of momentum modes. For validation, one can translate the regions into parameter space via eq.~(\ref{eq:region_scaling_relation_momentum_LPparameter}) and compare them with the region vectors from \texttt{pySecDec}. We have done so for 18,000+ graphs in table~\ref{table-validation_status_FRI}, with all results matched: for each case, we compared the complete lists of region vectors in both directions, and found neither missing nor extra regions. These graphs, together with the case-by-case comparison records, can also be accessed from the GitHub link above (the ``validation'' folder).
\begin{table}[t]
\centering
\begin{tabular}{|c||*{2}{c|}}
\hline
\makecell{\textbf{Class}} & \bf{hand-built graphs} & \bf{random graphs} \\ \hline\hline
\makecell{4 legs\\(in 6 chosen kinematics)}
  & 52 topologies (3--5 loops)
  & \makecell{1,200 (4-loop)\\+ 200 (5-loop) topologies} \\ \hline
\makecell{5 legs\\(in 4 chosen kinematics)}
  & 397 topologies
  & \makecell{1,000 (3-loop)\\+ 700 (4-loop)\\+ 300 (5-loop) topologies} \\ \hline
soft emission
  & \makecell{470 topologies (with soft lines\\attached to the topologies above)}
  & \\ \hline
\end{tabular}
\caption{The current status of validation through FRI, with 18,000+ cases in total.}
\label{table-validation_status_FRI}
\end{table}

Notably, in our tests, FRI can obtain region lists significantly faster than (the \texttt{find\_regions} module of) \texttt{pySecDec} at high loops, as briefly demonstrated in the README file of the package. Furthermore, we have successfully extended its application beyond the scope of this paper to include some cases with collinear external momenta. This suggests the potential for applying this method to general region analysis in the future.

A detailed description of this program will be given in a future publication.

\section{Conclusions and outlook}
\label{section-conclusion_outlook}

In this work, we have provided a complete momentum-space prescription for facet regions in generic massless wide-angle scattering, under the assumptions formulated in section~\ref{section-fundamental_facet_region_structure}. To this end, we have unified a broad class of asymptotic expansions---including various limits of external momenta---within a single framework termed the \emph{virtuality expansion}, defined in eq.~(\ref{eq:virtuality_expansion}). The diversity of external momentum types gives rise to rich and intricate structures of momentum modes in the corresponding regions.

A key conceptual advance of this paper is the algebraic characterization of momentum modes, developed in section~\ref{section-wedge_vee_operations}. We have noticed that every momentum scale relevant to the virtuality expansion can be expressed in the form $S^{m}C_i^{n}$ with $m,n\in\mathbb{N}$. This observation allows us to endow the set of modes with a natural partial order based on ``softness'', turning it into a \emph{partially ordered set} (poset). By further introducing the join ($\vee$) and meet ($\wedge$) operations, we elevate this poset to a \emph{non‑distributive lattice}. The algebraic properties of this lattice (for example, the identity $\mathscr{V}(X_1)+\mathscr{V}(X_2) = \mathscr{V}(X_1\wedge X_2) + \mathscr{V}(X_1\vee X_2)$) play a crucial role in the subsequent region analysis.

From a graphical viewpoint, each region corresponds to a specific assignment of modes to the edges and vertices of the Feynman graph $\mathcal{G}$. This assignment naturally partitions the graph into \emph{mode components} (specific subgraphs that share a common momentum scale), which serve as the fundamental building blocks of a region. The problem of classifying regions then reduces to understanding how these components can be connected. Our starting point is the \emph{fundamental pattern} introduced in section~\ref{section-fundamental_facet_region_structure}. This pattern rests on two well‑motivated physical assumptions: that infrared regions correspond to solutions of the Landau equations, and that the hard scattering subgraph is connected. We have argued why both assumptions are natural for wide‑angle kinematics.

The core of our analysis lies in identifying the necessary and sufficient conditions that a configuration of mode components must satisfy to yield a scaleful expanded integral and thus constitute a valid region. Remarkably, these conditions condense into only two concise requirements, summarized in section~\ref{section-summary_matching_results_known_expansions}: (1) a connectivity condition on certain unions of mode subgraphs, and (2) an \emph{infrared‑compatibility} condition on each individual mode component. The latter, a novel concept developed here, formalizes how a component acquires its infrared scaling. Together, these requirements impose strong constraints on the allowed momentum modes and, as a corollary, exclude the appearance of \emph{cascading modes} (unbounded number of soft hierarchies at the amplitude level) in the virtuality expansion. Applied to previously studied special cases, such as those in ref.~\cite{Ma23}, our framework reproduces known results with streamlined reasoning, replacing lengthy and technical proofs by surprisingly concise derivations.

The rigorous derivation of these subgraph requirements is presented in section~\ref{section-proof_subgraph_requirements}, the most technical part of the paper. The proof proceeds in the following logical sequence: (1) the First Connectivity Theorem, (2) generic structure of leading terms, (3) the Second Connectivity Theorem, (4) the sufficiency of the infrared‑compatibility condition, and finally (5) its necessity. Each step employs tailored graph-theoretical techniques that collectively provide a self-consistent foundation for the entire prescription.

Through two multiloop examples with highly nontrivial region structures, we have demonstrated how our framework works. In section~\ref{section-multiloop_examples}, two representative cases, a three‑loop four‑leg graph and a six‑loop five‑leg graph, were analyzed in detail. In both examples, we derived the full set of momentum‑space regions from first principles, translated them to the parametric representation, and found perfect agreement with the output of \texttt{pySecDec}. Moreover, as a first step toward extending the method to massive integrals, we examined a recent study of top‑mass corrections in $gg\to HH$~\cite{Jaskiewicz:2024xkd}. With a minor adaptation of the infrared‑compatibility condition to incorporate massive propagators, our analysis reproduces the regions identified in that work.

As a final highlight, we have implemented our prescription in a computer program called ``\texttt{Facet Region Interpreter}'' and validated the robustness of our approach through 18,000+ examples, as briefly illustrated in section~\ref{section-validation_through_codes}. We look forward to extending the method to broader kinematic regimes and upgrading the program in the future. The program is available from \href{https://github.com/yaomaphysics/FRI-public}{https://github.com/yaomaphysics/FRI-public}, and we welcome further testing of it by the community.

\bigbreak
The framework developed in this work already encompasses a broad class of asymptotic expansions in massless wide-angle scattering. The momentum-space prescription lends itself naturally to the development of efficient graph-finding algorithms to enumerate all relevant regions in arbitrary graphs, bypassing the need to construct high-dimensional polytopes in parameter space. For the special case of the on-shell expansion (see eq.~(\ref{eq:wideangle_onshell_kinematics})), such an algorithm has already been implemented in Maple~\cite{Maple} and is publicly available~\cite{git-Maple_file}. Extensions to more general virtuality hierarchies are expected to be similar.

Nevertheless, the present analysis is restricted to massless wide-angle kinematics. We expect the methodology introduced here---combining graph theory, convex geometry, and the algebraic structures of momentum modes---to serve as a foundation for systematic treatments of other asymptotic expansions, despite anticipated subtleties. Massive propagators, for instance, introduce quadratic terms in the Symanzik polynomials that fundamentally alter the structure of the associated polytopes and the corresponding region analysis. While section~\ref{section-massive_extension_revisiting_recent_work_ggHH} offers an initial exploration, a complete all-order prescription remains a distant goal. Possible threshold singularities~\cite{EdenLdshfOlvPkhn02book,Capatti:2022mly} could also affect the region structure by inducing the ``potential region''~\cite{BnkSmn97}. Further challenges arise in kinematics involving spacelike-collinear external momenta, where Glauber exchanges may appear, requiring dedicated treatment and potentially invalidating the fundamental pattern due to factorization-violating effects~\cite{Catani:2011st,Forshaw:2012bi,Schwartz:2017nmr,Cieri:2024ytf,Becher:2024kmk,Duhr:2025lyg}.

A further limitation of this work is the exclusive focus on facet regions. Extending the all-order prescription to hidden regions (though rare in massless wide-angle scattering) would fully resolve the region structure in this regime. Ref.~\cite{GrdHzgJnsMa24} conjectures that all hidden regions exhibit a Landshoff-type scattering pattern, a proposal verified for two-to-two kinematics up to three loops, and recently supported on a broader range of application~\cite{Gardi:2026mcx}. Confirming this conjecture at all orders and interpreting all hidden regions in momentum representation remains an open and technically demanding challenge.

Once some of these extensions are achieved, the prescription may find broad application across diverse computations. Beyond the standard loop integrals of the form (\ref{eq:Feynman_integarl_dim_reg}), to which many applications have been made in recent years~\cite{Davies:2022ram,Herzog:2023sgb,terHoeve:2023ehm,Niggetiedt:2023uyk,Guan:2024hlf,Hou:2025ovb,Stahlhofen:2025hqd,Capatti:2025gqj}, the EbR technique have recently been applied to conformal correlators~\cite{Li:2025glx}, cosmological correlators~\cite{Beneke:2023wmt}, $\mathcal{N}=4$ super Yang-Mills theory~\cite{BltskSmn23,BltskBorkPklnSmn23,BltskBorkSmn23,Bork:2025ztu}, among others. A particularly promising direction is the treatment of phase-space integrals~\cite{Bonocore:2014wua,Bahjat-Abbas:2018hpv,Ebert:2020lxs,Smirnov:2024pbj,Haag:2025ywj}, which opens pathways to fully differential cross sections. Another interesting phenomenon resides in a recent work of soft-anomalous-dimension calculation~\cite{Gardi:2025ule,Gardi:2025lws}, where eikonal propagators give rise to ultraviolet modes (loop momenta scaling to infinity), in addition to the hard and infrared modes discussed in this work.

From a theoretical perspective, this work focuses on region identification, one of the fundamental aspects of EbR. A full understanding of this topic provides crucial insights into even harder questions: is EbR always correct, and why? Sophisticated mathematical deductions may be needed to answer these questions~\cite{Jtz11,SmnvSmnSmv19,Smnv20}. Another key challenge involves performing expansions for specific region sets, particularly when certain regions yield integrals with unregulated divergences, even under dimensional regularization. Such cases possibly arise when some regions degenerate~\cite{Chen25divergence}, requiring additional regulators (e.g., the analytic regulator, see section~2.3 of ref.~\cite{BchBrgFrl15book} for a pedagogical example) to carry out the integration. It remains to understand in general how this works out systematically, and how the complete set of regions depends on the regulators one chooses. Resolving these issues would also advance the reliability and further development of computational packages that employ EbR~\cite{LiuMa23AMFlow,Hdg21DiffExp,AmdlBcnDvtRanaVcn23SeaSyde,Chen25AmpRed,Zhang:2024fcu}.

More broadly, the all-order prescription developed here may advance our understanding of infrared singularities in scattering amplitudes. A prominent example is the hard-collinear-soft factorization formula for wide-angle scattering~\cite{Sen1983,StmTjd-Yms03,DxnMgnStm08,FgeSwtz14,EdgStm15,Ma20},
\begin{align}
\label{eq:factorization_formula}
\mathcal{M} = \mathcal{H} \cdot \frac{\mathcal{J}_\text{parton}}{\mathcal{J}_\text{eikonal}} \cdot \mathcal{S},
\end{align}
which asserts that collinear singularities reside in the jet factor ratio $\mathcal{J}_\text{parton}/\mathcal{J}_\text{eikonal}$, soft singularities in the soft function $\mathcal{S}$, and that the hard function $\mathcal{H}$ is infrared finite. Although formally reminiscent of our region prescription, the two frameworks differ fundamentally: the factorization above applies locally (i.e., at the integrand level), whereas expansion by regions operates globally (i.e., at the integral level). Establishing a deeper connection between these structures could illuminate the interplay between ultraviolet and infrared singularities, potentially informing factorization-based local subtraction schemes, which have seen substantial recent development~\cite{AntsStm18,Ma20,Anastasiou:2020sdt,Bertolotti:2022aih,Anastasiou:2022eym,Magnea:2024jqg,Gambuti:2023eqh,delaCruz:2024xsm,Anastasiou:2024xvk,Anastasiou:2025cvy,Anastasiou:2026kpm}.

Finally, we highlight the broader potential of the algebraic structure of momentum modes uncovered in section~\ref{section-wedge_vee_operations}. On the one hand, further exploration of this lattice---including its poset structure, explicit join and meet operations, virtuality additivity, pairwise reduction properties, and especially its non-distributivity---could reveal deeper mathematical insights with connections to ordered sets and geometric interpretations relevant to asymptotic expansions. On the other hand, this framework offers a promising new perspective on Soft-Collinear Effective Theory (SCET) and related effective field theories~\cite{BchBrgFrl15book,BurFlmLk00,BurPjlSwt02-1,BurPjlSwt02-2,BnkChpkDhlFdm02,BnkFdm03,RstStw16,Beneke:2022pue}, potentially systematizing operator bases, multipole expansions, and mode interactions in multiscale processes. We hope that these directions will enrich the analytical understanding of Feynman integrals and precision phenomenology in the years ahead.

\acknowledgments
I wish to express my gratitude to Samuel Abreu, Thomas Becher, Martin Beneke, John Collins, Einan Gardi, Sebastian Jaskiewicz, and George Sterman for their insightful discussions, which greatly enriched this work. I am also grateful to the Mainz Institute for Theoretical Physics (MITP) of the Cluster of Excellence PRISMA+ (Project ID 390831469), for its hospitality and its partial support during the completion of this work. This work is supported by the Swiss National Science Foundation under the project funding scheme, grant number 10001706.

\appendix
\section{Some details in section~\ref{section-proof_subgraph_requirements}}
\label{appendix-details_proof}

In this appendix, we elaborate some details that are skipped in section~\ref{section-proof_subgraph_requirements}.

\subsection{Comparing the kinematic factors in proving (\ref{eq:connectivity_theorem_proof_config_I_weight})}
\label{appendix-comparing_kinematic_factors}

We shall demonstrate that $Q_{T_*^2}^2\gtrsim Q_{T^2(\r)}^2$ where $T^2(\r)$ and $T_*^2$ refers to the spanning 2-trees in figures~\ref{connectivity_theorem_proof_config_I} and~\ref{connectivity_theorem_proof_config_I_comparison}, respectively. To see this, recall the difference of these spanning 2-trees: the subtree $\mathfrak{t}=\gamma_*\cap t(\r;A)$, represented by the blob containing $v_A$ in figure~\ref{figure-connectivity_theorem_proof_I}, lies in the same component with $q_j$ in $T^2(\r)$, while in different components with $q_j$ in $T_*^2$. If no external momentum attaches to $\mathfrak{t}$, the momentum flowing between the components of $T^2(\r)$ and $T_*^2$ are identical. We thus have $Q_{T_*^2}^2 = Q_{T^2(\r)}^2$.

If there are some external momenta attached to $\mathfrak{t}$, we denote the total momentum entering $\mathfrak{t}$ as $\mathfrak{p}'$, and the total momentum entering $t(\r;B)$ as $\mathfrak{p}_B$. If $\mathscr{X}(\mathfrak{p}_B)$ is hard (same as $\mathscr{X}(q_j)$), then for both $T^2(\r)$ and $T_*^2$, the total momenta flow between their components are off shell, and we have $Q_{T_*^2}^2\sim Q_{T^2(\r)}^2\sim 1$. Otherwise, $\mathfrak{p}_B$ is in an infrared mode (collinear, soft, etc.), which must be identical to the mode of $Q_{T^2(\r)}$. Similarly, $\mathfrak{p}_B+\mathfrak{p}'$ is in the same mode as $Q_{T_*^2}$. Namely,
\begin{align}
    \mathscr{X}(Q_{T^2(\r)}) = \mathscr{X}(\mathfrak{p}_B),\qquad \mathscr{X}(Q_{T_*^2}) = \mathscr{X}(\mathfrak{p}_B)\vee \mathscr{X}(\mathfrak{p}').
\end{align}
It then follows that $\mathscr{X}(Q_{T_*^2})$ is harder than or equal to $\mathscr{X}(Q_{T^2(\r)})$, and $Q_{T_*^2}^2\gtrsim Q_{T^2(\r)}^2$.

\subsection{Solving for weights in figures~\ref{Sm_component_messenger_config2} and \ref{Sm_component_messenger_config3}}
\label{appendix-analyses_other_two_cases}

Here we show how to solve the weight $w_\gamma$ for figures~\ref{Sm_component_messenger_config2} and \ref{Sm_component_messenger_config3}, under the condition that the edge weights of $\gamma_1$ have been solved.

In figure~\ref{Sm_component_messenger_config2}, the $S^m$ component is relevant to $\gamma_1$ and $\gamma_2$, while the $S^mC_3^{n_3}$ component, which we denote as $\gamma'$, is relevant to $\gamma_1$, $\gamma_2$, and $\gamma_3$ simultaneously. Each mode $\mathscr{X}(\gamma_i)$ is of the form $S^{m-m_i}C_i^{m_i}$. Since the edge weights of $\gamma_1$ have been solved (in Step \emph{1}) as $w_{\gamma_1} = -\mathscr{V}(\mathscr{X}(\gamma_1))=2m_1-4m$, its external momentum $p_1$ must be of the mode $\mathscr{X}(\gamma_1)$. We therefore have
\begin{subequations}
\begin{alignat}{3}
    &\mathscr{X}(\gamma_1) = S^{m-m_1}C_1^{m_1}, &\quad&\mathscr{X}(\gamma_2) = S^{m-m_2}C_2^{m_2}, &\quad&\mathscr{X}(\gamma_3) = S^{m-m_3}C_3^{m_3+n_3};\\
    &\mathscr{X}(p_1) = S^{m-m_1}C_1^{m_1}, &\quad&\mathscr{X}(p_2) = S^{m-m_2}C_2^{m_2+n'_2}, &\quad&\mathscr{X}(p_3) = S^{m-m_3}C_3^{m_3+n_3+n'_3}.
\end{alignat}
\label{eq:Sm_component_messenger_config2_step2_general_modes}
\end{subequations}
From section~\ref{section-leading_terms}, the leading polynomial contains the following $\mathcal{U}^{(R)}$ terms
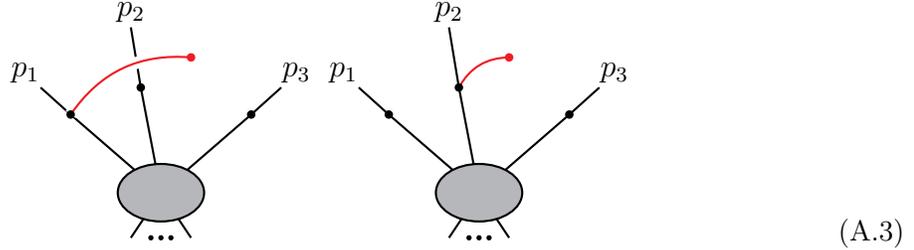
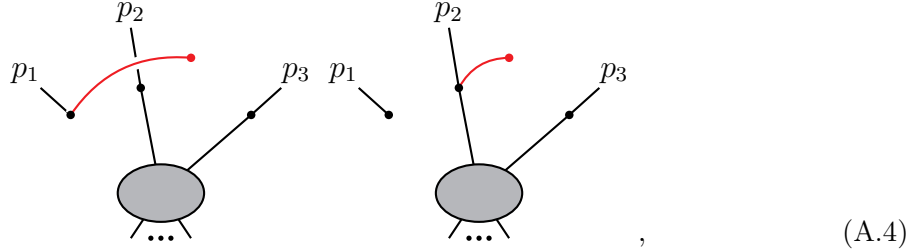
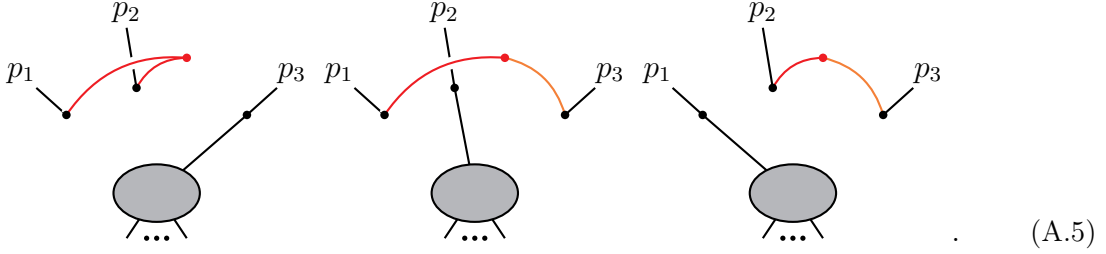
\begin{equation}
    \begin{tikzpicture}[line width = 0.6, scale=0.4, font=\large, mydot/.style={circle, fill, inner sep=.7pt}]
    \draw (5,2.5) edge [thick] (2,5.1) node [] {};
    \draw (2,5.1) edge [thick] (1,6) node [] {};
    \draw (5,2.5) edge [thick] (4.33,6) node [] {};
    \draw (4.33,6) edge [thick] (4,8) node [] {};
    \draw (5,2.5) edge [thick] (8,5.1) node [] {};
    \draw (8,5.1) edge [thick] (9,6) node [] {};
    \draw (6,7) edge [ thick, draw=white, double=white, double distance=3pt, bend right = 30] (2,5.1) node [] {};\draw (6,7) edge [ thick, Red, bend right = 30] (2,5.1) node [] {};
    \draw (5,2.5) edge [thick] (4,1) node [] {};
    \draw (5,2.5) edge [thick] (6,1) node [] {};
    \node () at (0.5,6.5) {$p_1$};
    \node () at (4,8.5) {$p_2$};
    \node () at (9.5,6.5) {$p_3$};
    \draw[fill, thick] (2,5.1) circle (3pt);
    \draw[fill, thick] (4.33,6) circle (3pt);
    \draw[fill, thick] (8,5.1) circle (3pt);
    \draw[fill, Red, thick] (6,7) circle (3pt);
    \node[ thick, draw, fill=Black!33, ellipse, minimum width=3em, minimum height=2em] () at (5,2.5){};
    \path (4,1)-- node[mydot, pos=.333] {} node[mydot] {} node[mydot, pos=.666] {}(6,1);
    \end{tikzpicture}
    \begin{tikzpicture}[line width = 0.6, scale=0.4, font=\large, mydot/.style={circle, fill, inner sep=.7pt}]
    \draw (5,2.5) edge [thick] (2,5.1) node [] {};
    \draw (2,5.1) edge [thick] (1,6) node [] {};
    \draw (5,2.5) edge [thick] (4.33,6) node [] {};
    \draw (4.33,6) edge [thick] (4,8) node [] {};
    \draw (5,2.5) edge [thick] (8,5.1) node [] {};
    \draw (8,5.1) edge [thick] (9,6) node [] {};
    \draw (6,7) edge [ thick, Red, bend right = 30] (4.33,6) node [] {};
    \draw (5,2.5) edge [thick] (4,1) node [] {};
    \draw (5,2.5) edge [thick] (6,1) node [] {};
    \node () at (0.5,6.5) {$p_1$};
    \node () at (4,8.5) {$p_2$};
    \node () at (9.5,6.5) {$p_3$};
    \draw[fill, thick] (2,5.1) circle (3pt);
    \draw[fill, thick] (4.33,6) circle (3pt);
    \draw[fill, thick] (8,5.1) circle (3pt);
    \draw[fill, Red, thick] (6,7) circle (3pt);
    \node[ thick, draw, fill=Black!33, ellipse, minimum width=3em, minimum height=2em] () at (5,2.5){};
    \path (4,1)-- node[mydot, pos=.333] {} node[mydot] {} node[mydot, pos=.666] {}(6,1);
    \end{tikzpicture}
    \label{eq:Sm_component_messenger_config2_leading_Uterms}
\end{equation}
$\mathcal{F}_\textup{I}^{(R)}$ terms
\begin{equation}
    \begin{tikzpicture}[line width = 0.6, scale=0.4, font=\large, mydot/.style={circle, fill, inner sep=.7pt}]
    \draw (2,5.1) edge [thick] (1,6) node [] {};
    \draw (5,2.5) edge [thick] (4.33,6) node [] {};
    \draw (4.33,6) edge [thick] (4,8) node [] {};
    \draw (5,2.5) edge [thick] (8,5.1) node [] {};
    \draw (8,5.1) edge [thick] (9,6) node [] {};
    \draw (6,7) edge [ thick, draw=white, double=white, double distance=3pt, bend right = 30] (2,5.1) node [] {};\draw (6,7) edge [ thick, Red, bend right = 30] (2,5.1) node [] {};
    \draw (5,2.5) edge [thick] (4,1) node [] {};
    \draw (5,2.5) edge [thick] (6,1) node [] {};
    \node () at (0.5,6.5) {$p_1$};
    \node () at (4,8.5) {$p_2$};
    \node () at (9.5,6.5) {$p_3$};
    \draw[fill, thick] (2,5.1) circle (3pt);
    \draw[fill, thick] (4.33,6) circle (3pt);
    \draw[fill, thick] (8,5.1) circle (3pt);
    \draw[fill, Red, thick] (6,7) circle (3pt);
    \node[ thick, draw, fill=Black!33, ellipse, minimum width=3em, minimum height=2em] () at (5,2.5){};
    \path (4,1)-- node[mydot, pos=.333] {} node[mydot] {} node[mydot, pos=.666] {}(6,1);
    \end{tikzpicture}
    \begin{tikzpicture}[line width = 0.6, scale=0.4, font=\large, mydot/.style={circle, fill, inner sep=.7pt}]
    \draw (2,5.1) edge [thick] (1,6) node [] {};
    \draw (5,2.5) edge [thick] (4.33,6) node [] {};
    \draw (4.33,6) edge [thick] (4,8) node [] {};
    \draw (5,2.5) edge [thick] (8,5.1) node [] {};
    \draw (8,5.1) edge [thick] (9,6) node [] {};
    \draw (6,7) edge [ thick, Red, bend right = 30] (4.33,6) node [] {};
    \draw (5,2.5) edge [thick] (4,1) node [] {};
    \draw (5,2.5) edge [thick] (6,1) node [] {};
    \node () at (0.5,6.5) {$p_1$};
    \node () at (4,8.5) {$p_2$};
    \node () at (9.5,6.5) {$p_3$};
    \draw[fill, thick] (2,5.1) circle (3pt);
    \draw[fill, thick] (4.33,6) circle (3pt);
    \draw[fill, thick] (8,5.1) circle (3pt);
    \draw[fill, Red, thick] (6,7) circle (3pt);
    \node[ thick, draw, fill=Black!33, ellipse, minimum width=3em, minimum height=2em] () at (5,2.5){};
    \path (4,1)-- node[mydot, pos=.333] {} node[mydot] {} node[mydot, pos=.666] {}(6,1);
    \end{tikzpicture},
    \label{eq:Sm_component_messenger_config2_leading_Fiterms}
\end{equation}
and $\mathcal{F}_\textup{II}^{(R)}$ terms
\begin{equation}
    \begin{tikzpicture}[line width = 0.6, scale=0.4, font=\large, mydot/.style={circle, fill, inner sep=.7pt}]
    \draw (2,5.1) edge [thick] (1,6) node [] {};
    \draw (4.33,6) edge [thick] (4,8) node [] {};
    \draw (5,2.5) edge [thick] (8,5.1) node [] {};
    \draw (8,5.1) edge [thick] (9,6) node [] {};
    \draw (6,7) edge [thick, draw=white, double=white, double distance=3pt, bend right = 30] (2,5.1) node [] {};\draw (6,7) edge [thick, Red, bend right = 30] (2,5.1) node [] {};
    \draw (6,7) edge [thick, Red, bend right = 30] (4.33,6) node [] {};
    \draw (5,2.5) edge [thick] (4,1) node [] {};
    \draw (5,2.5) edge [thick] (6,1) node [] {};
    \node () at (0.5,6.5) {$p_1$};
    \node () at (4,8.5) {$p_2$};
    \node () at (9.5,6.5) {$p_3$};
    \draw[fill, thick] (2,5.1) circle (3pt);
    \draw[fill, thick] (4.33,6) circle (3pt);
    \draw[fill, thick] (8,5.1) circle (3pt);
    \draw[fill, Red, thick] (6,7) circle (3pt);
    \node[ thick, draw, fill=Black!33, ellipse, minimum width=3em, minimum height=2em] () at (5,2.5){};
    \path (4,1)-- node[mydot, pos=.333] {} node[mydot] {} node[mydot, pos=.666] {}(6,1);
    \end{tikzpicture}
    \begin{tikzpicture}[line width = 0.6, scale=0.4, font=\large, mydot/.style={circle, fill, inner sep=.7pt}]
    \draw (2,5.1) edge [thick] (1,6) node [] {};
    \draw (5,2.5) edge [thick] (4.33,6) node [] {};
    \draw (4.33,6) edge [thick] (4,8) node [] {};
    \draw (8,5.1) edge [thick] (9,6) node [] {};
    \draw (6,7) edge [ thick, draw=white, double=white, double distance=3pt, bend right = 30] (2,5.1) node [] {};\draw (6,7) edge [ thick, Red, bend right = 30] (2,5.1) node [] {};
    \draw (6,7) edge [ thick, Orange, bend left = 30] (8,5.1) node [] {};
    \draw (5,2.5) edge [thick] (4,1) node [] {};
    \draw (5,2.5) edge [thick] (6,1) node [] {};
    \node () at (0.5,6.5) {$p_1$};
    \node () at (4,8.5) {$p_2$};
    \node () at (9.5,6.5) {$p_3$};
    \draw[fill, thick] (2,5.1) circle (3pt);
    \draw[fill, thick] (4.33,6) circle (3pt);
    \draw[fill, thick] (8,5.1) circle (3pt);
    \draw[fill, Red, thick] (6,7) circle (3pt);
    \node[ thick, draw, fill=Black!33, ellipse, minimum width=3em, minimum height=2em] () at (5,2.5){};
    \path (4,1)-- node[mydot, pos=.333] {} node[mydot] {} node[mydot, pos=.666] {}(6,1);
    \end{tikzpicture}
    \begin{tikzpicture}[line width = 0.6, scale=0.4, font=\large, mydot/.style={circle, fill, inner sep=.7pt}]
    \draw (5,2.5) edge [thick] (2,5.1) node [] {};
    \draw (2,5.1) edge [thick] (1,6) node [] {};
    \draw (4.33,6) edge [thick] (4,8) node [] {};
    \draw (8,5.1) edge [thick] (9,6) node [] {};
    \draw (6,7) edge [ thick, Red, bend right = 30] (4.33,6) node [] {};
    \draw (6,7) edge [ thick, Orange, bend left = 30] (8,5.1) node [] {};
    \draw (5,2.5) edge [thick] (4,1) node [] {};
    \draw (5,2.5) edge [thick] (6,1) node [] {};
    \node () at (0.5,6.5) {$p_1$};
    \node () at (4,8.5) {$p_2$};
    \node () at (9.5,6.5) {$p_3$};
    \draw[fill, thick] (2,5.1) circle (3pt);
    \draw[fill, thick] (4.33,6) circle (3pt);
    \draw[fill, thick] (8,5.1) circle (3pt);
    \draw[fill, Red, thick] (6,7) circle (3pt);
    \node[ thick, draw, fill=Black!33, ellipse, minimum width=3em, minimum height=2em] () at (5,2.5){};
    \path (4,1)-- node[mydot, pos=.333] {} node[mydot] {} node[mydot, pos=.666] {}(6,1);
    \end{tikzpicture}.
    \label{eq:Sm_component_messenger_config2_leading_Fiiterms}
\end{equation}
Again, the gray blobs above represent certain set of tree subgraphs. The kinematic coefficient for the three $\mathcal{F}_\textup{II}^{(R)}$ terms above are of the following scales:
\begin{align}
    (p_1+p_2)^2\sim \lambda^{2m-m_1-m_2},\quad (p_1+p_3)^2\sim \lambda^{2m-m_1-m_3},\quad (p_2+p_3)^2\sim \lambda^{2m-m_2-m_3}.
\end{align}
We then consider the characteristic equations of the $\mathcal{F}_\textup{II}^{(R)}$ terms and compare them with those of the $\mathcal{U}^{(R)}$ terms. To be specific, the differences between the first $\mathcal{U}^{(R)}$ and $\mathcal{F}_\textup{II}^{(R)}$ terms lie in the terms $w_{\gamma_1}$ and $w_{\gamma_2}$ (present in the $\mathcal{U}^{(R)}$-term equation but absent from the $\mathcal{F}_\textup{II}^{(R)}$-term equation), a $w_{\gamma}$ term (present in the $\mathcal{F}_\textup{II}^{(R)}$-term equation but absent from the $\mathcal{U}^{(R)}$-term equation), and the kinematic contribution (which is $0$ for the $\mathcal{U}^{(R)}$ term and $2m-m_1-m_2$ for the $\mathcal{F}_\textup{II}^{(R)}$ term). The sum over these contributions must be zero, which yields one equation. By comparing the $\mathcal{U}^{(R)}$ term and the other two $\mathcal{F}_\textup{II}^{(R)}$ terms, we obtain another two equations:
\begin{align}
\label{eq:Sm_component_messenger_config2_comparing_equations}
\begin{split}
    & \left.\begin{matrix}
    w_{\gamma_1}+w_{\gamma_2}-w_{\gamma}+ (2m-m_1-m_2) = 0\\
    w_{\gamma_1}+w_{\gamma_3}-w_{\gamma'}+ (2m-m_1-m_3) = 0\\
    w_{\gamma_2}+w_{\gamma_3}-w_{\gamma'}+ (2m-m_2-m_3) = 0
    \end{matrix}\right\}
    \quad\Rightarrow\quad w_{\gamma} = -2m.
\end{split}
\end{align}
Meanwhile, $w_{\gamma_3}$ and $w_{\gamma'}$ cannot be solved from these equations alone.

For figure~\ref{Sm_component_messenger_config3}, the three $\mathcal{F}_\textup{II}^{(R)}$ terms are
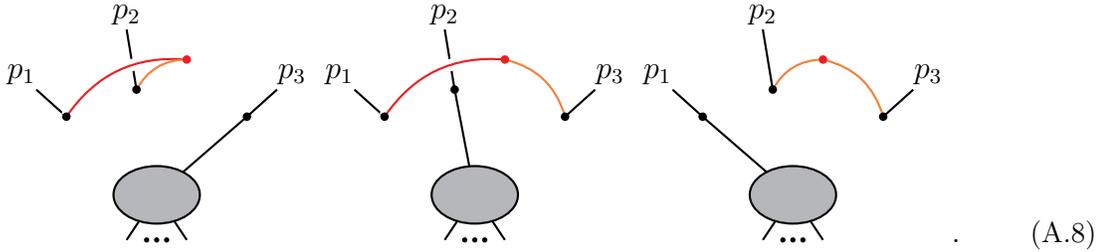
\begin{equation}
    \begin{tikzpicture}[line width = 0.6, scale=0.4, font=\large, mydot/.style={circle, fill, inner sep=.7pt}]
    \draw (2,5.1) edge [thick] (1,6) node [] {};
    \draw (4.33,6) edge [thick] (4,8) node [] {};
    \draw (5,2.5) edge [thick] (8,5.1) node [] {};
    \draw (8,5.1) edge [thick] (9,6) node [] {};
    \draw (6,7) edge [thick, draw=white, double=white, double distance=3pt, bend right = 30] (2,5.1) node [] {};\draw (6,7) edge [thick, Red, bend right = 30] (2,5.1) node [] {};
    \draw (6,7) edge [thick, Orange, bend right = 30] (4.33,6) node [] {};
    \draw (5,2.5) edge [thick] (4,1) node [] {};
    \draw (5,2.5) edge [thick] (6,1) node [] {};
    \node () at (0.5,6.5) {$p_1$};
    \node () at (4,8.5) {$p_2$};
    \node () at (9.5,6.5) {$p_3$};
    \draw[fill, thick] (2,5.1) circle (3pt);
    \draw[fill, thick] (4.33,6) circle (3pt);
    \draw[fill, thick] (8,5.1) circle (3pt);
    \draw[fill, Red, thick] (6,7) circle (3pt);
    \node[ thick, draw, fill=Black!33, ellipse, minimum width=3em, minimum height=2em] () at (5,2.5){};
    \path (4,1)-- node[mydot, pos=.333] {} node[mydot] {} node[mydot, pos=.666] {}(6,1);
    \end{tikzpicture}
    \begin{tikzpicture}[line width = 0.6, scale=0.4, font=\large, mydot/.style={circle, fill, inner sep=.7pt}]
    \draw (2,5.1) edge [thick] (1,6) node [] {};
    \draw (5,2.5) edge [thick] (4.33,6) node [] {};
    \draw (4.33,6) edge [thick] (4,8) node [] {};
    \draw (8,5.1) edge [thick] (9,6) node [] {};
    \draw (6,7) edge [ thick, draw=white, double=white, double distance=3pt, bend right = 30] (2,5.1) node [] {};\draw (6,7) edge [ thick, Red, bend right = 30] (2,5.1) node [] {};
    \draw (6,7) edge [ thick, Orange, bend left = 30] (8,5.1) node [] {};
    \draw (5,2.5) edge [thick] (4,1) node [] {};
    \draw (5,2.5) edge [thick] (6,1) node [] {};
    \node () at (0.5,6.5) {$p_1$};
    \node () at (4,8.5) {$p_2$};
    \node () at (9.5,6.5) {$p_3$};
    \draw[fill, thick] (2,5.1) circle (3pt);
    \draw[fill, thick] (4.33,6) circle (3pt);
    \draw[fill, thick] (8,5.1) circle (3pt);
    \draw[fill, Red, thick] (6,7) circle (3pt);
    \node[ thick, draw, fill=Black!33, ellipse, minimum width=3em, minimum height=2em] () at (5,2.5){};
    \path (4,1)-- node[mydot, pos=.333] {} node[mydot] {} node[mydot, pos=.666] {}(6,1);
    \end{tikzpicture}
    \begin{tikzpicture}[line width = 0.6, scale=0.4, font=\large, mydot/.style={circle, fill, inner sep=.7pt}]
    \draw (5,2.5) edge [thick] (2,5.1) node [] {};
    \draw (2,5.1) edge [thick] (1,6) node [] {};
    \draw (4.33,6) edge [thick] (4,8) node [] {};
    \draw (8,5.1) edge [thick] (9,6) node [] {};
    \draw (6,7) edge [ thick, Orange, bend right = 30] (4.33,6) node [] {};
    \draw (6,7) edge [ thick, Orange, bend left = 30] (8,5.1) node [] {};
    \draw (5,2.5) edge [thick] (4,1) node [] {};
    \draw (5,2.5) edge [thick] (6,1) node [] {};
    \node () at (0.5,6.5) {$p_1$};
    \node () at (4,8.5) {$p_2$};
    \node () at (9.5,6.5) {$p_3$};
    \draw[fill, thick] (2,5.1) circle (3pt);
    \draw[fill, thick] (4.33,6) circle (3pt);
    \draw[fill, thick] (8,5.1) circle (3pt);
    \draw[fill, Red, thick] (6,7) circle (3pt);
    \node[ thick, draw, fill=Black!33, ellipse, minimum width=3em, minimum height=2em] () at (5,2.5){};
    \path (4,1)-- node[mydot, pos=.333] {} node[mydot] {} node[mydot, pos=.666] {}(6,1);
    \end{tikzpicture}.
    \label{eq:Sm_component_messenger_config3_leading_Fiiterms}
\end{equation}
The three equations in (\ref{eq:Sm_component_messenger_config2_comparing_equations}) become
\begin{align}
\label{eq:Sm_component_messenger_config3_comparing_equations}
    &\left.\begin{array}{l}
    w_{\gamma_1} + w_{\gamma_2} - w_{\gamma'} + (2m - m_1 - m_2) = 0 \\
    w_{\gamma_1} + w_{\gamma_3} - w_{\gamma''} + (2m - m_1 - m_3) = 0 \\
    w_{\gamma_2} + w_{\gamma_3} - w_{\gamma'} - w_{\gamma''} + (2m - m_2 - m_3) = 0
    \end{array}\right\}
    \quad\Rightarrow\quad w_{\gamma} = -2m.
\end{align}
Meanwhile, $w_{\gamma_2}$, $w_{\gamma_3}$, $w_{\gamma'}$, and $w_{\gamma''}$ cannot be solved from these equations alone.

Summarizing from eqs.~(\ref{eq:Sm_component_messenger_config1_comparing_equations}), (\ref{eq:Sm_component_messenger_config2_comparing_equations}), and (\ref{eq:Sm_component_messenger_config3_comparing_equations}), we conclude that the edge weight of the $S^m$ component in a messenger $\Gamma^{[m]}$ can always be solved (provided it is relevant to some weight-solved mode component), whose value is the expected $-2m$. This justifies the corresponding statement in the weight-solving algorithm.

\subsection{Derivation of eqs.~(\ref{eq:harder_XaXb_possible_modes}) and (\ref{eq:softer_XaXb_possible_modes})}
\label{appendix-XaXb_possible_modes}

For $S^mC_i^n = \mathscr{X}(\gamma_a) \vee \mathscr{X}(\gamma_b)$, we aim to solve $\mathscr{X}(\gamma_a)$ and $\mathscr{X}(\gamma_b)$ under the condition that $\mathscr{X}(\gamma_a), \mathscr{X}(\gamma_b)$ are both softer than $S^mC_i^n$. A first indication is that $\mathscr{X}(\gamma_a)$ and $\mathscr{X}(\gamma_b)$ are overlapping.

Next, any mode that is softer than $S^mC_i^n$ must take one of the following forms:
\begin{enumerate}
    \item [(1)] $S^{m'}C_i^{n'}$, with $m'\geqslant m$ and $m'+n'\geqslant m+n$,
    \item [(2)] $S^{m+n}C_j^{n'}$, with $j\neq i$ and $n'\in \mathbb{N}$.
\end{enumerate}
Consider the case $\mathscr{X}(\gamma_a)$ and $\mathscr{X}(\gamma_b)$ are both of type (1), i.e.,
\begin{align}
    \mathscr{X}(\gamma_a) = S^{m'_a}C_i^{n'_a},\quad \mathscr{X}(\gamma_b) = S^{m'_b}C_i^{n'_b}.
\end{align}
Since $\mathscr{X}(\gamma_a)$ and $\mathscr{X}(\gamma_b)$ are overlapping, without loss of generality we can assume $m'_a>m'_b$ and $m'_a+n'_a < m'_b+n'_b$. From theorem~\ref{theorem-overlapping_modes_intersection_union_rules}, we have
\begin{align}
    \mathscr{X}(\gamma_a) \vee \mathscr{X}(\gamma_b) = S^{m'_b}C_i^{m'_a+n'_a-m'_b}.
\end{align}
In order that this is equal to $S^mC_i^n$, we must have $m'_b=m$ and $m'_a+n'_a = m+n$. Since $\mathscr{X}(\gamma_a)$ and $\mathscr{X}(\gamma_b)$ are both softer than $S^mC_i^n$, we have
\begin{align}
\label{eq:XaXb_possible_modes_case1}
    m'_a>m,\ \ m'_a+n'_a = m+n;\quad m'_b=m, \ \ n'_b>n.
\end{align}

Then, we consider the case $\mathscr{X}(\gamma_a)$ is of type (1) while $\mathscr{X}(\gamma_b)$ is of type (2), namely,
\begin{align}
    \mathscr{X}(\gamma_a) = S^{m'_a}C_i^{n'_a},\quad \mathscr{X}(\gamma_b) = S^{m+n}C_j^{n'_b}.
\end{align}
It follows immediately that $m'_a=m$; otherwise ($m'_a>m$) $\mathscr{X}(\gamma_a) \vee \mathscr{X}(\gamma_b)$ would be equal to or softer than $S^{m'_a}$, which is softer than $S^mC_i^n$. This implies $n'_a>n$. In other words,
\begin{align}
\label{eq:XaXb_possible_modes_case2}
    m'_a=m,\ \ n'_a>n;\quad n'_b\in \mathbb{N}.
\end{align}

Finally, eq.~(\ref{eq:harder_XaXb_possible_modes}) is a combination of eqs.~(\ref{eq:XaXb_possible_modes_case1}) and (\ref{eq:XaXb_possible_modes_case2}).

For $S^mC_i^n = \mathscr{X}(\gamma_a) \wedge \mathscr{X}(\gamma_b)$, we aim to solve $\mathscr{X}(\gamma_a)$ and $\mathscr{X}(\gamma_b)$ under the condition that $\mathscr{X}(\gamma_a), \mathscr{X}(\gamma_b)$ are both harder than $S^mC_i^n$. Again, $\mathscr{X}(\gamma_a)$ and $\mathscr{X}(\gamma_b)$ must be overlapping, and any mode harder than $S^mC_i^n$ must take the form of $S^{m'}C_i^{n'}$, with $m'\leqslant m$ and $m'+n'\leqslant m+n$. We then denote
\begin{align}
    \mathscr{X}(\gamma_a) = S^{m'_a}C_i^{n'_a},\quad \mathscr{X}(\gamma_b) = S^{m'_b}C_i^{n'_b},
\end{align}
and assume $m'_a>m'_b$ without loss of generality. From theorem~\ref{theorem-overlapping_modes_intersection_union_rules}, we have $\mathscr{X}(\gamma_a) \wedge \mathscr{X}(\gamma_b) = S^{m'_a}C_i^{m'_b+n'_b-m'_a}$. In order for this to equal $S^mC_i^n$, we obtain $m'_a=m$ and $m'_b+n'_b=m+n$. This is precisely eq.~(\ref{eq:softer_XaXb_possible_modes}), up to swapping $a$ and $b$.

\bibliographystyle{jhep}
\bibliography{refs}

\end{document}